\documentclass[twoside,12pt]{report}
\usepackage{mystile}
\usepackage[spanish,english]{babel}
\usepackage{geometry}
\usepackage{enumerate}
\usepackage{epigraph} 
\usepackage{graphicx}
\usepackage{rotating}   
\usepackage{fancyhdr}
\usepackage{amssymb}
\usepackage[Lenny]{fncychap}
\usepackage[small,normal,bf,up]{caption2}
\newcommand{\be}{\begin{equation}}
\newcommand{\ee}{\end{equation}}

\renewcommand{\textflush}{flushepinormal}
\renewcommand{\epigraphwidth}{6.5cm}
\newcommand{\mo}    {M_{\odot}}

\newcommand{\pri}   {${\rlap.}^{\prime \prime}$}
\newcommand{\rl}    {${\rlap.}^{\rm s}$}

\def\Ls{L_{s}^{\rm obs}}
\def\Lh{L_{h}^{\rm obs}}
\def\Ts{T_s}

\begin{document}

\setcounter{page}{1}
\renewcommand{\thepage}{\roman{page}}

\thispagestyle{empty}

\begin{center}

\selectlanguage{english}

\vspace*{1cm}

{\LARGE UNIVERSIDAD DE BUENOS AIRES}

\vspace{1cm}

{\LARGE Facultad de Ciencias Exactas y Naturales} 

\vspace{1cm}

{\LARGE Departamento de F\'{i}sica}

\vspace{3cm}

{\LARGE{\textit{Gamma-Ray Emission from Microquasars}}}

\vspace{1.5cm}

{\Large by \textit{Mar\'{i}a Marina Kaufman Bernad\'o}}

\vspace{3.2cm}

{\Large Supervisor: Dr. Gustavo E. Romero}

\vspace{1.5cm}

{\Large Working Place: Instituto Argentino de Radioastronom\'{i}a (I.A.R.)}

\vspace{2.0cm}

{\Large Thesis Work for PhD Degree issued by the}\\
{\Large University of Buenos Aires}

\vspace{1cm}

{\large \textit{December 15$^{th}$ 2004}}

\end{center}

\newpage
\thispagestyle{empty}
\phantom{.}

\newpage

\noindent {\Large\textbf{Abstract}}

\vspace{1cm}

\normalsize

\noindent Microquasars, X-ray binary systems that generate relativistic jets, were discovered in our Galaxy in the last decade of the $XX^{th}$ century. Their name indicates that they are manifestations of the same physics as quasars but on a completely different scale.\\ 

Parallel to this discovery, the EGRET instrument on board of the Compton Gamma Ray Observatory detected 271 point like gamma-ray sources 170 of which were not clearly identified with known objects. This marked the beginning of gamma-ray source population studies in the Galaxy.\\  

We present in this thesis models for gamma-ray production in microquasars with the aim to propose them as possible parent populations for different groups of EGRET unidentified sources. These models are developed for a variety of scenarios taking into account several possible combinations, i.e. black holes or neutron stars as the compact object, low mass or high mass stellar companions, as well as leptonic or hadronic gamma-ray production processes.\\ 
 
We also show that the presented models for gamma-rays emitting microquasars can be used to explain observations from well known sources that are detected in energy ranges other than EGRET's. Finally, we include an alternative gamma-ray producing situation that does not involve microquasars but a specific unidentified EGRET source possibly linked to a magnetized accreting pulsar. 

\vspace{2cm}

\noindent {\large{Keywords}}

\begin{itemize}
	\item Microquasars
	\item Gamma-Ray Sources
	\item Radiative processes: Non-Thermal
	\item Black Holes
	\item Accretion Disks
	\item Jets 
\end{itemize}

\newpage
\thispagestyle{empty}
\phantom{.}

\newpage

\noindent{\Large\textbf{Jury:}}\\

\vspace{0.5cm}

\noindent {\large Dr. P. Biermann (Max Planck Institute of Radioastronomy - Bonn)}\\
\\
{\large Dr. D. G\'omez (University of Buenos Aires)}\\
\\
{\large Dr. J.M. Paredes (University of Barcelona)}\\
\\
{\large Dr. F. Minotti (University of Buenos Aires)}\\

\newpage
\thispagestyle{empty}
\phantom{.}
\newpage

\textwidth 14.9cm
\textheight 23.5cm

\thispagestyle{empty}
\vspace*{16cm}

\epigraph {\normalsize\textit {Najini skupni prihodnosti, za katero sva se tako dolgo borila.}\vspace{0.5cm}}{\vspace{0.6cm}\normalsize\textit{A mi madre y \\a mi hermana}}

\newpage
\thispagestyle{empty}
\phantom{.}

\newpage

\selectlanguage{spanish}

\epigraph {[Jai Singh, observando las estrellas]: 
 
\vspace*{0.2cm}Medir, computar, entender, ser parte, entrar, morir menos pobre, oponerse pecho a pecho a esa incomprensibilidad tachonada, arrancarle un jir\'on de clave, hundirle en el peor de los casos la flecha de la hip\'otesis, la anticipaci\'on del eclipse, reunir en un pu\~no mental las riendas de esa multitud de caballos centelleantes y hostiles.}{ \textit{Prosa del observatorio}\\ \textsc{JULIO CORT\'AZAR}}
 
\vspace{1.5cm}

\begin{figure}[!h] 
\centering
\resizebox{15cm}{!}{\includegraphics{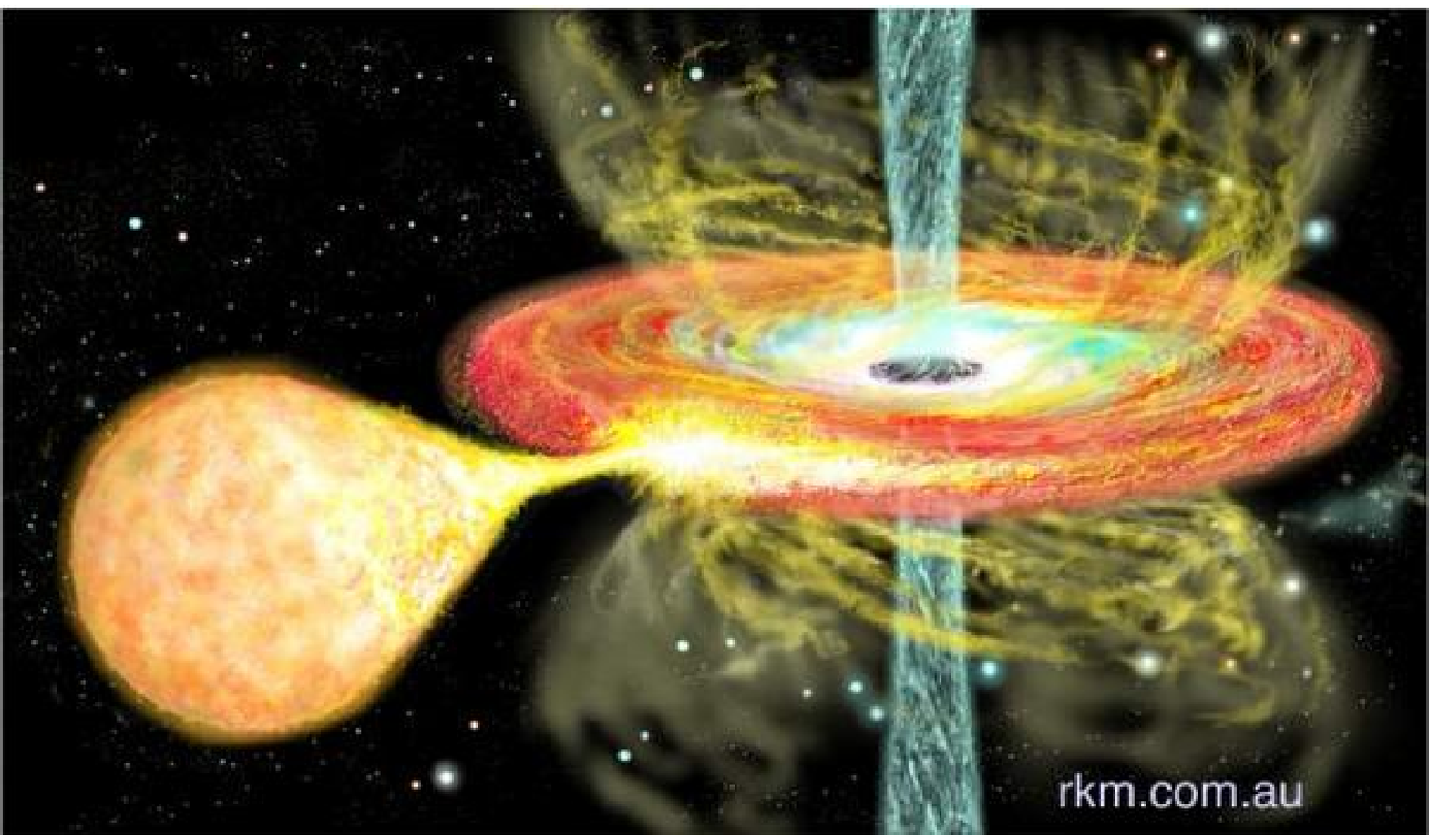}}
\end{figure}

\newpage
\thispagestyle{empty}
\phantom{.}

\newpage

\selectlanguage{spanish}

\hoffset 0cm
\LARGE\textbf{Agradecimientos}\\

\normalsize

En las siguientes l\'{i}neas quisiera agradecer a todas aquellas personas que directa o indirectamente contribuyeron en esta etapa de mi vida que comenz\'o en junio de 1999 con mi retorno a la Argentina luego de recibirme en Suecia y que culmina hoy, 15 de diciembre de 2004, con la obtenci\'on del t\'{i}tulo de \textit{Doctora en Ciencias F\'{i}sicas}.\\ 

Al llegar a la Argentina en junio de 1999, me acababa de recibir de Lic. en F\'{i}sica y hab\'{i}a realizado un trabajo de diploma que consisti\'o en una revisi\'on sobre las Explosiones de Rayos Gamma (GRBs, del ingl\'es \textit{Gamma-Ray Bursts}). La situaci\'on era dif\'{i}cil: no ten\'{i}a ninguna perspectiva de doctorado, no conoc\'{i}a a nadie que pudiese dirigirme en esa fase y tampoco ten\'{i}a trabajo. El tiempo y mucha gente posibilitaron que partiendo de dicho contexto llegase sin embargo hoy a coronar esta etapa con este doctorado que se titula \textit{Emisi\'on de Rayos Gamma en Microcu\'asares}.\\

Empezar\'e por mi director de doctorado, el Dr. Gustavo E. Romero. En primer lugar le agradezco su dedicaci\'on y sus  esfuerzos permanentes por integrarme a la comunidad de la astrof\'{i}sica de altas energ\'{i}as. Nunca dud\'o en hacerme un lugar, en brindarme oportunidades. Siempre me empuj\'o a que me moviera en forma independiente. Por sus ense\~nanzas, que excedieron lo estrictamente relacionado con lo acad\'emico, y por guiarme siempre con el af\'an de que creciera profesionalmente. Quisiera agradecerle tambi\'en el haberme dado un espacio para disentir, un espacio para la discusi\'on, un espacio para acordar. Por haberme tratado siempre como una colaboradora, d\'andole importancia desde un principio a los aportes que yo misma pudiera realizar. En fin Gustavo, gracias por ser mi director con todas las letras y un gran amigo.\\   

Quisiera agradecer a las autoridades del Instituto Argentino de Radioastronom\'{i}a (I.A.R.) y a todo su personal en general por siempre colaborar conmigo, por estar atentos a mis necesidades y por alegrarse ante mis logros.\\

Al CONICET por otorgarme la beca de doctorado con la que este proyecto se llev\'o a cabo. A la Asociaci\'on Antorchas por otorgarme una beca de finalizaci\'on de doctorado que me permiti\'o dedicarme con total exclusividad al cierre de este proyecto durante 1 a\~no.\\

A la Direcci\'on de Relaciones Internacionales de la SECyT, al \textit{Minist\`ere des Affaires Etrang\`eres} (DGCID) y al \textit{Minist\`ere de l'Education Nationale} (DRIC) por otorgarme en dos ocasiones subsidios para realizar estad\'{i}as de colaboraci\'on con grupos de investigaci\'on en Francia en el marco de los denominados proyectos ECOS. Esto me permiti\'o sumarle un ingrediente fundamental al desarrollo de mi doctorado: el contacto con grupos extranjeros. As\'{i} no solamente pude desarrollar una muy firme relaci\'on con el grupo de la Dra. Isabelle Grenier (\textit{Service d'Astrophysique}, CEA, Saclay, Francia) sino tambi\'en con el grupo del Dr. Josep Mar\'{i}a Paredes (Universidad de Barcelona) y con el grupo del Dr. Peter Biermann (\textit{Max Planck Institut f\"ur Radioastronomie} - Bonn).\\

Al Dr. Diego Harari por su respaldo y aval que fueron fundamentales en los comienzos del desarrollo de este doctorado.\\ 

Al Dr. Esteban Bajaja y al Dr. Anders Karlhede por brindarme su apoyo a trav\'es de sus cartas de referencia que formaron parte de mi postulaci\'on a la beca de doctorado con la que realic\'e este proyecto.\\

A Marta Pedernera por la infinita paciencia con la que gui\'o los \'ultimos pasos burocr\'aticos de este doctorado en su presentaci\'on oral y escrita. Por recibir cada una de mis consultas siempre con una sonrisa.\\ 

Al Dr. Jorge Devoto y a la Lic. Silvia Cederbaum por darme una mano enorme al llegar de Suecia. Gracias por confiar en m\'{i}, gracias por ayudarme a integrarme al mundo educativo universitario, gracias por permitirme dar mis primeros pasos como docente universitaria, gracias por recomendarme y por haberme dado oportunidades que fueron para m\'{i} fundamentales.\\ 

A Isabel, Nora, Ada y Susana, mis compa\~neras de trabajo en la Facultad de Ingeniería de la UBA, por ser tan agradable trabajar con ellas, por la paciencia que me tuvieron en mis comienzos, por ense\~narme tanto y por corregir muchos ex\'amenes en mi nombre cuando tuve que realizar viajes relacionados con mi doctorado. Por seguir tan de cerca la evoluci\'on de mi doctorado y por alegrarse siempre mucho ante cada paso que logr\'e dar. Por lo mucho que me cuidaron.\\

A Diego Torres por ese cartelito que dej\'o en aquel p\'oster que present\'e en 1999 al volver de Suecia. Por interesarse en mi trabajo, por darme la oportunidad de trabajar con \'el y ayudarme a dar mis primeros pasos luego de recibida. Por po\-nerme en contacto con el que fue mi director de tesis, Gustavo Romero. En fin por ubicarme y ofrecerme un lugar cuando todo se ve\'{i}a muy difuso e incierto. Por siempre confiar en m\'{i} y por sus muchas cartas de referencias que tan \'utiles me resultaron.\\

A Jorge Combi por ser un muy buen compa\~nero de trabajo. Un agradecimiento lleno de ternura por aquellos primeros pasitos dados junto a \'el. Por ense\~narme la existencia y el uso de las bases de datos astrof\'{i}sicas y por las clases de AIPS que tan gentilmente me dio al principio.\\  

A Ernesto Eiroa por ayudarme en mis comienzos en el uso del programa ``mate\-m\'atica'' (programa que utilic\'e durante todo mi doctorado) y por estar siempre dispuesto a aclarar cualquier duda que al respecto me surgiera.\\

A Mariana Orellana por ser una excelente colaboradora y compa\~nera de trabajo. Por acompa\~narme en mis corridas de \'ultimo momento y por participar tan activamente de uno de los cap\'{i}tulos de esta tesis.\\ 

A la Dra. Isabelle Grenier por recibirme en Saclay y por su colaboraci\'on que desde un principio estuvo llena de confianza y entusiasmo. Por las muchas discusiones de trabajo que tuvimos personalmente o por tel\'efono que  siempre esclarecieron dudas y me sirvieron para avanzar en el entendimiento de temas directa o indirectamente relacionados con mi tesis. Por los momentos compartidos de intenso trabajo. Gracias Isabelle por tu apoyo permanente y por trabajar conmigo de manera tan abierta y pareja.\\ 

Al Dr. Peter Biermann porque desde que lo conoc\'{i} me puso a prueba y confi\'o en m\'{i}. Por su entusiasmo y permanente apoyo. Por invitarme en dos oportunidades a realizar estad\'{i}as cortas en su lugar de trabajo (\textit{Max Planck Institut} - Bonn) que me permitieron entrar en contacto con otros investigadores. Por las discusiones que pude tener con \'el y que siempre fueron tan fruct\'{i}feras porque abarcan un amplio abanico de temas en los que \'el trabaja. Gracias Peter por estar siempre tan pen\-diente de mis pasos.\\

Al Dr. Josep Mar\'{i}a Paredes porque desde que lo conoc\'{i} me trat\'o como a una colega. Por las discusiones que abarcaron tanto lo profesional como las que lo excedieron y de las cuales siempre saqu\'e alguna conclusi\'on provechosa. Por invitarme en dos ocasiones a realizar estad\'{i}as cortas en su lugar de trabajo (Universidad de Barcelona). Por darme la oportunidad de dar una de mis primeras charlas en un \'ambito que no me fuera familiar. Por la confianza que esa experiencia me dio. Gracias Josep Mar\'{i}a por estar siempre dispuesto a darme una mano para avanzar.\\

Al Dr. Daniel G\'omez por aguantar la insistencia con la que le ped\'{i} una y otra vez que diera un curso de magnetohidrodin\'amica. De hecho, gracias por finalmente haberlo darlo. Por ser un excelente docente que me dedic\'o mucho tiempo, que siempre estuvo dispuesto a discutir consultas. Por el entusiasmo con el que transmite los conocimientos.\\

Al Dr. Fernando Minotti por haber recibido esta tesis a \'ultimo momento y as\'{i} y todo interesarse y leerla con detalle. Porque si bien particip\'o en los \'ultimos tramos de este proceso, lo hizo con un compromiso y una confianza que parecieron de larga data.\\

Al Dr. F\'elix Mirabel por varias colaboraciones que realizamos en trabajos conjuntos. Por darme la oportunidad de trabajar con \'el que tan fundamental fue y es en el campo en el cual esta tesis se centra, los microcu\'asares.\\

A Regis Terrier por estar tan pendiente de m\'{i} en las dos oportunidades que estuve trabajando en Saclay. Por su amabilidad, cordialidad y amistad. Por compartir conmigo ricos tes e interesante discusiones.\\ 

A Marc Rib\'o por sus consejos, por sus respuestas siempre instant\'aneas, por muchas informaciones muy \'utiles, por colaborar de manera clave en los \'ultimos momentos de la preparaci\'on del escrito de esta tesis, por estar siempre atento a brindar su ayuda.\\

A Valent\'{i} Bosch-Ram\'on por ser desde un principio un colaborador y un amigo. Por largas discusiones sobre los temas que en lo laboral nos competen y otros tantos m\'as tambi\'en. Por los infinitos \textit{e-mails} intercambiados con paciencia sobre aspectos muy detallistas de nuestro trabajo. Por la buena voluntad que compartimos por entendernos y avanzar. Por recibirme en su ciudad, Barcelona, con todo el entusiasmo y guiarme a trav\'es de la misma. Gracias Valenti por las palabras de catal\'an y por trabajar conmigo en un marco de amistad.\\ 

Al padre George V. Coyne por lo que significa para todos aquellos alumnos de los que se estila denominar ``pa\'{i}ses en vías de desarrollo''. Porque con su tes\'on incansable es el alma de las escuelas del Observatorio del Vaticano, que nos dan la posibilidad de entrar en contacto cotidiano con astrof\'{i}sicos de primera l\'{i}nea durante un tiempo apreciable y de dar nuestros primeros pasos y charlas en p\'ublico. Por su entusiasmo siempre intacto, por su energ\'{i}a y vitalidad, por su amistad. Por haberme otorgado todo el apoyo econ\'omico posible para que pudiera participar de la 8va escuela, VOSS 01, y del encuentro general de escuelas, SuperVOSS II. Por la experiencia inolvidable que esa escuela fue para m\'{i}. Porque no s\'olo marc\'o mis comienzos como astrof\'{i}sica sino mi vida en general. Gracias George por sus ense\~nanzas y cari\~no.\\  

A todo el staff que conform\'o la 8va escuela del Vaticano, VOSS01, y el encuentro denominado SuperVOSS II, por hacer de esos encuentros experiencias \'unicas y de un valor incalculable.\\ 

A Paula Benaglia por la amistad que fuimos desarrollando desde los comienzos de mi doctorado. Por su dulzura. Por acompa\~narme muy de cerca en cada uno de los pasos que fui dando hasta llegar a este t\'{i}tulo. Por ayudarme en m\'as de una oportunidad en la b\'usqueda de datos que fui necesitando para mi trabajo. Gracias Pauli por preocuparte y alentarme.\\

A Vero por ser una amiga que me cuida y que me tir\'o un salvavidas fundamental en las \'ultimas semanas de este proceso. Vero, !`!`!`sab\'es que me salvaste!!! Por las tan diversas experiencias compartidas. Gracias Veruchi.\\

A Christian porque es Christian, ese amigo del alma. Porque estuvo conmigo siempre desde que me conoce hace ya m\'as de una d\'ecada. Porque me alienta, me sostiene, me cuida y me quiere sin condiciones. Por lo mucho que deseaba que consiga este doctorado. Gracias Chris.\\

A mi hinchada sueca, Karin y Henrik, ah\'{i} siempre presentes y cerca a pesar de las mayores o menores distancias que nos separan. Por seguir mis pasos con extremo detalle. Porque son mi tesoro sueco. Gracias Karincita y Henriquito.\\

A Evi esa compa\~nera de estudio que se volvi\'o una amiga entra\~nable, siempre presente. Por haberme aguantado durante este \'ultimo a\~no el hecho de que estando tan cerca no me la haya podido cruzar m\'as que por casualidad. Gracias Evi.\\

A Graciela y a Laura esas amigas que tanto quiero y extra\~no. Por los consejos de hermanas mayores que siempre saben darme. Por seguir mis pasos y alegrarse conmigo cuando algo sale bien. Gracias Grace. Gracias Lauri.\\

A Leo por haber vuelto en medio de este doctorado. Por su amistad intacta. Por alegrarse siempre tanto y tan genuinamente por mis logros. Por alentarme a viva voz cuando lo necesito. Gracias Leo por haber vuelto a estar presente.\\ 

A Andrea y a Pablo por el espacio Arroyo-Quanta en el que conoc\'{i} tanta gente que me integr\'o a sus vidas y que me sigui\'o en cada uno de mis pasos con mucho cari\~no y alegr\'{i}a. Por crear un espacio donde he jugado y me he divertido tanto.\\ 

A Matej y a Miha, mis traductores eslovenos, por traducir tan gentilmente la dedicatoria que en ese idioma aparece al principio de esta tesis y que est\'a dirigida a mi novio, Simon.\\  

A Estela y L\'azaro por el cari\~no inmenso con el que me albergaron en las dos oportunidades que estuve trabajando en Francia. Por hacer de su casa la m\'{i}a. Por cuidarme y mimarme. Gracias a los dos por estar en mi vida desde siempre, tan presentes y atentos a mis necesidades.\\

A Marie-Ange y Richard por albergarme en su casa en mi segunda estad\'{i}a de trabajo en Francia. Por hacerme sentir tan c\'omoda. Gracias Richard por las eternas y divertidas discusiones. Gracias Marie-Ange por cuidarme y mimarme tanto. Por esas charlas que quisieran arreglar el mundo o al menos nuestros mundos.\\

A Marie-Th\'e por estar en los principios de mi educaci\'on y no haber parado nunca de incentivarme en los caminos que he seguido. Por acompa\~narme tan activamente desde que eleg\'{i} estos estudios y por interesarse tan seriamente en mis progresos.\\

A Marcela, Cecilia, Ver\'onica y Leo por todas las gestiones burocr\'aticas o de impresi\'on que tan importantes fueron en todo este proceso.\\ 

A Mariano por ese recuerdo entregado justito unos d\'{i}as antes de terminar esta etapa. Por el s\'{i}mbolo que eso signific\'o para m\'{i} en un momento tan l\'{i}mite e importante.\\

A mi t\'{i}a Lita, a Celia y a Norma por seguirme siempre tan de cerca y cuidarme. Por estar presentes una vez m\'as en un d\'{i}a tan importante.\\ 

A Val\v{c}i, Mitja, Marta, Klara y Matej por acompa\~narme tanto a pesar de la distancia. Por estar tan pendientes de m\'{i} y por haberme aceptado con tanto cari\~no en el seno de su familia.\\  

A Isadora por aquellas horas de estudio compartidas.\\

A Octavio por esa cajita de m\'usica que tanto me gusta.\\

A V\'{i}ctor por siempre valorar mi trabajo y mis esfuerzos tan expl\'{i}citamente. Por no dudar en repet\'{i}rmelo una y otra vez. Por defenderme y as\'{i} cuidarme.\\  

A mi Lolo siempre tan presente en mis pensamientos, por haber sido mi Lolo con todo lo que eso signific\'o y en particular aqu\'{i} por aquella l\'ampara que me regal\'o para estudiar hace 20 a\~nos volviendo a la Argentina y que volv\'{i} a usar luego de muchos a\~nos durante la escritura de esta tesis de doctorado. Por seguir as\'{i} estando tan cerca m\'{i}o.\\ 

A mi madre y a mi hermana por tanto y m\'as. Por estar conmigo en todos los momentos de mi vida y en este en particular. Por seguirme d\'{i}a y noche durante el \'ultimo a\~no que fue poni\'endose cada vez m\'as dif\'{i}cil. Por noches en vela, por compa\~n\'{i}as eternas, por dictados sin sentido, por entender lo incomprensible. En fin, por sostenerme y darme energ\'{i}as y \'animos. Por su emoci\'on en este d\'{i}a. Por su cari\~no tan pero tan grande. Gracias Marie por tantos dulces Torroncinos y por tan dulces momentos. Gracias mami por la educaci\'on y la fuerza.\\

A mi compa\~nero, Simon, por tanta cercan\'{i}a en medio de tanta lejan\'{i}a. Por su apoyo constante a toda hora. Por su paciencia y dedicaci\'on. Por sostenerme. Por quererme as\'{i} de mucho. Por elegir esta historia a pesar de todas sus dificultades. Por la paz y la calma que me transmite. Por trabajar a mi lado. Por sus cr\'{i}ticas constructivas. Gracias Simon por volver a elegir todos los d\'{i}as el ser mi compa\~nero en esta ruta sinuosa que hemos emprendido juntos. 

\vspace{0.7cm}
\begin{center}
$\clubsuit$
\end{center}
\vspace{0.7cm}

Et finalement ma si ch\`ere Nany, cette pens\'ee pour toi\ldots tu nous as quitt\'es quelques jours avant le 15 d\'ecembre\ldots quelques jours avant que je puisse te t\'el\'epho\-ner pour te raconter que finalement j'\'etais docteur\ldots toi, ma ch\`ere Nany qui as suivi mes pas de si pr\`es pendant 20 ans\ldots voil\`a cette pens\'ee pleine de cafard\ldots``cafard'', ce mot que tu m'as appris un jour ensoleill\'e \`a Apremont\ldots pour toi qui m'as si gentiment d\'edi\'e une de tes derni\`eres pens\'ees\ldots pour toi, la nouvelle de ce r\'esultat que je sais tu attendais anxieusement recevoir\ldots \`a toi ma pens\'ee et ma tendresse pour toujours.

\vspace{1cm}

\begin{flushright}
Mar\'{i}a Marina Kaufman Bernad\'o 
\end{flushright}

\newpage

\LARGE\textbf{Acknowledgements}\\

\selectlanguage{english}

\normalsize

In the following lines I would like to express my gratitude to all those people who have directly or indirectly contributed to the period of my life that started in June 1999 with my return to Argentina after finishing my undergraduate studies in Sweden, and that culminates today, December 15th  2004, with my Ph.D. in Physics.\\ 

When arriving to Argentina in June 1999, I had just finished my undergraduate studies in Physics with a diploma work which was a review of Gamma-Ray Bursts (GRBs). The situation was difficult: I did not have any perspective of Ph.D. project, I did not know anybody that could supervise me in that phase and I did not have any work either. Time itself and many people made it possible that starting from such a context I nevertheless managed to crown this period with this Ph.D., titled Gamma-Ray Emission from Microquasars.\\

I will begin with my Ph.D. supervisor, Dr. Gustavo E. Romero. In the first place I would like to thank him for his dedication and his permanent efforts to integrate me into the community of the astrophysics of high energies. He never doubted in making a place for me, in offering me opportunities. He always pushed me to move independently. For his lessons, which exceeded  strictly academic issues, and for always guiding me with the aim that I grew professionally. I would also like to thank him for giving me a space to dissent, a space to discuss, a space to agree. Because he always treated me as a collaborator, respecting my contributions from the very beginning to. In a few words, Gustavo, thank you for being my supervisor with capital letters and a great friend.\\   

I would like to thank to the authorities of the \textit{Instituto Argentino de Radioastronom\'{i}a} (I.A.R.) and to all its personnel in general for always collaborating with me, for being alert to my necessities and for rejoicing over my achievements.\\

To the CONICET for granting me the scholarship with which this project was carried out. To the \textit{Asociaci\'on Antorchas} for granting a complementary scholarship for a year which allowed me to dedicate myself with total exclusivity to finish this project.\\

To the \textit{Direcci\'on de Relaciones Internacionales} - SECyT, to the \textit{Minist\`ere des Affaires Etrang\`eres} (DGCID) and to the \textit{Minist\`ere de l´Education Nationale} (DRIC) for granting me for two working stayings in France to collaborate with research groups of that country within the framework of the denominated ECOS projects. This allowed me to add a fundamental ingredient to the development of my Ph.D.: the contact with foreign groups. This way I was not only able to develop a very firm relation with the group of Dr. Isabelle Grenier (\textit{Service d'Astrophysique}, CEA, Saclay, France) but also with the group of Dr. Josep Mar\'{i}a Paredes (\textit{Universidad de Barcelona}) and with the group of Dr. Peter Biermann (\textit{Max Planck Institut f\"ur Radioastronomie} - Bonn).\\

To Dr. Diego Harari for his support and endorsement that were fundamental in the beginnings of the development of this Ph.D.\\

To Dr. Esteban Bajaja and to Dr. Anders Karlhede for their support through the letters of recommendation that were part of the application to the Ph.D. scholarship with which I made this project.\\

To Marta Pedernera for her huge patience with which she guided the last bureaucratic steps of this Ph.D. in its oral and written presentation. For always receiving each of my questions with a smile.\\ 

To Dr. Jorge Devoto and to Lic. Silvia Cederbaum for giving me a hand when I arrived from Sweden. Thank you for trusting in me and for helping me to insert myself into the academic and university world as a teacher, thank you for recommending me and for giving me opportunities that were essential for my  professional development.\\ 

To Isabel, Nora, Ada and Susana, my fellow workers at the Faculty of Engineering of the \textit{Universidad de Buenos Aires} (U.B.A.), because it was so pleasant to work with them. For their patience at my beginnings, for teaching me so much and for correcting so many exams on my behalf when I had to make trips related to my Ph.D. For following the evolution of my Ph.D. in detail and for being always so happy at each step that I managed to give. In a word, for taking so much care of me.\\

To Diego Torres for that little note that he left in that poster that I presented in 1999 when returning from Sweden. For being interested in my work, for giving me the opportunity to work with him and helping me to take my first steps after finishing my undergraduate studies. For putting me in contact with Gustavo Romero, who became my thesis supervisor. For locating me and for offering me a place when everything looked so diffuse and uncertain. For always trusting in me and for writing many letters of recommendation that were so useful to me.\\

To Jorge Combi for being a very good fellow worker. An acknowledgment, full of tenderness, for those first little steps that I made next to him. For teaching me the existence and the use of the astrophysics data bases and for the classes of AIPS that he gave to me so kindly at the beginning.  

To Ernesto Eiroa for helping me in my beginnings with the use of the program ``mathematica'' (program that I used throughout my whole Ph.D.) and for being always willing to clarify any doubt that I would have on that matter.\\

To Mariana Orellana for being an excellent collaborator and companion of work. For being next to me in the last minute running and for participating so actively in one of the chapters of this thesis.\\

To Dr. Isabelle Grenier for receiving me in Saclay and for her collaboration full of confidence and enthusiasm from the beginning. For many working discussions that we had in person or by telephone which always helped me to clarify doubts and that were always very useful to advance in my understanding of the issues that were directly or indirectly related to my thesis. For sharing moments of intense work. Thank you, Isabelle, for your permanent support and for working next to me in such an open and even way.\\

To Dr. Peter Biermann because since I met him he tested me and he trusted me. For his enthusiasm and permanent support. For inviting me to make short working staying at his place of work (Max Planck Institut - Bonn) at two occasions, which allowed me to take contact with other researchers. For the working discussions that I could have with him and that were always so fruitful because they included the wide range of subjects in which he works. Thank you, Peter, for being always so attentive to my steps.\\

To Dr. Josep Mar\'{i}a Paredes because since I met him he treated me as a colleague. For the discussions that included professional issues and for those that exceeded them, from all of which I could always extract a useful conclusion. For inviting to make short working visits at his place of work (Universidad de Barcelona) on two occasions. For offering me the opportunity to give one of my first talks in an environment that was not familiar to me. For the self-confidence I gained from that experience. Thank you, Josep María, for being always ready to help me to advance.\\

To Dr. Daniel G\'omez for bearing the insistence with which I requested him once and again that he would give a course in magneto-hydrodynamics. In fact, thank you for finally giving it. For being an excellent teacher that dedicated to me a lot of his time and that was always willing to discuss any my doubts. For the enthusiasm with which he transmits the knowledge.\\

To Dr. Fernando Minotti for having received this thesis in the last moment and nevertheless showing interest and reading it in detail. Because although he participated in the last period of this process, his commitment and confidence seemed to be of long-term.\\

To Dr. F\'elix Mirabel for several collaborations in the works we did together. For giving me the opportunity to work with him who played and plays such a fundamental role in the field in which this thesis is focused, microquasars.\\

To Regis Terrier for being so attentive to my needs on the two occasions when I was working in Saclay. For his amiability, cordiality and friendship. For sharing good teas and interesting discussions with me.\\ 

To Marc Rib\'o for his advices, for his always immediate answers, for many very useful pieces of information, for his fundamental collaboration in the last moments of the preparation of the writing of this thesis, for being always attentive to offer his help.\\

To Valent\'{i} Bosch-Ram\'on for being a collaborator and a friend from the very beginning. For long discussions over the issues related to our work and so many over other topics as well. For the infinite e-mails on very detailed aspects of our work interchanged with patience. For the good will that we shared to understand each other and to advance. For receiving me in his city, Barcelona, with enthusiasm and for guiding me through it. Thank you, Valenti, for the words of Catalan and for working with me in a frame of friendship.\\ 

To father George V. Coyne for the whole he means to all those students that belong to the so called ``developing countries''. Because with his untiring tenacity he is the soul of the schools of the Vatican Observatory, which give us the possibility of taking a daily contact with very important  astrophysicists during an appreciable period of time and which also give us the opportunity to make our first steps and talks in public. For his always high enthusiasm, his energy and vitality, for his friendship. For having granted me with all the possible economic support so that I could participate in the 8th school, VOSS 01, and in the general encounter of schools, SuperVOSS II. For the unforgettable experience that school was for me, because it did not only influence my professional beginnings but my life in general. Thank you, George, for your lessons and affection.\\  

To all the staff that was part of the 8th Vatican School, VOSS01, and of the special meeting, so called  SuperVOSS II, for making those encounters unique experiences of an incalculable value.\\ 

To Paula Benaglia for the friendship that we developed from the beginnings of my Ph.D. For her gentleness. For being close to me in each of the steps that I made to arrive to this title. For helping me on many occasions in the search of data needed for my work. Thank you, Pauli, for caring and for encouraging me all the time.\\

To Vero for being a friend who takes care of me and who threw a life-vest to me in the last weeks of this process. Vero, you know that you saved me!!! For so many diverse shared experiences. Thank you, Veruchi.\\

To Christian because he is Christian, that soul mate. Because he was with me always since we met more than a decade ago. Because he encourages me, supports me, he takes care of me and he loves me unconditionally. For the strong wish he had that I obtain this Ph.D. Thank you, Chris.\\

To my Swedish staunchest supporters, Karin and Henrik, always present and close to me despite  greater or smaller distances that separate us. For following every detail of the evolution of my Ph.D. Because they are my Swedish treasure. Thank you, Karincita and Henriquito.\\

To Evi that companion of studies that became a close friend, always present. For bearing during the last year the fact that being so close I couldn't cross her more than by chance. Thank you, Evi.\\

To Graciela and Laura those friends that I love and miss so much. For the pieces of advices they are always giving to me like being my older sisters. For their interest in my progress and being glad when something comes out well. Thank you, Grace. Thank you, Lauri.\\

To Leo for coming back in the middle of this Ph.D. For his friendship that remained intact. For being always so genuinely glad about my achievements. For strongly encouraging me whenever I need it. Thank you, Leo, for being present again.\\ 

To Andrea and Pablo for the Arroyo-Quanta space in which I met so many people who integrated me to their lives and who showed much interest in each of my steps with much affection and joy. For creating a space where I have played and had much fun.\\ 

To Matej and Miha, my Slovenian translators, for their kind translation of my dedication to my boyfriend Simon, which appears in the Slovenian language at the beginning of this thesis.\\
  
To Estela and Lazaro for the immense affection with which they sheltered me in their house in the two occasions when I was working in France. For making their home mine. For taking care of me and for spoiling me. Thank you both for being in my life from ever, so present and kind to my needs.\\

To Marie-Ange and Richard for lodging me in their house at my second working staying in France. For making me feel so comfortable. Thank you, Richard, for the eternal and amusing discussions. Thank you, Marie-Ange, for taking care of me and spoiling me so much. For those chats that try to fix the world or at least our worlds.\\

To Marie-Th\'e for being in the beginning of my education and not having ever stopped stimulating me along the paths that I have been following. For accompanying me so actively since I have chosen these studies and for being so seriously interested in my progress.\\

To Marcela, Cecilia, Ver\'onica and Leo for their help in all the bureaucratic or printing managements that were  so important during all this process.\\ 

To Mariano for that souvenir that he gave me just some days before finishing this period. For the symbol it meant in a moment for me so important  in which I was so in the limit.\\

To my aunt, Lita, to Celia and to Norma for being always very close to me and for taking care of  me. For being present once again in such an important day.\\ 

To Val\v ci, Mitja, Marta, Klara and Matej for accompanying me so much in spite of the distance. For being so attentive and for having accepted me with so much affection in the bosom of their family.\\  

To Isadora for those shared hours of study.\\

To Octavio for that small music box that I like so much.\\

To V\'{i}ctor for always valuing my work and my efforts so explicitly. For never doubting in repeating it to me once and again. For defending me and thus taking care of me.\\
  
To my Lolo always so present in my thoughts, for being my Lolo with the whole this have meant and in particular here for that lamp he had given me to study 20 years ago when I had come back to Argentina and that I used again after many years during the writing of this Ph.D. thesis. For continuing this way to be so close to me.\\ 

To my mother and my sister for so much and more. For being with me in all the moments of my life and in this one in particular. For following me day and night in this last year that became each time more difficult. For working nights, for their eternal company, for dictations without sense, for  understanding the incomprehensible. In short, for their support and for giving me energy and cheering me up. For their emotion on this special day. For their so great affection. Thank you, Marie, for so many sweet Torroncinos and for so sweet moments. Thank you, mami, for the education and the strength.\\

To my companion, Simon, for so much closeness in the middle of so much distance. For standing by me at any hour. For his patience and dedication. For his support. For loving me so much. For choosing this story in spite of all its difficulties. For the peace and the calmness that he transmits to me. For working at my side. For his constructive critics and remarks. Thank you, Simon, for choosing every day, once and again, to be my companion in this winding route we have undertaken together.  

\begin{center}
$\clubsuit$
\end{center}
\vspace{0.3cm}

Et finalement ma si ch\`ere Nany, cette pens\'ee pour toi\ldots tu nous as quitt\'es quelques jours avant le 15 d\'ecembre\ldots quelques jours avant que je puisse te t\'el\'epho\-ner pour te raconter que finalement j'\'etais docteur\ldots toi, ma ch\`ere Nany qui as suivi mes pas de si pr\`es pendant 20 ans\ldots voil\`a cette pens\'ee pleine de cafard\ldots``cafard'', ce mot que tu m'as appris un jour ensoleill\'e \`a Apremont\ldots pour toi qui m'as si gentiment d\'edi\'e une de tes derni\`eres pens\'ees\ldots pour toi, la nouvelle de ce r\'esultat que je sais tu attendais anxieusement recevoir\ldots \`a toi ma pens\'ee et ma tendresse pour toujours.

\vspace{0.1cm}

\begin{flushright}
Mar\'{i}a Marina Kaufman Bernad\'o 
\end{flushright}

\selectlanguage{english}


\newpage
\large{\underline{\scshape List of acronyms appearing in the text}}

\vspace{1cm}

\normalsize

\begin{tabular}{lc}
Active Galactic Nuclei \hspace{1cm}& AGN\\
Advection-Dominated Accretion Flow model \hspace{1cm}& ADAF\\
All-Sky Monitor \hspace{1cm} & ASM\\
Annual Review of Astronomy and Astrophysics \hspace{1cm} & ARA\&A\\
Astro-rivelatore Gamma a Immagini LEggero \hspace{1cm} & AGILE\\
Astronomical Journal \hspace{1cm} & AJ\\
Astronomical Unit \hspace{1cm} & AU\\
Astronomische Nachrichten \hspace{1cm} & AN\\
Astronomy and Astrophysics \hspace{1cm} & A\&A\\
Astronomy and Astrophysics Supplement \hspace{1cm} & A\&AS\\
Astronomy with a Neutrino Telescope and Abyss \hspace{1cm} & \\
environmental RESearch \hspace{1cm} & ANTARES\\
Astroparticle Physics \hspace{1cm} & APh\\
Astrophysics and Space Science \hspace{1cm} & Ap\&SS\\
Australia Telescope Compact Array \hspace{1cm} & ATCA\\
Black Hole \hspace{1cm} & BH\\
Burst And Transient Source Experiment \hspace{1cm} & BATSE\\
Chinese Journal of Astronomy and Astrophysics \hspace{1cm} & ChJAA\\
Compton Gamma Ray Observatory \hspace{1cm} & CGRO\\
Energetic Gamma Ray Experiment Telescope \hspace{1cm} & EGRET\\
European VLBI Network \hspace{1cm} & EVN\\
External Compton \hspace{1cm} & EC\\
Gamma-ray Large Area Space Telescope \hspace{1cm} & GLAST\\
High Energy Stereoscopic System \hspace{1cm} & HESS\\ 
High-Mass Microquasars \hspace{1cm} & HMMQs\\
High-Mass X-ray Binaries \hspace{1cm} & HMXBs\\
Imaging Compton Telescope \hspace{1cm} & COMPTEL\\
International Astronomical Union Circular\hspace{1cm} & IAUC\\
Inverse Compton \hspace{1cm} & IC\\
Journal of Physics G \hspace{1cm} & JPhG\\
Low-mass microquasars \hspace{1cm} & LMMQs\\ 
Low Mass X-ray Binaries \hspace{1cm} & LMXBs\\
Major Atmospheric Gamma Imaging Cherenkov \hspace{1cm} & MAGIC\\
MagnetoHydroDynamic \hspace{1cm} & MHD\\
Microquasar(s) \hspace{1cm} & MQ(s)\\
Monthly Notices of the Royal Astronomical Society \hspace{1cm} & MNRAS\\
Multi-Element Radio-Linked Interferometer Network \hspace{1cm} & MERLIN\\
Neutron Star \hspace{1cm} & NS\\ 
New Astronomy Reviews \hspace{1cm} & New AR\\
Oriented Scintillation Spectrometer Experiment \hspace{1cm} & OSSE\\
Physics Reports \hspace{1cm} & PhR\\
Physical Review \hspace{1cm} & PhRv\\
Physical Review D \hspace{1cm} & PhRvD\\
\end{tabular}

\begin{tabular}{lc}
Physical Review Letters \hspace{1cm} & PhRvL\\
Publications of the Lick Observatory \hspace{1cm} & Publ. Lick Obs.\\
Radio Emitting X-ray Binaries \hspace{1cm} &  REXBs\\
Reports on Progress in Physics \hspace{1cm} & RPPh\\
Reviews of Modern Physics \hspace{1cm} & RvMP\\
Rossi X-Ray Timing Explorer \hspace{1cm} & RXTE\\
Synchrotron Self Compton \hspace{1cm} & SSC\\
Space Science Review \hspace{1cm} & SSRv\\
Spectral Energy Distribution \hspace{1cm} & SED\\
The Astrophysical Journal \hspace{1cm} & ApJ\\
The Astrophysical Journal Supplement \hspace{1cm} & ApJS\\
Torsional Alfv\'en Waves \hspace{1cm} & TAW\\
Very Energetic Radiation Imaging Telescope \hspace{1cm} & \\ 
Array System  \hspace{1cm} & VERITAS\\
Very Large Array \hspace{1cm} & VLA\\
Very Long Baseline Array \hspace{1cm} & VLBA\\
X-ray binary \hspace{1cm} & XRB\\
X-Ray Timing Explorer \hspace{1cm} & XTE\\
\end{tabular}

\newpage
\normalsize 
\voffset 0cm
\setcounter{page}{1}
\renewcommand{\thepage}{\arabic{page}}

\pagestyle{myhead}

\vspace*{4.4cm} 
\thispagestyle{empty}
\tableofcontents

\chapter{\label{Introduction}Introduction}
\thispagestyle{empty}

\newpage
\thispagestyle{empty}
\phantom{.}

\newpage

\vspace*{5.6cm} 

\noindent This thesis deals with the possibility to link two of the most intriguing and recent discoveries of the astrophysics from the last decade of the $XX^{th}$ century: {\scshape microquasars} and {\scshape the galactic population of gamma-ray sources}.\\

The first complete map of the Galaxy in gamma-rays was obtained with the ESA COS-B satellite\footnote{COS-B carried a single large experiment, the Gamma-Ray Telescope. The energy range covered was: 2 keV - 5 GeV. It was originally projected to last two years, but it operated successfully for 6 years and 8 months.}, launched on August $9^{th}$, 1975, and operating until April $25^{th}, 1982$. It was, however, only with the advent of the Compton Gamma Ray Observatory (CGRO) in the 1990s that population studies of gamma-ray sources properly started. EGRET instrument, on board of CGRO, detected 271 point like gamma-ray excesses with a significance of more than 5$\sigma$ on the galactic plane and more than 4$\sigma$ out of it. On the other hand, X-ray binary systems able to generate relativistic radio jets were baptized as ``microquasars'' with the discovery of the sources 1E1740.7-2942 and GRS 1758-258 in 1992, soon after the beginning of the ``Compton Era''.\\

We will present in this thesis models for gamma-ray production in microquasars (MQs). These models will be developed through the following chapters, where the link with gamma-ray sources will be discussed for different scenarios.\\

\section{\label{MQs}Microquasars}

When X-ray binary systems (XRB) are able to generate relativistic radio jets, we call them \textit{microquasars}. The name was given by Mirabel et al. (1992) as a clear reference to the already quite studied phenomena associated with quasars. Quasars and microquasars share three common basic features:\\

\noindent 1) A spinning compact object (in the case of MQs it can be either a neutron star (NS) or a black hole (BH)).\\
2) An accretion disk.\\
3) Collimated jets of relativistic particles.\\

However, as it is indicated by their names, quasars and microquasars, are different manifestations of the same physics at different scales. We can highlight some of their differences in the following table:

\vspace{0.5cm}

\begin{center}
\begin{tabular}{|c|c|c|}
\hline
Feature&Quasars&Microquasars\\
\hline
\hline
Black hole mass&Several million&Few solar masses\\
&solar masses&\\
\hline
\hline
Accretion disk size&$\sim 10^9$ km&$\sim 10^3$ km\\
\hline
\hline
Mean thermal $T$ of&Several thousand&Several million\\
the accretion disk&degrees&degrees\\
\hline
\hline
Characteristic $\lambda$&ultraviolet and&X-rays\\
of the disk radiation&optical&\\
\hline
\hline
Distance traveled&Millions&Few light-years\\
by the jets&of light-years&\\
\hline
\end{tabular}
\end{center}

\vspace{0.5cm}

\begin{figure}[!t] 
\centering
\resizebox{10cm}{!}{\includegraphics{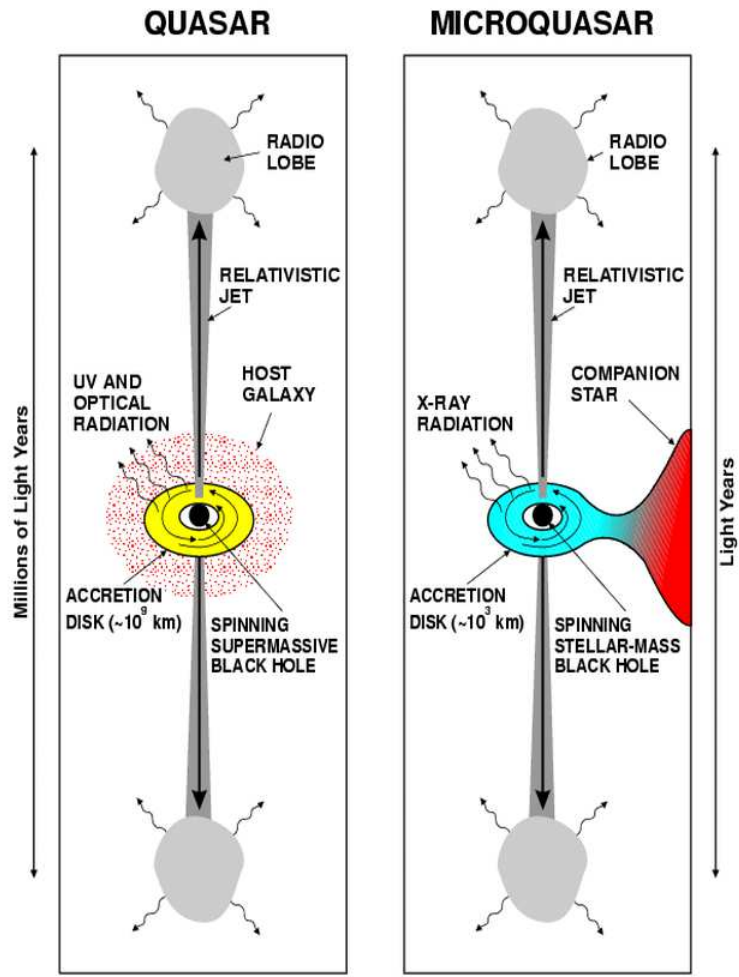}}
\caption{ {Quasars and Microquasars are supposed to be different scales of the same physics (from Mirabel \& Rodr\'{i}guez 1998).}}\label{mqvsq}
\end{figure}

All this information is summarized in a schematic way in Fig.~\ref{mqvsq}. We can see from the table that the linear and time scales are proportional to the black hole mass (Sams et al. 1996). In fact, quasars have super-massive black holes located in the center of distant galaxies (active galactic nuclei, AGN), while MQs have stellar-mass black holes forming binary stellar systems and can be found in the Galaxy. In quasars, the accretion disk feeds itself from disrupted stars or from the interstellar medium of the host galaxy; in MQs, instead, the matter supply is from the companion star in the binary system.\\ 

Taking into account the nature of the donor star, MQs are classified into high-mass microquasars (HMMQs) and low-mass microquasars (LMMQs). HMMQs have a young massive stellar companion and the mass transfer is done mainly through the stellar wind. On the other hand, LMMQs contain an old stellar companion that transfers mass by Roche lobe overflow.\\     

The presence of a relativistic jet is a fundamental characteristic in quasars as well as in MQs. The first evidence of a jet-like feature in astrophysics was discovered by Curtis (1918): he found the optical jet from the elliptical galaxy M87 in the Virgo cluster. The finding that jets can also be produced at smaller scales by binary stellar systems is much more recent. In the late 1950s several radio sources were matched with very dim optical objects that looked like stars, but had strange spectra with a strong ultraviolet excess. They were called quasars for ``quasi-stellar radio source'', and it was in the early 1960s that the first spectrum lines of these ``stars'' were identified showing a redshift of 0.158 (it was the case of 3C273). The discovery of such a large redshift established the cosmological nature of quasars.\\

Relativistic jets in galactic objects were first observed by
Margon et al. (1979) in SS433. This binary remained the only known object of this kind for more than a decade. As it was already mentioned above, supermassive black hole disks emit strongly at optical and UV wavelengths. The more massive the black hole, the cooler the surrounding accretion disk. For a black hole accreting at the Eddington limit, the characteristic blackbody temperature at the last stable
orbit in the surrounding accretion disk is given approximately by $T \sim 2 \times 10^7 M^{-1/4}$ (Rees 1984), with $T$ in Kelvin and the mass of the black hole, $M$, in solar masses. This explains why whereas accretion disks in quasars have strong emission in the optical and ultraviolet, BH and NS binaries usually are identified the first time through their X-ray emission. It is therefore understandable that the jet phenomenon was first discovered far away in very distant galaxies, and that, there was an impasse in the discovery of stellar size sources with relativistic jets in our galactic surroundings until the recent developments in X-ray astronomy. It has to be noted that among galactic jet sources, SS 433 is unusual, because of its broad optical emission lines and its brightness in the visible that made it detectable before such a technological development took place.\\

\begin{figure}[!t] 
\centering 
\resizebox{9cm}{!}{\includegraphics{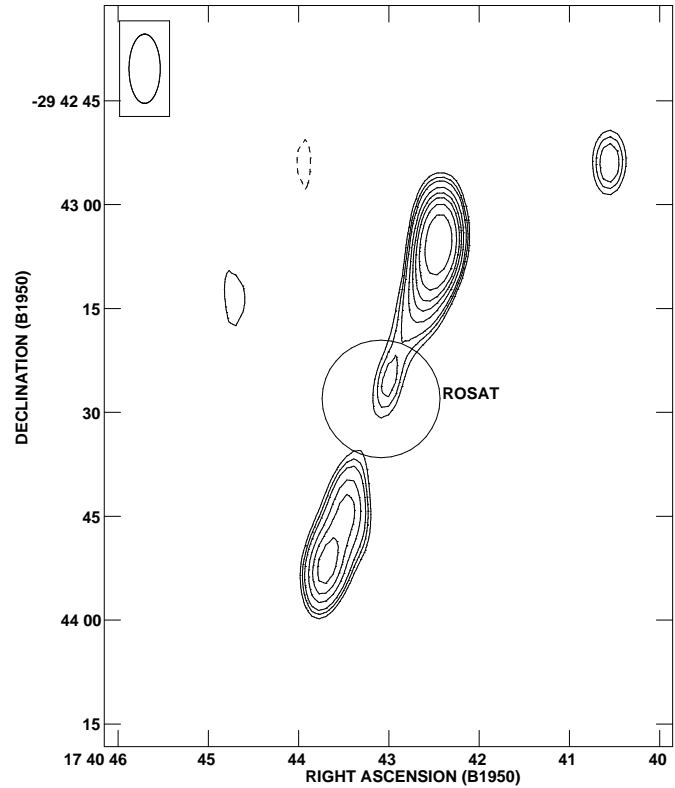}}
\caption{ {Radio counterpart from microquasar 1E1740.7-2942. Notice the clear two-sided 
jet feature (from Mirabel et al. 1992).}}\label{jet1E1740.7-2942}
\end{figure}

As we remarked some lines above, MQs are firstly detected by their X-ray emission as X-ray binaries. A subsequent multi-wavelength study of the system can establish the presence of a radio counterpart, which is the unmistakable signature of the non-thermal emission from a jet composed by relativistic particles. The confluence of observations made in hard X-rays and in radio wavelengths gave as a result the discovery, in 1992, of two such stellar sources of relativistic jets in the galactic center region: 1E1740.7-2942 and GRS 1758-258 (Mirabel et al. 1992, Rodr\'{i}guez et al. 1992). The radio counterpart of 1E1740.7-2942 is shown in Fig.~\ref{jet1E1740.7-2942}. Based on the spectral shape, it was assumed that the steady radio emission in X-ray binaries comes from jets (Rodr\'{i}guez et al. 1995; Fender et al. 1999). This assumption was corroborated recently with VLBA observations of GRS 1915+105 (Dhawan et al. 2000), Cyg X-1 (Stirling et al. 2001) and other sources. The resu
 lting VLBA images of GRS 1915+105 are in agreement with the model of a conical expanding jet emitting through synchrotron processes (Hjellming \& Johnston 1988; Falcke \& Biermann 1999). It is the high brightness temperature, the rapid variability, and the linear polarization which suggest that the observed radio emission is due to synchrotron processes, from the relativistic leptons in the jet.\\

Another very important parallelism can be established between quasars and MQs. Apparent \textit{superluminal}\footnote{See Appendix \ref{superluminal} for a detailed explanation of apparent superluminal motions.} transverse motions in the expansion of jets have been observed in quasars (Pearson \& Zensus 1987; Zensus 1997) with values up to ten or more times the speed of light. Shortly after the quasar -- microquasar analogy was proposed, the first superluminal source in the Galaxy was discovered: GRS 1915+105\footnote{Found with the satellite GRANAT (Castro-Tirado et al. 1994; Finogenov et al. 1994).} (Mirabel \& Rodr\'{i}guez 1994). Thereafter, this characteristic was observed in MQs like GRO J1655-40\footnote{Found with the satellite CGRO (Zhang et al. 1994).} (Tingay et al. 1995; Hjellming \& Rupen 1995) and XTE J1748-288\footnote{Found with the satellite X-Ray Timing Explorer (XTE)(Smith et al. 1998).} (Hjellming et al. 1998), among others.\\

\begin{figure} 
\centering 
\resizebox{8.5cm}{!}{\includegraphics{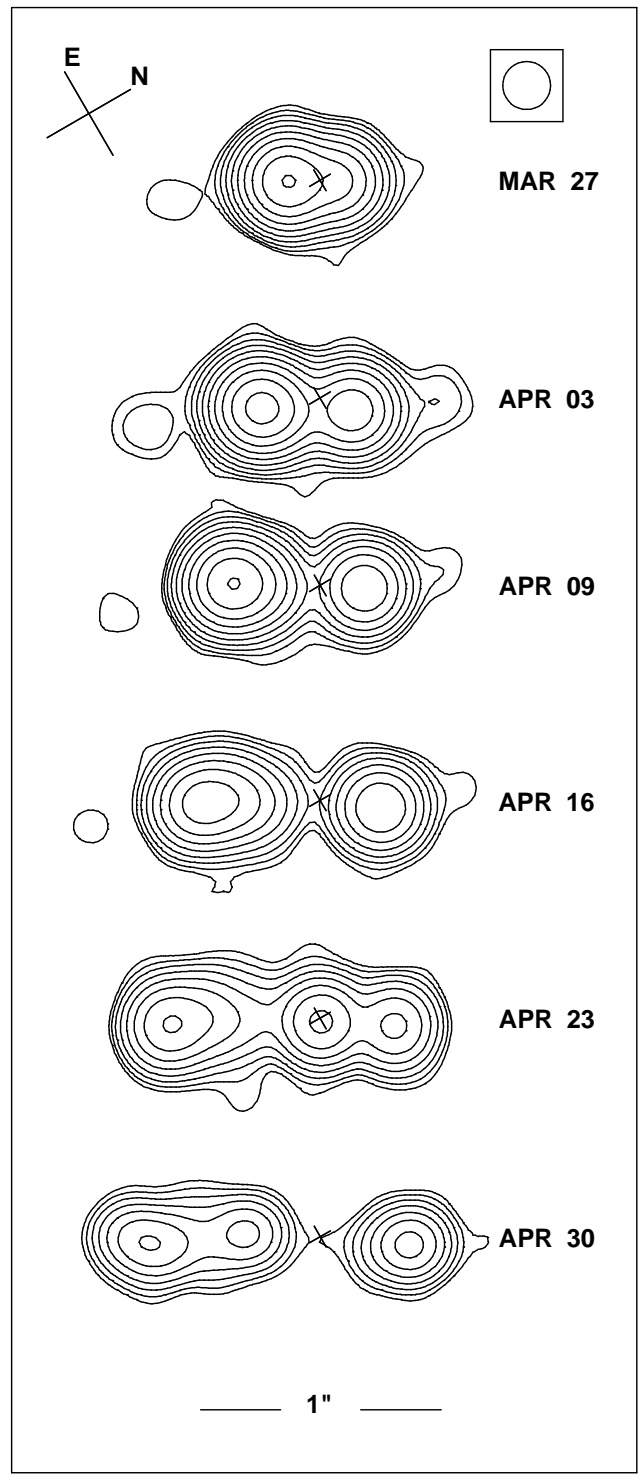}}
\caption{ {Bright radio condensations ejected in opposite directions. The approaching component (left) shows an apparent velocity on the plane of the sky of 1.25c while the receding one (right) seems to move at 0.65c. Observations of GRS 1915+105 with VLA in 1994 (from Mirabel \& Rodr\'{i}guez 1994).}}\label{GRS 1915+105}
\end{figure}

The ejection of two radiating components into opposite directions implies (if the viewing angle is smaller than $90^{\circ}$) that one of them will be approaching the observer and the other one, receding. The synchrotron emission from the ejecta will be affected by the Doppler effect. For example, in the case of GRS 1915+105, the approaching condensation shows an apparent velocity on the plane of the sky of 1.25c and the receding one of 0.65c with the first one appearing brighter than the second (see Fig.~\ref{GRS 1915+105}). Using VLA data, it was inferred that the ejecta moves with a speed of 0.92c at an angle $\phi=70^{\circ}$ to the line of sight. We can then assume that the existence of apparent superluminal movements implies relativistic bulk motions in the jets.\\  

As well as there are relativistic bulk motions, recently, X-ray synchrotron emission from the jet of XTE J1550-564 has been observed (Corbel et al. 2002, Kaaret et al. 2003), which implies the presence of extremely relativistic electrons with TeV energies and shock reacceleration in the outer jet (see Fig.~\ref{Fender1550}).\\

\section{\label{Gamma-Ray Emission}Gamma-Ray Sources}

The study of gamma-ray emission from unidentified sources was the first motivation of this thesis, searching for an understanding of the origin of some mysterious EGRET detections.\\ 

\begin{figure}[!t] 
\centering 
\resizebox{14.5cm}{!}{\includegraphics{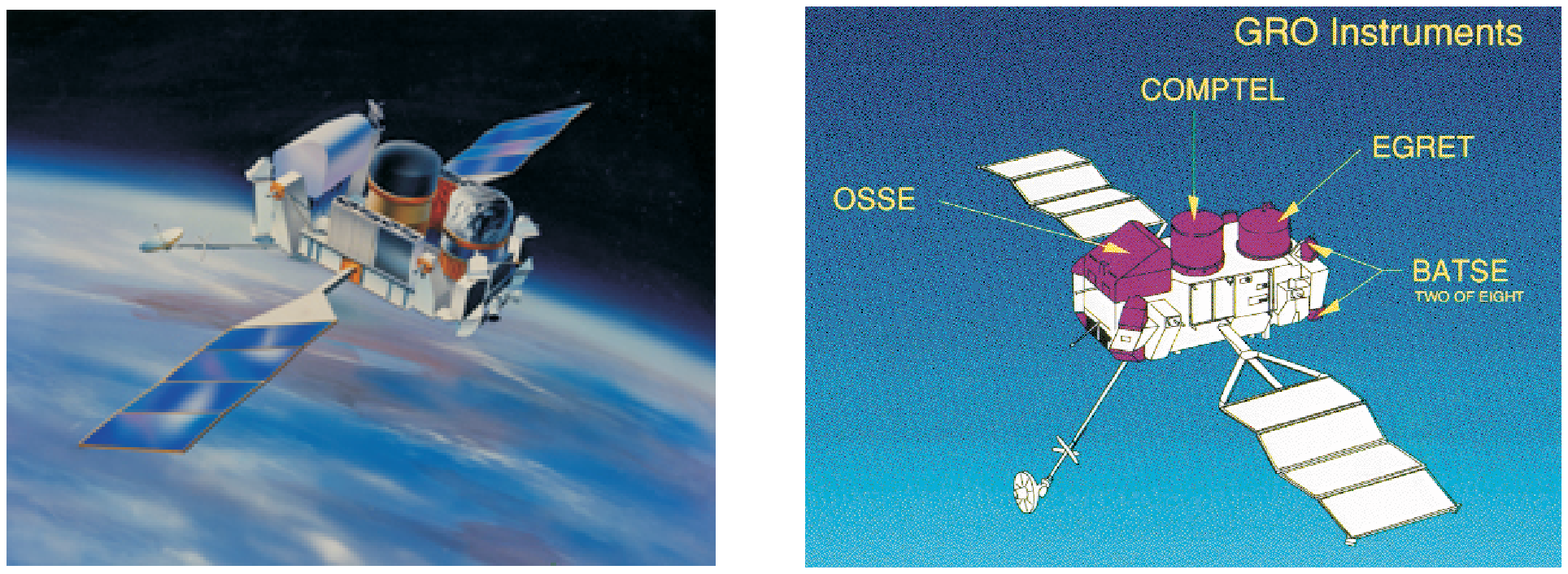}}
\caption{ {The Compton Gamma-Ray Observatory and its four instruments on board, BATSE, OSSE, COMPTEL, EGRET. The four scientific instruments on Compton were designed to operate simultaneously and cooperatively, to cover the gamma-ray energy range from 15 keV to 30 GeV.}}\label{CGRO}
\end{figure}

EGRET, which stands for Energetic Gamma Ray Experiment Telescope, was one of the four instruments on board of the Compton Gamma-Ray Observatory. This satellite was launched on April $5^{th}$ 1991 aboard the space shuttle Atlantis; on June $4^{th}$ 2000 it was safely deorbited and re-entered the Earth's atmosphere. The main characteristics of the four instruments (that can be seen on board in Fig.~\ref{CGRO}) were:

\begin{itemize}
	\item \textbf{BATSE} was optimized to measure brightness variations in gamma-ray bursts and solar flares on timescales down to microseconds, over the energy range \textbf{30 keV to 1.9 MeV}. It was also continuously monitoring all transient sources and bright persistent sources in the gamma-ray sky. It consisted of 8 detectors.
	\item \textbf{OSSE} was designed to undertake comprehensive spectral observations of astrophysical sources in the \textbf{0.05 to 10 MeV} range, with capability above 10 MeV for solar gamma-ray and neutron observations. Each of the 4 detectors had a single axis pointing system which enabled a rapid OSSE response to target of opportunities, such as transient X-ray sources, explosive objects, and solar flares. 
	\item \textbf{COMPTEL} has performed the first sky survey in the energy range from \textbf{1 to 30 MeV}. Source mapping was provided over a field of view of about 1 steradian. 
	\item \textbf{EGRET} was the highest energy instrument on Compton, and covered the broadest energy range, from \textbf{20 MeV to \textbf{$\sim$} 30 GeV}. It had a wide field of view, good angular resolution and very low background. Because it was designed for high-energy studies, the detector was optimized to detect gamma rays when they interact by  pair-production process, which forms an electron and a positron within the EGRET spark chamber. 
	 
\end{itemize}

As we have already mentioned, we will focus our attention on the last instrument.  The \textbf{Third EGRET Catalog (3EG)} (Hartman et al. 1999) of high-energy gamma-ray sources  includes data from April $22^{nd}$, 1991, to October $3^{rd}$, 1995. The 271 sources that conform the catalog were detected at $E>100$ MeV. They  include:

\begin{itemize}
	\item The single 1991 solar flare bright enough to be detected as a source.
	\item The Large Magellanic Cloud.
	\item 5 pulsars (their number has been recently extended up to 7).
	\item One probable radio galaxy detection - Cen A - (there are two other candidates at the moment).
	\item 66 high-confidence identifications of blazars (BL Lac objects, flat-spectrum radio quasars, or unidentified flat-spectrum radio sources).
	\item 27 lower-confidence potential blazar identifications.
	\item \large{\textbf{$\sim$ 170 sources not yet identified firmly with known objects.}}
\end{itemize}

\begin{figure}[!t] 
\centering 
\resizebox{14.7cm}{!}{\includegraphics{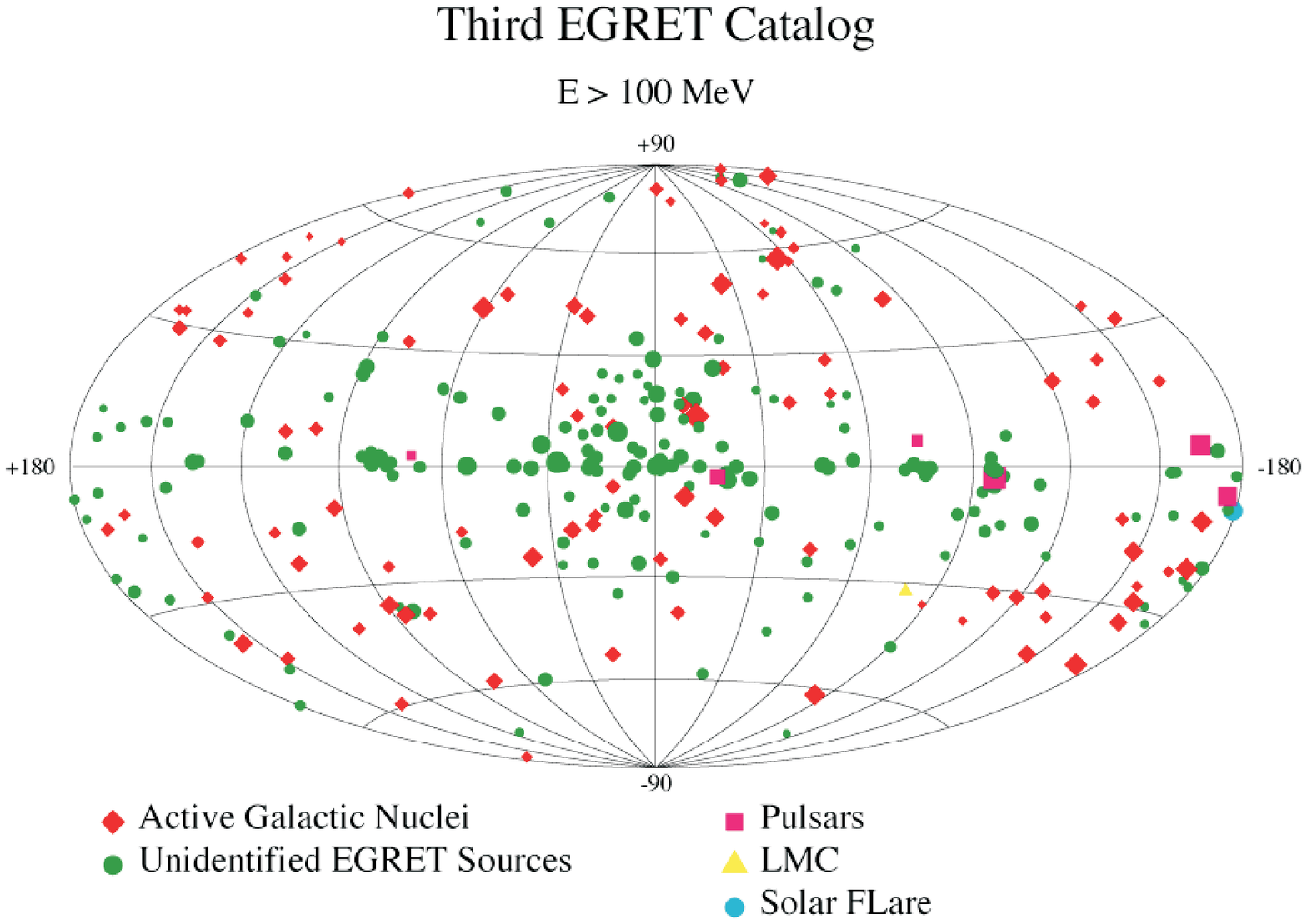}}
\caption{ {This diagram indicates the positions on the sky of EGRET sources of celestial gamma-rays at energies above 100 MeV. The source positions are plotted in Galactic coordinates with color-coded classifications. The sources are represented as point-like ones.}}\label{fuentes}
\end{figure}

Fig~\ref{fuentes} shows the positions in the sky of these sources.\\

The detections included have a significance of more than $5\sigma$ at less than 10 degrees from the galactic plane, where the background radiation is stronger, and more than  $4 \sigma$ otherwise. Most of the sources can be fitted by a single power-law spectra, and the corresponding spectral photon index usually ranges between 1.5 and 3.0.\\

\begin{figure}[!t]
 \begin{minipage}[t]{0.5\linewidth}
\centering
  \resizebox{7cm}{!}{\includegraphics{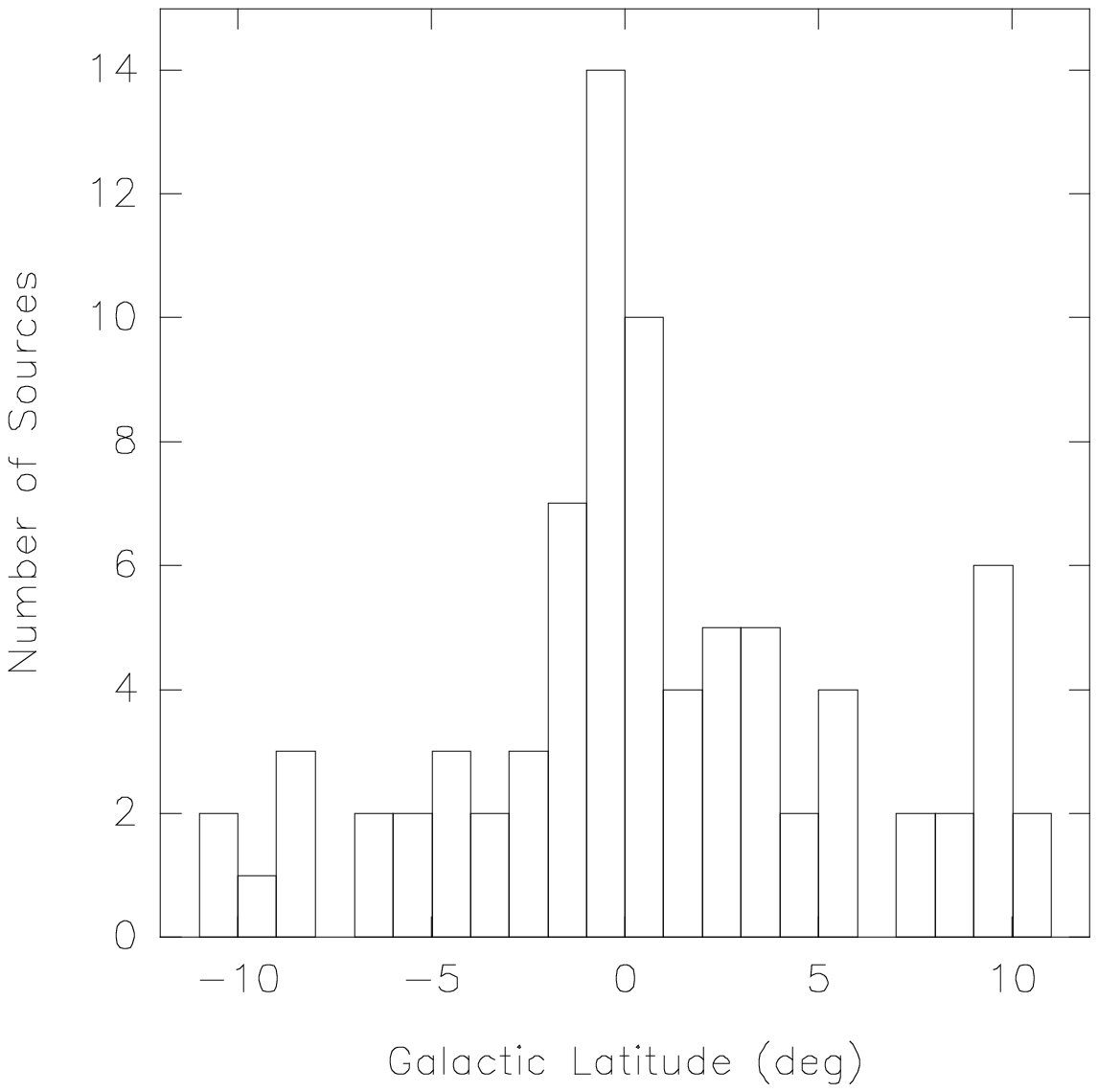}}
 \end{minipage}
 \begin{minipage}[t]{0.5\linewidth}
 \centering
 \resizebox{7cm}{!}{\includegraphics{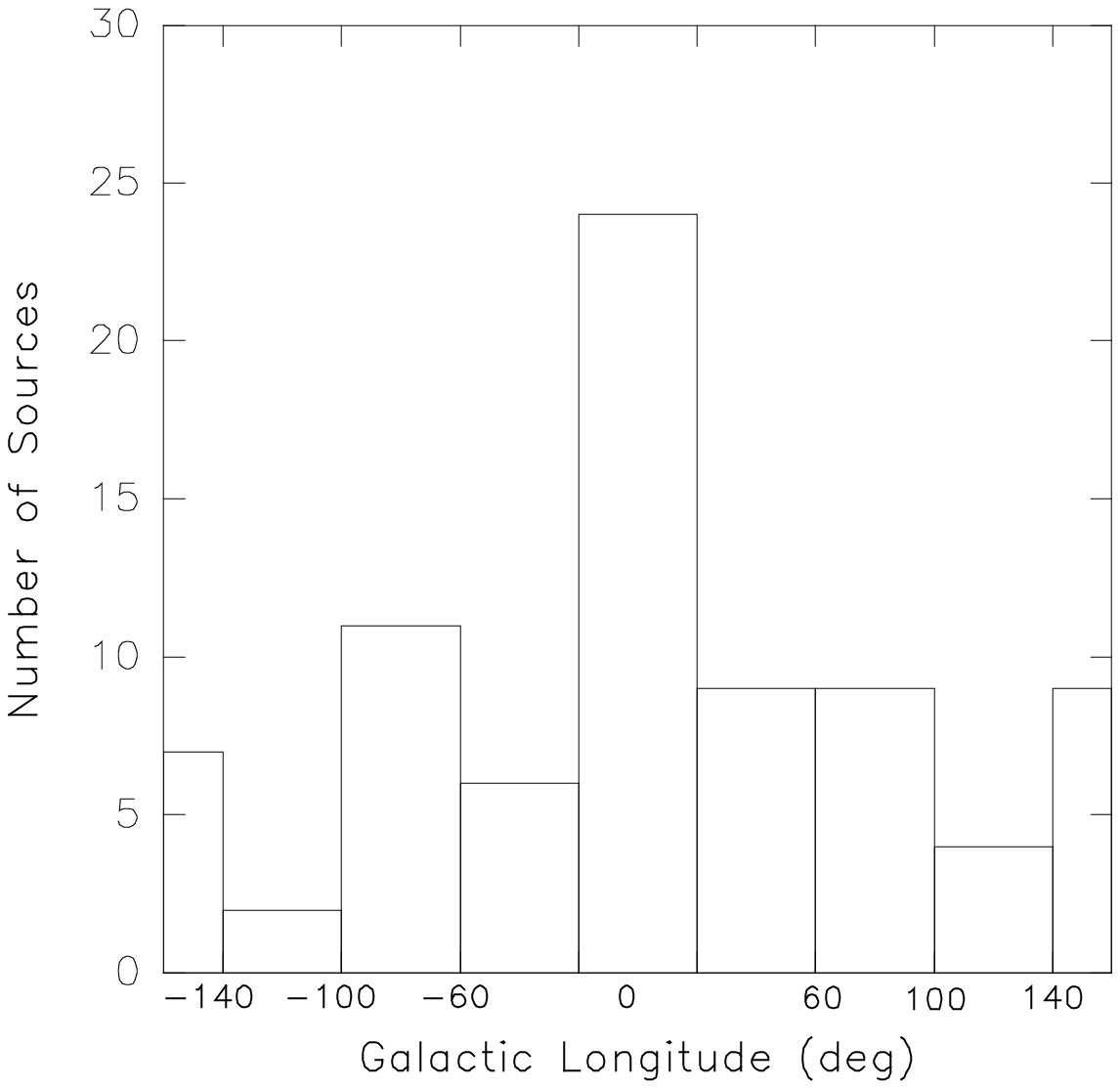}}
  \end{minipage}
 \caption{ {Distribution of EGRET sources with galactic latitude on the left panel and with galactic longitude on the right panel (from Romero et al. 1999).}}
 \label{distribution}
\end{figure}

Our interest here will be focused on the \textit{unidentified} sources. It has been shown in Romero et al. (1999) that the distribution of the unidentified EGRET sources with galactic coordinates has a clear concentration of sources on the galactic plane plus a concentration in the general direction of the galactic center (see Fig.~\ref{distribution}). This indicates a significant contribution from galactic sources.
A correlation ana\-lysis between 3EG-low-latitude sources and bright and giant HII regions, the usual tracers for the galactic spiral structure, shows that there is a strong correlation at $\sim 7\sigma$-level with the spiral arms of the Galaxy, the places where stars are formed (Romero 2001). This means that there is a significant number of Population I objects\footnote{A term used to describe stars and other objects, such as star clusters, that tend to be found in, or near to, the plane of a spiral galaxy and follow roughly circular orbits around the center. They are younger than Population II objects, have a relatively high heavy element content, and have probably been formed continuously throughout the lifetime of the disk. Extreme Population I objects are found in spiral arms and consist of young objects, such as T Tauri stars, O stars, B stars, and stars newly arrived on the zero-age main sequence. Older Population I objects include stars like the Sun. All Population I 
 stars are relatively rich in elements heavier than hydrogen and helium since they formed from clouds of gas and dust which contained the products of nucleosynthesis from previous generations of stars.} in the parent population of the low-latitude gamma-ray sources.\\

The large error boxes in the location of the EGRET sources (typically $\sim 1$ degree in diameter), make correlation studies to find secure counterparts very uncertain. This technique has been successful mainly to identify sources at high latitudes (such as bright blazars),  where the turmoil is significantly lower than in the galactic plane. Complementary tools like population studies using the known characteristics of the gamma-ray sources, like their spectra, variability and flux density had to be used also. The so-called $\log N - \log S$ \footnote{A $\log N - \log S$ plot displays the number of sources ($N$) with a flux ($S$) greater than a given value.} studies can be particularly useful on this respect (e.g. Reimer \& Thompson 2001, Bhattacharya et al. 2003, Bosch-Ramon at al. 2005). Combining this with studies including variability and source distribution models that take into account the non-uniform detection sensitivity across the sky (Gehrels et al. 2000; Grenier 2
 001, 2004; Grenier et al. 2004), it has been possible to establish a division of the sources into two well-defined  groups:\\

\begin{enumerate}[I)]
	\item One at low latitudes, $|b|<5^{\circ}$, on the galactic plane and spiral arms. These are bright sources, relatively hard (photon index $\Gamma\sim 2.18$). This group should be formed by young sources (a few million years at most) with isotropic luminosities in the range of $10^{34-36}$ erg/s. These sources contain a subgroup of clearly variable sources (Torres et al. 2001a, Nolan et al. 2003).
	\item The other group, at mid-high latitudes, includes 3 populations: 
\begin{itemize}
	\item 45 $\pm$ 6 sources spatially associated with the Gould belt (star forming region at $\sim$ 600 pc), with $\Gamma\sim2.25$, stable and weaker. In this group young sources are also expected but with luminosities in the range $10^{32-33}$ erg/s.
	\item 45 $\pm$ 9 sources with softer spectra ($\Gamma\sim2.5$) and high variability forming a kind of halo around the galactic center with a scale height of $\sim2$ kpc. The sources in this group might be formed in and ejected from globular clusters (however, there is no significant correlation with individual clusters) or from the galactic plane. These sources should be old, age measured in Gyrs, and very luminous, in the range $10^{35-37}$ erg/s. Significant variability between different EGRET viewing periods is observed in many of these sources (Nolan et al. 2003).   

	\item Around 35 potentially extragalactic sources (isotropically distributed). 
\end{itemize}
\end{enumerate}

The transition between both groups occurs at $|b|\sim 5$ degrees.\\ 

\section{Motivation of the Thesis: the Link}

\begin{figure}[!t] 
\centering 
\resizebox{10cm}{!}{\includegraphics{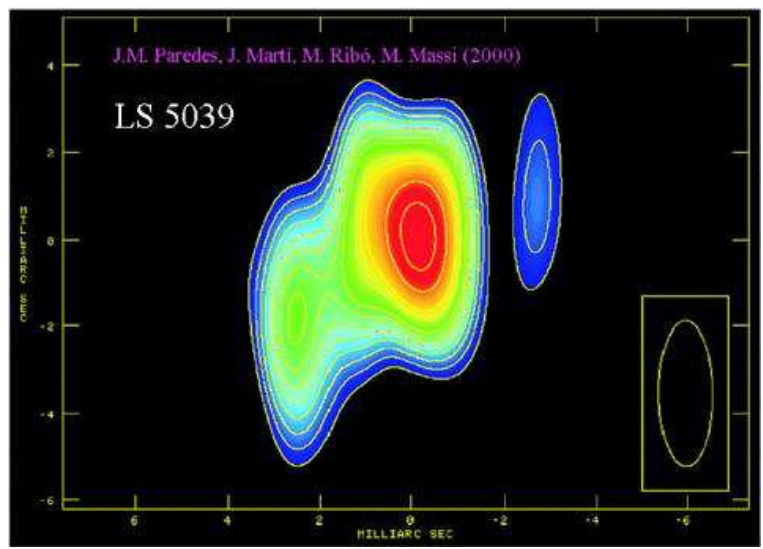}}
\caption{{VLBA observation where the relativistic radio jets of the microquasar LS 5039 can be seen (from Paredes et al. 2000).}}\label{LS5039}
\end{figure}

\begin{figure}[!t] 
\centering 
\resizebox{14.5cm}{!}{\includegraphics{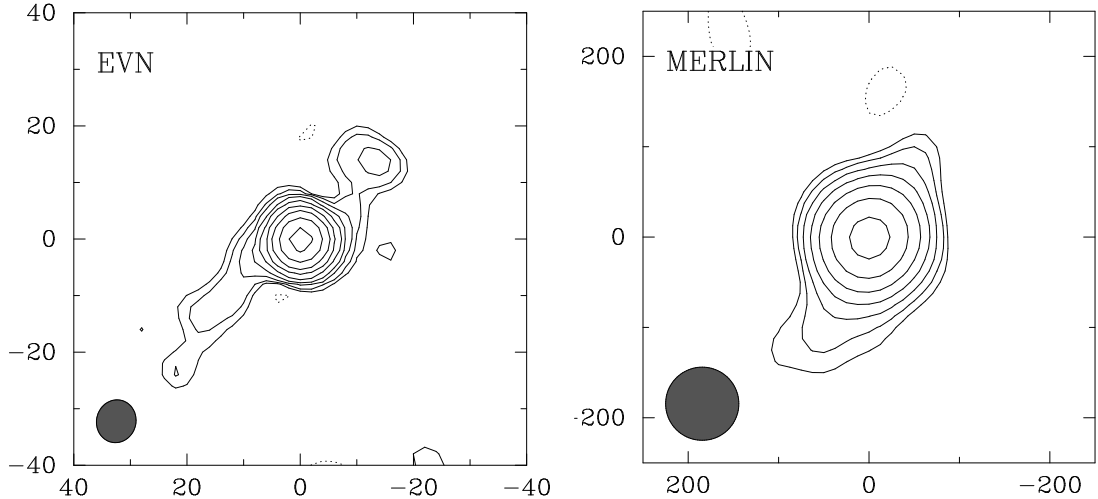}}
\caption{{Observations of LS 5939 made with EVN (\textit{left panel}) and MERLIN (\textit{right panel}) confirming the existence of a jet in that object. Axis units are in milliarcseconds (from Paredes et al. 2002a).}}\label{LS5039-2}
\end{figure}

\begin{figure}[!t] 
\centering 
\resizebox{10cm}{!}{\includegraphics{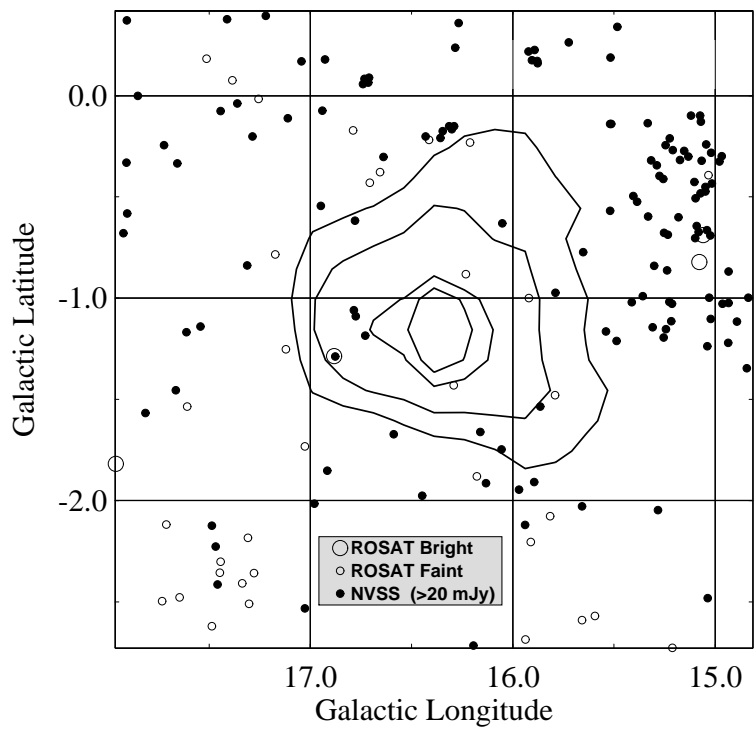}}
\caption{{Location map of the EGRET source 3EG J1824-1514. We can see the statistical probability that the gamma-ray source lies within the $50\%$, $68\%$, $95\%$ or $99\%$ contour. LS 5039, here represented by a filled circle inside an open cirle, has coordinates, l = $16.88^{\circ}$, and b =  $-1.29^{\circ}$, and lies in the $95\%$ contour. It's the only X-ray and radio source in the EGRET error box (from Rib\'o 2002).}}\label{LS5930vs3EG}
\end{figure}

A few words now about the link we will propose between MQs and gamma-ray emission. In the year 2000, Paredes et al. reported the discovery of a new microquasar, LS 5039\footnote{The presence of an asymmetric two-sided jet in LS 5039 was confirmed with observations from EVN and MERLIN (Paredes et al. 2002a). The longest jet arm was estimated to reach up to $\sim$ 1000 AU (see Fig~\ref{LS5039-2}).} (see Fig.~\ref{LS5039}). This discovery involved a very interesting issue: the microquasar LS 5039 is inside the error box of the unidentified EGRET source 3EG J1824-1514 (see Fig.~\ref{LS5930vs3EG}), and was therefore proposed to be its counterpart (Paredes et al. 2000).\\

Two similar cases followed. The discovery of radio jets in LS I +61 303 confirmed the microquasar nature of this binary system (Massi et al. 2001, 2004). There is an argument that seems to suggest that LS I +61 303 is also a gamma-ray emitting microquasar (Massi 2004a): the likely correlation of the short term variability of 3EG J0241+6103 with the orbital period of the binary. The other case has just been proposed by Combi et al. (2004) through the association between the microquasar candidate AX J1639.0-4642 and 3EG J1639-4702. This latter case is the most uncertain one, since there are other potential gamma-ray emitting objects inside the error box and the microquasar nature of AX J1639.0-4642 is not demonstrated.\\

The original proposal by Paredes et al. (2000) motivated the research project described in this thesis. We investigated the possibility that MQs could be the parent population of the gamma-ray sources distributed on the galactic plane and at high latitude forming a halo around the galactic center and whose spectral energy distribution peaks in the MeV-GeV range. \\  

Though our first aim was to propose a new parent population for different groups of EGRET sources, the model of microquasars emitting gamma-rays will also be used to explain observations from well-known sources  at other wavelengths than EGRET's. The models described in this thesis can also be applied outside this energy range, as it will be shown for different cases.\\  

\section{Outline of the Thesis} 

In the next chapter we will review the main features of each component of a microquasar. The compact object, BH or NS, the accretion disk, the corona, the stellar companion and the jet: main characteristics and some models will be presented. We will focus on those aspects that are more relevant from the point of view of the possible high-energy emission.\\

After the second chapter, we will start the presentation of our proposal of MQs generating gamma-ray emission. Different scenarios and combinations of parameters will be studied. In Chapter 3 we  will introduce a high-mass microquasar model, showing the fundamental role of a high mass companion. We will develop a leptonic model studying the interaction of the jet with the field of photons from the stellar companion (Inverse Compton, IC). The variability observed in some sources will also be considered.\\

The toy model of Chapter 3 will become more complex in Chapter 4 where the interaction of the jet with the accretion disk and the corona photon fields will be taken into account in order to analyze the possible contributions to the total gamma-ray production. We will also discuss the application of the model as an attempt to interpret some repeated soft gamma-ray events observed in Cyg X-1.\\

In Chapter 5 we present another kind of model to produce high-energy gamma-rays in HMMQs. It is based on hadronic interactions between multi-TeV protons of the jet and the ions present in the wind of the stellar companion. The gamma-ray emission results from the decay of neutral pions that are produced in the inelastic collisions between the protons from the jet and from the wind. This kind of scattering implies also the generation of neutrinos. We include estimations for this emission as well.\\

Low-mass microquasars are finally considered in Chapter 6. We come back to a leptonic model and we propose LMMQs as the parent population of one of the groups of the unidentified mid-high latitudes EGRET sources. We focus on the sources that are forming a kind of halo around the galactic center. In this scenario it will be fundamental to consider not only the interactions between leptons of the jet and photons from external sources, but also the IC interactions with the synchrotron photons (Synchrotron-self-Compton, SSC).\\

Four appendices complete the thesis. The first one is the presentation of another gamma-ray producing situation, also studied during the thesis working period. It does not involve MQs but a specific unidentified gamma-ray source, 3EG J0542+2610 that is proposed as the gamma-ray counterpart of the Be/X-ray transient A0535+26. This is the case of gamma-rays from magnetized accreting pulsars, where the high-energy photons are produced in the interaction of a beam of relativistic protons and a transient accretion disk.\\

Relevant radiative processes, that are used in the models presented in the former chapters of this work, are described in Appendix~\ref{radiative processes}. A detailed explanation about superluminal motions is presented in Appendix~\ref{superluminal}. The last appendix lists the publications completed during the development of the thesis.        

\newpage
\thispagestyle{empty}
\phantom{.}

\chapter{\label{Microquasars}Microquasars: Main Features}
\thispagestyle{empty}

\newpage
\thispagestyle{empty}
\phantom{.}

\newpage
\vspace*{6cm}

\noindent In the following sections we will review the main statistical and physical characteristics of MQs. MQs were discovered in recent times, and the number of objects recognized as such is still far from representing a statistically important figure. Questions like ``Can they be assumed as rather common objects in our Galaxy?'', are still lacking a clear answer.\\

On the other hand, we have to deal with the observational data we already have in order to start to understand which are the main features of the different components of MQs and which are the active physical processes that are taking place in them. Observations made at different wavelengths are being interpreted and different models are being proposed.\\

We need to ponder how significant are the chances to rise the number of confirmed MQs and also to understand the physics involved, if we want to be able to propose them as a parent population of EGRET unidentified sources.\\ 

\section{\label{SandP}Statistics and Projections}

Taking into account that MQs are until now considered as a particular subgroup of XRBs that generate jets, the first step is to wonder how many XRBs are hosted in the Galaxy. The most updated catalogs of XRBs are found in Liu et al. (2000) for High Mass X-ray Binaries (HMXBs) and Liu et al. (2001) for Low Mass X-ray Binaries (LMXBs). After an increase from $\sim 30$ known HMXBs in 1983 (van Paradijs 1983) to 69 in 1995 (van Paradijs 1995), the new catalog contains 130 of these sources. The number of registered LMXBs went from 33 in 1983 (van Paradijs 1983, Bradt \& McClintock 1983) to 119 in 1995 (van Paradijs 1995) to reach in the newest catalog, 150 confirmed of such XRBs. The two catalogs have then 280 XRBs in total (several new binaries have been discovered since then).\\

Next step is to check how many of those 280 XRBs are also radio emitters (Radio Emitting X-ray Binaries, REXBs), since this is the wavelength at which jets manifest themselves more clearly. Taking into account both catalogs, there are $\sim 75$ X-ray pulsars, which are suspected not to emit in radio. It is believed that the strong magnetic field of X-ray pulsars disrupts the accretion disk at some thousand kilometers from the neutron star, suppressing the possibility to generate a jet (Fender \& Hendry 2000). In fact, synchrotron radio emission has never been detected from any of these sources. Otherwise, the two catalogs contain 43 radio emitting sources (Paredes 2004b). Once detected, the radio emission has to be resolved in order to confirm that it is produced by a relativistic jet. Only when the presence of a jet is unquestionable we can call the XRB a microquasar.\\

Nowadays there are 16 systems identified as MQs. They are listed in Table~\ref{census} (from Paredes 2004a and 2004b with some modifications)\footnote{XTE J1118+480 is included in the list because even though its relativistic jet has not been resolved yet it is inferred from theoretical reasons.}. Different multi-wavelength cross-searchings have been carried out in X-ray and radio source catalogs in order to find more MQ candidates that could rise this number after being confirmed as such with VLBA, ATCA or VLBI observations (e.g Paredes et al. 2002b). The chances to increase the number of MQs radically would enhance if we could know that there are even more XRBs in the Galaxy. Grimm et al. (2002) have made some estimate in this respect. Using data of XRBs from RXTE/ASM\footnote{ASM, All-Sky Monitor, instrument on board of the satellite RXTE, Rossi X-Ray Timing Explorer, also known as XTE.} they obtained a $\log N - \log S$ diagram and the luminosity function. The total numbe
 r of XRBs brighter than $2 \times {10}^{35}$ erg s$^{-1}$ is $\sim$ 190 divided into $\sim 55$ high mass and $\sim 135$ low mass binaries. Extrapolating the luminosity functions towards low luminosities they estimated a total number of $\sim 705$ XRBs brighter than ${10}^{34}$ erg s$^{-1}$, distributed as $\sim 325$ LMXBs and $\sim 380$ HMXBs. As a consequence of these expectations the existence of MQs in a larger number becomes more likely. In fact, some authors sustain that jets are a quite common feature of X-ray binary systems, suggesting that up to $70$\% of XRBs might produce them (Fender 2004). In such a context it would be a matter of time to detect them, providing that new instruments with more sensitivity and resolution in both X-rays and radio-band become available. We will come back over the arguments for such a high percentage in Section~\ref{dynamics}.

\begin{sidewaystable}
{\footnotesize 
\begin{center}
\caption{\label{census} {\bf Microquasars in our Galaxy}}
\begin{tabular}{@{}l@{\hspace{0.07cm}}@{}l@{\hspace{0.07cm}}c@{\hspace{0.05cm}}cc@{\hspace{0.05cm}}c@{\hspace{0.05cm}}c@{\hspace{0.05cm}}c@{\hspace{0.05cm}}c@{\hspace{0.05cm}}c@{}c@{}}
             &                 &           &      &              &              &                           \\
\hline \hline \noalign{\smallskip}
Name  & Position &  System  & $D$ & $P_{\rm orb}$  & $M_{\rm compact}$   & \hspace{0.1cm} Activity &
\hspace{0.2cm} $\beta_{\rm app}$ & $\theta$$^{\rm (c)}$ & Jet size & Remarks$^{\rm (d)}$ \\
& (J2000.0)  & type$^{\rm (a)}$& (kpc) & (d)   &   $(\mo)$  & \hspace{0.3cm} radio$^{\rm (b)}$&   &  & (AU)\\
\noalign{\smallskip} \hline \hline \noalign{\smallskip}
\multicolumn{10}{c}{\bf High Mass X-ray Binaries}\\
\noalign{\smallskip} \hline \hline \noalign{\smallskip}

{\bf LS~I~+61~303} & $02^{\rm h}40^{\rm m}$31\rl 66 &B0V & 2.0 & 26.5 & $-$ & p & $\geq$0.4 & $-$ & 10$-$700 & Prec?\\
& $+61^{\circ}13^{\prime}$45\pri 6 & +NS? &  & & & & & \\

{\bf V4641~Sgr} & $18^{\rm h}19^{\rm m}$21\rl 48  & B9III & $\sim10$ & 2.8  & 9.6 & t & $\ge9.5$ & $-$&$-$ & \\
& $-25^{\circ}25^{\prime}$36\pri 0 &+BH & \\

{\bf LS~5039} & $18^{\rm h}26^{\rm m}$15\rl 05   &  O6.5V((f)) & 2.9 & 4.1 & 1$-$3 & p  & $\geq0.15$ &$<81^{\circ}$& 10$-$$10^3$ & Prec?\\
&$-14^{\circ}50^{\prime}$54\pri 24 & +NS?& &   & & & & &  \\

{\bf SS~433} & $19^{\rm h}11^{\rm m}$49\rl 6&evolved A?   & 4.8  &  13.1 &  11$\pm$5?& p & 0.26 &  $79^{\circ}$&$\sim10^4$$-$$10^6$ & Prec   \\
& $+04^{\circ}58^{\prime}58^{\prime\prime}$ &+BH?  & &   & & & & & &XRJ\\

{\bf Cygnus~X-1} & $19^{\rm h}58^{\rm m}$21\rl 68&O9.7Iab  & 2.5 & 5.6 & 10.1 &  p &$-$& 40$^{\circ}$& $\sim40$ &Prec?\\
& $+35^{\circ}12^{\prime}$05\pri 8& +BH & &   & & & & & \\

{\bf Cygnus~X-3} & $20^{\rm h}32^{\rm m}$25\rl 78 &WNe     &  9      &  0.2   & $-$& p  & 0.69 & 73$^{\circ}$ & $\sim10^4$    \\
&$+40^{\circ}57^{\prime}$28\pri 0 & +BH? & &   & & & & & \\

\noalign{\smallskip} \hline \hline \noalign{\smallskip}
\multicolumn{10}{c}{\bf Low Mass X-ray Binaries}\\
\noalign{\smallskip} \hline \hline \noalign{\smallskip}

{\bf XTE~J1118+480}   &$11^{\rm h}18^{\rm m}$10\rl 85 & K7V$-$M0V    & 1.9  &  0.17  & 6.9$\pm$0.9 &  t &$-$& $-$&$\le0.03$   \\
 &$+48^{\circ}02^{\prime}$12\pri 9 & +BH &       &    &    &  \\

{\bf Circinus~X-1}    & $15^{\rm h}20^{\rm m}$40\rl 9& Subgiant &  5.5     &  16.6  &$-$&  t   & $>15$& $<6^{\circ}$ & $>10^4$ \\
& $-57^{\circ}10^{\prime}01^{\prime\prime}$ & +NS \\

{\bf XTE~J1550$-$564} & $15^{\rm h}50^{\rm m}$58\rl 70 & G8$-$K5V & 5.3  & 1.5   & 9.4 &  t
&$>2$& $-$& ${\sim10^5}^{\ast}$  & XRJ   \\
& $-56^{\circ}28^{\prime}$35\pri 2& +BH & &   & & & & &  \\

{\bf Scorpius~X-1}  &  $16^{\rm h}19^{\rm m}$55\rl 1 & Subgiant    & 2.8      &  0.8  &  1.4 &  p &$ 0.68$& $44^{\circ}$& $\sim40$    \\
 & $-15^{\circ}38^{\prime}25^{\prime\prime}$& +NS  \\

{\bf GRO~J1655$-$40} & $16^{\rm h}54^{\rm m}$00\rl 25 & F5IV  & 3.2    &  2.6   & 7.02 & t & 1.1& $72^{\circ}$$-$$85^{\circ}$&  8 $10^3$  & Prec?   \\
& $-39^{\circ}50^{\prime}$45\pri 0 & +BH & &   & & & & &  \\

{\bf GX~339$-$4}   & $17^{\rm h}02^{\rm m}$49\rl 5 & $-$         & $>6$     &  1.76  & 5.8$\pm$0.5 & t& $-$&$-$&   ${\sim4.6\,10^4}^{\dagger}$  \\
& $-48^{\circ}47^{\prime}23^{\prime\prime}$& +BH \\

{\bf 1E~1740.7$-$2942}& $17^{\rm h} 43^{\rm m} 54$\rl 83&$-$  & 8.5? &  12.5?   &$-$ & p &$-$& $-$&$\sim10^6$  \\
& $-29^{\circ} 44^{\prime}$42\pri 60& +BH ?& &   & & & & & \\

{\bf XTE~J1748$-$288} & $17^{\rm h}48^{\rm m}$05\rl 06& $-$ &  $\geq8$   &  ?    &   $>4.5$? & t & 1.3&$-$ & $>10^4$         \\
& $-28^{\circ}28^{\prime}$25\pri 8 &+BH? \\

{\bf GRS~1758$-$258}  &$18^{\rm h}01^{\rm m}$12\rl 40  & $-$ & 8.5?  & 18.5?  &$-$   & p &$-$& $-$&$\sim10^6$\\
& $-25^{\circ}44^{\prime}$36\pri 1 &+BH ?\\

{\bf GRS~1915+105} &  $19^{\rm h}15^{\rm m}$11\rl 55 &  K$-$M III   & 12.5 & 33.5  & 14$\pm$4 &  t &1.2-1.7& 66$^{\circ}$-70$^{\circ}$&$\sim10$$-$$10^4$ & Prec?\\
&$+10^{\circ}56^{\prime}$44\pri 7 &+BH & &   & & & & & \\
\hline
\hline
\end{tabular}
\end{center}
{\begin{center}
\small Notes: $^{\rm (a)}$ NS: neutron star; BH: black hole. $^{\rm (b)}$ p: persistent; t: transient. $^{\rm (c)}$ jet inclination.\\ 
$^{\rm (d)}$ Prec: precession; XRJ: X-ray jet. $^{\ast}$Reported by Corbel et al. 2002. $^{\dagger}$Recently reported by Gallo et al. 2004.
\end{center}}}
\end{sidewaystable}

\section{Physical Components}

\begin{figure}[!t] 
\centering 
\includegraphics[scale=0.6,angle=270]{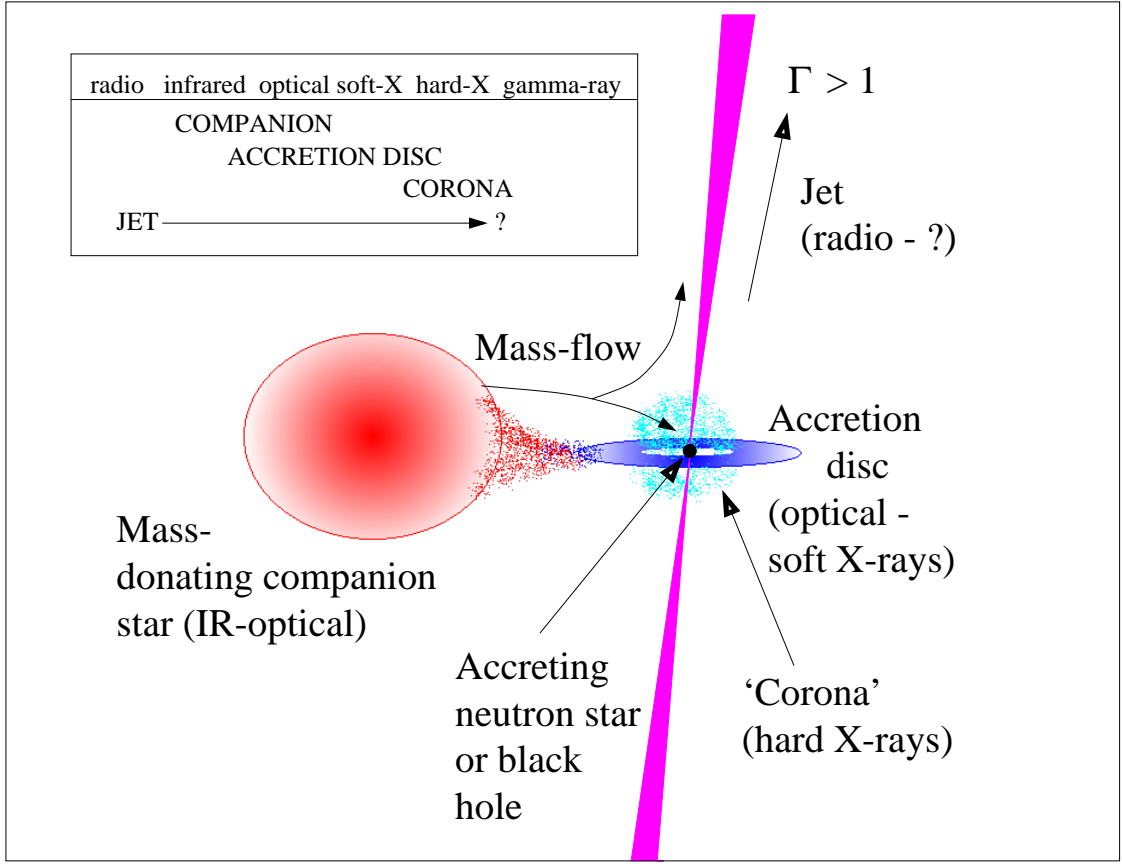}
\caption{{Sketch of the physical components of a microquasar. We can see the binary system formed by a compact object (NS or BH) that accretes material from the stellar companion, forming an accretion disk and corona. The potential accretion energy is converted into high energy emission and mechanical energy of the jet, that emits from radio wavelengths to gamma-rays, as we will propose in the forthcoming chapters (from Fender \& Maccarone 2004).}}\label{components}
\end{figure}

In the next four subsections we will make an overview of the main components of a microquasar, studying their physical properties, the classifications that arise from them and some of the models that attempt to describe the observational data. These components are: the companion star, the disk, the corona, and the jet. Fig.~\ref{components} sketches a microquasar system showing the range of wavelengths corresponding to each emitting component.\\

\subsection{\label{CSandCO}The Companion Star and the Compact Object}

According to the mass of the stellar companion, XRBs are divided in two groups: high mass and low mass X-ray binaries. HMXBs are systems where the donor star is a young massive star of $\sim 8-20\;M_{\odot}$ and orbital periods are several days. The nature of the star determines a further division of HMXBs in two subgroups: those in which the stellar companion is a Be star (Be/X-ray binary) and those in which it is an O or B supergiant (SG/X-ray binary). In general, these high mass companion stars transfer mass to the compact object through strong stellar winds, though in some cases it can also happen via Roche Lobe overflow. X-ray outbursts are expected to happen in the Be/X-ray sources during the periastron passage.\\

In the case of LMXBs, the companion is an older low-mass type of star, with $M \leq 2 \;M_{\odot}$; it can be a white dwarf, a late-type main-sequence star, an A-type star or an F-G sub-giant (in which case, it may be the remnant of a star that originally was of intermediate mass $\sim$ 1.5 to 4 $M_{\odot}$). The orbital periods are in the range of 0.2 to 400 hours and the mass transfer occurs through Roche Lobe overflow.\\

\begin{figure}[!t] 
\centering 
\resizebox{14.8cm}{!}{\includegraphics{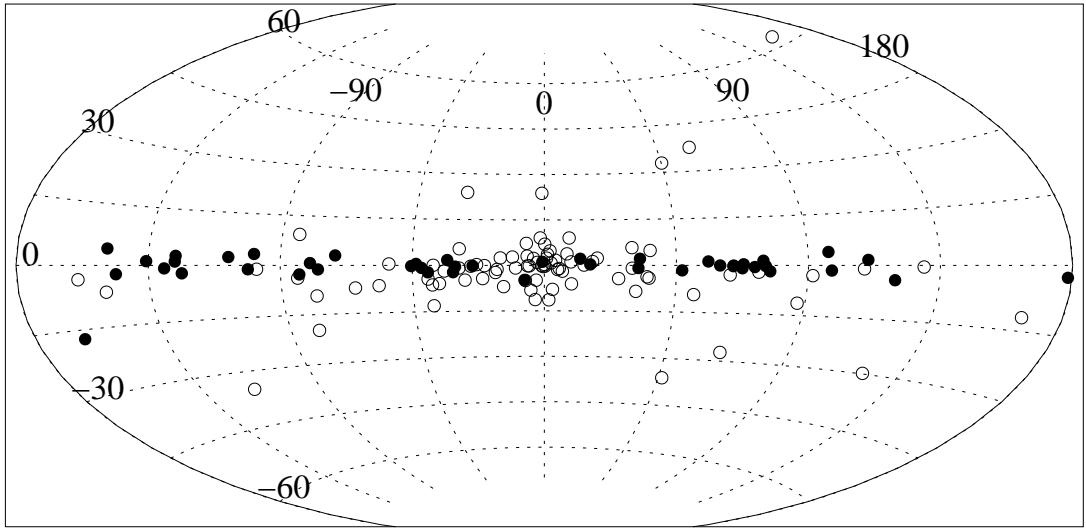}}
\caption{{All-sky map where open circles representing 86 LMXBs and filled circles, 52 HMXBs. The two spatial distributions in the Galaxy can be noted: HMXBs are significantly concentrated on the galactic plane whereas LMXBs are mainly forming a halo in the galactic bulge (from Grimm et al. 2002).}}\label{XRBs}
\end{figure}

\begin{figure}[t]
 \begin{minipage}[t]{0.5\linewidth}
\centering
  \resizebox{7cm}{!}{\includegraphics{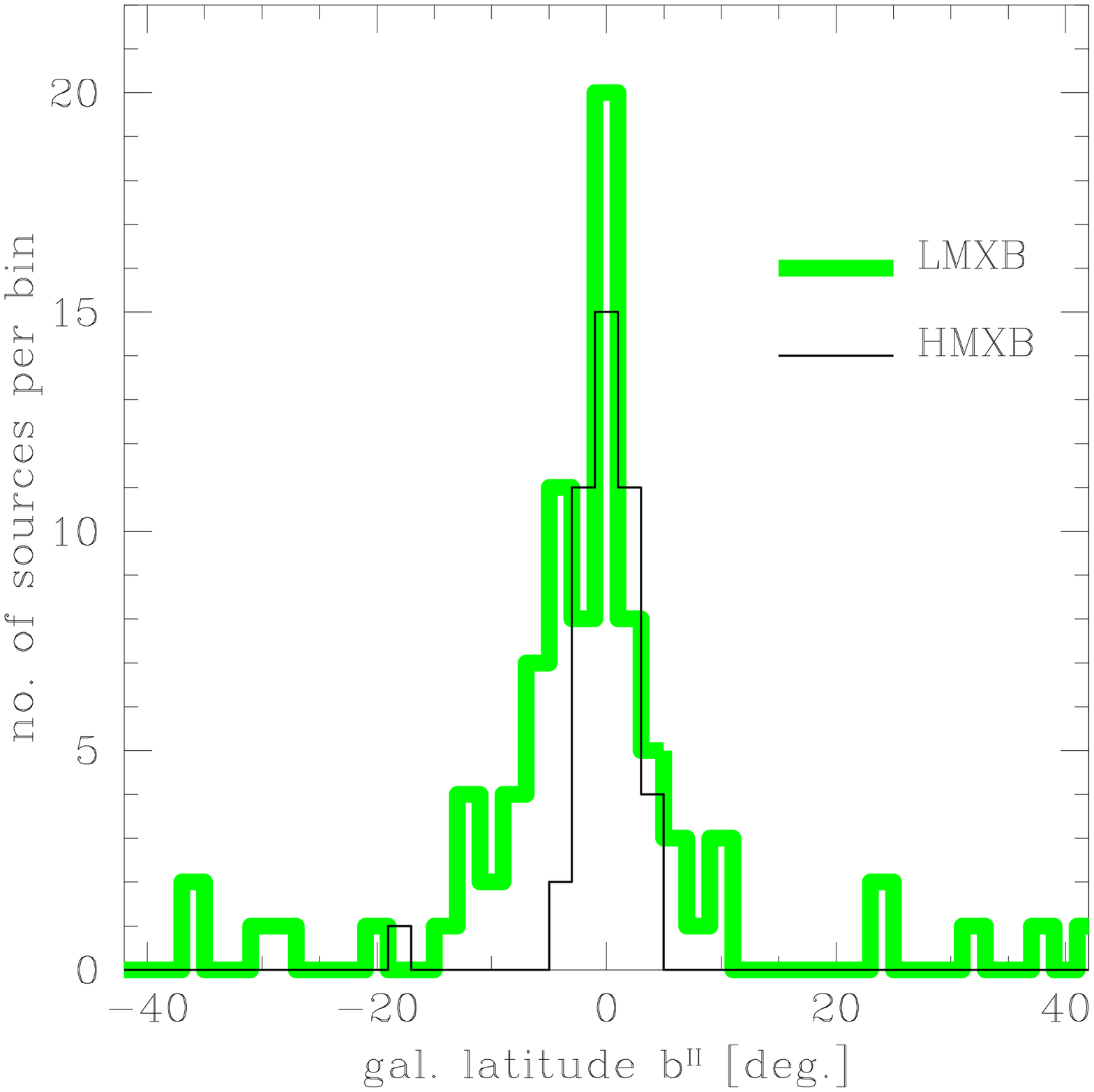}}
 \end{minipage}
 \begin{minipage}[t]{0.5\linewidth}
 \centering
 \resizebox{7cm}{!}{\includegraphics{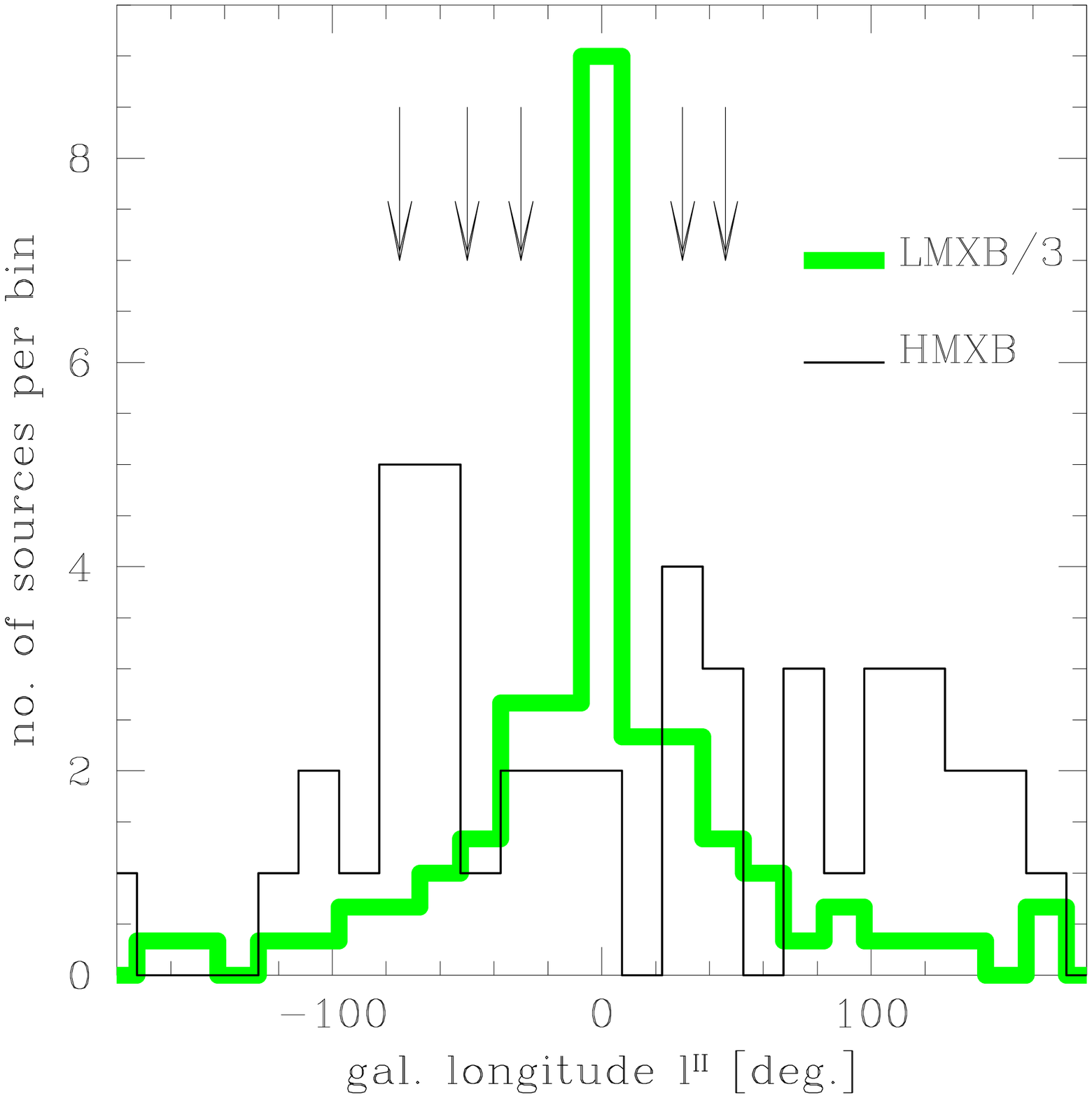}}
  \end{minipage}
 \caption{ {Angular distribution of galactic HMXBs (solid lines) and LMXBs (thick green lines) against the galactic latitude (left panel) and longitude (right panel). These two graphs illustrate the well-known fact that HMXBs are strongly concentrated towards the galactic plane. An important difference in the longitude distributions of HMXBs and LMXBs can be noticed, with the second ones significantly concentrated towards the galactic center/bulge and the former distributed in clumps approximately coinciding with the location of tangential points of the spiral arms (whose position is marked by arrows in the right panel). The LMXBs number is divided by 3 on the right panel (from Grimm et al. 2002).}}
 \label{XRBs2}
\end{figure}

The already mentioned ASM catalog includes 340 sources being 217 galactic, 112 extragalactic, and 10 unidentified. Studying the galactic sources, Grimm et al. (2002) found significant differences in the 3D spatial distribution of HMXBs and LMXBs in our Galaxy as it is shown in Fig.~\ref{XRBs} (in good agreement with theoretical expectations and earlier results -- van Paradijs \& White 1995; White \& van Paradijs 1996; Koyama et al. 1990; Nagase 1989). Whereas HXMBs are more concentrated towards the galactic plane with a vertical scale height of 150 pc, and clear indications of a distribution following the spiral structure, LMXBs have a strong tendency to concentrate towards the galactic bulge and their vertical distribution has a scale height of 410 pc. This important difference in the angular distribution of high and low mass XRBs is also illustrated by Fig.~\ref{XRBs2} where the angular distribution is represented against the galactic latitude and longitude. 

\vspace{0.7cm}
\begin{center}
$\clubsuit$
\end{center}
\vspace{0.9cm}

The compact object that is present in an X-ray binary system can be either a neutron star or a black hole. While stars are on the \textit{main sequence}, the thermal pressure of hot gas maintains an equilibrium with gravity. The source of energy that provides the pressure comes from nuclear reactions generated in their cores. The life of the star evolves out from the main sequence through the so-called \textit{giant branch} towards the final phases when the outer layers of the star are ejected. During all that path the nuclear processing occurs using all the available nuclear energy resources of the star. In the most massive stars, $M\geq 10M_{\odot}$, it is quite probable that the nuclear burning continues till the formation of iron. At the end of this phase when the core of the star runs out of nuclear fuel, gravity overtakes and the star collapses until another form of pressure appears to establish a new kind of equilibrium. The new possible configurations in this new stag
 e after the star collapses are \textit{are known as white dwarfs, neutron stars and black holes}. These are called ``dead stars''.\\

Statistics made with observed cases seems to indicate that low mass stars, $1\leq M \leq 4M_{\odot}$, evolve into white dwarfs, whereas more massive stars, $4\leq M \leq 10M_{\odot}$, towards NSs. Many of these latter cases are formed in supernova explosions. Supernova explosions associated with even more massive stars, $M\geq10M_{\odot}$, are assumed to end up in the formation of a stellar-mass BHs.\\

As we stated, the compact objects that are of our interest are NSs and BHs. In NSs, the pressure that provides the force opposed to gravity is the quantum mechanical pressure associated with the fact that neutrons are fermions and therefore only one particle is allowed to occupy one quantum state at a time. NSs are then held up against gravitational collapse by neutron degeneracy pressure and have masses in the range of $\sim 1.5 M_{\odot}$ to $3 M_{\odot}$ with a diameter that goes from 10 to 20 km. The corresponding escape velocity is about half the speed of light and hence the path of light is strongly curved by gravity. Theories support the concept that NSs should have a solid crust and a liquid core with properties of superfluidity, having almost no viscosity.\\

NSs are the last stable-star form. In the case of more compact objects the attractive force of gravity can not be balanced by any known physical force and the last collapse to a physical singularity is unavoidable. Even electromagnetic radiation cannot escape from this singularity since the escape velocity exceeds the speed of light, hence the origin of the name \textit{black holes}. The escape velocity from the surface of a star of mass $M$ and radius $r$ can be obtained through classical calculations: $v=(2GM/r)^{1/2}$. If we set $v$ as the speed of light ($c$) the radius of such a star would be $r=2GM/c^2$. This is the so-called \textit{Schwarzschild radius}, $r_{\rm Sch}$, of a black hole of mass $M$. For non-rotating BHs there is a last stable circular orbit around this point mass at $3r_{\rm Sch}$. At radii less than this value the particles start spiraling into the singularity, located at $r=0$.\\  

Setting the values of the constant in the expression of the Schwarzschild radius one obtains a very practical expression: $r_{\rm Sch}=3(M/M_{\odot})$ km. This straightforwardly implies that for the Sun, $r_{\rm Sch}=3$ km, which is negligible compared to the solar radius, $R_{\odot} \sim 7 \times 10^5$ km. Instead, in the case of a NS with $M \approx 3 M_{\odot}$, we find $r_{\rm Sch}\approx10$ km which, being of the order of the neutron star radius, means that general relativity becomes important in determining the stability of this case.\\

To distinguish whether a binary system harbors a neutron star or a black hole as the compact object is not a trivial task since the gravitational field near a NS is nearly as strong as near a BH. Differences in the X-rays emitted by infalling matter in the two cases are expected to be subtle. Though regular pulsations are a sure proof that an X-ray emitter is a neutron star\footnote{See e.g. Begelman \& Rees (1998) or Longair (1997), for a detailed explanation on the physics of pulsars.}, they are not always present. Many sources like Sco X-1, are suspected to hold a NS despite the fact that they show no sign of regular pulsations.\\

On the other hand there are severe differences between NSs and BHs that should imply certain differences in the radiation spectra involving one or the other. NSs have real surfaces whereas BHs have horizons, which represent the complete absence of a physical surface. The characteristic magnetic field present in NSs would also be expected to mark a distinction. Nonetheless, up to now, no one has succeeded in predicting the spectral differences only from theoretical considerations.\\     

There is however a decisive discriminant that can at least ensure the absence of a NS from some XRB systems: if the present compact object has a mass greater than $3M_{\odot}$, then one could assume the presence of a BH. Such a mass estimation can be obtained by measuring the orbital period and the parameters of the normal star in the binary system. In this way the X-ray source in the binary system Cygnus X-1 was inferred to have a mass of at least $6 M_{\odot}$, becoming in this way the first example of a black hole candidate.\\

By definition microquasars have relativistic jets. Bright radio detections associated with the production of a relativistic jet are by now a common property of accreting black holes whereas they are not so commonly associated with accreting neutron stars. Nonetheless the most relativistic flow observed so far in the Galaxy has been very recently discovered by Fender et al. (2004a) in the source Circinus X-1, whose compact object is a neutron star (Tennant et al. 1986). It was established that the apparent superluminal velocity is $\beta_{\rm app} > 15$. Knowing that for an intrinsic velocity $\beta$ at an angle $\theta$ to the line of sight, $\beta_{\rm app}$ has a maximum value of $\beta_{\rm app}=\Gamma \beta$ when $\beta = \cos \theta$ (see Appendix~\ref{superluminal}), then $\Gamma \beta > 15$, and it follows that the minimum velocity solution is $\Gamma > 15$, with $\beta > 0.998$. In order to have a comparison, let us mention MQs with outflowing components are estimated
  to have bulk Lorentz factors in the range of 2 $\leq \Gamma \leq$ 5; in the case of active galactic nuclei (AGN) typical Lorentz factors are $\Gamma \sim 10$.\\

An important comment is that the lower limit to $\Gamma$ may correspond to a perturbation (shock) propagating along the flow instead of the underlying jet. We should then consider the relation $\Gamma_{\rm bulk}= \Gamma_{\rm shock}/\sqrt{2}$, that is valid in the approximation of a 1D-relativistic shock propagating along a relativistic fluid, in the limit $\beta_{\rm shock} \rightarrow 1$. This would imply $\Gamma_{\rm bulk} \geq 10$.\\

It seems that Sco X-1 is another neutron star binary that may be generating ultrarelativistic jets (Fomalont et al. 2001). The discovery of ultrarelativistic flows in XRBs with a NS instead of a BH as the compact object, leads to the conclusion that the generation of such jets is not related to any property that would be exclusively of BHs, like the event horizon or the ergosphere. It rather
highlights the role of the accretion flow (which is a common feature of both NSs and BHs) as the main element that might lead to the formation of the jet.\\      

\subsection{The Accretion Disk}\label{accret-disk}

We will devote this section to the physical description of the accretion disk that is present in MQs as it can be inferred from X-ray observations. The accretion disk component is usually modeled by the so-called \textit{``Standard Disk Model''} in which the accreting gas forms a geometrically thin and optically thick disk, producing a quasi--blackbody spectrum due to thermal emission. The effective temperature of the accreting gas is in the range $10^5 - 10^7$ K, depending on the compact object mass and the accretion rate,  $T_{\rm eff} \propto M^{-1/4}\dot{M}^{1/4}$.\\

Accretion processes around compact objects imply the presence of rotating gas flows. The situation is therefore modeled by hydrodynamic equations of viscous differentially-rotating flows. The ``standard disk model'' is the name of the most famous solution to this set of equations. The model was developed by Shakura \& Sunyaev (1973) and Novikov \& Thorne (1973).\\

If a particle is in a circular orbit around a central gravitating body, it will stay in that orbit. If then energy and angular momentum are extracted from the particle, it will spiral slowly inwards. In the context of the standard disk model, the \textit{viscosity} will cause the gas to lose energy and will have the effect of transporting angular momentum outwards, allowing the accreting gas to spiral in towards the central mass. The amount of energy that can be extracted by such a process is equal to the binding energy of the innermost stable orbit. Around a neutron star, $\sim 10$\% of the rest mass can be extracted and up to $40$\% for orbits around a black hole (Pringle 1981). Viscosity also acts as a source of heat; some or all of this heat is radiated, leading to the observed spectrum. The accretion process can then be an efficient converter of rest mass to radiation.\\

It is known already from the 1920s (Jeffreys 1924) that the action of viscosity on an initial ring is to spread it out. Most of the mass moves inwards losing energy and angular momentum, but a tail of matter moves out to larger radii in order to conserve the angular momentum.\\

Let us develop this idea in a more quantitative way\footnote{Skipped steps in any of the calculations included in this section can be followed in Shapiro \& Teukolsky (1983) and Frank et al. (2002).}. A thin disk means that the matter lies very close to the plane, which in cylindrical coordinates ($R$, $\phi$, $z$) means $z=0$. We assume that the gas moves with angular velocity $\Omega$ in circles around the accreting compact object which has mass $M$ and radius $R_{\ast}$\footnote{For the case where the compact object is a BH, $R_{\ast}$ should be replaced by $R_{\rm in}$ in all the following calculations, with $R_{\rm in}$ the innermost stable orbit.}. When the angular velocity has the Keplerian value then,

\begin{equation}
	\Omega = \Omega_{\rm K}(R)= \left(\frac{GM}{R^3}\right)^{1/2},
\end{equation}
\\
\noindent which means that the circular velocity is,

\begin{equation}
	v_{\phi} = R \Omega_{\rm K}(R).
\end{equation}
\\
\noindent Since the gas is being accreted, its radial velocity, $v_{\rm R}(R,t)$, near the compact object, is negative. Another important magnitude is the disk surface density, $\Sigma(R,t)$, which is the mass per unit surface area of the disk, obtained by integrating the gas density, $\rho$, in the $z$-direction.\\

An annulus of the disk material lying between $R$ and $R+\Delta R$ has a total mass $2 \pi R \Delta R  \Sigma$ and total angular momentum $2 \pi R \Delta R  \Sigma R^2 \Omega$. Taking into account that the rate of change of both quantities is given by the net flow from the neighboring annuli and making the limit $\Delta R \rightarrow 0$, we get the \textit{mass conservation} equation:

\begin{equation}\label{eq:mass-cons}
	R\frac{\partial \Sigma}{\partial t}+\frac{\partial}{\partial R}(R\Sigma v_R)=0\,.
\end{equation}

\vspace{0.6cm}

\noindent To obtain the conservation equation for angular momentum the procedure is the same except that we have to include the transport due to the net effects of the viscous torques, $G(R,t)$:

\vspace{0.3cm}

\begin{equation}\label{eq:ang-mom}
	R\frac{\partial}{\partial t}(\Sigma R^2 \Omega)+\frac{\partial}{\partial R}(R \Sigma v_R R^2 \Omega)= \frac{1}{2\pi}\frac{\partial G}{\partial R}\;,
\end{equation}
\\
\noindent where ,
\begin{equation}
	G(R,t)= 2 \pi R \nu \Sigma R^2 \Omega^{'},
\end{equation}
\\
\noindent with $\Omega^{'} = {\rm d} \Omega / {\rm d} r$, and $\nu$, the kinematic viscosity.\\

Assuming $\partial \Omega / \partial t = 0$, which holds for orbits in a fixed gravitational potential, performing some algebraic combinations with all these formulas, we obtain:

\vspace{0.4cm}

\begin{equation}\label{eq:sigma}
	\frac{\partial\Sigma}{\partial t}= \frac{3}{R}\frac{\partial}{\partial R}\left\{R^{1/2}\frac{\partial}{\partial R}\left[\nu \Sigma R^{1/2}\right]\right\},
\end{equation}

\begin{figure}[!t] 
\centering 
\resizebox{11cm}{!}{\includegraphics{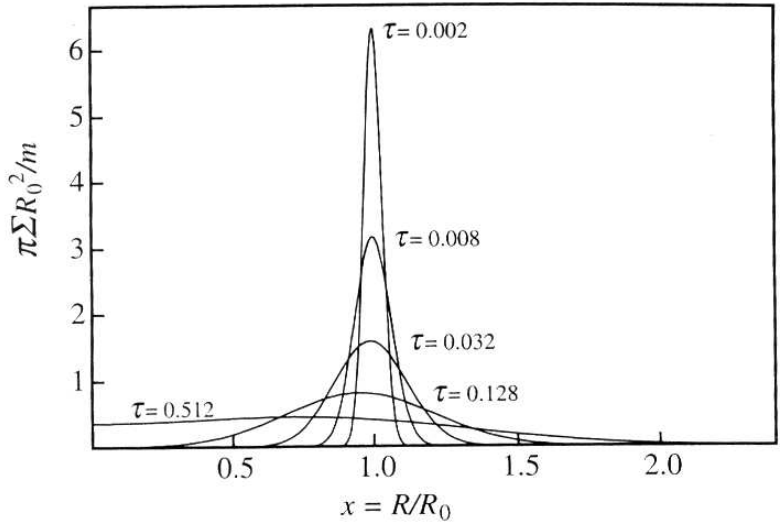}}
\caption{{A matter ring of mass $m$ placed in a Keplerian orbit at $R=R_0$ spreads out under the action of viscous torques. Surface density, $\Sigma$, vs $x=R/R_0$, the dimensionless radius, for different values of the dimensionless time, $\tau=12\nu t R^{-2}_0$, where $\nu$ is the viscosity.}}\label{sigma}
\end{figure}

\vspace{0.4cm}

\noindent that is the basic equation governing the time evolution of surface density in a Keplerian disk. In general it is a nonlinear diffusion equation of $\Sigma$, because $\nu$ may be a function of local conditions in the disk, like $\Sigma$, $R$ and $t$. Therefore, in order to solve (\ref{eq:sigma}), we need some prescription for $\nu$. Unfortunately, the nature of viscosity is still quite unclear, and as a consequence, its mathematical form is still uncertain and not unique. We gain some insight into the evolution of disks by choosing forms for $\nu$ that makes (\ref{eq:sigma}) solvable, which is just a mathematical help to show how the disk's dynamics works, rather than a physically based prediction of the theory. For instance, assuming $\nu$ = constant, (\ref{eq:sigma}) can be solved by separation of variables as:
\\

\begin{equation}
	\Sigma(x,\tau)=\frac{m}{\pi R^2_0}\;\tau^{-1}x^{-1/4}exp\left\{-\frac{(1+x^2)}{\tau}\right\}I_{1/4}(2x/\tau),
\end{equation}
\\

\noindent where $I_{1/4}(2x/\tau)$ is a modified Bessel function. Taking as the initial matter distribution a ring of mass $m$ at $R=R_0$, $x=R/R_0$ and $\tau=12\nu t R^{-2}_0$ are dimensionless radius and time variables. Fig~\ref{sigma} shows $\Sigma(x,\tau)$ as a function of $x$ for different values of $\tau$. We can see in the figure that viscosity has the effect of spreading the original ring in radius.\\

The asymptotic behavior of $I_{1/4}(2x/\tau)$,

\begin{eqnarray}
	I_{1/4}(2x/\tau) & \propto & \left(\frac{2x}{\tau}\right)^{-1/2} e^{(2x/\tau)},\; \frac{2x}{\tau} \gg 1\nonumber\\
	          & \propto & \left(\frac{2x}{\tau}\right)^{1/4}, \; \frac{2x}{\tau} \ll 1
\end{eqnarray}

\noindent implies that 

\begin{equation}
	v_R \sim \frac{3 \nu}{R_0}\left\{\frac{1}{4x}+\frac{2x}{\tau}-\frac{2}{\tau}\right\}>0 \qquad \textrm{for 2$x\gg\tau$},
\end{equation}
\\
\noindent and

\begin{equation}
	v_R \sim \frac{-3 \nu}{R_0}\left\{\frac{1}{2x}-\frac{2x}{\tau}\right\}<0 \qquad \textrm{for 2$x\ll\tau$}.
\end{equation}

\vspace{0.8cm}

\noindent Thus, we confirm that the outer parts of the matter distribution ($2x\gg\tau$) move outwards, taking away the angular momentum of the inner parts, which move inwards towards the accreting compact object. 

\vspace{0.7cm}
\begin{center}
$\clubsuit$
\end{center}
\vspace{0.7cm} 
 
In the steady state\footnote{We impose the steady state condition by setting $\partial/\partial t = 0$ in (\ref{eq:mass-cons}) and (\ref{eq:ang-mom}), obtaining in this way (\ref{eq:steady-mass}) and (\ref{eq:steady-angmom}). The steady assumption will be valid whenever the external conditions change on timescales longer than the time $t_{\rm visc}\sim \frac{R^2}{\nu}$, in which changes in the radial structure of the disk are registered, i.e. the timescale of the spreading because of  viscosity effects.} the disk structure is determined by solving simultaneously four conservation equations: conservation of mass, angular momentum, energy, and vertical momentum (where an expression for the pressure is needed). In addition a prescription for $\nu$ must be specified as well as a law describing the vertical energy transport from the center to the surface (where an expression for the opacity is needed). The set of equations is the following:

\begin{enumerate}

	\item {\scshape{rest-mass conservation}}
		
		
	\begin{equation}\label{eq:steady-mass}
	 \dot{M} = 2 \pi R \Sigma v_R = {\rm constant}.
	\end{equation}
	
  \vspace{0.2cm}
	
	\item {\scshape{angular momentum conservation}}
 
  \vspace{0.2cm}

	\begin{equation}\label{eq:steady-angmom}
		\nu \Sigma = \frac{\dot{M}}{3\pi}\left[1-\left(\frac{R_{\ast}}{R}\right)^{1/2}\right].
	\end{equation}
	
	\vspace{0.4cm}
	
  \item {\scshape{energy conservation}}
  
  \vspace{0.2cm}
  
  The kinematic viscosity, $\nu$, generates dissipation in the disk at a rate $D(R)$ per unit area per unit time, such that:
  
  \begin{eqnarray}\label{eq:dissipation}
	D(R) & = & \frac{1}{2}\nu \Sigma (R \Omega^{'})^2 \nonumber\\
	     & = & \frac{3GM\dot{M}}{8\pi R^3}\left[1-\left(\frac{R_{\ast}}{R}\right)^{1/2}\right].
\end{eqnarray}

\vspace{0.5cm}

An interesting issue is that expression (\ref{eq:dissipation}) is independent of the major uncertainty of the disk's theory, the viscosity. This is at the expense of the assumption that the chosen viscosity prescription can be adjusted to provide the steady mass flux $\dot{M}$.\\

Using (\ref{eq:dissipation}) we obtain the total disk luminosity:

\begin{eqnarray}\label{eq:lum-disk}
	L_{\rm disk}& = & \int^{\infty}_{R_{\ast}}D(R)\;2\pi R \ dR \nonumber\\
	     & = & \frac{1}{2}\frac{GM\dot{M}}{R_{\ast}}.
\end{eqnarray}

\vspace{0.3cm}

This is only one half of the total available accretion energy, since the total potential drop from infinity to $R_{\ast}$ is $GM/R_{\ast}$. The point is that the matter just outside the boundary layer still retains as kinetic energy one half of the potential energy it has lost in spiraling in. Hence, the rest of the accretion luminosity is emitted in the boundary layer. This has a practical implication: the study of the details of emission from the inner edge of an accretion disk can be just as important as studying the emission from the disk itself.

\item {\scshape{vertical momentum conservation}}
  
\vspace{0.2cm}

  Since there is no net motion of the gas in the vertical direction, momentum conservation along the $z$-direction reduces to a hydrostatic equilibrium condition. Equating the component of the gravitational force of the compact object along the $z$-direction to the vertical pressure gradient in the disk,
  
  \vspace{0.2cm}
  
  \begin{equation}
		\frac{1}{\rho}\frac{dP}{dz} = -\frac{GM}{R^2}\frac{z}{R} \qquad (z \ll R).
	\end{equation}
	
	\vspace{0.4cm}
	
Replacing the differentials by finite differences, setting $\Delta P \approx P$ and $z=H$, yields:

  \begin{equation}
		H=c_{\rm s} \frac{R^{3/2}}{(GM)^{1/2}},
	\end{equation}
	
	\vspace{0.4cm}
	
where $c_s$ is the sound speed, ($c_{\rm s}^2 = P/\rho$), $P$, pressure, and $\rho$, density of the disk.

\item {\scshape{pressure}}

\vspace{0.2cm}

The total pressure of the disk material is the sum of gas and radiation pressures:

 \begin{equation}
		P = \frac{\rho k T_{\rm c}}{\mu m_{\rm p}}+\frac{4 \sigma}{3c}T_{\rm c}^4,
 \end{equation}
	
\vspace{0.4cm}
	
where $k$ is the Boltzmann constant, $\mu$ is the mean molecular weight, $m_{\rm p}$ the proton mass, $\sigma$ the Stefan-Boltzmann constant, and $T_{\rm c}$ the central temperature: $T(R,z)=T(R,0)$.	

\item {\scshape{radiative transport}}

\vspace{0.2cm}

The temperature $T_{\rm c}$ can be given by an energy equation relating the energy flux in the vertical direction to the rate of generation of thermal energy by viscous dissipation. The heat generated internally by viscosity dissipation is transported vertically\footnote{Since we are working under the geometrically thin disk condition ($H\ll R$), the disk medium is then essentially ``plane-parallel'', so that the temperature gradient is effectively in the $z$-direction, and we need only to consider the energy transport perpendicular to the disk.} through the disk before being radiated at the surface. The vertical energy transport mechanism can be either radiative or convective. We will here assume the radiative one (see Frank et al. 2002, Section 5.8, for a justification of this choice).\\

We then equate the heat input per unit area, $D(R)$, with the heat loss per unit area given by the radiative transport,
\vspace{0.4cm}
\begin{eqnarray}\label{eq:transport}
	\frac{4\sigma}{3\tau}T_{\rm c}^4& = & D(R) \nonumber\\
	     & = & \frac{3GM\dot{M}}{8\pi R^3}\left[1-\left(\frac{R_{\ast}}{R}\right)^{1/2}\right],
\end{eqnarray}
   
\vspace{0.4cm}   
   
where $\tau$ is the \textit{optical depth} that is defined through

\begin{equation}
	\tau=\kappa_{\rm R} \rho H =\kappa_{\rm R}\Sigma,
\end{equation}

\vspace{0.4cm}

with $\kappa_{\rm R}=\kappa_{\rm R}(\rho,T_{\rm c})$, the total Rosseland mean opacity. The opacity in the disk is related with the photon absorption (mainly due to ``free-free'' transitions) and the photon scattering (Thomson scattering in this case).\\ 

The optically thick condition of the standard disk model implies that $\tau\gg1$. If $\tau$ becomes $\lesssim1$ the radiation escape directly; the radiative transport is  then not required and equation (\ref{eq:transport}) is no longer valid. 

\item {\scshape{viscosity}}

\vspace{0.2cm}

The standard model uses for the viscosity the so-called \textit{$\alpha$-prescription} of Shakura \& Sunyaev (1973). Suppose that the dominant process for redistributing angular momentum is a \textit{turbulent} viscosity. The effective kinematic viscosity of a turbulent process is given by $\nu \sim l v$, where $l$ the size and $v$ the velocity of the largest eddies in the flow. In an accretion disk, we can assume that the scale of the eddies is less than the disk thickness, $H$, and the turbulence is subsonic\footnote{In the case that the turbulence would be supersonic, the turbulent motions would probably be thermalized by shocks.}.  Consequently,
\begin{equation}\label{eq:alfa}
	\nu = \alpha \; c_{\rm s} \; H,
\end{equation}

\vspace{0.4cm}

with $\alpha \leq 1$. It is important to realize that (\ref{eq:alfa}) is a parametrization that helps to isolate in $\alpha$ the present lack of knowledge on the viscosity. We have at least obtained a constraint over the new parameter, $\alpha \leq 1$.
Though just a parametrization, the $\alpha$-prescription has been very useful and has promoted a semi-empirical approach to the viscosity issue, which looks for estimating the magnitude of $\alpha$ by comparison of theory and observation.\\

An alternative to the turbulent viscosity process and in fact one of the more plausible, is the \textit{magnetic} viscosity (Lynden-Bell 1969, Shakura \& Sunyaev 1973), where the transfer of angular momentum occurs through magnetic stresses due to a field anchored in the disk. This kind of viscosity can also be contained within the $\alpha$-prescription, with $\alpha\sim v_{\rm A}^2/c_{\rm s}^2$, where $v_{\rm A}$ is the Alfv\'en speed in the disk.\\

A more radical approach to the idea of using magnetic fields to transfer away angular momentum has been proposed by Blandford (1976) and by Lovelace (1976). The energy and angular momentum of the disk can be carried away in the form of a magnetized relativistic wind (See also Blandford \& Znajek 1977; Blandford \& Payne 1982).

\end{enumerate}

Solving the above listed set of equations we obtain the magnitudes $\rho$, $\Sigma$, $H$, $v_R$, $P$, $T_{\rm c}$, $\tau$, and $\nu$ as functions of $\dot{M}$, $M$, $R$ and $\alpha$. It can be shown that for fixed values of $M$ and $\dot{M}$, the disk can be divided into three distinct regions, depending on $R$:

\begin{itemize}
	\item An \textit{outer} region, at large $R$, in which gas pressure dominates over radiation pressure and in which the opacity is mainly due to free-free absorption.
	\item A \textit{middle} region in which gas pressure still dominates over radiation pressure but the opacity is mainly due to electron scattering.
	\item An \textit{inner} region, at very small $R$, in which radiation pressure dominates over gas pressure and again the opacity is dominated by electron scattering. 
\end{itemize}

\vspace{0.4cm}
\begin{center}
$\clubsuit$
\end{center}
\vspace{0.7cm}
 
Let us now concentrate on the emitted disk spectrum. A very important consequence of the assumption that the disk is optically thick is that each element of the disk radiates roughly as a blackbody with a temperature $T(R)$. This temperature is given by equating the dissipation rate $D(R)$ per unit area to the blackbody flux:

\begin{equation}
\sigma T^4(R)= D(R). 
\end{equation}

\vspace{0.4cm}

\noindent Using (\ref{eq:dissipation}) for $D(R)$ we obtain,
\vspace{0.2cm}
\begin{equation}
T(R)= \left\{\frac{3GM\dot{M}}{8\pi R^3 \sigma}\left[1-\left(\frac{R_{\ast}}{R}\right)^{1/2}\right]\right\}^{1/4}. 
\end{equation}

\vspace{0.5cm}

\noindent This temperature is analogous to the effective temperature of a star. We  approximate the emitted spectrum from each element of area of the disk as: 

\vspace{0.5cm}
\begin{equation}
I_{\nu}=B_{\nu}[T(R)]=\frac{2h\nu^3}{c^2(e^{h\nu/kT(R)}-1)}(\rm erg\; s^{-1}\; cm^{-2}\;Hz^{-1}\;sr^{-1}).
\end{equation}

\vspace{0.7cm}

\noindent For an observer at a distance $D$ whose line of sight makes an angle $\theta$ to the normal to the disk plane ($\theta$ is the binary inclination), the flux at frequency $\nu$ from the disk is
\vspace{0.4cm}
\begin{equation}
S_{\nu}=\frac{2\pi\cos\theta}{D^2}\int^{R_{\rm out}}_{R_{\ast}}I_{\nu}\;R\;dR,
\end{equation}

\vspace{0.7cm}

\noindent where $R_{\rm out}$ is the outer radius of the disk. With the blackbody assumption, we get

\vspace{0.4cm}

\begin{equation}\label{eq:diskspectrum}
S_{\nu}=\frac{4\pi h \nu^3\cos\theta}{c^2 D^2}\int^{R_{\rm out}}_{R_{\ast}}\frac{R\;dR}{e^{h\nu/kT(R)}-1}.
\end{equation}

\vspace{0.4cm}

\begin{figure}[!t] 
\centering 
\resizebox{10cm}{!}{\includegraphics{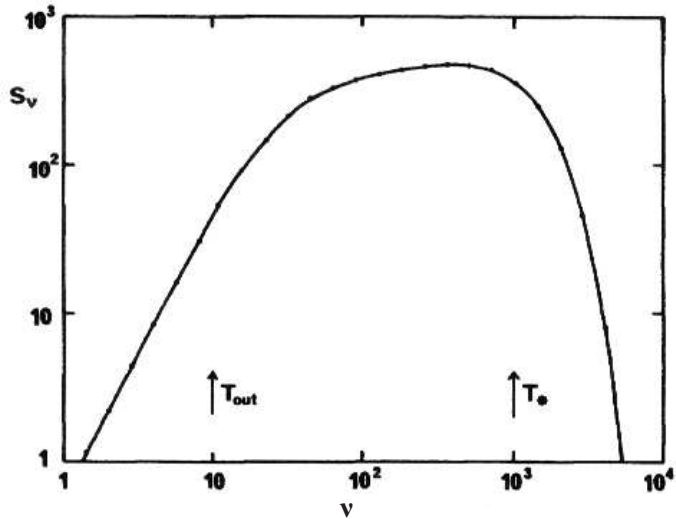}}
\caption{{The integrated spectrum of a steady accretion disk that radiates a local blackbody spectrum at each point. The units are arbitrary, but the frequencies corresponding to $T_{\rm out}=T(R_{\rm out})$ and $T_{\ast}$ are marked (from Pringle 1981).}}\label{disco}
\end{figure}
 
\noindent Notice that the expression we obtain for $S_{\nu}$ is independent of the disk viscosity. This is a consequence of the assumptions of steady and optically thick disk. This implies that if there is any important difference between the spectrum obtained with (\ref{eq:diskspectrum}) and the observed ones, it will not be due to our ignorance of the viscosity but to the assumptions made.\\

The integrated spectrum given by (\ref{eq:diskspectrum}) is shown in Fig.~\ref{disco} and its functional form is the following:
\vspace{0.4cm}
\begin{displaymath}
\left\{ \begin{array}{ll} \nu \ll  kT(R_{\rm out})/h \hspace{1cm}& S_{\nu} \propto \nu^2 \\
kT(R_{\rm out})/h\ll \nu \ll kT_{\ast}/h \hspace{1cm} & S_{\nu}\propto \nu^{1/3}\\
\nu\gg kT_{\ast}/h \hspace{1cm} & \textrm{the spectrum drops exponentially}
\end{array} \right.
\end{displaymath}

\vspace{0.4cm}
  
The integrated spectrum is therefore a \textit{stretched-out} blackbody. The flat part corresponding to $S_{\nu}\propto \nu^{1/3}$ characterizes the disk spectrum: unless $T_{\rm out}=T(R_{\rm out})$ is significantly smaller than $T_{\ast}$ this part of the curve may be quite short and the spectrum is not very different from a blackbody.\\

Finally, some remarks with respect to how the nature of the compact accreting object may affect the disk emission. If it is a star with a strong magnetic field, we already mentioned in previous section that the disk flow can be disrupted at some distance from the compact object itself. In this case most of the accretion luminosity comes from the matter when it actually strikes the compact object surface and not from the disk. If the accretion disk is around a black hole, the motion near the inner regions deviates from the simple Keplerian law. It is assumed that the disk extends down to the innermost stable circular orbit $R_{\rm in}$. Within the radius  $R_{\rm in}$ the gas spirals into the hole without radiating (see Stoeger 1980 for a justification). The energy released by the accreting matter corresponds then to the binding energy at the radius $R_{\rm in}$. For a Schwarzschild black hole this corresponds to about the $6$\% of the rest mass energy and, for a maximal Kerr 
 hole, to about $42$\%.\\ 

We end this section mentioning that the major success of the standard disk model is that the surface disk temperature it predicts approximately corresponds to the observed UV emission in AGNs and soft X-ray emission in galactic black holes.\\

\subsection{\label{corona}The Corona}

A strong motivation for much of the work on nonthermal and optically thin emission from accretion disks was provided by the observations of the X-ray binary source Cyg X-1. There is reasonable evidence that the X-ray source is powered by disk accretion onto a black hole, but the observed X-ray spectrum is too hard to be produced by the standard, optically thick disk model that we just described in the previous section. With that model a maximum disk surface temperature is $T_{\ast} \sim 10^5$ K for AGNs ($M \sim 10^7 - 10^9 M_{\odot}$) and $T_{\ast} \sim 10^7$ K for stellar-mass black holes ($M \sim 10 M_{\odot}$). A hard X-ray component would, however, correspond to a temperature $\sim 10^9$ K. The standard disk model may yield such high temperatures only if the accretion rate approaches the critical value, $\dot{M}_{\rm crit}$, that implies a high radiative efficiency of accretion and the situation then corresponds to the critical Eddington luminosity. Near-critical disks c
 an strongly deviate from the blackbody state provided $\alpha$ is large (Shakura \& Sunyaev 1973). An overheating then occurs in the inner region of the disk where the inflow time-scale is shorter than the time-scale for relaxation to thermodynamic equilibrium.\\ 

The region supposed to be responsible for emission that gives rise to the observed hard X-rays is named ``Corona'' (see Fig.~\ref{components}). Since the observations of Cyg X-1, different models appeared, based on different assumptions and physics, but having the same aim to interpret and fit those observed spectra that could not  be fully explained with the standard disk model. We will focus on two of them, that are still being developed and tested: the \textit{``Disk - Corona''} model and the \textit{``Advection - Dominated Accretion Flow''} model or ADAF.\\ 
 
The ADAF model originates as an alternative solution to the same hydrodynamic equations of viscous differentially-rotating flows that were solved in the standard disk model. Instead the Disk - Corona model implies to add a new component to the standard model.

\vspace{0.6cm}
\begin{center}
$\clubsuit$
\end{center}
\vspace{0.9cm}

The ADAF model is a solution of the hydrodynamic equations of viscous differen\-tially-rotating flows (listed in Section~\ref{accret-disk}) that works with low, sub-Eddington, accretion rates (Ichimaru 1977; Rees et al. 1982; Narayan \& Yi 1994, 1995a, 1995b; Abramowicz et al. 1995)\footnote{ADAF solutions are also possible at very high accretion rates (see e.g. the review by Reynolds \& Nowak 2003).}. In this solution, the accreting gas has a very low density which implies on one hand that it is optically thin and on the other that it is unable to cool efficiently within the accretion time. The viscous energy is therefore stored in the gas as thermal energy instead of being radiated, and is advected onto the central compact object. Since most of the viscously generated energy in the ADAF model is stored in the gas as internal energy, rather than being radiated, the gas temperature is quite high. This causes the gas to swell, then $H \sim R$ and therefore, geometrically, ADAF
 s are quasi-spherical. In this model the gas adopts a two-temperature configuration that will be explained later on.\\

There is a compact way to express the just stated definition:
\begin{eqnarray}
	q^{\rm adv}=q^+ - q^-,
\end{eqnarray}
 
\vspace{0.4cm} 
 
\noindent where $q^{\rm adv}$ represents the advective transport of energy, $q^+$ is the energy generated by viscosity per unit volume, and $q^-$ is the radiative cooling per unit volume.\\

Depending on the relative magnitudes of the terms in this equation, three regimes of accretion may be identified:

\begin{itemize}
	\item $q^+ \cong q^- \gg q^{\rm adv}$: this corresponds to a cooling-dominated flow where all the energy released by viscous stress is radiated; the amount of energy advected is negligible. The thin disk solution corresponds to this regime.

	\item $q^{\rm adv} \cong q^+ \gg q^-$: this corresponds to an ADAF where almost all the viscous energy is stored in the gas and is deposited into the black hole. The amount of cooling is negligible compared with the heating. For a given $\dot{M}$, an ADAF is much less luminous than a cooling-dominated flow.

	\item $- q^{\rm adv} \cong q^- \gg q^+$: this corresponds to a flow where energy generation is negligible, but the entropy of the inflowing gas is converted to radiation. Examples are Bondi accretion, Kelvin-Helmholtz contraction during the formation of a star, and cooling flows in galaxy clusters.
\end{itemize}

It is important to point out that models based on the two-temperature ADAF solution make certain important assumptions. The validity of these assumptions is not yet completely proved and is currently under testing.

\begin{enumerate}

	\item \textit{Equipartition Magnetic Fields}: 
It is assumed that magnetic fields contribute a constant fraction $(1 - \beta)$ of the total pressure:
\begin{eqnarray}
	p_{\rm m} = \frac{B^2}{24 \pi}=(1 - \beta)\rho {c_{\rm s}}^2,
\end{eqnarray}

where $p_m$ is the magnetic pressure due to an isotropically tangled magnetic field. The assumption of a constant $\beta$ is fairly innocuous since, in general, we expect equipartition magnetic fields in most astrophysical plasmas. In particular, Balbus \& Hawley (1991) have shown that differentially rotating disks with weak magnetic fields develop a strong linear MHD instability which exponentially increases the field strength to near equipartition values. ADAF models assume $\beta= 0.5$, which means equipartition between gas and tangled magnetic pressure.
	
	\item \textit{Thermal Coupling Between Ions and Electrons}:
ADAF models assume that ions and electrons interact only through Coulomb collisions and that there is no non-thermal coupling between the two species. In this case the plasma is two-temperature, with the ions much hotter than the electrons.
	
	\item \textit{Preferential Heating of Ions}:
The two-temperature ADAF model assumes that most of the turbulent viscous energy
goes into the ions (Shapiro et al. 1976; Ichimaru 1977; Rees et al. 1982; Narayan \& Yi 1995b), and that only a small fraction $\delta \ll 1$ goes to the electrons. The parameter $\delta$ is generally set to $\sim 10^{-3} \sim m_e/m_p$, but none of the results depend critically on the actual value of
$\delta$, as long as it is less than a few percent. There have been several theoretical investigations which consider the question of particle heating in ADAFs like Bisnovatyi--Kogan \& Lovelace (1997), Blackman (1999), Gruzinov (1998), and Quataert (1998).
	
	\item \textit{$\alpha$ Viscosity}:
The viscosity parameter $\alpha$ of Shakura \& Sunyaev (1973) is used to describe angular momentum transport; $\alpha$ is assumed to be constant, independent of radius. Some authors have proposed that it may vary as a function of ($H/R$). Since ADAFs have $H \sim R$, no radial dependence is expected, and a constant $\alpha$ appears to be a particularly good assumption (Narayan 1996a). 

\end{enumerate}

The optically thin ADAF has an important constraint: it exists as a solution only for $\dot{m}$\footnote{$\displaystyle \dot{m}=\frac{\dot{M}}{\dot{M}_{\rm Edd}}$, $\displaystyle \dot{M}_{\rm Edd}=\frac{L_{\rm Edd}}{\eta_{\rm eff} c^2}$, with $\eta_{\rm eff}$, the efficiency of converting matter to radiation.} less than a critical value $\dot{m}_{\rm crit}$ (Ichimaru 1977; Rees et al. 1982; Narayan \& Yi 1995b; Abramowicz et al. 1995), such that for $\dot{m}<\dot{m}_{\rm crit}$, we have $q^+ > q^-$ and a consistent ADAF solution, whereas for $\dot{m}>\dot{m}_{\rm crit}$ no ADAF is possible. At low densities Coulomb coupling between protons and electrons is very weak and the amount of viscous energy that is transferred to the electrons is very small. Coulomb coupling therefore restricts the amount of energy that can be lost by radiation. With increasing $\dot{m}$, Coulomb coupling becomes more efficient, and at a critical density the coupling is so efficient that a large fract
 ion of the viscous energy is transferred to the electrons and is radiated. Above this accretion rate, the flow ceases to be an ADAF and becomes a standard cooling-dominated thin disk. The critical accretion rate can be estimated by determining the $\dot{m}$ at which the viscous heating, $q^+$, equals the rate of energy transfer from the ions to the electrons, $q^{ie}$. Observations suggest that the two-temperature ADAF solution exists up to $\dot{m}_{\rm crit} \sim 0.05 - 0.1$. This suggests that $\alpha \sim 0.2 - 0.3$ in ADAFs.\\

It is interesting to remember, at this point, that in the case of the standard model there is also a critical value, $\dot{m}_{\rm crit}$, that implies a deviation from the thin disk feature to an emission model like for instance the ADAF model. The confluence of the two possible relations between $\dot{m}$ and $\dot{m}_{\rm crit}$ gives:

\vspace{0.4cm}

\begin{displaymath}
\left\{ \begin{array}{ll}\textrm{Thin disk to ADAF} & \dot{m} < \dot{m}_{crit}\\
\textrm{ADAF to thin disk} & \dot{m} > \dot{m}_{crit}.
\end{array} \right.
\end{displaymath}

\vspace{0.4cm}

The spectrum from an ADAF around a black hole ranges from radio frequencies $\sim 10^9$ Hz to gamma-ray frequencies $\geq 10^{23}$ Hz, and can be divided into two parts based on the emitting particles: 

\begin{enumerate}
	\item The radio to hard X-ray radiation is produced by electrons via synchrotron, Bremsstrahlung and IC processes (Mahadevan 1997). Comptonization process is over the locally produced synchrotron and Bremsstrah\-lung radiation and dominates the spectrum.
	\item The gamma-ray radiation results from the decay of neutral pions created in proton-proton collisions (Mahadevan et al. 1997)\footnote{This remains speculative since gamma-ray emission has never been observed from an accretion disk. As we will see most models for gamma-ray production in MQs invoke processes occurring in the relativistic jets.}.
\end{enumerate}

\begin{figure}[!t]
\begin{center}
\resizebox{12cm}{!}{\includegraphics{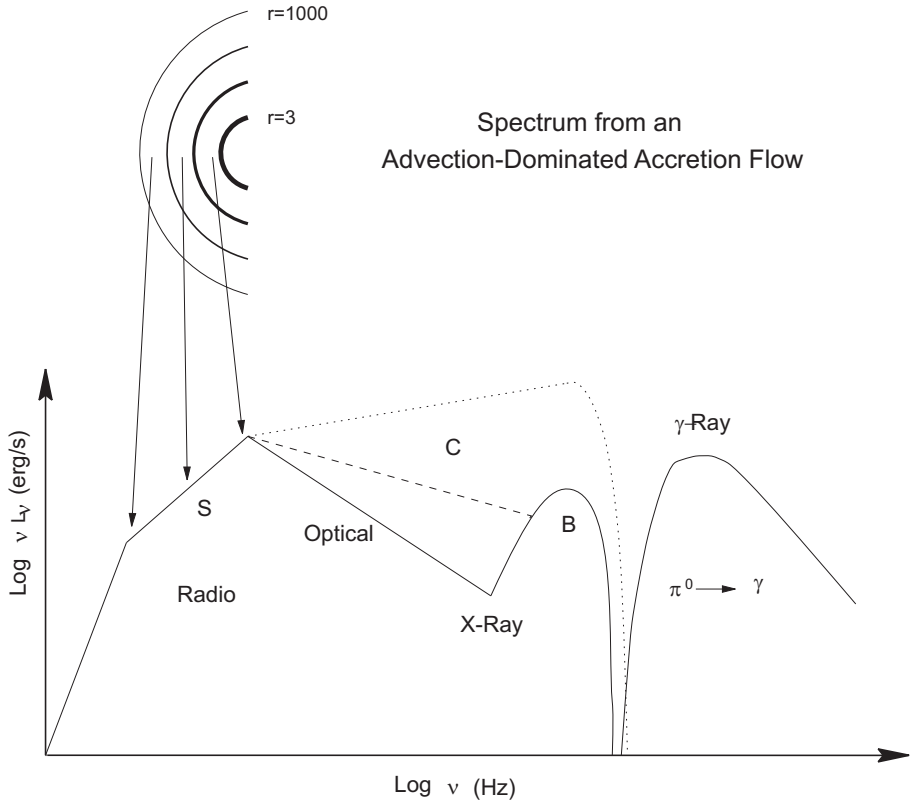}}
\caption{{Schematic spectrum of an ADAF around a black hole. S, C, and B refer to electron emission by synchrotron radiation, IC scattering, and Bremsstrahlung, respectively. The solid line corresponds to a low $\dot{m}$, the dashed line to an intermediate $\dot{m}$, and the dotted line to a high $\dot{m} \sim  \dot{m}_{crit}$. The gamma-ray spectrum is due to the decay of neutral pions created in proton-proton collisions (from Narayan, Mahadevan \& Quataert 1998).}}\label{adaf-emission}
\end{center}
\end{figure}

Fig.~\ref{adaf-emission} sketches the spectrum obtained with the ADAF model, showing the places where the emission is generated and the radiative processes that causes it for each frequency.\\

The low energy end of the spectrum, is due to \textit{synchrotron} cooling by semi-relativistic thermal electrons (Mahadevan et al. 1996). The synchrotron emission is highly self-absorbed and is very sensitive to the electron temperature (Mahadevan 1997). The emission at the highest (peak) frequency comes from near the black hole, while that at lower frequencies comes from further out. The peak frequency varies with the mass of the black hole and the accretion rate, roughly as $\nu_{\rm peak}^S\propto m^{-1/2}\dot{m}^{1/2}$ (Mahadevan 1997).\\

The soft synchrotron photons \textit{inverse Compton} scatter off the hot electrons in the ADAF and produce harder radiation extending up to about the electron temperature $\sim$ 100 keV ($h\nu_{\rm max}^C \approx kT_{\rm e}$). The relative importance of this process depends on the mass accretion rate. At high $\dot{m}$, because of the increased optical depth, the IC component dominates the spectrum.\\

As $\dot{m}$ decreases, Comptonization becomes less efficient and the IC component of the spectrum becomes softer and less important. At low $\dot{m}$, the X-ray spectrum is dominated by \textit{Bremsstrahlung} emission, which again cuts off at the electron temperature ($h\nu_{\rm max}^B \approx kT_{\rm e}$).\\

Gamma-ray emission from an ADAF is via the decay of neutral pions produced in \textit{proton-proton collisions}. The results depend sensitively on the energy spectrum of the protons. If the protons have a thermal distribution, the gamma-ray spectrum is sharply peaked at $\sim$ 70 MeV, and the luminosity is not very high. If the protons have a power-law distribution, the gamma-ray spectrum is a power-law extending to very high energies, and the luminosity is much higher. The photon index of the spectrum is equal to the power-law index of the proton distribution function.\\

Another outstanding feature is that ADAFs are much less luminous than thin disks at low values of $\dot{m}$. This is because most of the energy in an ADAF is advected, rather than radiated, leading to a low radiative efficiency. In fact, the luminosity of an ADAF scales roughly as $\sim \dot{m}^2$. In a thin disk, on the other hand, the luminosity scales as $\dot{m}$ (Narayan, Mahadevan \& Quataert 1998).\\

In determining the radiation processes in ADAFs, the electrons are assumed to be
thermal, while the protons could be thermal or non-thermal. It is clear that the spectrum will depend significantly on the energy distribution of the particles. Mahadevan \& Quataert (1997) considered two possible thermalization processes in ADAFs: (1) Coulomb collisions and (2) synchrotron self-absorption. In the case of the protons they found that, for all accretion rates of interest, neither Coulomb collisions nor synchrotron self-absorption lead to any significant thermalization. The proton distribution function is therefore determined principally by the characteristics of the viscous heating mechanism, and could therefore be thermal or non-thermal.\\

ADAF models have been applied to a number of accreting black hole systems. They
give a satisfying description of the spectral characteristics of several quiescent black hole binaries (Narayan et al. 1996; Hameury et al. 1997) and low luminosity galactic nuclei (Narayan et al. 1998; Manmoto et al. 1997) which are known to experience low efficiency accretion. ADAF models have also been applied successfully to more luminous systems which have higher radiative efficiencies (Esin et al. 1997, 1998).

\vspace*{0.5cm}
\begin{center}
$\clubsuit$
\end{center}
\vspace*{0.9cm}

As we already mentioned, the Disk - Corona model is not just another solution of the set of  equations that the standard disk or the ADAF fulfill; an extra component is added to the already defined thin disk, ``the corona''.  Likely a low density corona is heated by reconnecting magnetic loops emerging from the disk (Galeev et al. 1979). This implies that the corona is coupled to the disk by the magnetic field (for alternative models, where the corona accretes from above the disk, see, e.g., Esin et al. 1998). According to the model for the corona formation by Galeev et al. (1979), a seed magnetic field is exponentially amplified in the disk due to a combination of the differential Keplerian rotation and the turbulent convective motions. The amplification time-scale at a radius $r$ is given by $t_{\rm G} \sim r/3v_{\rm c}$ where $v_{\rm c}$ is a convective velocity. They showed that inside luminous disks the field is not able to dissipate at the rate of amplification. Then bu
 oyant magnetic loops elevate to the corona where the Alfv\'enic velocity is high and the magnetic field may dissipate quickly. The magnetic reconnection-heated corona process has therefore two phases (Liu et al. 2002a): a mass evaporation happens at the bottom of the magnetic flux tube and consequently builds the corona up to certain density; this is followed by Compton scattering cooling that radiates away the  previous magnetic heating. Liu et al. (2002a) showed the dependency of this forming and heating mechanism of the corona with the accretion rate. They concluded that for low luminosities ($L<0.1L_{\rm Edd}$) most of the accretion energy is transferred to the corona.\\

An alternative mechanism is proposed by Liu et al. (2002b) as a frictionally heated corona. The existence of the corona is assumed in their model. Heat released by friction in the corona flows down into lower and cooler parts. If there the density is low, cool matter is heated up and joins the coronal gas. Liu et al. (2002b) found that a larger value of the viscosity implies an increase of  the evaporation efficiency. The evaporated gas loses angular momentum because of the friction and drifts towards the central object. This is compensated by a steady mass evaporation flow from the underlying disk. Energy from the gravitational potential is released by friction in the form of heat in the corona.\\

Before going deeper into corona models it is convenient to mention some relevant spectral features that were the primary motivation for their formulation. We will also introduce some basic but useful notions of plasma physics in this context.\\

Broad-band X/$\gamma$-ray spectra of galactic black holes can be explained in terms of successive Compton scatterings of soft photons - \textit{Comptonization}- in a hot electron cloud. Besides, both galactic black holes and Seyfert galaxies (a specific case of AGN) show a hardening of the spectra at $\sim$ 10 keV, which is attributed to Compton reflection (combined effect of photo-electric absorption and Compton down-scattering) of hard radiation from a cold material (White, Lightman \& Zdziarski 1988; George \& Fabian 1991). This implies the presence of that close cold material. Hard radiation, reprocessed in the cold matter, can form a significant fraction of the soft seed photons for Comptonization. The energy balance of the cold and hot phases determines their temperatures and the shape of the emerging spectrum (Poutanen \& Svensson 1996).\\

The situation becomes more complex when a notable fraction of the total luminosity escapes at energies above $\sim$ 500 keV. Then hard photons can produce $e^+ e^-$--pairs which will be added to the background plasma. Electrons (and positrons) Comptonize soft photons up to gamma-rays and produce even more pairs. Thus, the radiation field, in this case, has an influence on the optical depth of the plasma, which in turn produces this radiation. This makes the problem non-linear.\\

Another complication appears when the energy distribution of particles starts to deviate from a Maxwellian. In the so-called non-thermal plasma models, relativistic electrons are injected to the soft radiation field.\\

Let us first assume a thermal pair plasma. We consider properties of an electron-positron plasma cloud in energy and pair equilibria. There are four parameters that describe the properties of hot thermal plasmas: 

\begin{enumerate}
	\item $l_{\rm h} \equiv L_{\rm h} \sigma_T /(m_ec^3r_{\rm c})$,
the hard compactness which is the dimensionless cloud heating rate. $L_{\rm h}$, is the heating rate of the hot cloud, $\sigma_T$, the Thomson scattering cross-section and $r_{\rm c}$, the cloud size.   
	\item The soft photon compactness, $l_{\rm s}$, which represents the cold disk luminosity that enters the hot cloud (corona). $L_{\rm s}$ is the luminosity of the seed soft photons that cool the plasma.
	\item $\tau_p$, the proton (Thomson) optical depth of the cloud (i.e., the optical depth due to the background electrons).
	\item The characteristic temperature of the soft photons, $T_{\rm bb}$.
\end{enumerate}
 
Pietrini \& Krolik (1995) proposed a very simple analytical formula that relates the observed X-ray spectral energy index to the amplification factor, $l_{\rm h}/l_{\rm s}$:

\begin{eqnarray}
	\alpha \approx 1.6 \left(\frac{l_{\rm h}}{l_{\rm s}}\right)^{-1/4}.
\end{eqnarray}

\vspace{0.4cm}

\noindent When $l_{\rm h}/l_{\rm s}$ increases, the source becomes more ``photon starved'' and the observed spectrum becomes harder.\\

Let us now consider various geometrical arrangements of the hot plasma cloud and the source of soft photons. If soft seed photons for Comptonization are produced by reprocessing hard X/$\gamma$-ray radiation, then the geometry will define the amplitude of feedback effect and the spectral slope (Liang 1979). What would be the most probable geometry taking into account all those concepts and observation features?. The main possibilities are:\\

``\textit{Sandwich}'': The simplest solution is to assume that a hot corona covers most of the cold disk (a sandwich, or a slab-corona model). The radiative transfer in such a geometry was considered by Haardt \& Maraschi (1993) who showed that in the extreme case, when all the energy is dissipated in the corona, the emitted spectra resemble those observed in Seyfert galaxies. Dissipation of energy in the cold disk (with subsequent additional production of soft photons) would produce too steep spectra in disagreement with observations. Even harder spectra observed in Galactic BHs cannot be reconciled with the slab-corona model (Gierli\'nski et al. 1997), and alternative models with more photon starved conditions and smaller feedback of soft photons are sought.\\

``\textit{Magnetic flares}'': A patchy corona (Galeev et al. 1979; Haardt et al. 1994), where the cold disk is not covered completely by hot material, has certainly a smaller feedback, and the resulting spectra are harder. A patchy corona can be described by a number of active regions above the cold accretion disk. Both patchy and slab-corona models predict an anisotropy break (i.e. a break in the power-law spectrum due to the anisotropy of the seed photons) that should appear at the energy corresponding to the second scattering order.\\

``\textit{Cloudlets}'': Another possible solution of the photon starvation problem is to assume that the cold disk within the hot corona is disrupted into cold dense optically thick clouds (Lightman 1974; Kuncic et al. 1997) that are able to reprocess hard X/$\gamma$-ray radiation and produce soft seed photons for Comptonization. If the height-to-radius ratio of the hot cloud is small, we can approximate this geometry by a plane-parallel slab. We assume further that the cold material is concentrated in the central plane of the hot slab and has a covering factor $f_c$. Compton reflection comes from these cold clouds (cloudlets) as well as from the outer cooler disk. The seed soft photon radiation is much more isotropic and the emerging high energy spectrum does not have an anisotropy break. The covering factor defines the amplitude of the feedback effect. The total soft seed luminosity (with corresponding compactness, $l_{\rm s}$) is the sum of the reprocessed luminosity and t
 he luminosity intrinsically dissipated in the cold disk (with corresponding compactness, $l^{\rm intr}_{\rm s}$).\\

``\textit{Sombrero}'': In this model, the cold disk penetrates only a short way into the central coronal region (see, e.g., Bisnovatyi-Kogan \& Blinnikov 1977, and Poutanen et al. 1997). We can assume that the X/$\gamma$-ray source can be approximated by a homogeneous spherical cloud of radius $r_{\rm c}$ situated around a black hole (probably, a torus geometry for a hot cloud would be more physically realistic, but then it would be more difficult to compute the radiative transfer). The inner radius, $r_{\rm in}$, of the cold geometrically thin, infinite disk is within the corona ($r_{\rm in}/r_{\rm c} \leq 1$). This geo\-metry is also similar to the geometry of the ADAF model already discussed. Spectra from the sombrero models are almost identical to the spectra expected from the cloudlets model, with the only difference that the amount of Compton reflection would be a bit larger for the same configuration of the outer cooler disk. From the observational point of view, these
  models are almost indistinguishable.\\
 
Let us add a new ingredient, by assuming that the plasma can be a ``hybrid thermal/non-thermal pair plasma''. There are reasons to believe that in a physically realistic situation, the electron distribution can notably deviate from a Maxwellian. A significant fraction of the total energy input can be injected to the system in form of relativistic electrons (pairs). In this hybrid model, the injection of relativistic electrons is allowed in addition to the direct heating of thermal electrons and then the particle energy distribution is approximately a Maxwellian plus a power law tail.\\

The most important input parameters of this version of the model are: 

\begin{enumerate}
	\item The thermal compactness, $l_{\rm th}$, which characterizes the heating rate of electrons (pairs).

	\item The analogous non-thermal compactness, $l_{\rm nth}$, which characterizes the rate of injection of relativistic electrons.

	\item The soft photon compactness, $l_{\rm s}$.

	\item $\Gamma_{\rm inj}$, the power-law index of the non-thermal electron injection spectrum.

	\item $\tau_p$, the proton (Thomson) optical depth.

	\item $T_{\rm bb}$
\end{enumerate}

By $l_{\rm h} = l_{\rm th} + l_{\rm nth}$, we now denote the total hard compactness.\\

\begin{figure}[!t]
\begin{center}
\resizebox{12cm}{!}{\includegraphics{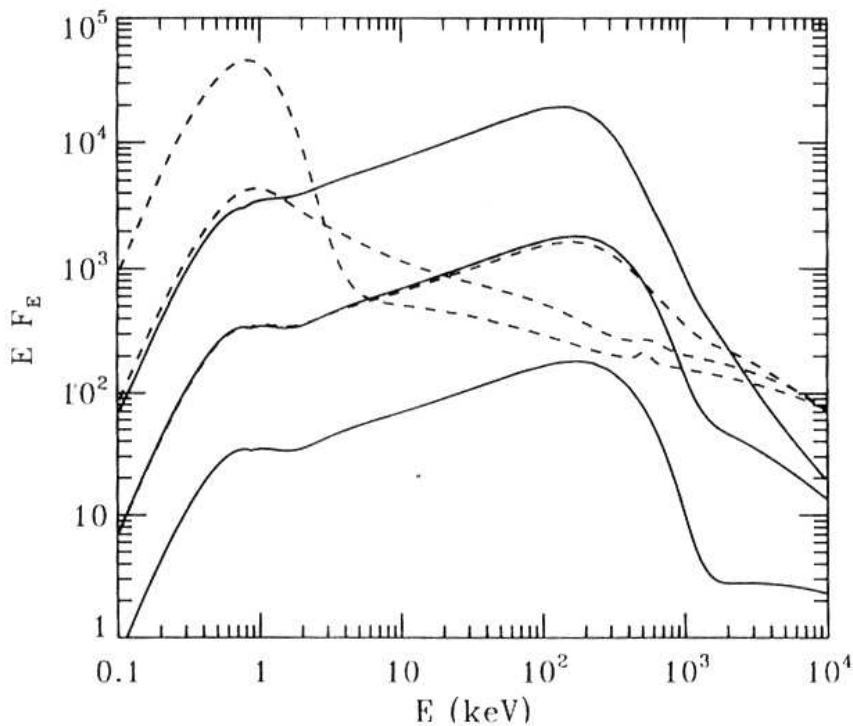}}
\caption{{Spectra from the hybrid pair plasmas. Solid curves show dependence on hard compactness $l_{\rm h}$. Other parameters: $l_{\rm h}/l_{\rm s} = 10$, $l_{\rm nth}/l_{\rm h} = 0.1$, $\tau_p = 1$, $\Gamma_{\rm inj} = 2.5$, $T_{\rm bb} = 0.2$ keV. The resulting electron temperature and optical depth are ($kT_e, \tau_T$ )=(126 keV, 1.0002), (123 keV, 1.02), and (82 keV, 1.47) for $l_{\rm h}$ = 1, 10, 100, respectively ($l_{\rm h}$ increases from the bottom to the top of the figure). For a higher compactness, the spectrum has a sharper cut-off at energies above 1 MeV due to larger optical depth for photon-photon pair production. These spectra are similar to the spectra of Galactic BHs in their hard state. Dashed curves show dependence on $l_{\rm h}/l_{\rm s}$. Here we fixed $l_{\rm h} = 10$, $l_{\rm nth}/l_{\rm h} = 0.5$. The resulting electron temperature and optical depth ($kT_e, \tau_T$) are (104 keV, 1.07), (34 keV, 1.02), and (5 keV, 1.01) for $l_{\rm h}/l_{\rm s} = 10,
  1, 0.1$, respectively. Increase in $l_{\rm h}$ results in a more pronounced blackbody part of the emerging spectrum. The blackbody is modified by Comptonization on thermal electrons (from Poutanen 1998).}}\label{corona-emission}
\end{center}
\end{figure}

Following this model, the spectrum of escaping radiation then consists of the incident blackbody, the soft excess due to Comptonization by a thermal population of electrons and a power-law like tail due to Comptonization by a non-thermal electron (pair) population (see Fig.~\ref{corona-emission}).\\

For large $l_{\rm h}/l_{\rm s}$, most of the spectrum is produced by Comptonization of a thermal population of electrons (pairs), while the tail at energies above $m_ec^2$ is produced by non-thermal electrons. For low $l_{\rm h}/l_{\rm s}$, the resulting spectrum is produced by a single Compton scattering off non-thermal electrons.\\
  
\subsection{\label{the jet}The Jet}

\begin{figure}
\begin{center}
\resizebox{10.5cm}{!}{\includegraphics{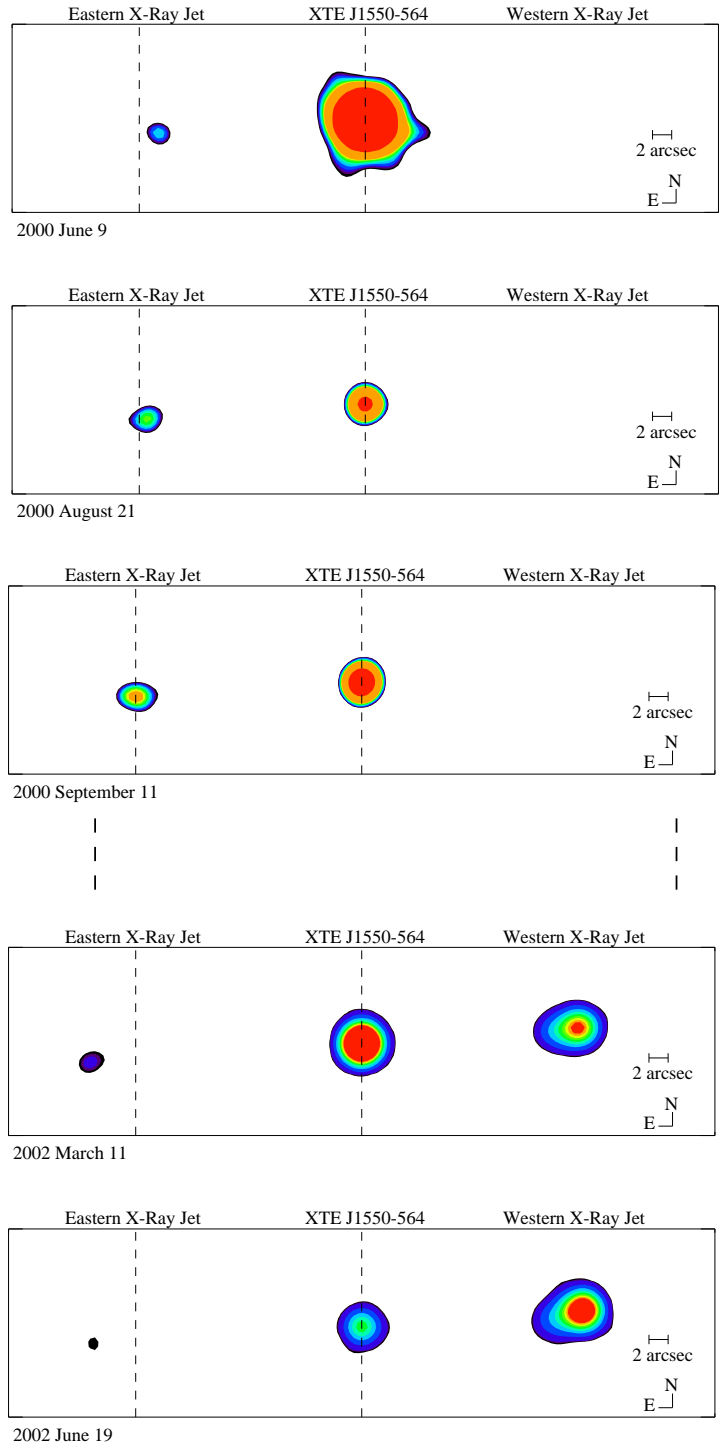}}
\caption{{X-ray emission from large-scale jets produced during an outburst of the black hole transient XTE 1550-564. The leptons radiating in soft X-rays should have TeV energies (from Corbel et al. 2002).}}\label{Fender1550}
\end{center}
\end{figure}

The main observational aspect that determines the identification of jets as such is their radio emission (see Fig.~\ref{jet1E1740.7-2942} and Fig.~\ref{GRS 1915+105}): high brightness temperatures, ``non-thermal'' spectra and high degree of linear polarization measurements indicate an origin as synchrotron emission from relativistic electrons. However, CHANDRA images of moving X-ray jets from the black hole transient XTE J1550-564 (Corbel et al. 2002 - see Fig~\ref{Fender1550}) made clear that the non-thermal electromagnetic radiation from X-ray binary jets may extend to at least the X-ray band\footnote{This observation corresponds to a transient relativistic jet. No direct evidence of X-ray non-thermal electromagnetic radiation has yet been found in the case of steady jets.}. The spectrum of these moving jets ranges from radio to X-ray wavelengths and can be fitted by a single power law with spectral index $\alpha=-0.660\pm0.005$, consistent with optically thin synchrotron e
 mission. Corbel et al. (2002) deduced that under minimum energy conditions, the magnetic field in the jets is of 0.3\,mG, which implies that the leptons emitting in soft X-rays have been accelerated to TeV energies through the interchange of bulk kinetic energy to particles via shocks.\\   

Detailed investigations of the jets from X-ray binaries, both in the radio band and at shorter wavelengths, have revealed certain features that seem to indicate a strong coupling between the accretion and the outflows occurring close to relativistic objects. Let us review some of the main characteristics;  interpretations will be developed in the next section.\\

Observed bright events reveal an optically thin spectrum above some frequency, from which the underlying distribution of the electron synchrotron emitting population can be derived. If the electron distribution is a power law of the form 

\vspace{0.4cm}

\begin{equation}
N(E)dE \propto E^{-p}dE,
\end{equation}

\vspace{0.4cm}

\noindent then determinations of the spectral index in the optically thin part of the synchrotron spectrum, $\alpha$, 

\vspace{0.2cm}

\begin{equation}
S_{\nu} \propto \nu^{\alpha},
\end{equation}

\vspace{0.6cm}

\noindent can be used through the relation $p = 1 - 2\alpha$ to find the index of the particles. Observed optically thin spectral indices $-0.4 \geq \alpha \geq -0.8$, indicate $1.8 \leq p \leq 2.6$. This is the same range derived for the majority of AGN jets and also for synchrotron emission observed in other astrophysical scenarios (e.g. supernova remnants) and is consistent with an origin for the electron distribution in diffusive shock acceleration (e.g. Longair 1997; Gallant 2002).\\

Observations of apparent superluminal motion in galactic jets is a characteristic that we already mentioned and developed in Section~\ref{MQs} (and with more details in Appendix~\ref{superluminal}). Apparent superluminal motions, $\beta_{\rm T}>1$, require intrinsic velo\-cities $\beta\geq0.7$, indicating the presence of relativistic bulk motions. The associated relativistic Doppler shift is given by 

\vspace{0.2cm}

\begin{equation}\label{eq:doppler}
D= \Gamma^{-1}(1\mp \beta\:\cos\phi)^{-1},
\end{equation}  

\vspace{0.6cm}

\noindent where $\phi$ is the angle to the line of sight and $\Gamma$ is the bulk Lorentz factor of the flow; the $\mp$ signs refer to approaching and receding components respectively.\\

In addition to the proper motions and Doppler-shifting of frequencies, there is a boosting effect (or de-boosting depending if we refer to the approaching or receding component) due to the relativistic Doppler factor expressed in (\ref{eq:doppler}). An object moving at angle $\phi$ to the line of sight with velocity $\beta$ (and resultant Lorentz factor $\Gamma$) will have an observed surface brightness $D^{\rm k - \alpha}$ times brighter (or fainter in the receding component case) than in the proper system, where 2 $\leq$ k $\leq$ 3 and k = 2 corresponds to the average of multiple events in e.g. a continuous jet, and k = 3 to an emission dominated by a singular evolving event, and $\alpha$ as defined above.\\

Until now it has been assumed, quite reasonably in the absence of other information, that the jet inclination is perpendicular to the plane of the binary. However, at least two jet sources, GRO J1655-40 and V4641 Sgr, appear to show significant misalignments (Maccarone 2002). The clearest example of a precessing jet is SS 433. The $\sim$ 162.5-day precession of these jets has been assumed to reflect the precession period of the accretion disk (see e.g. Ogilivie \& Dubus 2001). Hjellming \& Rupen (1995) suggested a precession period for GRO J1655-40 which was very close to the subsequently determined orbital period; similarly there seems to be marginal evidence for precession in the jets of GRS 1915+105 (Fender et al. 1999), LS I +61 303 (Massi et al 2004), and Cyg X-1 (Stirling et al. 2001; Pooley et al. 1999; Romero et al. 2002).\\

Because in X-ray binary jets only the synchrotron emission from the leptonic (electrons and/or positrons) component has been detected, there is little direct information on the baryonic jet content (or lack of it). This is also valid in the case of AGN. The exception  is SS 433, whose jets are associated with a variety of emission lines in optical, infrared and X-rays (e.g. Margon 1984), from where the presence of ions can be inferred.\\ 

Finally, a comment about the distinction between jets from neutron stars and those from black holes. Fender \& Kuulkers (2001) have found that defining a quantity ``radio loudness'' as the peak radio flux of transients (in mJy), divided by their peak X-ray flux (in Crab), black hole transients are more radio loud than neutron stars. Black holes seem to be about one or two orders of magnitude more radio loud than neutron star systems.

\vspace{0.7cm}
\begin{center}
$\clubsuit$
\end{center}
\vspace{0.7cm}

\begin{figure}[!t]
\begin{center}
\resizebox{14cm}{!}{\includegraphics{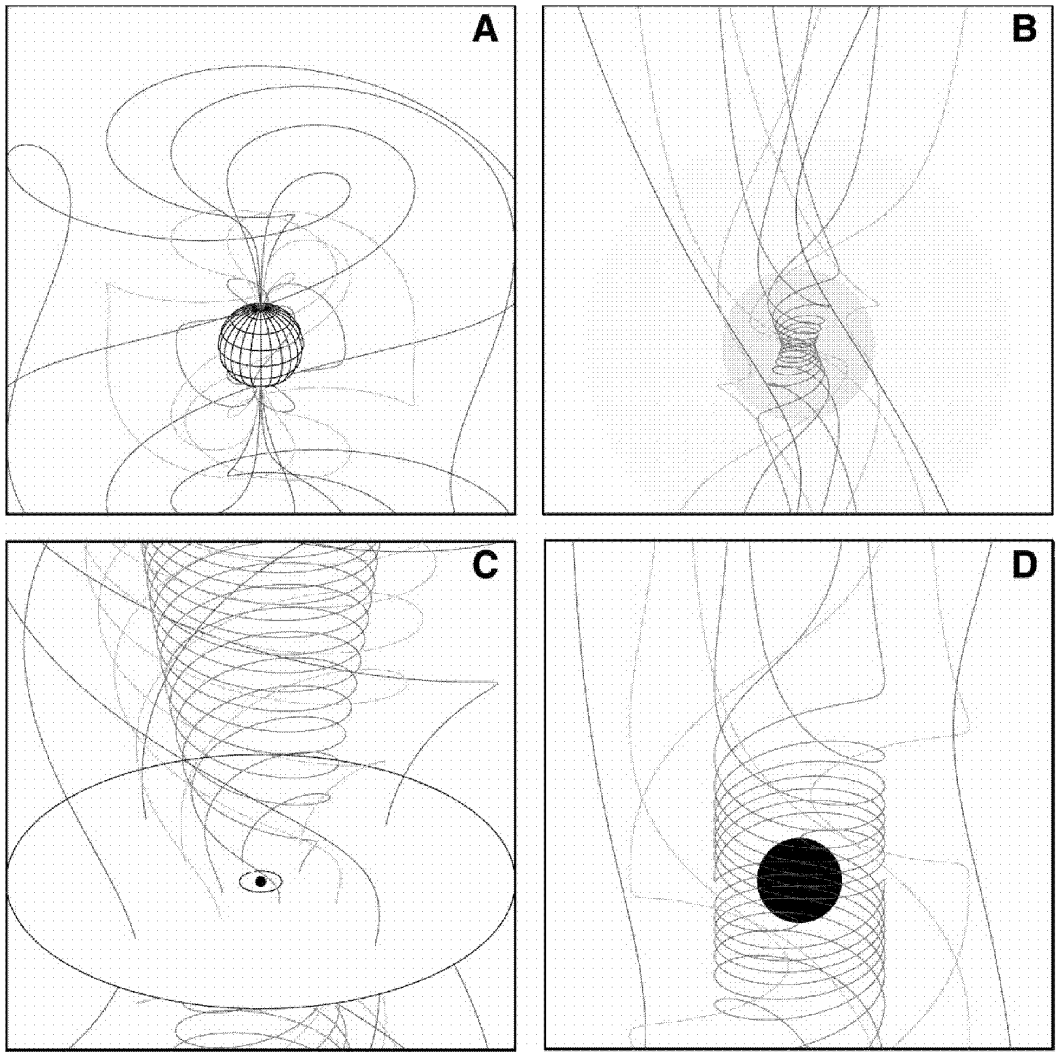}}
\caption{{Four ways to make jets with magnetic fields. A: dipole field of a rotating neutron star. B: A collapsing object drawing and winding up an initially uniform field. C: Poloidal magnetic field from a magnetized accretion disk. D: Frame-dragging near a rotating black hole resulting in strong coiling of the magnetic field lines. Types C and D (possibly also A) may be relevant for X-ray binaries; type A for isolated pulsars; types C and D for AGN (from Meier et al. 2001).}}\label{jet-origin}
\end{center}
\end{figure}

A very active research field related with jets is the investigation of their origin, how they may be generated. Blandford \& Payne (1982) provided a seminal model in which magnetic fields anchored in an accretion flow may produce ``radio'' jets by magneto centrifugal effects. Fig.~\ref{jet-origin} shows four different configurations of accretion with magnetic fields which may induce the formation of jets. In all these theoretical models a magnetized accretion flow is the basis for jet formation by magnetohydrodynamic (MHD) processes. The likely key role of magnetic fields in the coupled accretion - outflow system have been reviewed by Meier et al. (2001).\\

There are two ways to study these MHD jet formation processes:

\begin{itemize}
\item One is to straightforwardly set up and solve the full set of eight partial differential MHD equations: continuity equation (1), Navier-Stokes equation (3), induction equation (3) and energy equation (1), making as few assumptions as possible. Such demanding simulations run for several hours or days on large supercomputers, calculating the evolution of a rotating disk or stellar core of the magnetized plasma and showing the effects of the field on the flow and the flow on the field. Time evolution of multidimensional and supersonic processes, such as jets and shock waves, can be followed in detail. 
\item The other approach is the semi-analytic one. The main idea is to make several simplifying assumptions in order to reduce the set of eight partial differential equations that can then be solved on a modest computer. The most widely used assumptions are: i) Plasma flow and magnetic field structure do not change with time - \textit{steady state assumption}. Such solutions clearly focus only on a long-term structure that forms after transient effects due to initial conditions have died away. ii) Simplification of the geometry - \textit{self-similarity}, i.e. the structure scales with the distance from the compact object, so that it looks similar at every radius.
\end{itemize}

Historically speaking, the semi-analytic was the first approach to be considered; even now, in the era of modern computers, semi-analytic models still complement supercomputer numerical simulations. The simplifications of the semi-analytic approach may lead to less realistic solutions. For example, it is more difficult to generate highly relativistic narrow jets with such approach (Meier et al. 2001). However, because the results are expressed in terms of simple parameters (central object mass, magnetic field strength, etc.), they are often applicable to many different situations. On the other hand, whereas semi-analytic studies often ignore the effects of an inner accretion disk boundary or of a central black hole horizon, and they usually do not treat the time-dependent nature of the flow. In the case of numerical simulations, boundary conditions are taken into account, but careful attention should be put when choosing them, because they can cause spurious effects on the nu
 merical results. Though having a careful perfect control of the initial and/or boundary conditions, numerical simulations lack of a monitoring of the physics developed step by step, whose control is the main philosophy of the semi-analytic methods.\\

Besides the differences between these two approaches, that are definitely to be used in a  complementary way, they have a common essence with respect to the basic steps existing in the MHD jet generation approach, 

\begin{enumerate}
\item Generation from a seed magnetic field of a specific structure, always containing a poloidal component, $B_{\rm p}$, and a toroidal component, $B_{\phi}$, which state the existence of rotational symmetries.
\item Magnetic braking of the existent angular momentum (of the disk, of the compact object or of the young star).
\item Transfer of the extracted angular momentum and rotational energy through torsional Alfv\'en waves (TAWs).
\item Transfer of the magnetic energy from the TAWs (Poynting flux) to some of the present charged particles, pushing them upwards and outwards.
\item Collimation of those particles extracted through pinch effect.
\end{enumerate}

A rather different but still magnetically working model is proposed in Tagger \& Pellat (1999). Radiative acceleration (O'Dell 1981) as an alternative to magnetic acceleration, is unlikely to be able to push jets to the highest observed bulk velocities (Phinney 1982) but may still be operating in the case of SS 433 (Shapiro et al. 1986).\\

It seems quite unlikely that the jet formation region in X-ray binaries will ever be directly imaged. Junor et al. (1999) report the direct imaging of jet formation around the (low-luminosity) AGN M87 at a distance of $\sim$ 100 Schwarzschild radii from the black hole. Comparing M87 to X-ray binaries in our own galaxy, the ratio of distances is so much smaller than the ratio of black hole masses and therefore than the one between Schwarzschild radii, that such imaging will not be possible. For example, a structure of size 100 Schwarzschild radii around a 10$M_{\odot}$ black hole at a distance of 5 kpc would have an angular size of $\sim 10^{-9}$ arcsec (Fender 2004). Therefore the key for studying jet formation in X-ray binaries will remain in careful multiwavelength studies at the highest time resolution, such as those performed with such a success on GRS 1915+105 (e.g. Mirabel et al. 1998; Klein-Wolt et al. 2002).\\

Another family of mechanism for launching jets invoke one-shot acceleration in large differential drops established close to the BH (e.g. Blandford \& Znajek 1977). However, it has been recently argued that the mechanism is not efficient when the MHD equations are solved in a regime of strong gravity (Punsly 2001). There is an open discussion about the validity of these models.\\

\section{\label{dynamics}The Dynamics: Different Spectral States - A Cycle?}

\begin{figure}[!t] 
\centering 
\resizebox{12cm}{!}{\includegraphics{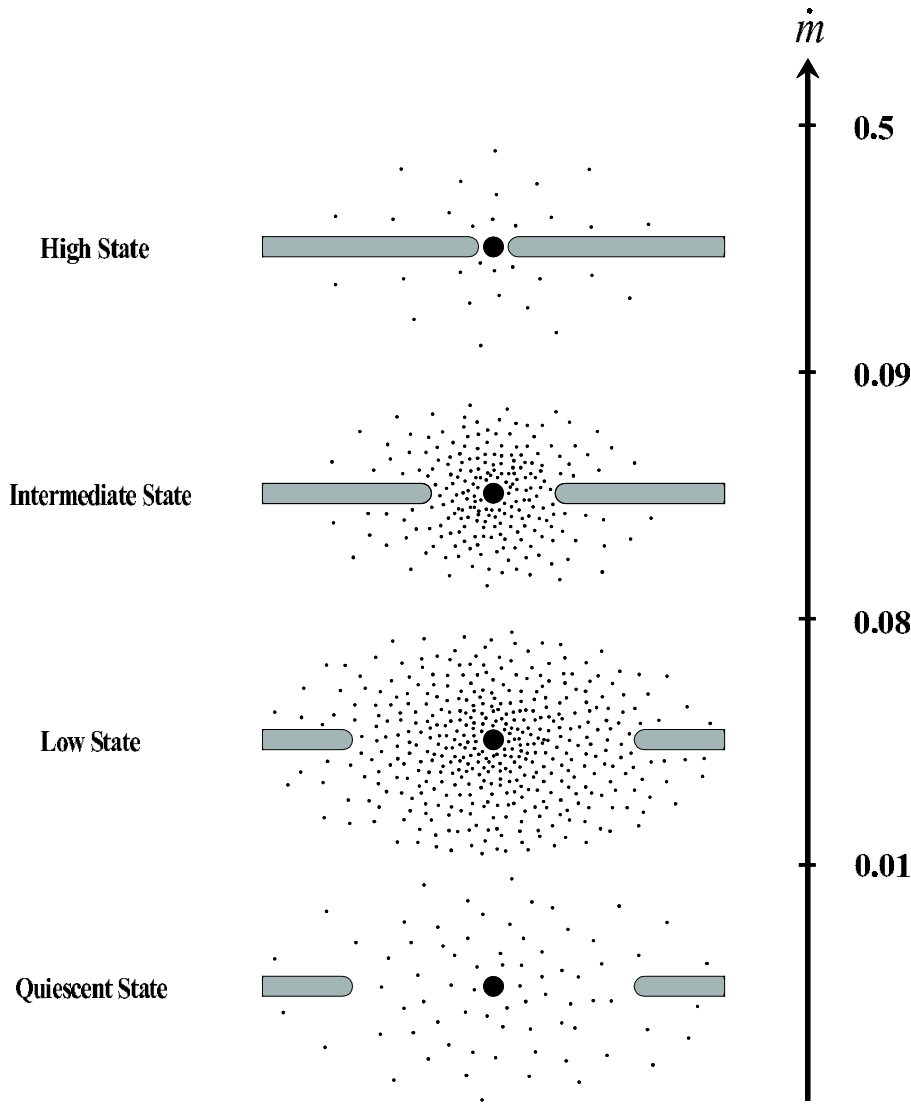}}
\caption{{The configuration of the accretion flow in different spectral states shown schematically as a function of the total mass accretion rate $\dot{m}$. The coronal region is indicated by dots and the thin disk by horizontal bars. (Adapted from Esin et al. 1997)}}\label{estados}
\end{figure}

XRBs are observed in different spectral states. The two most distinctive states (and somehow opposed in their characteristics) are the so-called ``Low/Hard'' and ``High/Soft'' states. These and the other existing states are sketched in Fig.~\ref{estados} as a function of the total mass accretion rate, $\dot{m}$. It has been proposed by different authors (Narayan 1996b; Esin et al. 1997, 1998; Fender et al. 2004b with new variants) that different spectral states observed in XRBs can be understood as a sequence of \textit{thin disk plus coronal region} models with varying $\dot{m}$ and $r_{\rm in}$ (inner thin disk radius).\\
 
A brief description of each state follows:

\begin{itemize}
	\item \textit{Quiescent state}: This is the lowest luminosity state and has $\dot{m}\lesssim 10^{-2}$. Due to the low accretion rate, Comptonization is weak, and the X-ray flux is much lower than the optical flux. The radiative efficiency is very low and the systems are extremely dim.
	
	\item  \textit{Low/Hard state}: For $\dot{m}$ above $10^{-2}$ and up to $\sim 10^{-1}$ the geometry of the accretion flow is similar to that of the quiescent state, but the luminosity and radiative efficiency are larger (and increase rapidly with $\dot{m}$). Comptonization becomes increasingly important, giving rise to a very hard power law spectrum which peaks around 100 keV. Even though in this state the coronal emission predominates, an excess at energies $\leq$ 1 keV is interpreted as radiation from the retracted thin disk (Baluci\'nska-Church et al. 1995) with a characteristic temperature, $kT_{\rm bb}$, of order 0.1 - 0.3 keV (usually quite poorly determined, due to strong interstellar absorption in that spectral range). 

\item  \textit{Intermediate state}: At still higher accretion rates, $\dot{m}$ approaches $\dot{m}_{\rm crit} \sim 0.1$, the coronal region progressively shrinks in size, the inner radius decreases, and the X-ray spectrum changes continuously from hard to soft. This occurs at roughly constant bolometric luminosity. In this state, the thin disk becomes radiatively comparable to, or even brighter than, the coronal region emission.

\item \textit{High/Soft state}: At still higher accretion rates, $\dot{m} > \dot{m}_{\rm crit}$, the coronal region cannot exist as an independent entity at any radius, and the thin disk comes all the way down to the last stable orbit; there is, however, a weak coronal region (with $\dot{m} \leq \dot{m}_{crit}$). A typical high state spectrum resembles a standard thin disk spectrum with a characteristic temperature $kT_{\rm bb}\approx 0.5 - 1$ keV, and a power-law tail due to the weak coronal emission.
Although signatures of Compton reflection are also observed in this state (e.g. Tanaka 1991), its amplitude is much more difficult to determine since it depends on the assumed distribution of ionization and detailed modeling of the continuum, which is rather curved in the spectral region around the iron edge. 

\end{itemize}

This would then be a sequence through which the coronal region gains importance and then looses it in favor of the thin disk and so on. While the corona is in its main activity period, the cold disk slowly regenerates and its edge moves inwards up to the last stable orbit, gradually becoming the main site of emission. The emission is then mainly soft with a little hard contribution. When the matter in the innermost part of the disk is engulfed by the BH there is a quick change of state usually accompanied by a mayor radio outburst. There is some evidence that seems to indicate a disk/jet coupling, i.e a relation between the inflow and the outflow of matter (we will come back to this point with more detail further on in this section). The disappearance of the inner part of the thin disk is then followed by the formation of the corona and the generation of the jet\footnote{Similar sequences have been proposed in the context of AGNs, e.g. Donea \& Biermann (1996).}.\\

It is important to remark that in addition to the explanation it gives of the different possible spectral states, the combination of a cold disk with a Comptonizing corona provides also a satisfactory explanation to other observed spectral characteristics, i.e. a fluorescent iron line at $\sim$ 6.4 keV, an iron edge at $\sim$ 7 keV, and the reflection ``hump'' at 10 - 30 keV. During the hard state, some of the photons from the coronal region are intercepted by the disk and Compton reflected by the colder matter back to the observer, producing this way the mentioned features. In the High/Soft state both, the iron line and the iron edge, appear to be smeared due to probably gravitational redshift and Doppler effect, implying that the cold disk extends very close to the central black hole (see Reynolds \& Nowak 2003 for a comprehensive discussion).

\vspace{0.3cm}
\begin{center}
$\clubsuit$
\end{center}
\vspace{0.8cm}

Two important issues can be mentioned at this point. On one hand, the presence of jets or the lack of them, as well as their characteristics, seem to be also related with the sequence of spectral states we have described above. On the other hand, the coupling between the thin accretion disk and the jet generation is apparently supported by a strong correlation between radio and X-ray flux.\\ 

Hannikainen et al. (1998) were the first to notice a broad correlation between the radio and X-ray fluxes for the black hole binary GX 339-4 on its Low/Hard state. A similar correlation between radio and X-ray fluxes was found for Cyg X-1 (Brocksopp et al. 1999), and Fender (2001) suggested that the magnitude of the radio/X-ray flux ratio was similar for all Low/Hard state black holes. In the past years the understanding of this coupling between radio and X-ray emission has advanced significantly. Corbel et al. (2000, 2003) and Corbel \& Fender (2002) in a detailed long-term study of GX 339-4, have found that the radio emission in the Low/Hard state scales as 

\begin{equation}\label{eq:coupling}
L_{\rm radio}\propto L^b_{\rm X}, 
\end{equation}

\vspace{0.6cm}

\noindent where b $\sim$ 0.7 for X-rays up to at least 20 keV. This relation holds over more than three orders of magnitude in soft X-ray flux. In fact, by compiling data for ten Low/Hard state sources, it was found that in the luminosity range $10^{-5}L_{\rm Edd}\leq\,L_{\rm X}\leq 10^{-2}L_{\rm Edd}$, all systems are consistent with the same coupling with a very small scatter (less than one order of magnitude in radio flux). Jets in the Low/Hard state are steady and self-absorbed (radio spectral index, $\alpha\geq0$). Bulk Lorentz factors of these jets were deduced by Gallo et al. (2003) to be $\Gamma<2$.\\

The radio/X-ray coupling that appears in the Low/Hard state under the form (\ref{eq:coupling}) induced some authors to suggest that both, radio and hard X-rays, have the same origin: synchrotron emission from the compact jet. This jet model was proposed by Markoff et al. (2001)\footnote{The formalism of this proposal is based on previous works developed by Falcke \& Biermann (1995, 1999) for AGNs.}, where it was applied to the MQ XTE J1118+480. Since then it was improved and applied to GX 339-4 in Markoff et al. (2003) and further generalized in Falcke et al. (2004). The radio emission is well known to be generated through optically thick synchrotron from the compact jet. It has been pointed out by Corbel \& Fender (2002; though apparently firstly noted by Motch et al. 1985) that the infrared/optical bands show a spectral break from optically thick to optically thin synchrotron emission which seems to indicate that the X-ray spectrum is an extension of the second.\\

At the beginning, this approach to the hard X-ray emission was received as an alternative explanation to the by then established idea that this kind of X-rays arise via thermal Comptonization of softer photons in a hot corona (such models are described in Section~\ref{corona}; a different interpretation of the here discussed correlation, but in the context of the coronal models, can be found in Zdziarski et al. 2003). Nowadays the need of both mechanisms of hard X-ray production starts to be suggested (Markoff et al. 2003, Falcke et al. 2004, Bosch-Ramon et al. 2005 and Fender et al. 2004b) since they seem to be able to account for different characteristics that are present in the spectra of MQs.\\

A third alternative to mimic this high-frequency part of the observed MQs spectra was proposed by Georganopoulos et al. (2002). In that scenario the hard X-rays would be due to EC scattering of photons from the companion star or the accretions disk by the leptons of the jet (more on this proposal will be said along the forthcoming chapters).\\

The next step in the sequence of spectral states is the transition from the Low/Hard to the High/Soft state is the so-called ``intermediate'' or recently also called ``very high'' state, VHS/IS, which seems to be associated with the formation of discrete, powerful jets. These jets have important differences with the middling relativistic steady optically thick ones present in the Low/Hard state: they are optically thin, transient and have $\Gamma>2$ (Fender et al. 2004b).\\

The VHS/IS is itself divided into hard--VHS/IS and soft--VHS/IS. Fender et al. (2004b) showed through radio observations how the compact jet persists into the hard--VHS/IS before the outcome of the outbursts that occur in the transition to the soft--VHS/IS while X-rays are softening. These outbursts or very relativistic transient jets are optically thin at radio as well as X-ray wavelengths. After these flares happen the fact that the radio emissions disappear is taken as a proof that the jet does not exist anymore in the soft--VHS/IS and nor will it exist in the following High/Soft state.\\

The components associated with radio flares have higher bulk Lorentz factors than the previously ejected steady jet. These components might be due to internal shocks that could arise when the fastest discrete ejections catch up the slower steady jet. Fender et al. (2004b) proposed that the very relativistic jets that are produced in the middle of the VHS/IS are likely to be formed by the coronal matter that is ejected, which, in this way, will also disappear in the transition to the High/Soft state.\\

Before 1998, radio monitoring of GX 339-4 in the Low/Hard state had already
established the existence of a weak, mildly variable radio counterpart (Hannikainen et al. 1998). When in 1998 the source spent a year in the High/Soft X-ray state, no radio counterpart was detected despite multiple observations (Fender et al. 1999). The source subsequently returned to the Low/Hard X-ray state and the weak radio counterpart reappeared (Corbel et al. 2000). This was the strongest evidence that in ``soft'' disk-dominated states the radio jet was either non-existent or more than an order of magnitude weaker. This conclusion was supported, among other observations of different XRBs, by the detailed studies of GRS 1915+105 reported by Klein-Wolt et al. (2002), in which steady ``soft'' X-ray states are never associated with bright radio emission.\\

\begin{figure}[!t] 
\centering
\resizebox{14.8cm}{!}{\includegraphics{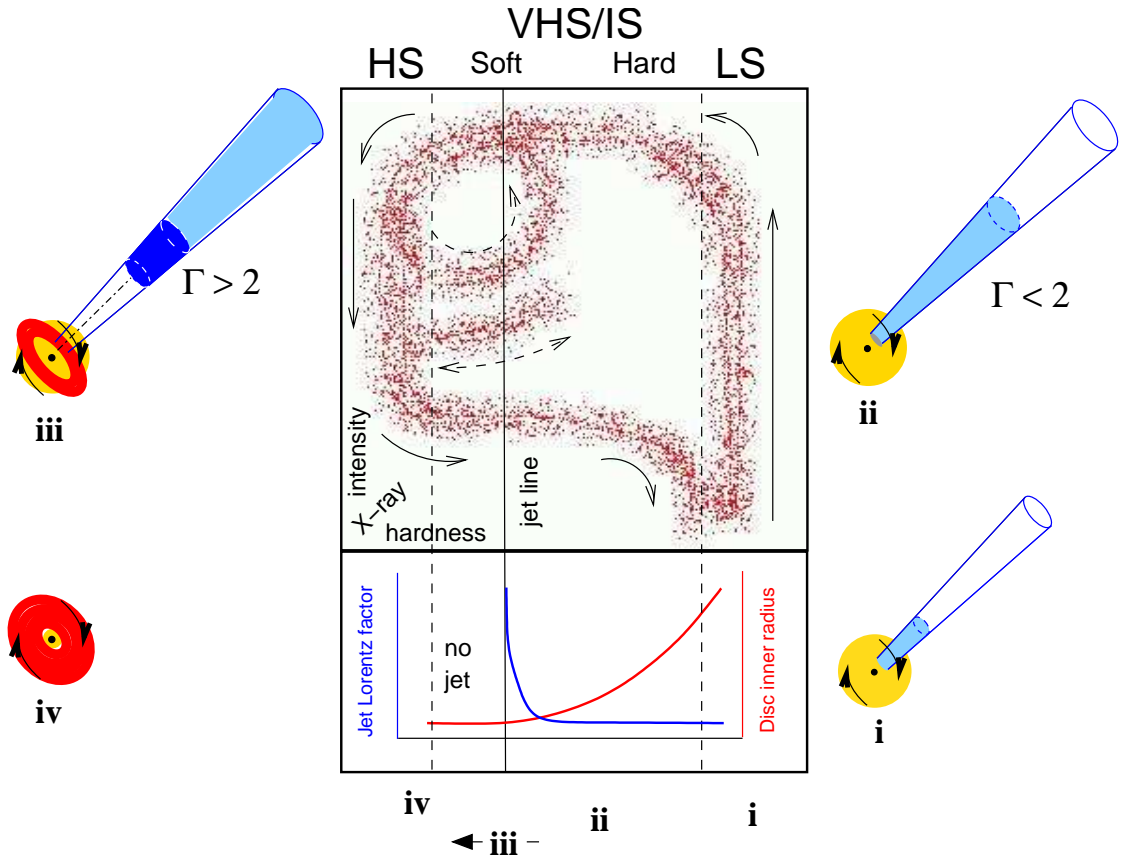}}
\caption{{The central box panel represents an X-ray
  hardness-intensity diagram. HS = high/soft state, VHS/IS = very high/intermediate state, LS = low/hard state. The lower panel indicates the variation of the bulk Lorentz factor of the outflow with hardness. The dashed loop and
  dotted track indicate the paths that GRS 1915+105 and some other
  transients take in repeatedly hardening and then crossing zone {\bf
  iii} -- the ``jet line'' -- from left to right, producing further optically thin radio
  outbursts. Sketches around the outside illustrate the
  relative contributions of jet (blue), corona (yellow) and
  accretion disk (red) at these different stages of the spectral states cycle (from Fender et al. 2004b).}}\label{cycle}
\end{figure}

This cycle of spectral states and the jet characteristics in each step are represented in a schematic way in Fig~\ref{cycle}.\\

\begin{figure}[!t] 
\centering
\resizebox{14cm}{!}{\includegraphics{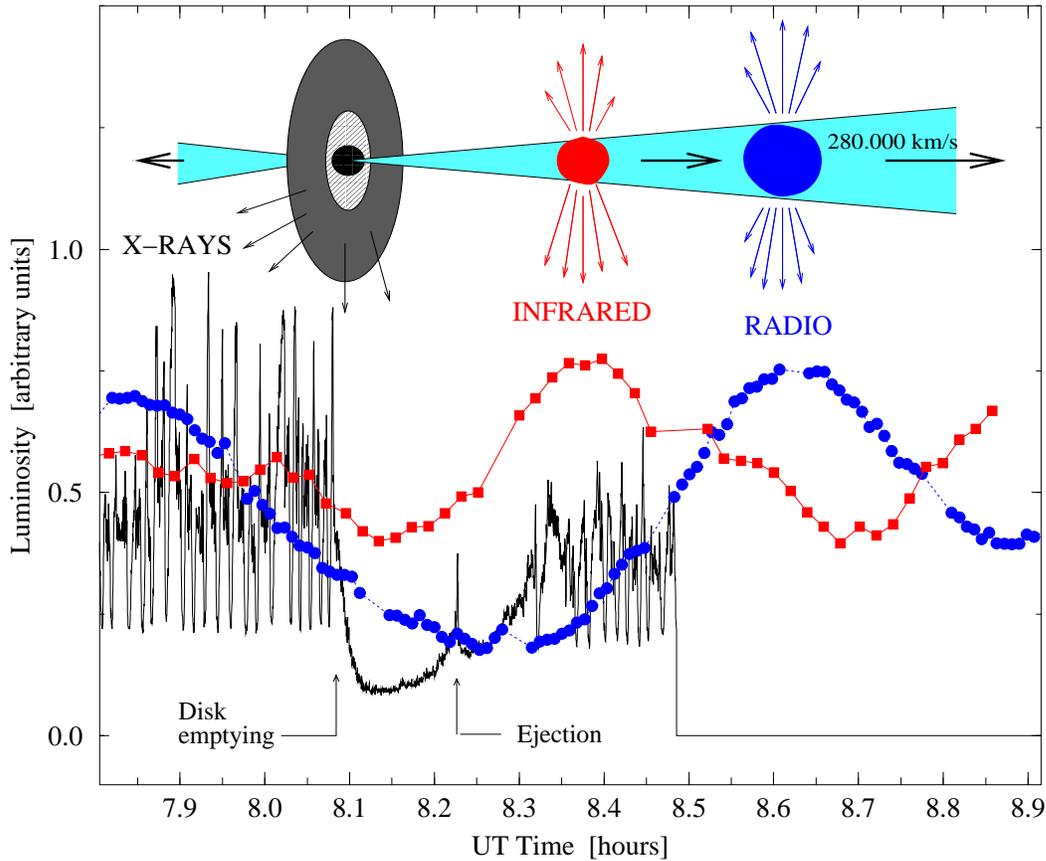}}
\caption{{Multi-wavelength behavior of the microquasar GRS 1915+105 as observed in
September 8th 1997 (Mirabel et al. 1998). The radio data at 3.6 cm (blue squares) were
obtained with the VLA interferometer; the infrared observations at 2.2 micron (red
squares) are from the UKIRT; the continuous line is the X-ray emission as observed by
RXTE in the 2–50 keV range. These observations show the disk/jet coupling, i.e a relation between the inflow seen in X-rays and the outflow of relativistic plasma
clouds observed as synchrotron emission at infrared wavelengths first and later at
radio wavelengths (from --original version-- Mirabel et al. 1998, modified in Paredes 2004b).}}\label{radio-X}
\end{figure}

A second issue of the relation between accretion disks and generation of
jets has been observed many times in the collimated ejections of GRS 1915+105,
which provides one of the best studied cases supporting the proposed disk/jet connection.
In Fig.~\ref{radio-X}, simultaneous observations realized on September 8th 1997 (Mirabel et al. 1998) are presented at radio, infrared and X-ray wavelengths. The data show the development of a radio outburst, with a peak flux density of about 50 mJy, as a result of a bipolar ejection of plasma clouds. Previous to the radio outburst there was a
former outburst in the infrared. The simplest interpretation is that both
flaring events were due to synchrotron radiation generated by the same relativistic electrons of the ejected plasma. The adiabatic expansion of the jets causes losses of energy of these electrons and as a result the spectral maximum of their synchrotron radiation is progressively shifted from the infrared to the radio wavelengths.
At the same time of the emergence of plasma clouds that produce the infrared and radio flares, a sharp decay and hardening of the X-ray emission occur. This can be noticed at 8.08 –- 8.23 h UT in the figure. The X-ray dip is interpreted as the disappearance of the inner regions of the accretion disk (Belloni et al. 1997). Part of the matter content in the disk is then ejected into the jets, while the rest is finally captured by the central black hole. The recovery of the X-ray emission level at 8.23 h UT is interpreted as the
progressive refilling of the inner accretion disk with a new supply of matter that will gradually reach the last stable orbit around the black hole. Similar effect has been observed in time scales of years in the extragalactic source 3C 120 (Marscher et al. 2002).

\vspace{0.4cm}
\begin{center}
$\clubsuit$
\end{center}
\vspace{0.7cm}

At this point the thesis sustained by some authors that most likely the 70\% (Fender 2004) of the galactic X-ray binary systems are able to generate jets looks more than reasonable considering that:

\begin{itemize}
\item All known BH systems which are either in the Low/Hard X-ray state, or in the intermediate, are associated with the formation of a jet (though of different kind). Then the majority of known binary BHs are, or have been in the past, associated with a jet.

\item The six brightest low magnetic field neutron star systems, the ``Z sources'', are all associated with jets. The lower luminosity, low magnetic field systems, which are dubbed as  ``Atoll'' sources, may be associated with radio emission (although
as in the case of black holes there may be bright soft states without jets), implying that the lack of radio detections of the majority is a sensitivity issue (for detailed description on these two kind of sources, see Fender 2004).

\item The high magnetic field neutron stars, including all but two of the accreting X-ray pulsars, are not associated with radio emission.
\end{itemize}

This basically means that all the systems except the high magnetic field X-ray pulsars, and a small number of black hole and neutron star systems which are in persistent ``soft'' states, are expected to have or to have had jets. 

\chapter{\label{High-mass}High-Mass Microquasars as Gamma-Ray Sources}
\thispagestyle{empty}

\newpage
\thispagestyle{empty}
\phantom{.}

\newpage
\vspace*{4.5cm}

\section{Unidentified EGRET Sources: the Galactic Plane Group}

As it is mentioned in Section~\ref{Gamma-Ray Emission}, about 170 out of 271 point like gamma-ray sources included in the third EGRET catalog have not yet been clearly identified with objects detected at lower energies. Many of these sources form a group that is concentrated towards the galactic plane and well correlated with the spiral arms of the Galaxy; something that indicates a significant contribution
from Population I objects. Several types of objects have been proposed as possible
counterparts of these gamma-ray sources, including early-type
stars (both isolated and forming binary systems), accreting
neutron stars, radio-quiet pulsars, interacting supernova
remnants, and black hole candidates (see Romero 2001, 2004 and
references therein).\\

In this chapter we will present a model for galactic variable
gamma-ray sources based on the idea of precessing high-mass\footnote{Let us recall the reader from Section~\ref{CSandCO} that HMXBs are distributed along the galactic plane.} \textit{microblazars}\footnote{Aharonian \& Atoyan (1998) and Mirabel \&
Rodr\'{\i}guez (1999) proposed that microquasars with jets forming
a small angle with the line of sight --by analogy with the unified
model for AGNs-- could appear as {\textit{microblazars}},
sources with highly variable and enhanced nonthermal flux due to
Doppler boosting.}. In our model, the jet precession will be induced by the gravitational
torque of the high-mass stellar companion, which is in a non-coplanar orbit,
on the accretion disk of the compact object (BH or NS). We propose that gamma-rays can be produced by external Compton (EC) scattering of UV stellar photons of the massive companions by relativistic leptons (electrons and/or positrons) far from the
base of the jet. We shall discuss cases where the particles are in
the form of a ``blob'' ejected from the central engine and where
the particles are continuously injected in the form of a
persistent jet. In the case of high-mass binaries, this mechanism
for generation of gamma-rays is more efficient than Compton
scattering upon disk photons or self-synchrotron jet emission
(Georganopoulos et al. 2002) and, in combination with the
gravitational effects of the stellar companion, it can be used to explain gamma-ray sources variable on time scales from weeks to years.\\ 

\section{\label{IC}Inverse Compton: the Jet - Photon Field Interaction}

Let us consider a binary system where accretion onto the
compact object results in the production of twin,
relativistic $e^+e^-$-pair jets propagating in opposite
directions. A sudden variation in the injection rate can result in
the formation of a blob (for which we adopt spherical geometry)
which propagates down the jet with a bulk Lorentz factor $\Gamma$.
We shall follow the general treatment derived by Georganopoulos et
al. (2001) for EC scattering in extragalactic
blazars, adapting it to microblazars.\\

In the blob frame the relativistic leptons are considered to have an isotropic power-law
density distribution,

\begin{equation}
    n'(\gamma\:')=\frac{k}{4\pi}{\gamma\:'}^{\;-p}P(\gamma_{1},\gamma_{2},\gamma\:'),
\end{equation}    
    
\vspace{0.4cm}

\noindent where $\gamma\:'$ is the Lorentz factor of the leptons, \textit{k}
is a constant, and 

\vspace{0.2cm}

\begin{displaymath}
P(\gamma_{1},\gamma_{2},\gamma\:')= \left\{ \begin{array}{ll}1 & \rm{for}\ \gamma_{1}\leq\gamma\:'\leq\gamma_{2}\vspace{0.3cm}\\

0 & \rm{otherwise.} \end{array} \right.
\end{displaymath}

\vspace{0.2cm}

Using the Lorentz invariant $n/\gamma^{2}$ and the relation
$\gamma=D\gamma\:'$ between the Lorentz factor of the
electrons/positrons in the lab frame and the blob frame
respectively, we get that the electron density in the lab frame is

\begin{eqnarray}
    n(\gamma)=\frac{k}{4\pi}D^{2+p}\gamma^{-p}P(\gamma_{1}D,\gamma_{2}D,\gamma) \label{den},
\end{eqnarray}

\vspace{0.4cm}

\noindent where $ D=\left[\Gamma\left(1\mp\beta\cos\phi\right)\right]^{-1}$ is
the usual Doppler factor: $\beta$ and $\phi$ are the bulk velocity
in units of $c$ and the viewing angle, respectively.\\

The energy distribution of the effective number of electrons is

\vspace{0.2cm}

\begin{equation}
N_{\rm eff}(\gamma)= n(\gamma)V_{\rm eff}, 
\label{eq:ele} 
\end{equation} 

\vspace{0.4cm}

\noindent where $V_{\rm eff}$ is the effective observed volume of the blob defined by,

\begin{eqnarray}
	V_{\rm eff}&=&
\int_{\rm blob}d^3x\int_{-\infty}^{+\infty} dt\; 
\delta\left(t-{\bf x}\cdot{\bf\hat{n}}/c\right).
\label{vobsdef}
\end{eqnarray}

\vspace{0.4cm}

\noindent To relate $V_{\rm eff}$ with $V$, the volume of the blob in its frame, we have to rewrite (\ref{vobsdef}) in terms of the spatial coordinates in the blob
frame, keeping the time $t$ in the lab frame. Knowing that $dx=dx'/\Gamma$,
$dy'=dy$, $dz'=dz$ and $x'=\Gamma(x-c\beta t)$, we perform the integral over $t$ and we obtain\footnote{To solve the time integral we use:\begin{displaymath}
\delta(g(x))= \sum_{i}\frac{1}{|g'(x_i)|}\delta(x-x_i),\qquad x_i \ \ \textrm{such that}\ \ g(x_i)=0 \ \ \textrm{and} \ \ g'(x_i)\neq0.
\end{displaymath}}: 

\vspace{0.2cm}

\begin{equation}\label{eq:voleff}
V_{\rm eff}={1\over\Gamma(1-\beta \cos\phi)}\int_{\rm blob}d^3x'=V D.
\end{equation}

\vspace{0.6cm}

\noindent Using at this point expressions (\ref{den}) and (\ref{eq:voleff}) into $N_{\rm eff}$, (\ref{eq:ele}), we get,

\vspace{0.3cm}

\begin{equation}
N_{\rm eff}(\gamma)=\displaystyle{kV \over 4\pi} D^{3+p} \gamma^{-p}\,P(\gamma_1 D,\gamma_2 D,\gamma). 
\label{eq:ele2} 
\end{equation} 

\vspace{0.4cm}

The blob is injected into an isotropic monoenergetic\footnote{In the coming sections and chapters we will apply the model to specific photon fields. Some of them will have blackbody distributions and we will work with the peak of such distribution, $\epsilon_0$, adapting the situation in a first approximation to a monoenergetic photon field.} photon
field and will then produce the inverse Compton upscattering of some of these
photons. The lab frame
rate of IC interactions per final photon energy is,

\begin{eqnarray}
    \frac{dN_{p}}{dtd{\epsilon}}=\frac{3\sigma_{T}c}{4\epsilon_{0}\gamma^{2}}f(x),
    \label{rate}
\end{eqnarray}

\vspace{0.4cm}

\noindent where $\sigma_{T}$ is the Thomson cross section, $\epsilon$ and $\epsilon_{0}$ are the energies of the scattered photons and seed photons in units of $m_e c^2$. For the case of
\textbf{Thomson scattering} and assuming isotropic scattering in the
electron frame (Blumenthal \& Gould 1970),

\begin{eqnarray}
    f(x)=\frac{2}{3}(1-x)P(1/4\gamma^{2},1,x),\ \ \
    x=\frac{\epsilon}{4\gamma^{2}\epsilon_{0}}.
\end{eqnarray}

\vspace{0.4cm}

The specific luminosity is then obtained by integrating the scattering rate
(\ref{rate}) over the particle energy distribution (\ref{eq:ele2}), and multiplying by the observed photon energy $\epsilon m_{e}c^{2}$ and the number density
$n_{p}=U/\epsilon_{0}m_{e}c^{2}$ (where \textit{U} is the energy
density of the photon field). The limits of integration are $\gamma_{\rm min}=(\epsilon/4\epsilon_0)^{1/2}$ and $\gamma_{\rm max}=\gamma_{2}D$. The final expression is:

\begin{eqnarray}\label{lumtot}    L_{\epsilon}&=&\frac{dL}{d{\epsilon}d\Omega}=D^{3+p}\frac{kV\sigma_{T}c\:U}{8\pi\epsilon_{0}}\frac{\epsilon}{\epsilon_{0}}\times\nonumber\\
&&\times\left\{(\gamma_{2}D)^{-(1+p)}\left(\frac{\epsilon}{4\epsilon_{0}(3+p)(\gamma_{2}D)^{2}}-\frac{1}{1+p}\right)+\right.\\
&&+\left.\left(\frac{\epsilon}{4\epsilon_{0}}\right)^{-(1+p)/2}\frac{2}{(1+p)(3+p)}\right\}.\nonumber
\end{eqnarray}

\vspace{0.6cm}

\noindent When $p>-1$ and $\gamma_{\rm min}\ll\gamma_{\rm max}$, the specific luminosity reduces to:

\vspace{0.4cm}

\begin{eqnarray}
    L_{\epsilon}=\frac{dL}{d{\epsilon}d\Omega}\approx D^{3+p}\frac{kV\sigma_{T}c\:U2^{p-1}}{\pi
    \epsilon_{0}(1+p)(3+p)}\left(\frac{\epsilon}{\epsilon_{0}}\right)^{-(p-1)/2}.
    \label{lumred}
\end{eqnarray}

\vspace{0.8cm}

\noindent The luminosity per sr is given by $L=\epsilon L_{\epsilon}$ and the scattered photons have:

\begin{eqnarray}
	\left\{\begin{array}{l}\epsilon_{\rm min,T}=4\epsilon_0\gamma_{1}^2 D^2 \vspace{0.2cm}\\
	\epsilon_{\rm max,T}=4\epsilon_0\gamma_{2}^2 D^2 \end{array} \right.
\end{eqnarray}

\vspace{0.6cm}

In the \textbf{Klein-Nishina regime}\footnote{When $\displaystyle \frac{E_e E_{\rm ph}}{(m_e c^2)^2} > 1$, which is equivalent to $\gamma\epsilon_0>1$.} the expression for $f(x)$ in (\ref{rate}) becomes\footnote{Obtained by Jones (1968).}

\vspace{0.3cm}

\begin{eqnarray}
\begin{array}{c}\displaystyle f(x)=\left[2x \ln x+x+1-2x^2+\frac{(4\epsilon_0\gamma x)^2}{2( 1+4 \epsilon_0 \gamma x)}\right] \,P(1/4\gamma^2,1,x),\vspace{0.6cm}\\

\displaystyle x= \frac{\epsilon }{4 \epsilon_0 \gamma^2 (1- \epsilon/ \gamma)}\end{array}
\end{eqnarray} 

\vspace{0.6cm}

\noindent and hence the procedure to calculate $L_{\epsilon}$ requires numerical integrations instead of working with an analytical expression like in the Thomson case (\ref{lumtot}). In this context:
\begin{eqnarray}
	\left\{\begin{array}{l}\displaystyle \epsilon_{\rm min,KN}=\frac{4\epsilon_0\gamma_{1}^2 D^2}{(1+4\epsilon_0\gamma_1 D)} \vspace{0.2cm}\\
	\displaystyle \epsilon_{\rm max,KN}=\frac{4\epsilon_0\gamma_{2}^2 D^2}{(1+4\epsilon_0\gamma_2 D)}\end{array} \right.
\end{eqnarray}

\vspace{0.7cm}
\begin{center}
$\clubsuit$
\end{center}
\vspace{0.7cm}

In the case of a single blob, the interaction time with
the photon field, and hence the duration of the gamma-ray flare,
will be strongly limited by the apparently superluminal speed of
the feature. For instance, in the case of an O7 star and a blob
with $\beta=0.98$ ($\Gamma=5$), the flux will decrease to $1/e$ of
its original value in $\sim 1$ s. Gamma-ray production through EC in blobs in
microblazars are then transient phenomena difficult to detect on
Earth. This is not the case, however, of microblazars with
persistent jets.\\

The amplification due to the Doppler factor in (\ref{lumred}) is given by $D^{3+p}$. This factor changes to $D^{2+p}$ when we consider a continuous relativistic flow instead of a blob ejection. The change comes from the fact that in the jet-like case the observed flux is obtained by integrating over the area (and thereafter the luminosity as well) of the feature instead of integrating over the volume as we did in the blob case\footnote{See Kembhavi \& Narlikar (1999) and references therein for the calculations that justify this change.}.\\

\section{External Photon Field: the High-Mass Companion Star}
\subsection{Gamma-Ray Emission}

Let us now consider some concrete examples of binary systems, where the jet is injected into the photon field due to the high-mass companion star.\\

\begin{figure}[!t]
\centering
\resizebox{12cm}{!}{\includegraphics{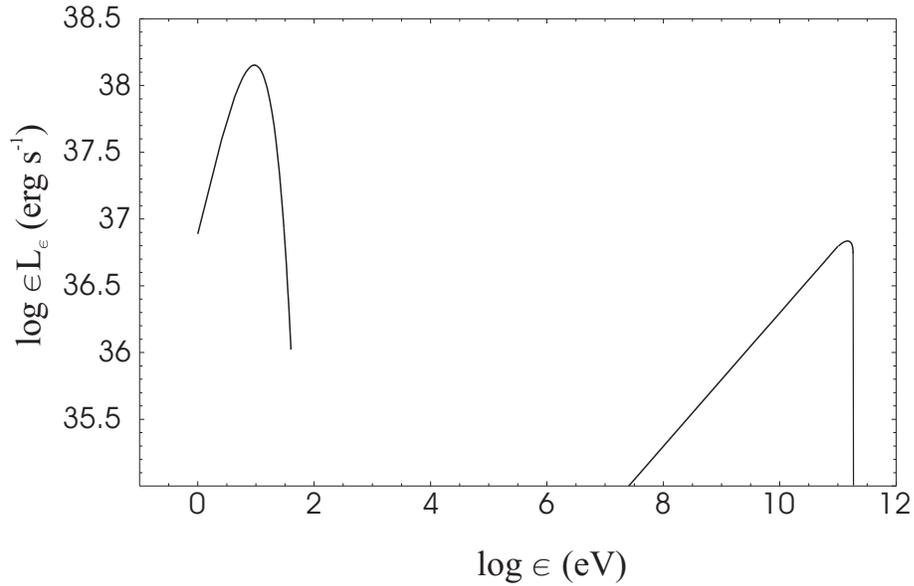}}
\caption{ Spectral high-energy distribution of the scattered photons for a
microquasar injecting a power-law spectrum of electrons in the
photon field of the high-mass stellar companion (an O7 star in
this example). A cut-off at Lorentz factors $\gamma\sim10^3$ has
been assumed. The spectral energy distribution of the star is also
shown (left top corner).} \label{fig1-toymodel}
\end{figure}

We have calculated the
spectral energy distribution for a specific model with $p=2$, bulk
Lorentz factor $\Gamma=5$, viewing angle $\phi=10$ deg, photon
field of an O7 star (photon energies $\sim 10$ eV), and a
high-energy cut-off of $\gamma_2=10^3>>\gamma_1$. The results are
shown in Fig.~\ref{fig1-toymodel}, where we also show the energy distribution of
the stellar photons. The important point is that luminosities of $\sim 10^{36-37}$
erg s$^{-1}$ can be obtained in the observer frame at EGRET's energy range, i.e. 100 MeV - 20 GeV.\\ 

\begin{figure}[!t]
\centering
\resizebox{12cm}{!} {\includegraphics{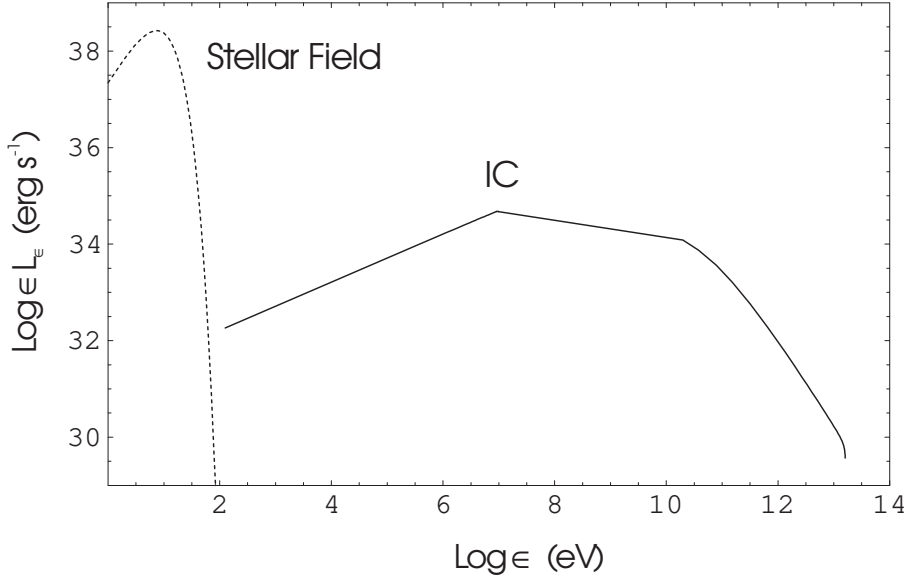}}
\caption{Inverse-Compton spectral energy distribution for a microquasar with a massive stellar companion. Leptons in the jet are assumed to have a power-law energy distribution with an index $p=2$ and a high-energy cut-off at multi-TeV energies. Notice the softening of the spectrum at high-energies due to the Klein-Nishina effect. \label{fig2-integral}}
\end{figure}

Another similar example can be seen in Fig.~\ref{fig2-integral} where the important difference with the case of Fig.~\ref{fig1-toymodel} is that the leptons of the jet are assumed to have a high energy cut-off in the TeV range\footnote{We saw in Section~\ref{the jet} that the existence of TeV electrons in microquasar jets (at least at large scale) have been recently inferred from the detection of synchrotron X-ray emission in XTE J1550-564 (Corbel et al. 2002).}. This situation corresponds to an electron Lorentz factor of $\gamma=10^{7.5}$. The stellar companion is also a high-mass type corresponding to an O9 I star.\\

The spectral energy distribution of the IC emission has a peak at MeV energies, due to the effect of losses on the particle spectrum\footnote{This physical effect will be described in Chapter~\ref{External fields}.}. At energies above $10^{10}$ eV the Klein-Nishina effect becomes important and the spectrum softens significantly. The Klein-Nishina spectral energy distribution resulting from a power law electron energy distribution is not a power law, and it is not possible to assign a unique spectral index to it. The photon spectral index at the MeV-GeV band is $\Gamma\sim 2$. At TeV energies, where the spectral index is $\Gamma>3$, the luminosity is $L(E>1\;{\rm TeV})\sim 10^{32}$ erg s$^{-1}$. Hence, one of these microquasars located at a few kpc might be a detectable TeV gamma-ray source. For instance some of the unidentified MeV sources detected in the galactic plane by COMPTEL (Zhang et al. 2002, 2004) might correspond to sources of this type, and they are potential targe
 ts for high-energy Cherenkov imaging telescopes like HESS or MAGIC.\\

\subsection{Precession vs Variability}

The companion star in a high-mass microblazar system not only
provides a photon field for inverse Compton interactions, but also
a gravitational field that can exert a torque onto the accretion
disk around the compact object. The effect of this torque, in a
non-coplanar system, is to induce a Newtonian precession of the
disk. If the jets are coupled to the disk, as it is usually
thought, then the precession will be transmitted to them. This
situation, which should not be confused with the geodetic
precession (a purely General Relativity effect), has been
extensively studied in the case of SS433 (Katz 1980) and
extragalactic sources like 3C 273 and A0235+16 (e.g. Romero et al. 2000, 2003b).\\

\begin{figure}[!t]
\centering
\resizebox{13cm}{!}{\includegraphics{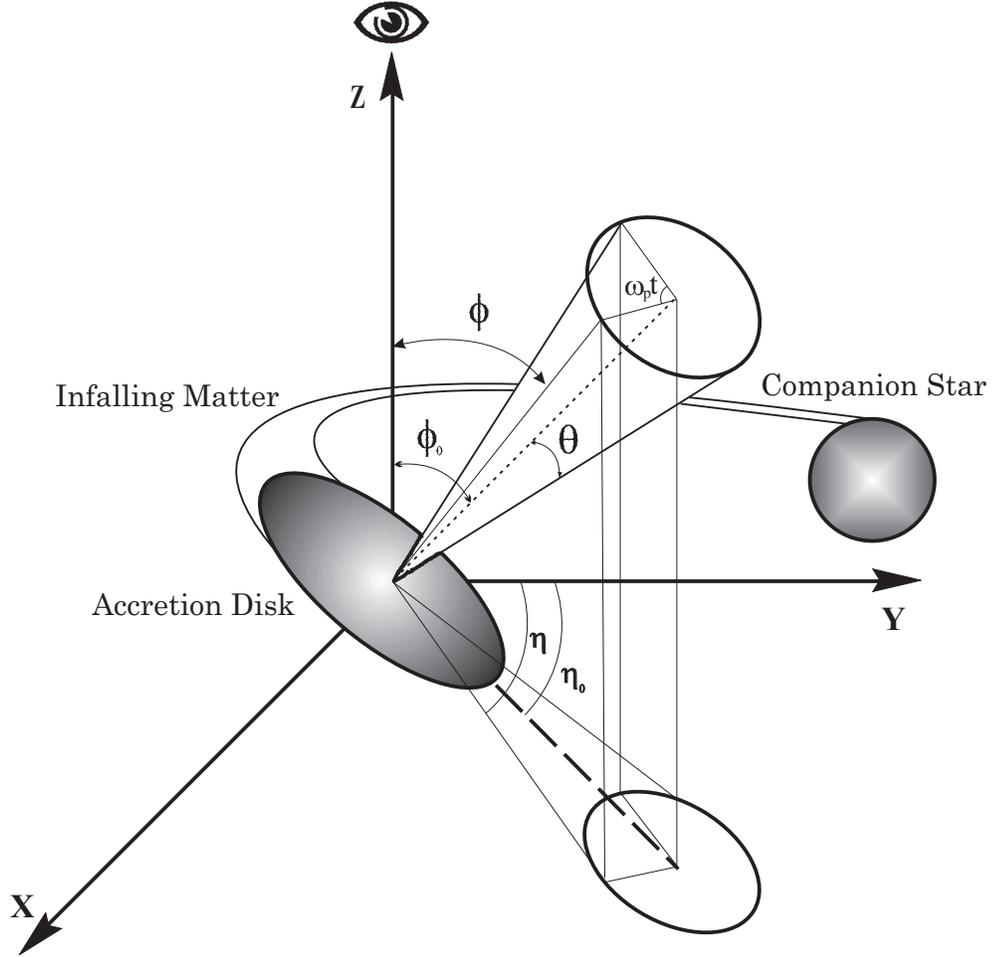}} 
\caption{Precessing jet model.} \label{fig2-toymodel}
\end{figure}

A sketch of the situation is presented in Fig.~\ref{fig2-toymodel}. The disk
Keplerian angular velocity is,

\begin{equation}
 \omega_{\rm d}=\left(\frac{GM}{r_{\rm d}^3}\right)^{1/2},
\end{equation} 

\vspace{0.4cm}
 
\noindent where $M$ is the mass of the compact object and
$r_{\rm d}$ is the radius of the precessing part of the disk. The
orbital period is $T_{\rm m}$ and the orbital radius is given by
Kepler's law, with $m$ the mass of the star: 

\begin{equation}
r_{\rm m}^3= \frac{G}{4\pi^2}(m+M) T_{\rm m}^2\;.
\end{equation}

\vspace{0.4cm}

The angular velocity of the tidally induced precession can then be
approximated by,

\begin{equation}\label{eq:omega-prec}
 \omega_{\rm p}\approx - \frac{3Gm\cos\theta}{4 r_{\rm m}^3 \omega_{\rm d}},
\end{equation}

\vspace{0.4cm}

\noindent where $\theta$ is the half-opening angle of the precession cone (Katz 1980, Romero et al. 2000). We can now introduce a time-parametrization of the jet's viewing angle as (see Abraham \& Romero 1999):

\begin{eqnarray}
    \phi(t)&=&\arcsin\left[x^{2}+y^{2}\right]^{1/2}\label{para}
\end{eqnarray}
\begin{eqnarray}   x&=&(\cos\theta\sin\phi_{0}+\sin\theta\cos\phi_{0}\sin\omega_{p}t)\cos\eta_{0}-\nonumber\\
    &&-\sin\theta\cos\omega_{p}t\sin\eta_{0}\nonumber \\
    y&=&(\cos\theta\sin\phi_{0}+\sin\theta\cos\phi_{0}\sin\omega_{p}t)\sin\eta_{0}+\nonumber\\
    &&+\sin\theta\cos\omega_{p}t\cos\eta_{0}\nonumber,
\end{eqnarray}

\vspace{0.4cm}

\noindent where all angles are defined in Fig.~\ref{fig2-toymodel}.\\

\begin{figure}[!t]
\centering
\resizebox{10cm}{!}{\includegraphics{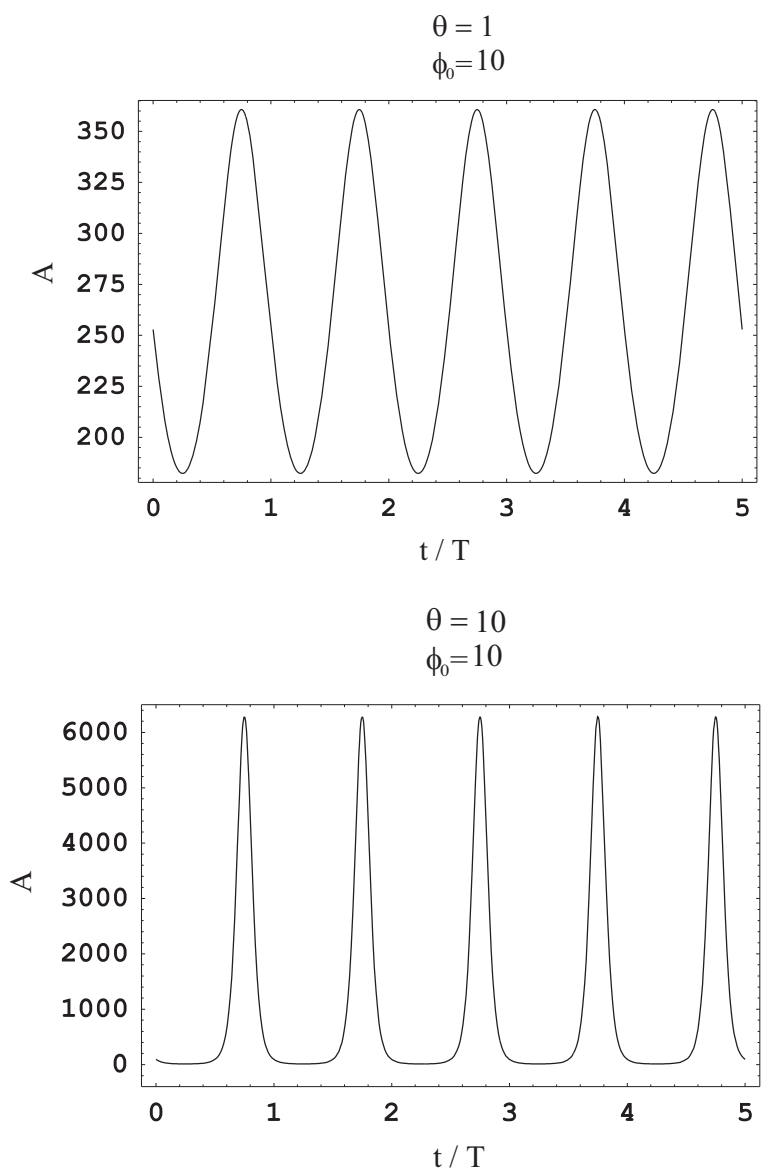}} 
\caption{ Variation of the amplification factor for continuous jet emission
as a function of time in the precessing microblazar model (for two
different opening angles). Time units are normalized to the
precessing period.} \label{fig3-toymodel}
\end{figure}

In Fig.~\ref{fig3-toymodel} we show the time evolution of the boosting
amplification factor of the gamma-ray emission for the case of a
continuous jet for two different sets of geometrical parameters
(viewing angle of 10 deg and precession half-opening angle of 1
and 10 deg). Time units are normalized to the precession period
$T$.\\

We see that the flux density can change by a factor of
$\sim6 \times 10^3$ in a single period. A very weak, otherwise
undetected gamma-ray microblazar, can increase its flux due to the
precession and then enter within the sensitivity of an instrument
like EGRET, producing a variable unidentified gamma-ray source.
The duty cycle (i.e. the fraction of time in which the source is
highly variable) in this example is $\sim 0.2 T$, so the source
could appear in several EGRET viewing periods.\\

Just to give a
feeling of the magnitudes involved, we mention that for a black
hole of 4 $M_{\odot}$, an O7Ia stellar companion, an orbital period
of 10 days, a half-opening angle of $10^{\circ}$ for the
precession, and an accretion disk of $\sim5\;10^{11}$ cm ($\approx
5\;10^{-2}$ a.u.), we get a precession period $T\sim100$ days. The
source could then be detectable $\sim70$ days per year. This is
consistent with some EGRET observations of highly variable sources
(e.g., the case discussed by Punsly et al. 2000).\\

\section{\label{discussion3.4}Discussion}

Taking into account that there exist more than $\sim130$ high-mass
X-ray binaries detected so far\footnote{See Section~\ref{SandP}.} and that
this number should be a small fraction of the total number of
these objects in the Galaxy, it is not unreasonable to expect the
existence of a few tens of microblazars at mid and low galactic
latitudes that could be responsible for the variable galactic
gamma-ray sources detected by EGRET. In particular, we already mentioned that
Paredes et al. (2000) have proposed that the microquasar LS 5039,
a massive X-ray binary with persistent non-thermal radio emission,
could be physically associated with the gamma-ray source 3EG J1824-1514.
In fact they suggested that the observed gamma-ray flux is the result of
EC scattering of UV photons from the high-mass
stellar companion. Our calculations support that suggestion (see also the more recent treatment by Bosch-Ramon \& Paredes 2004).\\ 

The recently discovered X-ray transient V4641 Sgr, which seems to harbor a
$\sim 9$ $M_{\odot}$-black hole and displays extreme superluminal
velocities, could also be a microblazar (Orosz et al 2001). Since the
stellar companion is a late B-type star, external Compton
gamma-ray production is not expected to be very efficient in this
case, but its high-energy emission could fall within GLAST
sensitivity. Another interesting candidate, from the theoretical
point of view, is the high-mass X-ray binary LS I +61 303, which presents a one-sided jet at milliarcsecond
scales and evidence for a precessing accretion disk (Massi et al.
2001, 2004). The intrinsic jet velocity, however, seems not to be very
high: $\sim0.4c$, but the source location is consistent with its
identification with a highly variable gamma-ray source 3EG
J0241+6103 (Tavani et al. 1998, Torres et al. 2001a). A recent discussion of this object as a potential gamma-ray source can be found in Massi (2004a, b)\\

Perhaps the best way to identified precessing gamma-ray
microblazars is through the detection of the electron-positron
annihilation feature in their spectra. This annihilation signature
should appear as a broad, blueshifted (by a factor $D$) line in
the spectrum at a few MeV, exactly within the energy range of IBIS
imager of the INTEGRAL satellite. Due to the
precession of the jet, the Doppler factor will change periodically
with time, and hence the position of the annihilation peak should
oscillate in energy in the lab frame around a mean value (for a
detailed discussion of the phenomenon see Abraham et al. 2001).
Chandra X-ray observations of non-thermal radio sources within the
EGRET location error boxes (a complete list of these sources is
given by Torres et al. 2001a) could help to find candidates to new
microquasars (through the detection of X-ray disk emission), and
then INTEGRAL exposures could be used in an attempt to find the
annihilation line. Such a detection would be a remarkable
discovery, since it would establish, at a same time, the matter
content of the jets in microquasars, and would help to clarify the
nature of some variable unidentified high-energy gamma-ray
sources. Unfortunately, INTEGRAL sensitivity seems not to fulfill pre-launch expectations and the detection of the annihilation line might require prohibitively long integration times in most cases.\\

To finish this chapter, a comment that concerns the fact that not all high-mass microblazars should be confined to the galactic plane.
Recent direct measurements of proper velocities in high-mass microquasars (Rib\'o et al. 2002) show that these objects can present very high
velocities and then could be ejected from the galactic plane. This
is also supported by the discovery of V4641 Sgr at $b\sim -4.8$
deg (estimated distance: $7.4\leq d \leq 12.3$ kpc). In the case
of microquasars with high-mass companions, which are young
objects, we could expect to find them up to distances $\sim100$ pc
(Rib\'o et al. 2002) or even more (e.g. V4641 Sgr) from the
galactic plane. 

\newpage
\thispagestyle{empty}
\phantom{.}

\chapter{\label{External fields}External Fields: Jet Interactions}
\thispagestyle{empty}

\newpage
\thispagestyle{empty}
\phantom{.}

\newpage
\vspace*{5.3cm}

\section{External Photon Fields: the Disk and the Corona}

We briefly recall the reader from Chapter~\ref{Microquasars} that the soft X-ray blackbody component in the SED of MQs is usually understood as emission from a cold, optically thick accretion disk, whereas the power law
component is thought to be originated in an optically thin hot
corona by thermal Comptonization of photons from the disk --disk/corona model-- or locally produced by synchrotron and Bremsstrahlung radiation --ADAF model. The hot corona would fill the inner few tens of gravitational radii around the black hole. The accretion
disk penetrates only marginally in the coronal region. We already saw that in the hard
state the thermal X-ray emission is dominated by the corona, with
typical luminosities of a few times $10^{37}$ erg s$^{-1}$. In the
soft states, the disk approaches to the black hole and then most
of the energy dissipation occurs through it.\\

\begin{figure}
\centering
\resizebox{12cm}{!}{\includegraphics{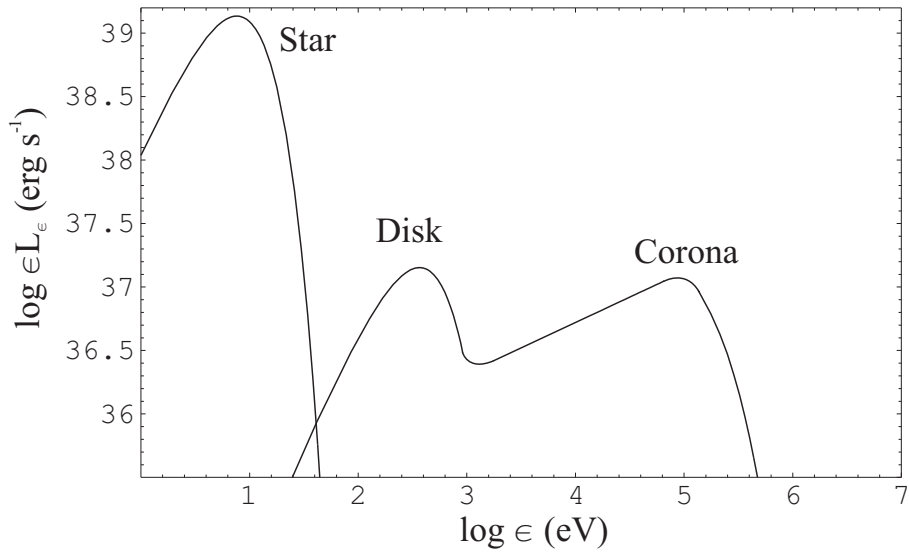}} 
\caption{ External photon fields to which the jet is exposed.} \label{fig2-CygX1}
\end{figure}

All photon
fields that interact with the jet are shown in Fig.~\ref{fig2-CygX1}. The
photon field of the corona is assumed to be isotropic. External
inverse Compton interactions with these photons can be treated in the Klein-Nishina regime since $\gamma \epsilon_{0} \gg 1$, with $\epsilon_{0}$ the peak of the spectral energy distribution of the coronal photons in units of the electron rest energy
($\epsilon_{0} m_{\rm e} c^2\sim 100$ keV).\\

In the case of the photons from the accretion disk and the
companion star we can make calculations in the Thomson regime. Taking into
account that the photons from the disk come from behind the jet,
an additional factor $(1-\cos\phi)^{(p+1)/2}$ that
reduces the effects of beaming has to be introduced (Dermer et al. 1992). The stellar
photons, instead, can be treated as an isotropic field, at least as a first approximation as we already saw when we studied the interaction of the jet with this photon field in Chapter~\ref{High-mass}.\\

The emerging spectrum of the specific luminosity can be approximated by a power law of index $(p-1)/2$ in the Thomson regime, whose complete expression is given by (\ref{lumred}). In the Klein-Nishina regime, where numerical
integrations are necessary, the results significantly depart from
a power law, resulting in a softer spectrum (see Fig.~\ref{fig2-integral}).\\

The Compton losses in the different regions will modify the
injected electron spectrum, introducing a break in the luminosity power law
at the energy at which the cooling time equals the escape time.
This will occur at (e.g. Longair 1997): 

\begin{equation}
\gamma_{\rm b}=\frac{3m_{\rm e} c^2}{4\sigma_{\rm T} \Gamma^2 U t_{\rm esc}}\hspace{0.2cm}.
\end{equation}

\vspace{0.4cm}

\noindent Here $t_{\rm esc}$ is the average time spent by the particles in the field region
(typically $t_{\rm esc}\sim l/c$, with $l$ the linear size), and
the rest of the symbols have their usual meanings. The spectrum
will steepen from an index $p$ to $p+1$ for energies higher than
$\gamma_{\rm b}$. After the interaction of the jet with the photons from the 
disk and the corona, the modified spectrum will be injected in the
stellar photon field region, suffering further losses and
modifications.\\

\section{Application to Cygnus X-1}

\subsection{Main Properties of the System and Observations}

Cygnus X-1 is the most extensively studied black hole candidate in
the Galaxy. It is a very bright X-ray binary with a compact object
of $\sim 10.1$ $M_{\odot}$ and a companion O9.7 Iab star of
$\sim17.8$ $M_{\odot}$ (Herrero et al. 1995), at an estimated
distance of $\sim 2$ kpc (e.g. Gierli\'nski et al. 1999 and
references therein). As in other sources of this type, the X-ray
emission switches between soft and hard states, being most of the
time in the latter. The spectrum in both states can be
approximately represented as the sum of a blackbody plus a power
law with exponential cut-off (e.g. Poutanen et al. 1997). During
the soft state the blackbody component is dominant and the power
law is steep, with a photon spectral index $\Gamma\sim 2.8$ (e.g.
Frontera et al. 2001). During the hard state more energy is in the
power law component, which is then harder, with photon index
$\sim 1.6$ (e.g. Gierli\'nski et al. 1997).\\

\begin{figure}[t!]
\begin{center}
\includegraphics[width=10cm,height=10cm]{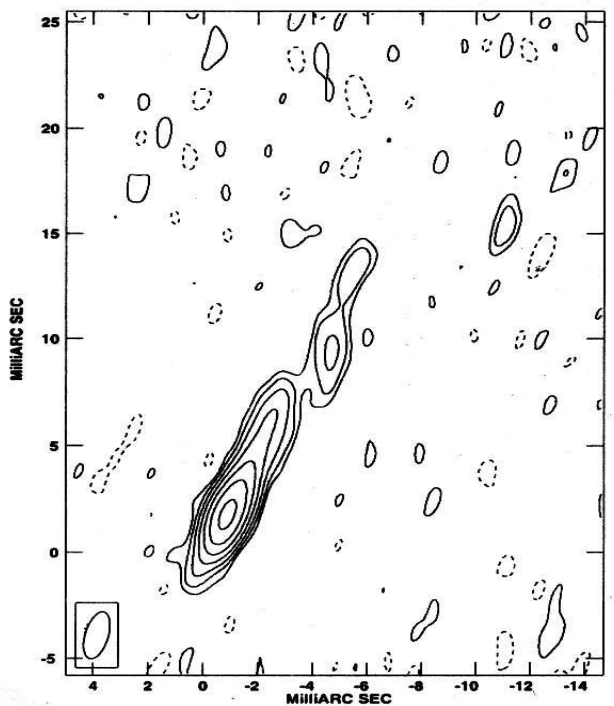}
\caption{\label{Cyg-jet} The radio image of the continuous jet of Cygnus X-1 at 8.4 GHz obtained by Stirling et al. (2001) when the source was in the Low/Hard state.}
\end{center}
\end{figure}

Cygnus X-1 has a persistent, mildly variable, compact continuum
counterpart of flat spectrum (e.g. Pooley et al. 1999). During
many years, evidence for non-thermal radio jets in Cygnus X-1 was
lacking, despite the efforts of the observers (e.g. Mart\'{\i} et
al. 1996). Finally, the jet was detected by Stirling et al. (2001)
at milliarcsecond resolution using VLBA observations -  see Fig~\ref{Cyg-jet}. The jet-like
feature extends up to $\sim 15$ mas with an opening angle of less
than 2 degrees. The spectrum seems to be flat, and no counterjet is
observed. The total radio emission at 8.4 GHz is $\sim 11$ mJy,
with variations of $\sim2$ mJy over timescales of 2 days (Stirling
et al. 2001). The average angle with the line of sight, if the jet
is perpendicular to the disk, seems to be $\sim 30^{\circ}$
(Fender 2001).\\

Recently, the Interplanetary Network detected a transient
soft-gamma ray event from the general direction of Cygnus X-1
(Golenetskii et al. 2002). Analysis of previous data indicates
that at least other two events were observed during 1995. These
latter events were also detected by BATSE instrument on the
Compton Gamma-Ray Observatory, suggesting that they were
originated in Cygnus X-1 (Schmidt 2002). The luminosities above 15
keV of the outbursts were in the range $1-2\times 10^{38}$ erg
s$^{-1}$, much higher than the typical thermal luminosity in the
hard state.\\

\subsection{Spectral Energy Distributions}

In this section we suggest that these soft-gamma ray flaring events can be
interpreted in terms of non-thermal microblazar activity. We study the effects of the interaction of the relativistic jet with the ambient photon fields from the
accretion disk, the corona, and the companion star, and we
calculate the expected non-thermal contribution to the keV-MeV
spectrum. In this context the recurrent character of the events can be explained
through variable Doppler boosting originated in the precession of
the jet.\\

\begin{figure}[!t]
\centering
\resizebox{12cm}{!}{\includegraphics{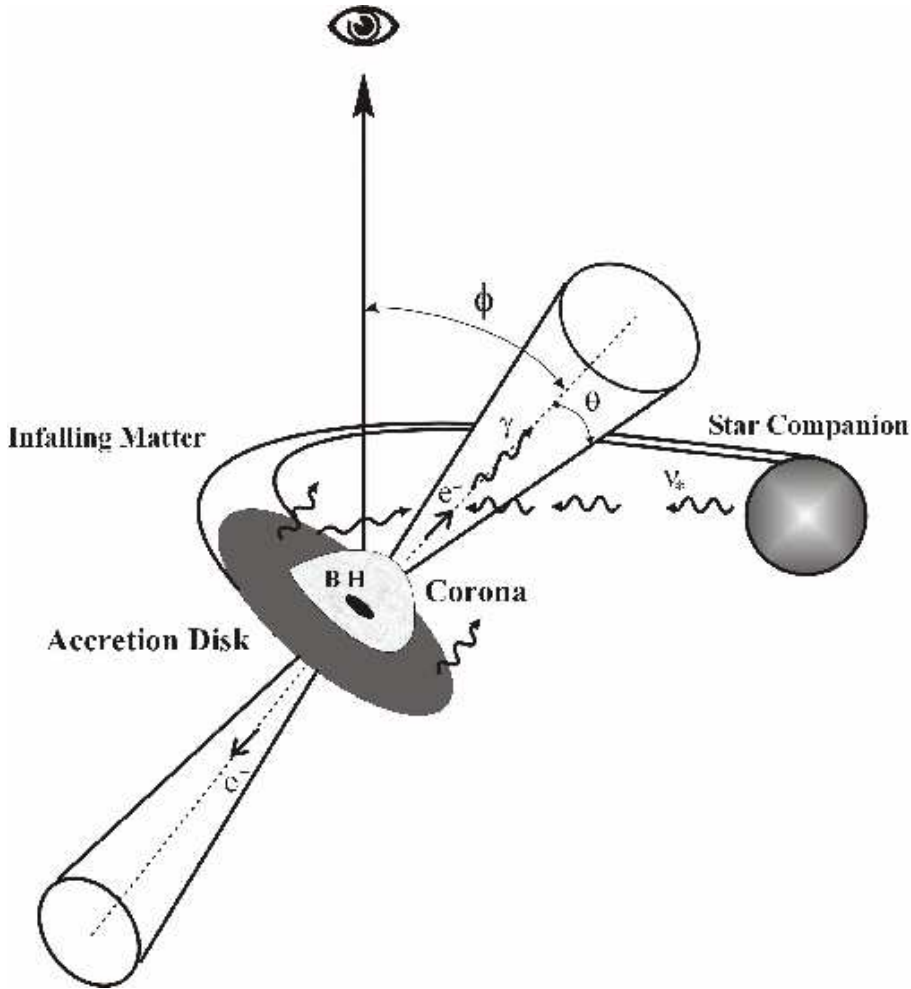}} 
\caption{ Sketch of the general situation discussed in this chapter. A
relativistic jet is injected close to the black hole in Cygnus
X-1. This jet must traverse photon fields created by the cold
accretion disk, the hot corona, and the stellar companion. Inverse
Compton up-scattering of some of these photons is unavoidable. Here $\theta$ is the half-opening angle of the cone defined by a possibly precessing jet.}
\label{fig1-CygX1}
\end{figure}

A sketch of the injection of a relativistic
leptonic jet at a few Schwarzschild radii from the central black
hole and its subsequent propagation through the ambient photon
fields is shown in Fig.~\ref{fig1-CygX1}. The individual
electrons have Lorentz factors $\gamma$ in the lab frame and the
flow is assumed to have a bulk Lorentz factor $\Gamma$. In
accordance to the disk/jet symbiosis model (e.g. Falcke \&
Biermann 1999, Markoff et al. 2001) and the estimated accretion
rate of Cygnus X-1, $\dot{M}\sim 10^{-8} M_{\odot}$ yr$^{-1}$ (e.g.
Poutanen et al. 1997), we adopt a mean particle density
$n=10^{14}$ cm$^{-3}$ for the jet at $10 r_{\rm Sch}$ from the black
hole. In our calculations we have adopted two different values for the
original electron energy index\footnote{Let us remember that the electron energy distribution is assumed to be a power law (see Section~\ref{IC}), 
\begin{eqnarray}
n(\gamma)=\frac{k}{4\pi}D^{2+p}\gamma^{-p}P(\gamma_{1}D,\gamma_{2}D,\gamma)\nonumber.
\end{eqnarray}
}: a hard index $p=1.5$ and a steeper
index $p=2.3$. The jet was assumed forming an angle of 30 degrees
with the line of sight, and a bulk Lorentz factor $\Gamma=5$ was
adopted. The initial part of the jet was modeled as a cylindrical
structure.\\

The jet will traverse first the coronal region. The parameters
that characterize this region and its photon field in the hard
state change with time. We shall assume here the typical values
given in Table~\ref{tab-cyg}. Following Poutanen et al. (1997), the coronal region was considered as a spherical region of $\sim 500$ km in radius.\\

\begin{table}[!t]

\begin{center}
\caption{\label{tab-cyg}Observational characteristics of Cyg X-1 
in its soft and hard states (from Poutanen et al. 1997).}
\vspace{0.4cm}
\begin{tabular}{lcc}
\hline
\hline
{Parameter} \hspace{4cm}& {Hard state} & \hspace{3cm}{Soft state}\\
\hline
\hline 
$\Ls$$^{\;\;\rm a}$  \hspace{4cm}& $1 \times 10^{37}$ erg/s& \hspace{3cm}$4 \times 10^{37}$ erg/s \\
$\Lh$$^{\;\;\rm b}$  \hspace{4cm}&  $4\times 10^{37}$ erg/s &  \hspace{3cm}$1 \times 10^{37}$ erg/s \\ 
$\alpha$$^{\rm c}$  \hspace{4cm}& 0.6& \hspace{3cm} 1.6   \\ 
$\Ts$$^{\rm d}$ \hspace{4cm}& 0.13 keV& \hspace{3cm} 0.4 keV \\  
$E_c$$^{\rm e}$  \hspace{4cm}&  150 keV& \hspace{3cm} $\gtrsim 200$ keV \\ 
$C$$^{\rm f}$    \hspace{4cm} &   0.4& \hspace{3cm}    0.55       \\ 
\hline
\hline
\end{tabular}
\end{center}
$^{\rm a}${Observed soft luminosity
 (total luminosity below $\sim 1$ keV in the hard state
and below $\sim 3$ keV in the soft state)} \\
$^{\rm b}${Observed hard luminosity} \\
$^{\rm c}${Energy spectral index} \\
$^{\rm d}${Temperature of the soft component}\\
$^{\rm e}${Cut-off energy of hard component} \\
$^{\rm f}${Covering factor of the cold matter}

\end{table}

\begin{figure}[!t]
\centering
\resizebox{12cm}{!}{\includegraphics{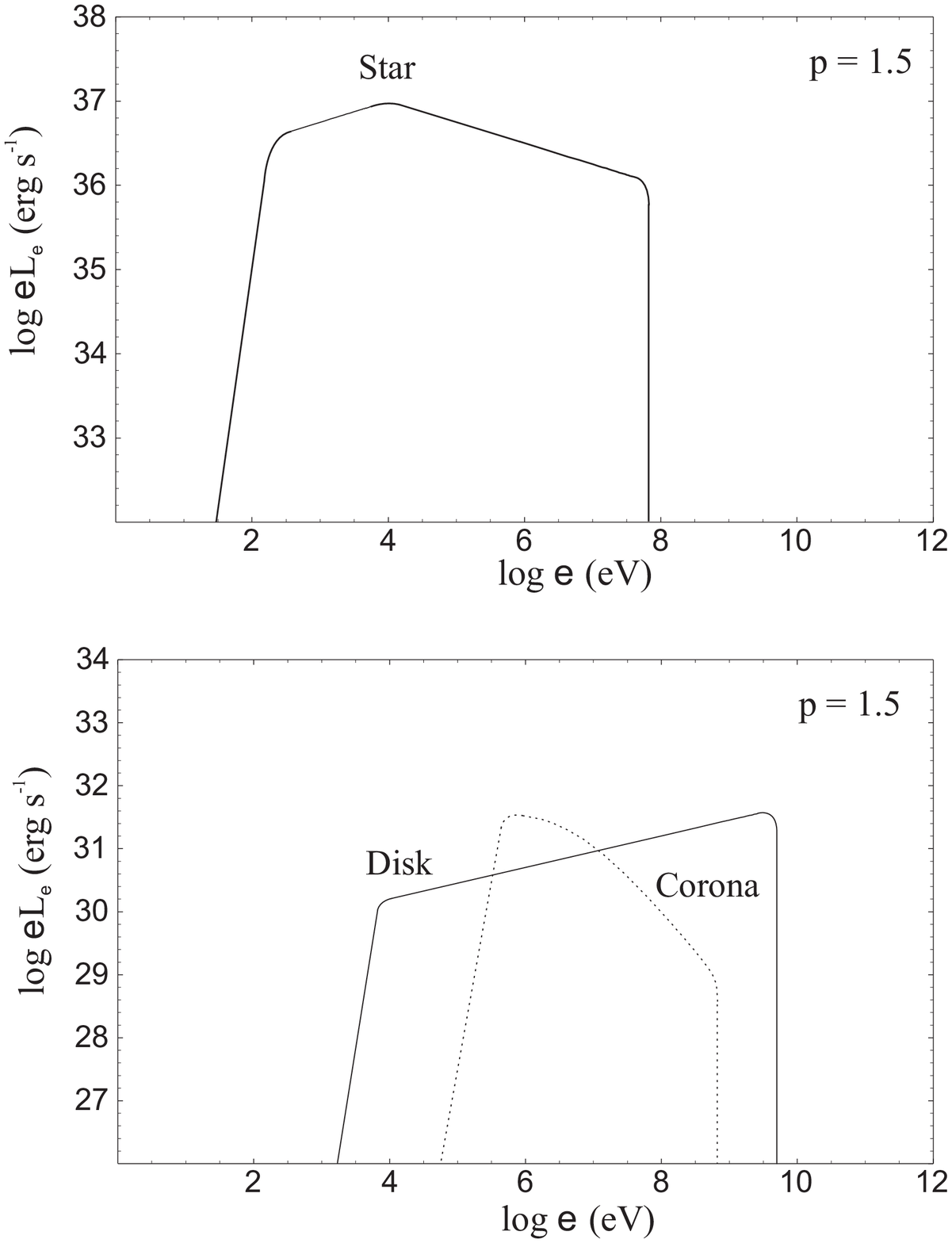}}
\caption{Results of the model for an injection electron spectrum with index
$p=1.5$ in a cylindrical jet forming a viewing angle of 30
degrees. The bulk Lorentz factor is $\Gamma=5$ and the electron
power law extends from $\gamma_1=2$ to $\gamma_2=10^3$. Three
different components are shown, resulting from the up-scattering
of photons from the star (top panel), the disk and the corona (bottom panel, solid and dashed lines respectively). Notice that
the contribution from the coronal photons is not a power law
because of the Klein-Nishina effect.} \label{fig3-CygX1}
\end{figure}

\begin{figure}[!t]
\centering
\resizebox{12cm}{!}{\includegraphics{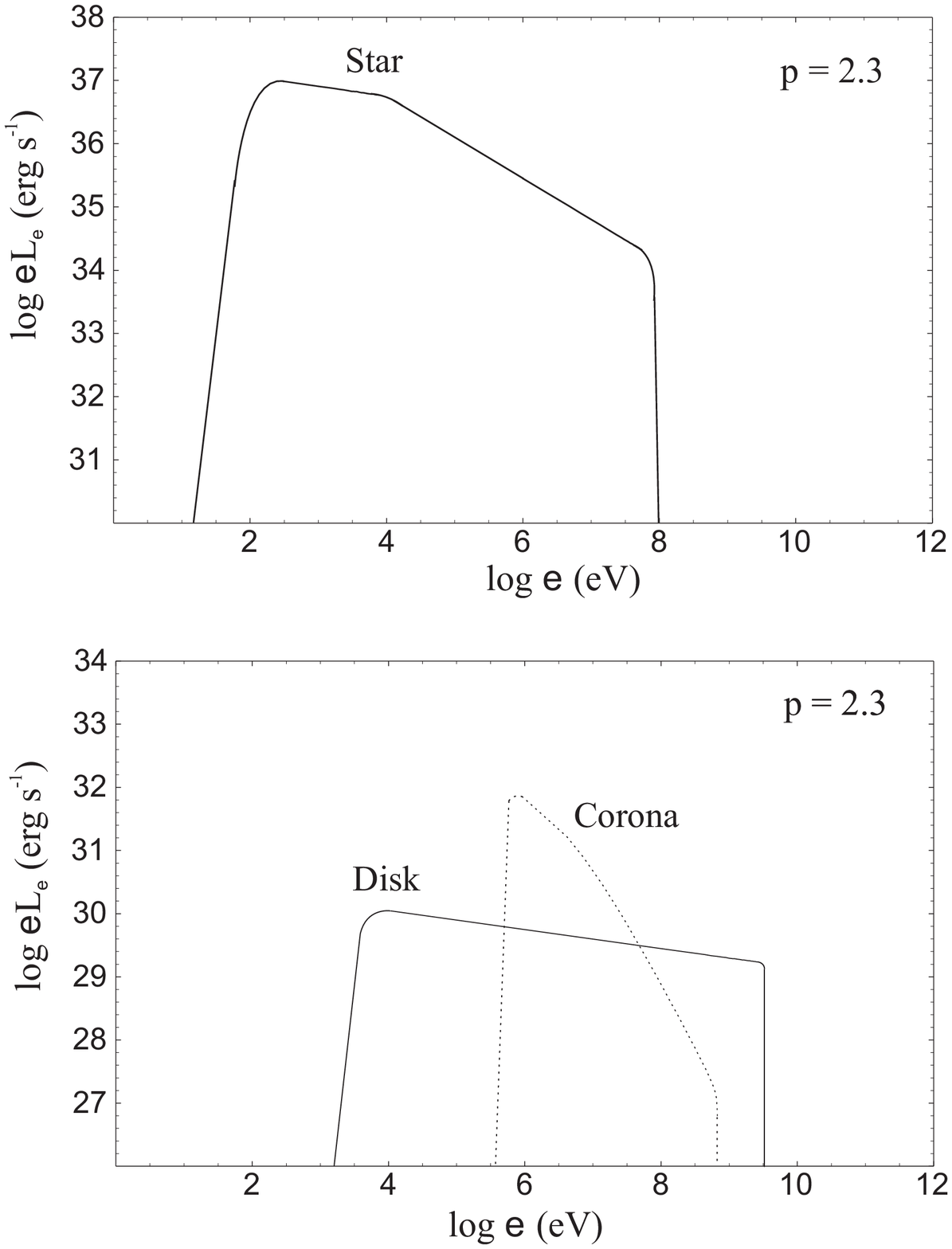}} 
\caption{ Idem Fig.~\ref{fig3-CygX1} for $p=2.3$.} \label{fig4-CygX1}
\end{figure}

The results obtained for the two assumed electron distributions are shown in Fig.~\ref{fig3-CygX1} and Fig.~\ref{fig4-CygX1}. There, we show all the components due to external Compton scattering of the different photon fields by the injected electrons in the lab frame. This non-thermal contribution should be added to the
thermal components shown in Fig.~\ref{fig2-CygX1} in order to recover the total
emission. We can see that the non-thermal emission is dominated by
the up-scattering of the stellar photons. Moreover, gamma-rays produced within the coronal region will be mostly absorbed in
the local field through pair creation. Using the simple formulae
by Herterich (1974) along with the adopted parameters for the
coronal region, we estimate an optical depth $\tau\sim 1$ for
photons of 5 MeV. The probability for a 10 MeV photon to escape
from the corona is only 0.1.\\

\begin{figure}[!t]
\centering
\resizebox{12cm}{!}{\includegraphics{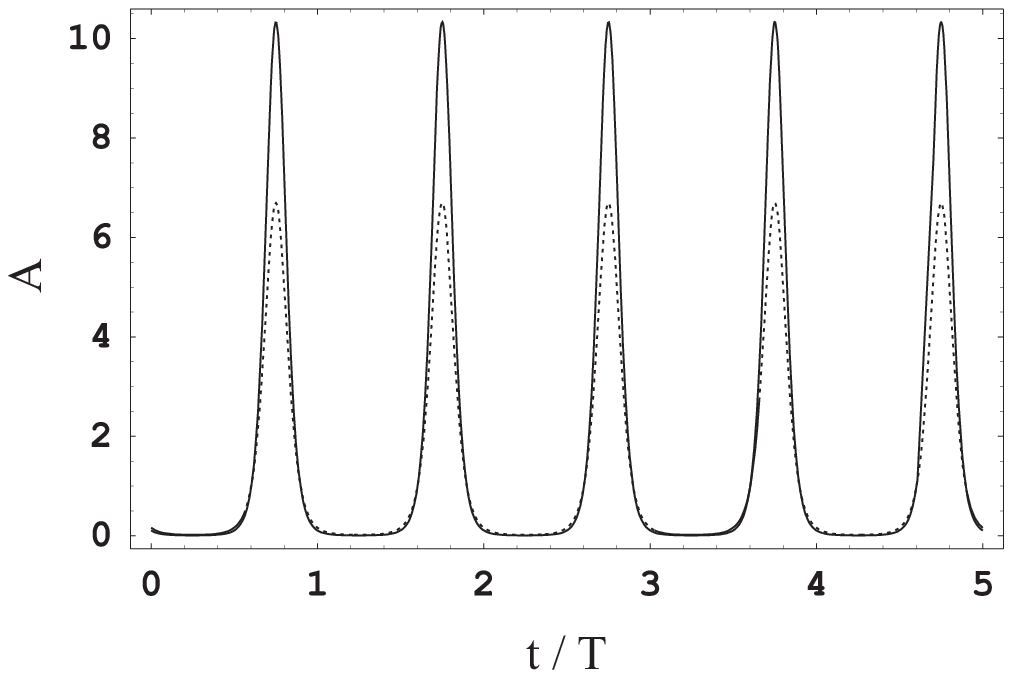}}
\caption{ Variation of the amplification factor for the external inverse
Compton emission from the jet (electron power-law indices of
$p=1.5$ --dots-- and $p=2.3$ --solid line--) as a function of time
in the precessing microblazar model for Cygnus X-1 (precessing
opening angle of 16.5 degrees). Time units are normalized to the
precessing period.} \label{fig5-CygX1}
\end{figure}

For a viewing angle of 30 degrees, we still have the thermal
emission dominating by a factor of 5. But if we introduce the
gravitational effects of the companion star, then precession of
the disk should occur. In Fig.~\ref{fig5-CygX1} we show the modification of the  beaming
amplification factor of the external inverse Compton emission for
a precessing angle of 16.5 degrees. The time axis is normalized in
units of the period $T$. We see that there is a variation of about
1 order of magnitude in the emission measured in the observer's
frame because of the precession. This means that when the jet is
closer to the line of sight, the non-thermal luminosity can reach
values of $\sim 10^{38}$ erg s$^{-1}$, as observed in the
recurrent outbursts detected by the Interplanetary Network. The transit through the peak of flux magnification can be very fast in
the observer frame, leading to quick and transient states when
the total flux is dominated by the non-thermal contribution.\\

The angular velocity of the tidally induced precession can be
approximated by (\ref{eq:omega-prec}). For a half-opening angle of the precession cone,  $\theta=16.5$ deg as
we have assumed, the observed orbital period of 5.6 days, and a
precessing period of $\sim 140$ days (Brocksopp et al. 1999, see
below), we get a disk size of $\sim 3.9\times 10^{11}$ cm, quite
reasonable for a wind-accreting system like Cygnus X-1.\\

\section{\label{discussioncap4}Discussion}

The above outlined model incorporates the different known
components of Cygnus X-1, accretion disk, corona, stellar
companion, and relativistic jet, in an integrated picture where
transient non-thermal outbursts are a natural and expected result.
The amplitude of these outbursts can be similar to what has been
recently observed in some intriguing flaring episodes in this
source.\\

Our model is different from the model proposed by
Georganopoulos et al. (2002) not only because we incorporate the
effects of precession, but also because we do not attempt to
explain the bulk of X-ray emission as non-thermal {\sl all the
time}. This emission is normally dominated by thermal
Comptonization in the hot corona around the black hole, except
during the {\sl microblazar} phase, and in this case we
incorporate the effects of the interaction of the jet with the
corona in our calculations. We emphasize that, as it is shown in
Fig.~\ref{fig3-CygX1} and Fig.~\ref{fig4-CygX1}, during the transient microblazar phase the X- and
soft gamma-ray spectrum will be softer than in the normal hard
state, when the coronal emission dominates. This is an unavoidable
consequence of the steepening produced by Compton losses in the
injected electron spectrum and can be used to test our proposal,
not only through new observations of Cygnus X-1, but also of other
potentially precessing MQs as LS 5039 and LS I +61 303 (Paredes et al. 2002a; Massi et al. 2004).\\

Brocksopp et al. (1999) have found multiwavelength
evidence for the presence of a $142.0\pm 7.1$ days period in
Cygnus X-1. The optical and X-ray period seem to originate in
the precession of the accretion disk (Brocksopp et al. 1999),
whereas the modulation at radio wavelengths is probably produced
by the associated precession of the jet (see Pooley et al. 1999).
The morphology of the extended radio jet, with a clear bend, is
also consistent with a precession of the inner beam (Stirling et
al. 2001). The periodic signal in the radio lightcurve, however,
is not expected to be as strong as at high energies since the
magnification factor for the synchrotron emission goes as
$D^{(3+p)/2}$, whereas as we already saw for the external Compton component, it goes
as $D^{2+p}$.\\

The time lag between the two high-energy flares observed in 1995
is $\sim 75$ days, about a half of the value reported by Brocksopp
et al. (1999), but since Cygnus X-1 is a wind-accreting system
variations in the period along a span of several years are
possible. Certainly, more observations on longer time spans are
necessary to constrain the dynamical models. In the model
presented here the duty cycle of the microblazar phase is rather small,
$\sim 10\%$. Future X-ray observations of non-thermal flares can
be used for a better determination of the geometric parameters.\\

We saw that in the case of Cyg X-1 an interesting resulting characteristic is that most of the gamma-rays produced within the coronal region will be absorbed by pair
production. Sooner or later these pairs will annihilate producing
a broad, blueshifted feature in the MeV spectrum that might be detected by the INTEGRAL satellite. It is expected that this satellite will probe Cygnus X-1
spectrum and its temporal evolution at this energy range, helping
to test and constrain the model here proposed\footnote{It is expected though not certain, taking into account the remarks we made in Section~\ref{discussion3.4} about the after-launched resulting INTEGRAL sensitivity.}.  

\newpage
\thispagestyle{empty}
\phantom{.}

\chapter{\label{Hadron}Hadronic Gamma-Ray Emission}
\thispagestyle{empty}

\newpage
\thispagestyle{empty}
\phantom{.}

\newpage
\vspace*{5.2cm}

\section{Components of the Model}

The model for gamma-ray production in MQs proposed in Chapters~\ref{High-mass} and \ref{External fields} was based on the process of IC upscattering of seed photons from the high-mass stellar companion, the disk and the corona. In that context the resulting gamma-ray emission is dominated by the EC of the stellar photon field. Besides, pair creation absorption mechanism in the disk and coronal X-ray field might quench the contributions from these two regions to the total gamma-ray emission.\\

In this chapter we present a new mechanism for the generation
of high-energy gamma-rays in MQs that is based on
hadronic interactions occurring outside the coronal region.
The gamma-ray emission arises from the decay of neutral pions
created in the inelastic collisions between relativistic protons in the jet and the ions of the stellar wind. The requisites for the model are a windy high-mass stellar
companion and the presence of multi-TeV protons in the
jet\footnote{Interactions
of hadronic beams with moving clouds in
the context of accreting pulsars have been previously discussed in
the literature by Aharonian \& Atoyan (1996). For an early
discussion in a general context see Bednarek et al (1990).}. The
presence of hadrons in MQ jets like those of
SS 433 has been inferred from iron X-ray line observations (e.g.
Kotani et al. 1994, 1996; Migliari et al. 2002), although direct and clear
evidence exists only for this source so far.
In what follows we describe the model and
present the results of our calculations.\\

\subsection{Hadronic Jet}

\begin{figure}
\centering
\resizebox{12cm}{!}{\includegraphics{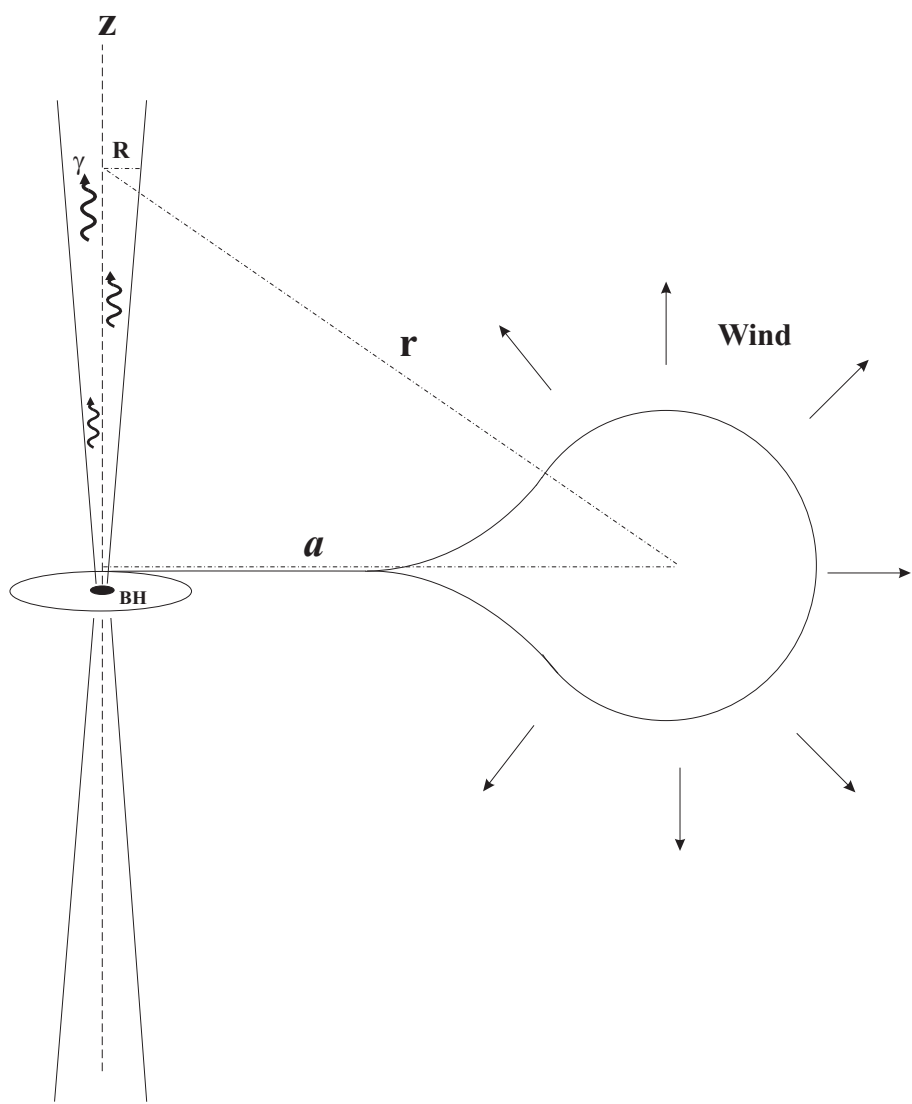}}
\caption{Sketch of the general situation in the hadronic windy MQ:
relativistic $e-p$ jet is injected close to the black hole in a
MQ with a high-mass stellar companion. The stellar wind penetrates on the jet from the sides. The
resulting interaction produces gamma-ray emission. In
the figure perpendicularity is assumed between the jet and the orbital
plane, but this particular assumption can be relaxed in a more general situation.} \label{fig1-Hadron}
\end{figure}

The general situation discussed in this chapter is shown in
Fig.~\ref{fig1-Hadron}. A binary system is formed by a black hole and a
high-mass early-type star. A relativistic $e-p$ jet is ejected perpendicularly to the accretion disk plane. For simplicity, we shall assume that this is also the orbital
plane, but this condition can be relaxed to allow, for instance, a
precessional motion or more general situations as discussed by Maccarone (2002), Butt et al. (2003), and Romero \& Orellana (2004).\\

The jet axis, $z$, is assumed to be normal to the orbital radius $a$. We shall
allow the jet to expand laterally, in such a way that its radius
is given by 

\vspace{0.2cm}

\begin{equation}
R(z)=\xi z^{\epsilon},
\end{equation}

\vspace{0.4cm}

\noindent with $\epsilon\leq 1$ and
$z_0\leq z\leq z_{\rm max}$. For $\epsilon=1$ we have a conical
beam. The jet starts to expand at a height $z_0\sim$ a few hundred
km above the black hole, outside the coronal region. The particle
spectrum of the relativistic $e-p$ flow is assumed to be a power
law,

\begin{equation}
 N'_{e,\;p}(E'_{e, \;p})= K_{e, \;p}\; E_{e,
\;p}'^{-\alpha_{\rm p}},
\end{equation} 

\vspace{0.8cm}

\noindent valid for $ E_{e, \;p}'^{\;\rm min}\leq  E'_{e,
\;p} \leq  E_{e, \;p}'^{\;\rm max}$, in the jet frame (denoted by the prime). The
corresponding particle flux will be,

\vspace{0.3cm}

\begin{equation}
J'_{e, \;p}( E'_{e, \;p})=\frac{c}{4\pi} N'_{e,\;p}(E'_{e,\;p}).
\end{equation}

\vspace{0.6cm}

\noindent Since the jet expands, the proton flux can be written as:

\vspace{0.4cm}

\begin{equation}
J'_p(E'_p)=\frac{c}{4 \pi} K_0
\left(\frac{z_0}{z}\right)^{\epsilon n} {E'_p}^{-\alpha_{\rm p}},
\label{Jp}
\end{equation}

\vspace{0.6cm}

\noindent where $n>0$ (a value $n=2$ corresponds to the conservation of the
number of particles, see Ghisellini et al. 1985). Using relativistic invariants, it can be
proven that the proton flux, in the observer (or lab) frame,
becomes (e.g. Purmohammad \& Samimi 2001)

\vspace{0.1cm}

\begin{equation}
J_p(E_p,\theta)=\frac{c K_0}{4 \pi} \left(\frac{z_0}{z}\right)^
{\epsilon n} \frac{\Gamma^{-\alpha_{\rm p}+1} \left(E_p-\beta
 {\displaystyle{\sqrt{E_p^2-m_p^2c^4}}} \cos \theta\right)^{-\alpha_{\rm p}}}{\left[\sin ^2
\theta + \Gamma^2 \left( \cos \theta -{\displaystyle{\frac{\beta
E_p}{\sqrt{E_p^2-m_p^2 c^4}}}}\right)^2\right]^{1/2}}\ \ ,
\label{Jp_lab}
\end{equation}

\vspace{0.6cm}

\noindent where $\Gamma$ is the bulk Lorentz factor, $\theta$ is the angle
subtended by the emerging photon direction and the jet axis, and $\beta$
is the corresponding velocity in units of $c$. The exponential dependence of the cross
section on the transverse momentum ($p_{\rm t}$) of the incident protons beams the
gamma-ray emission into an angle $\phi<cp_{\rm t}/m_{\rm p} \Gamma\sim 0.17/\Gamma$
along the proton direction, hence justifying the assumption that both directions are similar. Note that only photons emitted with angles similar to that of the inclination
angle of the jet will reach a distant observer, and thus $\theta$
can be approximated by the jet inclination angle.\\

In order to determine the matter content of the jet we will adopt
the jet-disk coupling hypothesis proposed by Falcke \& Biermann
(1995) and applied with success to AGNs (see also Falcke \&
Biermann 1996), i.e. the total jet power scales with the accreting
rate as: 

\begin{equation}
Q_{\rm j}=q_{\rm j} \dot{M}_{\rm disk} c^2,
\end{equation}

\vspace{0.7cm}

\noindent with $q_{\rm
j}=10^{-1}-10^{-3}$. The number density $n'_0$ of particles
flowing in the jet at $R_0=R(z_0)$ is then given by,

\begin{equation}
 c\pi R_0^2 n'_0= \frac{Q_{\rm j}}{m_p c^2}\;,
\end{equation} 
 
\vspace{0.6cm} 
 
\noindent where $m_p$ is the proton rest mass. From here we can obtain $n'_0$: 

\vspace{0.4cm}

\begin{equation}
n_0'=\frac{q_{\rm j} \dot{M}_{\rm disk}}{\pi c  R_0^2 m_p}.
\label{n01}
\end{equation}

\vspace{0.4cm}

\noindent Additionally,

\vspace{0.2cm}

\begin{equation}
n_0'=\int^{E_p'^{\rm max}}_{E_p'^{\rm min}}
N'_p(E'_p, \;z_0)\; d{E'_p}.
\end{equation}

\vspace{0.8cm}

\noindent Then, if $E_p'^{\rm max}>>E_p'^{\rm min}$, which is always the
case, we have,

\vspace{0.4cm}

\begin{equation}
K_0=n'_0 (\alpha_{\rm p}-1) (E_p'^{\:\rm min})^{\alpha_{\rm p}-1},\label{n02}
\end{equation}

\vspace{0.7cm}

\noindent which gives the constant in the power-law spectrum at $z_0$.\\

\subsection{The Wind from the High-Mass Companion Star}

Early-type stars, like OB stars, lose a significant fraction of
their masses through very strong supersonic winds. Typical mass
loss rates and terminal wind velocities for O stars are of the
order of $10^{-5}$ $\dot{M_{\odot}}$ yr$^{-1}$ and 2500 km
s$^{-1}$, respectively (Lamers \& Cassinelli 1999). At the base of
the wind, the density can easily reach $10^{-12}$g\,cm$^{-3}$.
Such strong winds provide a field of matter dense enough as to
produce significant hadronic gamma-rays when they penetrate on the jet from the sides.\\

The structure of the matter field will be determined essentially
by the stellar mass loss rate and the continuity equation:
$\dot{M_*}=4\pi r^2 \rho (r) v(r)$, where $\rho$ is the density of
the wind and $v$ is its velocity. Hence,

\begin{equation}
\rho(r)=\frac{\dot{M_*}}{4\pi r^2 v(r)}.
\end{equation}

\vspace{0.4cm}

\noindent The radial dependence of the wind velocity is given by (Lamers \& Cassinelli 1999):

\vspace{0.3cm}

\begin{equation}
v(r)=v_{\infty}\left(1-\frac{r_*}{r}\right)^{\zeta},
\end{equation}

\vspace{0.7cm}

\noindent where $v_{\infty}$ is the terminal wind velocity, $r_*$ is the
stellar radius, and the parameter $\zeta$ $\sim 1$ for massive
stars. Hence, using the fact that $r^2=z^2+a^2$ and assuming a gas
dominated by protons, we get the particle density of the medium
along the jet axis:

\begin{equation} 
n(z)=\frac{\dot{M_*}}{4\pi m_p v_{\infty}
(z^2+a^2)}\left(1-\frac{r_*}{\sqrt{z^2+a^2}}\right)^{-\zeta}\label{n(z)}
\end{equation}

\vspace{0.8cm}

\noindent The stability of a relativistic jet under the effects of an
external wind has been recently investigated by Hardee \& Hughes
(2003) through both theoretical analysis and numerical
simulations. Their results indicate that jets surrounded by
outflowing winds are in general more dynamically stable than those
surrounded by a stationary medium. Note that protons pertaining to
the wind can diffuse into the jet medium. The wind penetration into the jet outflow
depends on the parameter 

\vspace{0.2cm}

\begin{equation}
\varpi \sim v R(z)/D,
\end{equation}

\vspace{0.4cm}

\noindent where $v$ is velocity of wind, $R(z)$ is
the radius of the jet at a height $z$ above the compact object,
and $D$ is the diffusion coefficient. $\varpi$ measures the ratio
between the convective and the diffusive timescale of the
particles. In the Bohm limit, with typical magnetic fields $B_0\sim 1-10$ G,
$\varpi \leq 1$, and the wind matter penetrates the jet by diffusion.\\

\newpage

\section{\label{GREHM}Gamma-Ray Emission in the Hadronic Model}

The $p-p$ interaction results in the production of pions whose decay chain leads to gamma-ray and neutrino generation:\\ 

\begin{displaymath}
\begin{array}{ccccc}
	p+p & \longrightarrow & p + p + a \pi^0 + b(\pi^+ + \pi^-),& \textrm{a and b} \in \mathbb{N}& \\
	\pi^0 & \longrightarrow & 2  \gamma&&\\
 \pi^+ & \longrightarrow & \nu_{\mu} + \mu^+ & \longrightarrow &\nu_{\mu}+ e^+ + \nu_e + \bar{\nu}_{\mu}\\
 \pi^-& \longrightarrow & \bar{\nu}_{\mu} + \mu^- & \longrightarrow & \bar{\nu}_{\mu}+e^-+ \bar{\nu}_e + \nu_{\mu}
\end{array}
\end{displaymath}

\vspace{0.4cm}

The differential gamma-ray emissivity from $\pi^0$-decays is:

\vspace{0.4cm}

\begin{equation}
q_{\gamma}(E_{\gamma})= 4 \pi \sigma_{pp}(E_p)
\frac{2Z^{(\alpha_{\rm p})}_{p\rightarrow\pi^0}}{\alpha_{\rm p}}\;J_p(E_{\gamma},\theta)
\eta_{\rm A}. \label{q} 
\end{equation} 

\vspace{0.8cm}

\noindent Here, the parameter
$\eta_{\rm A}$ takes into account the contribution from different
nuclei in the wind and in the jet (for standard composition of
cosmic rays and interstellar medium  $\eta_{\rm A}=1.4-1.5$,
Dermer 1986b). $J_p(E_{\gamma})$ is the proton flux distribution
evaluated at $E=E_{\gamma}$. The cross section $\sigma_{pp}(E_p)$
for inelastic $p-p$ interactions at energy $E_p\approx 10
E_{\gamma}$ can be represented above $E_p\approx 10$ GeV by
$\sigma_{pp}(E_p)\approx 30 \times [0.95 + 0.06 \log (E_p/{\rm
GeV})]$ mb.  Finally, $Z^{(\alpha_{\rm p})}_{p\rightarrow\pi^0}$ is the
so-called spectrum-weighted moment of the inclusive cross-section.
Its value for different spectral indices $\alpha_{\rm p}$ is given, for
instance, in Table A1 of Drury et al. (1994). Notice that
$q_{\gamma}$ is expressed in ph s$^{-1}$ erg$^{-1}$ when we adopt
CGS units.\\

The spectral gamma-ray intensity (photons per unit of time per
unit of energy-band) is:

\begin{equation}
I_{\gamma}(E_{\gamma},\theta)=\int_V n(\vec{r'})
q_{\gamma}(\vec{r'}) d^3\vec{r'}, \label{I}
\end{equation}

\vspace{0.6cm}

\noindent where $V$ is the interaction volume.\\

Since we are interested here in a general model and not in the
study of a particular source, the spectral energy distribution

\vspace{0.4cm}

\begin{equation}
L^{\pi^0}_{\gamma}(E_{\gamma},\theta)=E_{\gamma}^2
I_{\gamma}(E_{\gamma},\theta)
\end{equation}

\vspace{0.6cm}

\noindent is a more convenient quantity than the flux. Using eqs. (\ref{Jp_lab}),
(\ref{n(z)}),  (\ref{q}) and (\ref{I}), we get:

\begin{eqnarray}
\lefteqn{L^{\pi^0}_{\gamma}(E_{\gamma},\theta)\approx  \;\frac{ q_{\rm
j} z_0^{\epsilon (n-2)} Z^{(\alpha_{\rm p})}_{p\rightarrow\pi^0}}{ 2\pi
m_p^2 v_{\infty}} \;\frac{\alpha_{\rm p}-1}{\alpha_{\rm p}}\;(E_p'^{\rm
min})^{\alpha_{\rm p}-1}\times {}} \nonumber \\
& & {} \times \dot{M}_* \dot{M}_{\rm
disk} \; \sigma_{pp}(10\; E_{\gamma})\;
 \frac{\Gamma^{-\alpha_{\rm p}+1} \left(E_\gamma-\beta
{\displaystyle{\sqrt{E_\gamma^2-m_p^2c^4}}} \cos
\theta\right)^{-\alpha_{\rm p}}}{\left[\sin ^2 \theta + \Gamma^2 \left(
\cos \theta - {\displaystyle{\frac{\beta E_\gamma}{\sqrt{E_\gamma^2-m_p^2
c^4}}}}\right)\right]^{1/2}} \times \nonumber \\
& & {} \times \int_{z_0}^{\infty}
\frac{z^{\epsilon (2-n)}}{(z^2+a^2)}
\left(1-\frac{r_*}{\sqrt{z^2+a^2}}\right)^{-\zeta} dz.
\end{eqnarray}

\vspace{1cm}

\noindent This expression gives approximately the $\pi^0$-decay gamma-ray
luminosity for a windy MQ at energies $E_{\gamma}>1$ GeV,
in a given direction $\theta$ with respect to the jet axis.\\

Depending on the characteristics of the primary star and the geometry of the system, high-energy gamma rays can be absorbed in the anisotropic stellar photon field through pair production, and then inverse Compton emission from these pairs can initiate an $e^{\pm}$-pair cascade. This effect has been studied in detail by Bednarek (1997) and  Sierpowska \& Bednarek (2005). The main effect of these cascades is a degradation of TeV gamma-rays into a form of softer MeV-GeV emission. Very close systems ($a\sim 10^{11}$ cm), with O stars and perpendicular jets, can be optically thick for gamma-rays between $\sim 10^{-2}-10$ TeV. If the jet is inclined towards the star, the effect can be stronger (Bednarek 1997). For systems with larger separations, the opacity rapidly falls below unity.\\

\subsection{Specific Spectral Energy Distributions}

In order to make some numerical estimates, we shall adopt the
specific MQ model presented in Table \ref{t1}. The values
chosen for the different parameters are typical for MQs
with O stellar companions, like Cygnus X-1. We shall consider a
conical jet ($\epsilon=1$) with conservation of the number of
protons ($n=2$) and a high-energy cut-off for the population of
relativistic protons of $E_p'^{\rm max}=100$ TeV. The minimum distance from the jet to the primary star is $a=70\;R_{\odot}\sim 5 \times 10^{12}$ cm.\\

\begin{table} 
\centering
\caption{Basic parameters of the model}
\vspace{0.4cm}
\begin{tabular}{lll}
\hline
\hline
Parameter & Symbol  & Value  \\
\hline
\hline
Type of jet &  $\epsilon$ & 1 \\
Black hole mass & $M_{\rm bh}$ & 10 $M_{\odot}$\\
Injection point & $z_0$ & 50 $R_{\rm g}^{\;\ast}$  \\
Initial radius & $R_0$ & 5 $R_{\rm g}$ \\
Radius of the companion star & $r_*$ &35 $R_{\odot}$ \\
Mass loss rate & $\dot{M}_*$ & $10^{-5}$ $M_{\odot}$ yr$^{-1}$ \\
Terminal wind velocity & $v_{\infty}$ & 2500 km s$^{-1}$\\
Black hole accretion rate & $\dot{M}_{\rm disk}$ & $10^{-8}$ $M_{\odot}$
yr$^{-1}$ \\
Wind velocity index & $\zeta$ & 1 \\
Jet's expansion index & $n$ & 2 \\
Jet's Lorentz factor & $\Gamma$ & 5 \\
Minimum proton energy & ${E'}_p^{\rm min}$ & 10 GeV \\
Maximum proton energy & ${E'}_p^{\rm max}$ & 100 TeV \\
Orbital radius & $a$ & 2 $r_*$\\
\hline
\hline
\\
\multicolumn{3}{l} {$^\ast$$R_{\rm g}=GM_{\rm bh}/c^2$.}\cr
\end{tabular}
\label{t1}
\vspace{1cm}
\end{table}

\begin{figure}[!t]
\centering
\resizebox{10cm}{!}{\includegraphics{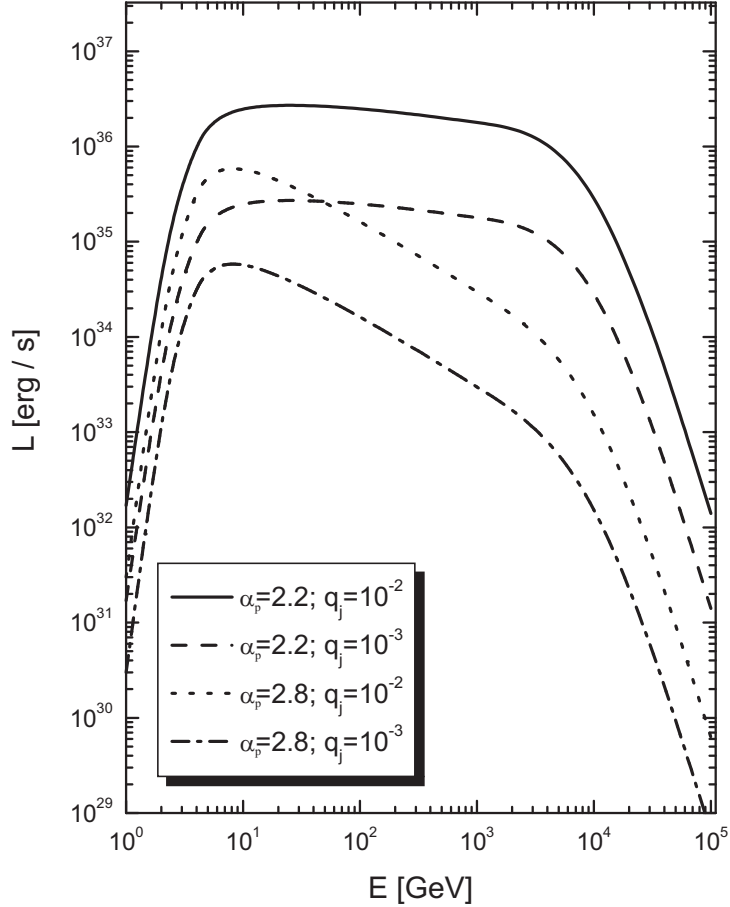}} 
\caption{Spectral high-energy distribution for windy MQs with
proton index  $\alpha_{\rm p}=2.2$ and $\alpha_{\rm p}=2.8$, for different
jet/disk coupling constants ($q_j$). The jet inclination with
respect to the line of sight is assumed to be 10 degrees. An angle
of 30 degrees reduces the luminosity for about two orders of
magnitude.} \label{fig2-Hadron}
\end{figure}

In Fig.~\ref{fig2-Hadron} we show the spectral high-energy distribution
for models with proton index $\alpha_{\rm p}=2.2$ and $\alpha_{\rm p}=2.8$, and
for different values of the jet/disk coupling parameter $q_{\rm
j}$, with the results obtained using a numerical integration
routine. We have added an exponential-like cut-off at $E_{\gamma}\sim 0.1
E_p'^{\rm max}$. Since the forward momentum of the protons in the jet is so great,
the gamma-rays will be highly beamed, within an angle $\Theta\sim$
arctg $(R/z)$. Hence, only hadronic microblazars would be
detected (see, nonetheless, Romero \& Orellana 2004). We assumed a jet inclination with
respect to the line of sight of 10 degrees.

\section{Production of Neutrinos and Secondaries}

Neutrinos are also generated by pion decay chains (see Section~\ref{GREHM}). The signal-to-noise (S/N) ratio for the detection of such a $\nu$-signal can be
obtained analyzing the event rate of atmospheric $\nu$-background
and comparing it with the event rate from the source (see e.g.
Anchordoqui et al. 2003 for details). The S/N ratio in a km-scale detector (like IceCube) in the 1--10 TeV band (including the effects of neutrino oscillations) is $\sim
3$ for one year of operation, assuming an inclination angle of 30
degrees, $\alpha_{\rm p}=2.2$ and that the neutrino spectrum roughly
satisfies (e.g. Dar \& Laor 1997): 

\begin{equation}
\frac{dF_{\nu}}{dE_{\nu}}\simeq 0.7\; \frac{dF_{\gamma}}{dE_{\gamma}}. 
\end{equation}

\vspace{0.4cm}

\noindent This S/N is high enough as to justify speculations on the possibility
of detecting a hadronic microblazar first from its neutrino signal
(a serendipitous discovery in a detector like IceCube), and only
later from its gamma-ray emission (through a pointed observation).\\ 

The here presented hadronic model contemplates the production of gamma-rays in MQs as the final stage that follows the decay of neutral pions previously generated in $p-p$ interactions. However, neutral pions can also be produced by $p-\gamma$ interactions (also called ``photo-meson'' processes), which as an alternative model also gives rise to the generation of TeV photons and neutrinos\footnote{This kind of model is similar to the AGN proton initiated cascade model of Mannheim \& Biermann (1992), Mannheim et al. (1992) and Mannheim (1993). See also Biermann \& Strittmatter (1987) for  the presentation of pioneering ideas in this sense.} (Levinson and Waxman 2001, Distefano et al. 2002). The site where the emission is generated in this case is within the coronal region in the innermost part of MQs' jets. This implies a fundamental difference of this model with the one used in this chapter. Pions are then produced in collisions of hadrons from the jet with the strong external
  X-ray photons (e.g. from the hot corona) and with the synchrotron photons produced by leptons inside the jet. In this process protons are required to be accelerated up to very high energies ($\sim 10^{16}$ eV) in the inner jet, in order to produce a significant flux of multi-TeV neutrinos. This requirement might represent an important difficulty for the photo-meson model considering that the possibility for particles as protons to be accelerated up to such energies in MQs is still under strong discussion. Besides, neutrinos are supposed to extract more energy from the original protons in the $p-p$ process than in the $p-\gamma$ one: $E_{\nu}^{\rm max}= \frac{1}{12} E_p^{\rm max}$ and $E_{\nu}^{\rm max}= \frac{1}{20} E_p^{\rm max}$ respectively (Alvarez-Mu\~niz \& Halzen 2002).\\ 

It is not possible to separate in the neutrino signal the contributions from the photo-meson and $p-p$ channels, but simultaneous X-ray observations can help to determine the characteristics of the relevant photon fields to which the inner jet is exposed, making then possible estimates of each contribution in particular cases. That one channel might dominate over the other will depend on the relative target density of the photons and protons in the source region where the protons are accelerated (Halzen \& Hooper 2002). In any case the detection of TeV neutrinos from MQs could be used as an important diagnostic of jets' content. This kind of detection would immediately imply the presence of hadrons in the jets. Therefore it would help to discriminate $e-p$ jets from pair dominated ones (see Bednarek et al. 2004 for additional discussion).\\

Another interesting aspect of the hadronic microblazar described in this chapter is that
$e^{\pm}$ are injected outside the coronal region through
$\pi^{\pm}$ decays (see Section~\ref{GREHM}). These leptons do not experience the severe
IC losses that affect to primary electrons and pairs when crossing the disk and coronal photon fields as we mentioned in Chapter~\ref{External fields}. These secondaries will mainly cool through synchrotron radiation (at X-rays, in the case of TeV particles)
and IC interactions with the stellar seed photons (that would
result into an additional source of MeV-GeV gamma-rays). The
spectrum of secondary pairs roughly mimics the shape of the proton
spectrum. Hence, synchrotron emission from these particles will
present indices $\alpha_{\rm syn}\simeq (\alpha_{\rm p}-1)/2$, which for
values of $\alpha_{\rm p}\sim2$ are similar to what is observed in
MQs' jets at radio wavelengths. The losses of the
primaries in the inner source lead to a soft particle spectrum that is then injected in
the region where the particles produce IC gamma-rays through
interactions with stellar UV photons.  Pure leptonic models for
the gamma-ray flux (e.g. Kaufman Bernad\'o et al. 2002, Romero et al. 2002) then require particle re-acceleration in order to explain the flat
radio-spectrum of sources like LS 5039 far from the core. In the
hadronic model, the leptons might be injected {\em in situ} with the
right spectrum through hadronic decays, avoiding the problem.\\

At TeV gamma-ray energies hadronic microblazars which are optically thin to pair production can be detected as
unidentified, point-like sources with relatively hard spectra.
This kind of sources could display variability.
In the near future, new ground-based Cherenkov telescopes like HESS
and MAGIC\footnote{The VERITAS project is also expected to contribute to the detection of this kind of sources.} might detect the signatures of such sources on the
galactic plane. Hadronic microblazars might be part of this
population, as well as of the parent population of low latitude
unidentified EGRET sources.

\chapter{\label{Low-mass}Low-Mass Microquasars and Halo Sources}
\thispagestyle{empty}

\newpage
\thispagestyle{empty}
\phantom{.}

\newpage
\vspace*{5.2cm}

\section{Variable Gamma-Ray Sources off the Galactic Plane}

\begin{figure}[!t]
\centering
\resizebox{14.8cm}{!}{\includegraphics{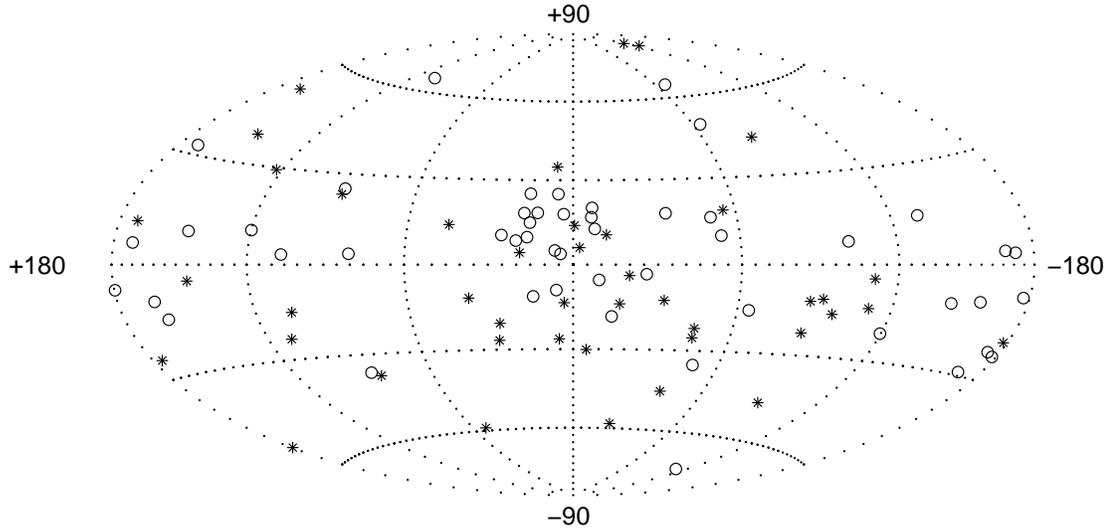}} 
\caption{All-sky plot, in galactic coordinates, of the unidentified EGRET sources at latitudes $|b|>3^{\circ}$. The steadier sources associated with the Gould
Belt and the variable sources are marked as circles and stars, respectively.} \label{fig1-Lowmass}
\end{figure}

\begin{figure}[!t]
\centering
\resizebox{13cm}{!}{\includegraphics{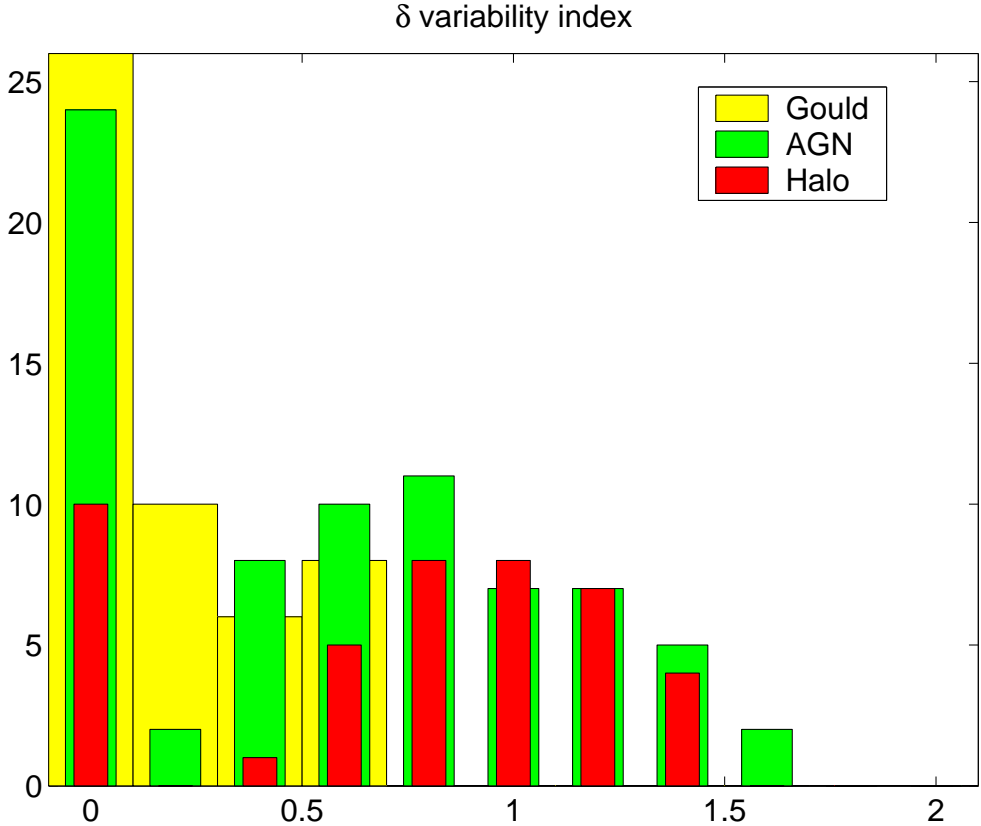}} 
\caption{Distributions of the $\delta$ variability
indices from Nolan et al. (2003) for the firm AGN, Gould Belt, and halo sources at latitudes
$|b|>3^{\circ}$.} \label{fig2-Lowmass}
\end{figure}

A brief description of the mid-high latitude EGRET sources was included in Chapter~\ref{Introduction}. In what follows, we will describe in some more detail the properties that induce to the classification of these sources in three subgroups. The 93 unidentified sources detected away from the Galactic plane are displayed in Fig~\ref{fig1-Lowmass}. Their concentration at $3^{\circ} < |b| < 30^{\circ}$ and in the inner half steradian indicates (at a 7$\sigma$ confidence level) that 70 to 100 \% of them have a Galactic origin, depending on the choice of
Galactic scale height. Their temporal and spatial characteristics reveal an heterogeneous sample. Fig.~\ref{fig2-Lowmass} shows their variability index distribution, which closely follows that of the variable AGN sources (Nolan et al. 2003).
The average $\overline{\delta}$ index is 0.79 $\pm$ 0.08, 0.66 $\pm$ 0.06, and 0.42 $\pm$ 0.06 for the halo, AGN, and Belt sources, respectively. The halo sources are clearly variable. Taking into account the detection biases (Grenier 2000), the spatial distribution of the variable sources implies an origin in a thick galactic disk with a scale height of 1.3 $\pm$ 0.6 kpc. At typical distances of 5 to 10
kpc, the luminosities of the halo sources range from 2 to $30 \times 10^{34}$ erg s$^{-1}$ sr$^{-1}$ above 100 MeV. They exhibit large luminosity $L_{\gamma}/L_X$ ratios of a few hundred.\\

The compact objects likely to power sources located high above the plane include millisecond pulsars and microquasars with a low-mass star companion, both having migrated away from the galactic plane or escaped from globular clusters. In the first case we refer to MQs that could have been ejected by kicks imparted in the natal supernova explosions to galactic centric orbits (Mirabel \& Rodrigues 2003). In the second case they might have been born in globular clusters from where they might have been injected in large galactic orbits (Mirabel et al. 2001). Until now no gamma-ray source is found to be positionally  coincident with a globular cluster.\\ 

Pulsed gamma-rays have been detected from the ms pulsar PSR J0218+4232, in phase with the radio and X-ray peaks (Kuiper et al. 2002). This object shares many traits with the halo sources: a distance of 5.7 kpc and an altitude of 1.6 kpc, a luminosity of
$1.6 \times 10^{34}$ erg s$^{-1}$ sr$^{-1}$ and a spectral index of 2.6 above 100 MeV. PSR J0218+4232 does not belong to a globular cluster either. Yet, no long-term variability is expected from theory (Zhang \& Cheng 2003).\\

In the next section, we explore whether low-mass microquasars can produce the halo
sources despite their intrinsic faintness and softness compared with the young high-mass systems. The stellar luminosity in LMMQs is reduced by 4 to 6 orders of magnitude with respect to the case of a high-mass stellar companion. There is also a reduction of 1 order of magnitude in the emission from the accretion disk. The thermal emission in both components peaks a decade or two lower in energy. Compare Fig.~\ref{fig2-CygX1} with Fig.~\ref{fig3a-Lowmass} to check these differences luminosities and peak energies.\\

Only leptonic models can be considered for halo sources since hadronic ones require the presence of an early type of companion star in order to ensure the existence of strong stellar winds (Romero et al. 2003a).\\

\section{Emission from Low-Mass Microquasars}

XTE J1118+480 is a LMMQ at a high galactic latitude above the plane, $b = 62^{\circ}$,  and a distance of $1.8 \pm 0.6$ kpc. It lies at 1.6 $\pm$ 0.5 kpc
above the galactic plane (McClintock et al. 2001a). The central object accretes matter through a disk from a low-mass star via Roche lobe overflow. McClintock et al. (2001a) and Wagner et al. (2001) constrained the companion spectral type to be between K5 V and M1 V. A large mass function, $f(M) \approx$ 6 M$\odot$, strongly
suggests that the compact object is a black hole in a fairly compact binary system with a short orbital period of 4.1 hr (Wagner et al. 2001). The disk may be precessing under the stellar tidal influence (Torres et al. 2002).\\

The rapid and correlated UV-optical-X-ray variability in the low-hard state is interpreted as a signature of the strong coupling between a hot corona and a jet emitting synchrotron radiation up to, at least, the UV band (Hynes et al. 2003; Chaty et al. 2003; Malzac et al. 2004).  The outflow has remained steady through the outburst evolution (Chaty et al. 2003).\\

The coronal emission extends up to
$\sim 150$ keV (McClintock et al. 2001b). The optical to hard-X-ray data can be modeled by the Comptonization in a hot corona or in the inner accretion flow of the soft photons emitted by the outer cold disk with an inner
radius of $\sim 55 R_{\rm Sch}$ and a temperature of $\sim 24$ eV (Esin et al. 2001; McClintock et al. 2001b; Malzac et al. 2004).\\

\begin{table}[!t]
\begin{center} 
\caption{Parameter set corresponding to the MQ XTE J1118+480 and used in the model}\label{parset}
\vspace{0.8cm}
\begin{tabular}{ll}
\hline
\hline
\\
Black hole mass           & $M_{\rm bh} = 6.5$ M$\odot$\\
Mass accretion rate       & $\dot{M} = 3 \times 10^{-8}$ M$\odot$ yr$^{-1}$\\
K-M star bolometric luminosity & $L_{\rm KM} = 4 \times 10^{32}$ erg/s\\
F star bolometric luminosity & $L_{\rm F} = 1.5 \times 10^{34}$ erg/s\\
Star temperature          & $kT_{\rm KM} = 1$ eV and $kT_{F} = 1.8$ eV\\
Star orbital radius       & $D_* = 1.7 \times 10^{11}$ cm\\
Jet/accretion power ratio & $q_{\rm jet} = P_{\rm jet}/\dot{M}c^2 = 10^{-3}$ to $10^{-2}$\\
Corona luminosity         & $L_{\rm cor} = 7.8 \times 10^{34}$ erg/s\\
Corona outer radius       & $R_{\rm cor} = \, 10^8$ cm\\
Corona photon index ($dN_X/dE \propto E_X^{-\alpha}$) & $\alpha_{\rm cor} = 1.8$\\
Corona cut-off energy     & $E_{\rm cor} = 150$ keV\\
Disk luminosity           & $L_{\rm disk} = 8.6 \times 10^{35}$ erg/s\\
Disk temperature          & $kT_{\rm disk} = 24$ eV \\
Initial jet radius        & $R_{\rm jet} = 1.9 \times 10^7$ cm\\
Jet bulk Lorentz factor   & $\Gamma_{\rm jet} = 3$ to $10$\\
Jet viewing angle         & $\phi = 1^{\circ}$ to $30^{\circ}$\\
Jet electron index ($dN_e/dE \propto E_e^{-p}$) & $p = 2$ to 3\\
Maximum electron energy   & $E_{e\,\rm max} = 5$ GeV to 5 TeV\\
Minimum electron energy   & $E_{e\,\rm min} = 1$ to 5 MeV\\
\\
\hline
\hline
\end{tabular}
\end{center}
\end{table}

XTE J1118+480 therefore serves as a good example for a low-mass microquasar and we adopt its characteristics as input to our model (see table \ref{parset}). SED of the thermal
stellar, disk and coronal components, taken from the aforementioned publications, are displayed in Fig.~\ref{fig3a-Lowmass}. The case of an F star companion is also considered in Fig.~\ref{fig3b-Lowmass}. We calculate the EC emission from interactions of the jet with the three radiation fields following the model developed in Chapters~\ref{High-mass} and \ref{External fields}.
We therefore assume a population of $e^+-e^-$ pairs in a persistent, cylindrical jet, with a power-law distribution in number density per unit energy, $N(E_e)=k E_e^{-p}$, between $E_e^{\rm min}$ and $E_e^{\rm max}$. The jet is assumed to be parallel to the disk axis, at an angle $\phi$ to the line of sight. It moves with a bulk Lorentz factor $\Gamma$ and carries a total power $P_{\rm jet} = q_{\rm jet} \dot{M}c^2$ (adopting the jet-disk coupling proposed by Falcke \& Biermann 1995).\\

\begin{figure}[!t]
\centering
\resizebox{13cm}{!}{\includegraphics{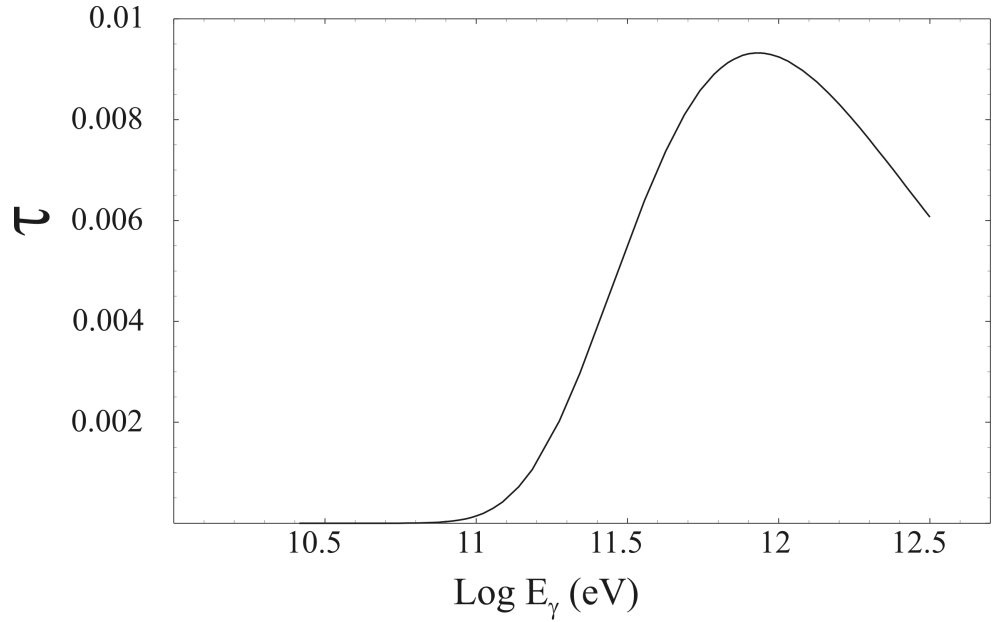}} 
\vspace{0.4cm}
\caption{The optical depth for a photon with energy $E_{\gamma}$ in the ambient radiation from the companion star. The stellar field is clearly optically thin to the gamma-rays since $\tau<1$ for the whole range of $E_{\gamma}$.}\label{tau-estrella}
\end{figure}

\begin{figure}
\centering
\resizebox{11cm}{!}{\includegraphics{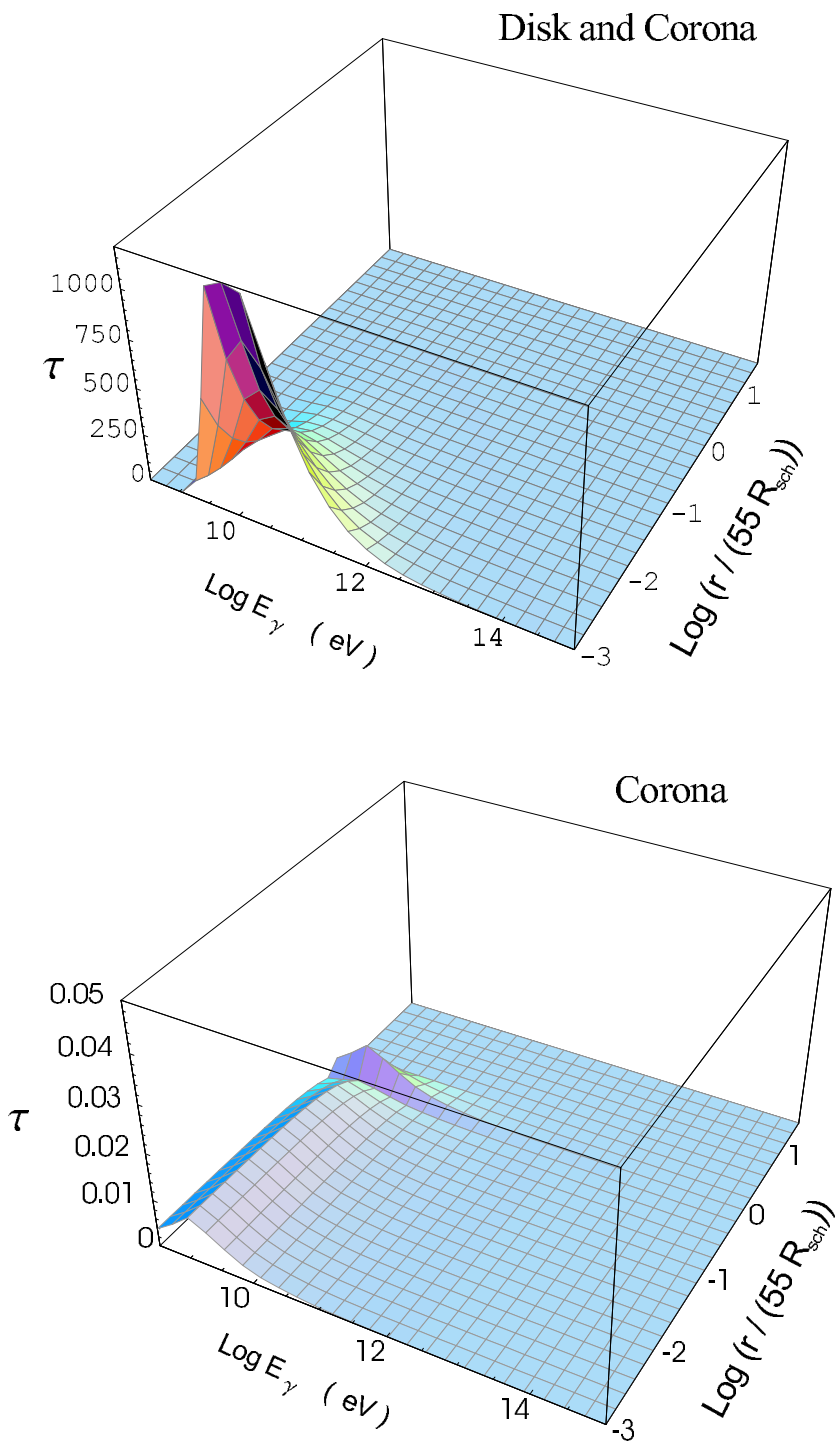}} 
\vspace{0.4cm}
\caption{Optical depth for gamma-rays in the inner region of a LMMQ. In the upper plot both photon densities, from the disk and the corona, are included in the calculation. On the bottom, only the coronal photon density was considered showing that the opacity is completely due to the seed photons from the disk. The opacity goes down progressively and the last layers from $r/55 R_{\rm Sch}\sim 10^{-1}$ are already optically thin to the gamma-rays of any energy.}\label{tau-disco+corona}
\end{figure}

The coronal region is assumed to fill a sphere inscribed in the inner disk radius. Let us recall that the Compton losses in the different regions can modify the injected electron spectrum, introducing a break in the power-law
from an index $p$ to $p+1$ at the energy at which the cooling time equals the escape time. In the scenario studied here, this occurs with the disk radiation field. We have seen from previous chapters that another important ingredient can be the absorption from two-photon pair creation in the ambient radiation. It turns out to be quite effective near the disk ($\tau>1$) whereas the coronal and stellar fields are optically thin ($\tau<1$) to the gamma-rays. These conclusions are supported by Fig~\ref{tau-estrella} and Fig~\ref{tau-disco+corona} where we show the computed opacity, $\tau$, of the stellar photon field on one hand and of the disk and corona fields on the other.\\

The ambient radiation from the companion star, $E_{\star}$, has an optical depth for a photon with energy $E_{\gamma}$ given by (Jauch \& Rohrlich 1976): 

\vspace{0.2cm}

\begin{equation}
\tau(E_\gamma)=\langle r \rangle \int N(E_{\star})\,\sigma(E_{\star},E_\gamma) dE_{\star},
\end{equation}

\vspace{0.4cm}

\noindent where $\langle r \rangle$ is the mean dimension of the pair production region and $\sigma(E_{\star},E_\gamma)$ is the photon-photon pair creation cross section given by:

\vspace{0.2cm}

\begin{equation}
\sigma_{e^+e^-}(E_{\star}, \;E_{\gamma})=\frac{\pi
r_e^2}{2}(1-\xi^2)\left[2\xi(\xi^2-
2)+(3-\xi^4)\ln\left(\frac{1+\xi}{1-\xi}\right) \right],
\end{equation}

\vspace{0.6cm}

\noindent where $r_e$ is the classical radius of the electron and $\displaystyle \xi=\left[1-\frac{(m_e c^2)^2}{E_{\star} E_{\gamma}}\right]^{1/2}$. The seed photon field from the star is assumed to be a blackbody type and its differential number density, $dn(E)=N(E)dE$, is (Leung et al. 1993):

\vspace{0.2cm}

\begin{equation}
N(E_{\star})dE_{\star} = \frac{E_{\star}^2 dE_{\star}}{\pi^2 (\hbar c)^3(e^{E_{\star}/kT}-1)}.
\end{equation}

\vspace{0.6cm}

The coronal region was considered as a spherical region of $\sim 55 R_{\rm Sch}$ in radius. Considering that this area is filled by seed photons from the disk and from the hot corona, we calculate the total opacity taking into account its dependency on $r$:

\begin{eqnarray}
\displaystyle{\lefteqn{ \tau(E_\gamma,r)=\int_r^\infty \int_{0}^\infty [N_{\rm disk}(E_{\rm X}, r') + N_{\rm corona}(E_{\rm X}, r')]\times}\hspace{7cm}}\nonumber\\
	& & \displaystyle{\times\, \sigma_{e^-e^+}(E_{\rm X},E_\gamma)\, dE_{\rm X}\, dr'}.
\end{eqnarray}
  
\vspace{0.4cm}

\noindent The disk photon distribution is that of a blackbody,

\vspace{0.6cm}

\be
N_{\rm disk}(E_{\rm X}, r)= \left(\frac{\pi B(E_{\rm
X})}{hc\,E_{\rm X}}\right)\frac{R_{\rm disk}^2}{r^2},    
\ee
  
\vspace{0.2cm}

\noindent with $R_{\rm disk}=55 R_{\rm Sch}$, and
\begin{eqnarray}
B(E_{\rm X})= \frac{2 E_{\rm X}^3}{(hc)^2\,(\displaystyle e^{E_{\rm X}/kT_{\rm disk}}\, -1)}.
\end{eqnarray}

\vspace{0.6cm}

\noindent The corona photon distribution is a power law such that:

\begin{eqnarray}
N_{\rm corona}(E_{\rm X}, r)= \left\{\begin{array}{ll}
 \displaystyle{\frac{L_{\rm corona}\,E_{\rm
X}^{-1.8}}{4\pi c\,R_{\rm corona}^2\,\,e^{E_{\rm X}/E_{\rm cut-off}}}} & \textrm{if $r\leq55 R_{\rm Sch}$}\vspace{0.5cm}\\
  \displaystyle{\frac{L_{\rm corona}\,E_{\rm
X}^{-1.8}}{4\pi c\,r^2\,\,e^{E_{\rm X}/E_{\rm cut-off}}}} & \textrm{if $r>55 R_{\rm Sch}$}
\end{array}\right.
\end{eqnarray}

\vspace{0.5cm}

\begin{figure}[!t]
\centering
\resizebox{14.5cm}{!}{\includegraphics{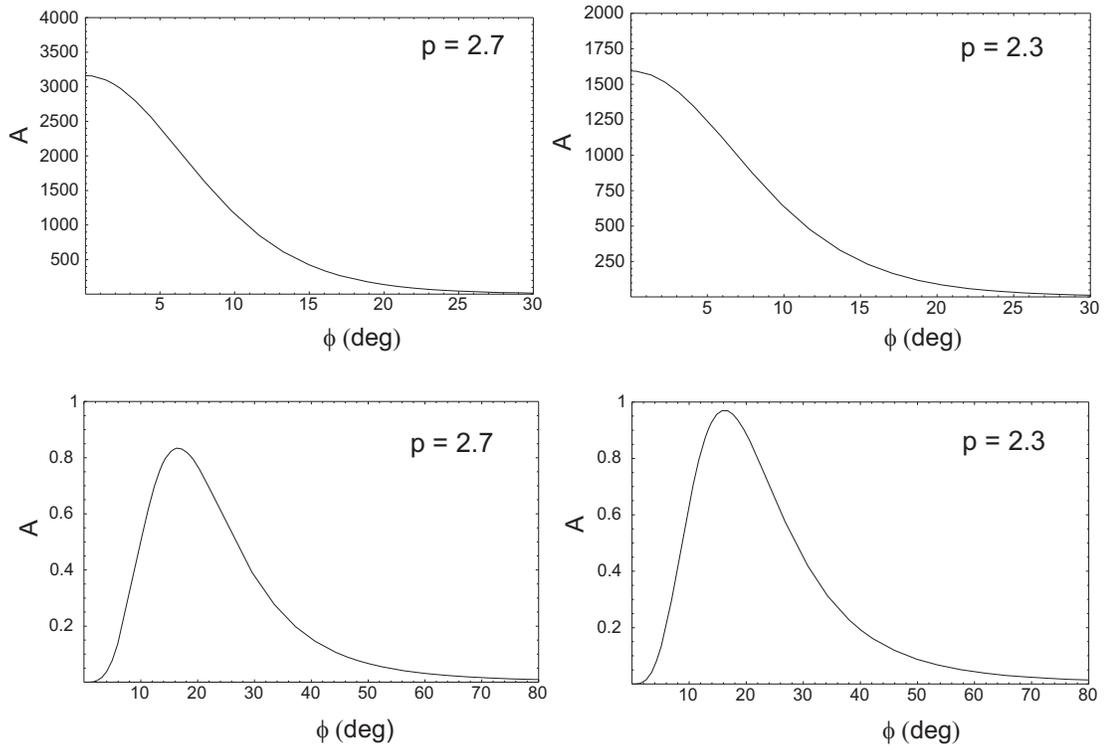}} 
\caption{Amplification factors on the SEDs: $A=D^{2+p}$, in the cases of the coronal and stellar contributions to the SED (\textit{top}); $A=D^{2+p}(1-\cos \phi)^{(p+1)/2}$ for the disk one (\textit{bottom}). $\phi$: viewing angle.}\label{amplif}
\end{figure}

\begin{figure}[!t]
\centering
\resizebox{13cm}{!}{\includegraphics{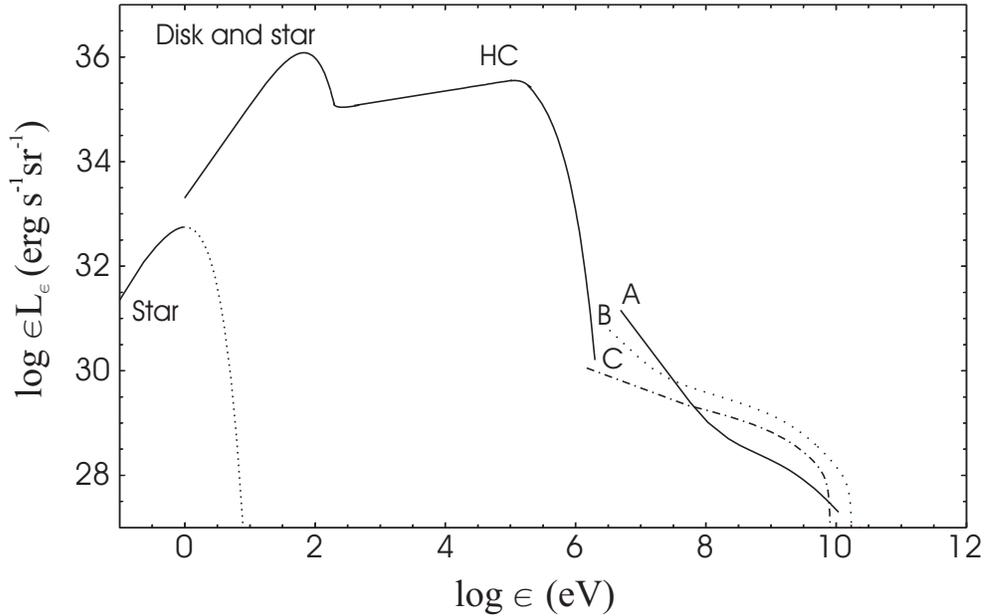}} 
\caption{Spectral energy distribution of the EC emission from the jet of a microquasar with a K-M star companion, seen at angles of $5^{\circ}$ (A), $15^{\circ}$ (B), and $30^{\circ}$ (C) from its
axis, for $\Gamma=3$, $q_{\rm jet}=10^{-2}$, and a jet electron index p=2.3 with $\gamma_{e}^{\rm min}=2$ and $\gamma_{e}^{\rm max}=10^4$.}\label{fig3a-Lowmass}
\end{figure}

\begin{figure}[!t]
\centering
\resizebox{13cm}{!}{\includegraphics{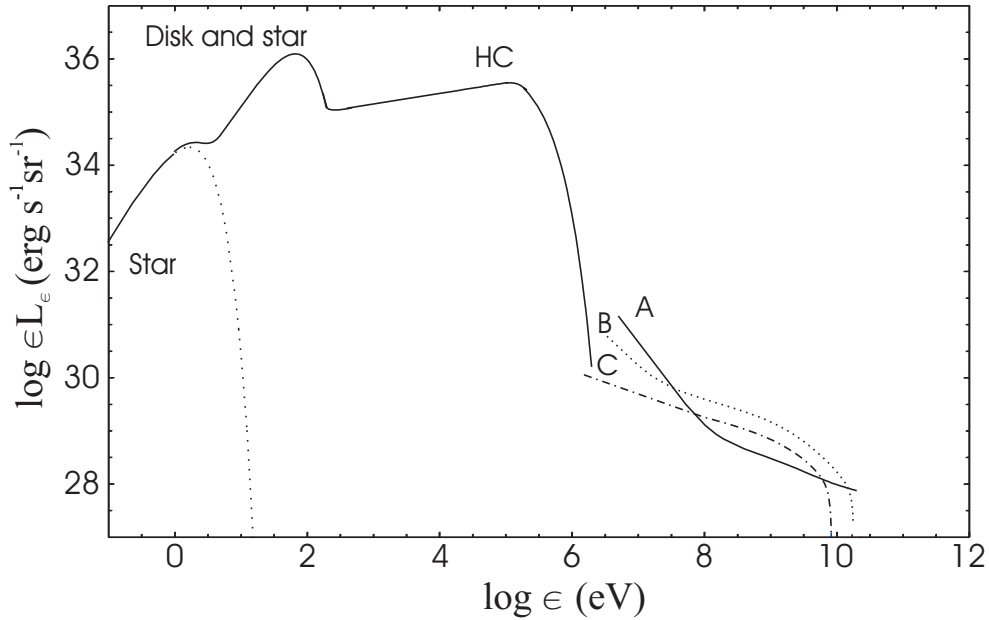}} 
\caption{Spectral energy distribution of the EC emission from the jet of a microquasar with an F star companion, seen at angles of $5^{\circ}$ (A), $15^{\circ}$ (B), and $30^{\circ}$ (C) from its axis, for $\Gamma=3$, $q_{\rm jet}=10^{-2}$, and a jet electron index p=2.3 with $\gamma_{e}^{\rm min}=2$ and $\gamma_{e}^{\rm max}=10^4$.}\label{fig3b-Lowmass}
\end{figure}

As it has been discussed, in the observer frame, the IC contributions from interactions with the stellar and coronal photons to the SED are amplified by $D^{2+p}$ whereas the disk one by $D^{2+p}(1-\cos\phi)^{(1+p)/2}$. As it can be seen in Fig.~\ref{amplif}, the amplification factor corresponding to the star and corona peaks along the jet axis in $\phi = 0^{\circ}$, whereas one for the disk peaks in $\phi \sim 15^{\circ}$ to $20^{\circ}$, for $\Gamma \sim 3$ and $p = 2$ to 3. \\

In Section~\ref{discussioncap4} we clarified the differences between the application of our leptonic model to the HMMQ Cygnus X-1 and the work presented by Georganopoulos et al. (2002) on the same object. Now we have to point out that with respect to the LMMQ XTE J1118+480, the scenario we are discussing also differs from that of Georganopoulos et al. (2002) who have imposed a much lower energy cut-off to the electrons so that the hard X-ray emission results from EC interactions with the stellar and disk
photons rather than from a hot inner accretion flow or from a hot coronal plasma energized by magnetic flares above the disk, elements that are absent in their model.\\

Figures \ref{fig3a-Lowmass} and \ref{fig3b-Lowmass}  show the SED obtained per steradian in the laboratory frame for a LMMQ with the parameters listed in Table~\ref{parset}, adding the contributions from the three external photon fields. The EC coronal component (in the Klein-Nishina regime) is dominant in the COMPTEL 3-30 MeV band whereas the EC disk component (in the Thomson regime) takes over above 100 MeV. This is why the maximum gamma-ray luminosity is reached for $\phi$ close to 15$^{\circ}$, reflecting the angle dependence of the disk amplification factor. This emission has a photon spectral index of 2.5 between 0.1 and 1 GeV. The IC stellar component is small at all viewing angles. These results show that even though the spectral index in the EGRET band matches that of the unidentified sources, the maximum predicted luminosity, $L_{\rm max} \sim 4 \times 10^{29} (E/100\, \rm MeV)^{-0.5}$ erg s$^{-1}$ sr$^{-1}$, is 5 orders of magnitude too faint to account for the typ
 ical halo source fluxes at distances of 5 to 10 kpc.\\

The relative contributions of the three EC components change in the case of an extreme microblazar where the bulk Lorentz factor $\Gamma$ reaches 10, electron energies extend up to 5 TeV, and the jet axis is close to the line of sight ($\phi = 1^{\circ}$). The generation of such a highly relativistic outflow, with
$\Gamma \geq 10$, has been recently observed from Circinus X-1, a neutron star with a stellar-mass companion as it is discussed in Section~\ref{CSandCO}. Particle energies as high as 10 TeV have been inferred for the large-scale jets of the
low-mass microquasar XTE J1550-564 (see Section~\ref{the jet}), 0.1 pc away from the black hole (Corbel et al. 2002).\\ 

\begin{figure}[!t]
\centering
\resizebox{13cm}{!}{\includegraphics{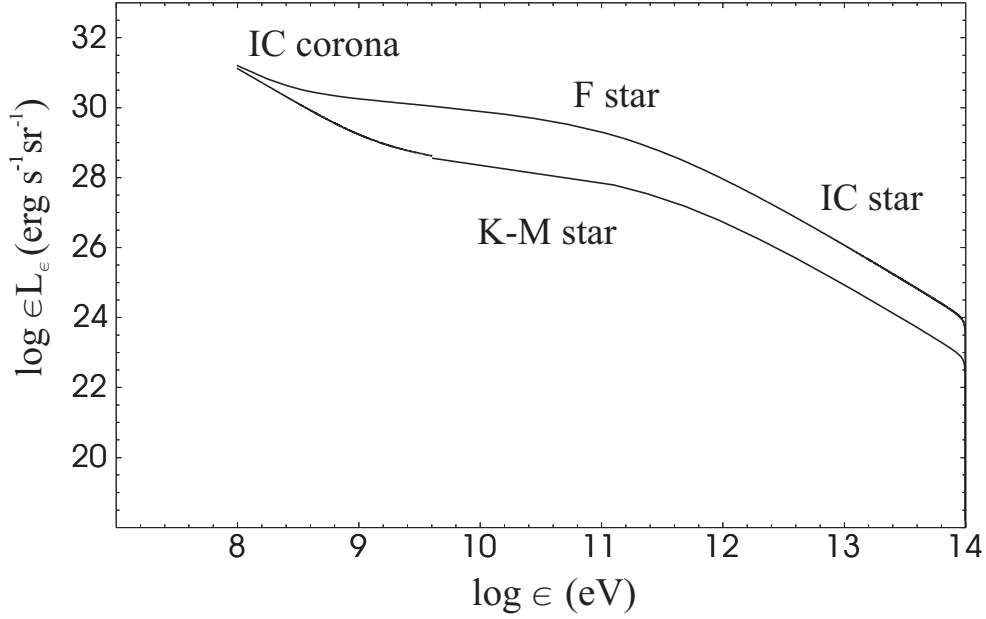}}
\caption{Spectral energy distribution of the EC emission from the jet of an extreme microblazar with an F or K-M star companion, seen at $1^{\circ}$
from its axis, for $\Gamma=10$, $q_{\rm jet}=10^{-2}$, and a jet electron index p=2.3 with  $\gamma_{e}^{\rm min}=10$ and $\gamma_{e}^{\rm max}=10^7$.} \label{fig4-Lowmass}
\end{figure}

Fig.~\ref{fig4-Lowmass} shows the result of our calculations for a low-mass microblazar. The assumed parameters are indicated in the caption. The corona EC emission predominates in a short range of energies up to several hundred MeV for an F star and to several GeV for a K-M star, beyond which the harder, stellar EC component takes over. The disk EC
emission is negligible because of the lesser amplification at small viewing angle. The $\sim$ 35 times larger energy density that the jet encounters around an F star compared with the case of a K-M star results in a modest luminosity increment by a factor $<$\,5 at the beginning of the EGRET band up to a maximum factor $\sim$ 30 at the end of that range. Anyhow, the very low luminosities, as well as the generally too soft spectra, with photon indices ranging from 3 to 4, cannot account for the typical characteristics of the EGRET sources. Nor would these systems be detected above 100 GeV by the new-generation Cherenkov telescopes since the predicted luminosity falls orders of magnitude below their sensitivity in all cases, even when multi-TeV electrons are present in the jet.\\

\section{Discussion}

LMMQs can remain quiescent for years or decades before brightening by as much as a factor $10^7$ in X-rays in a week. The accreting keV luminosity of $\sim 10^{35}$ erg s$^{-1}$sr$^{-1}$ adopted in the previous section corresponds to such a flaring state, i.e. to a bright X-ray source with a $\nu F_{\nu}$ flux of $2.6 \times 10^{-4}$ $(D/5\, {\rm kpc})^{-2}$ MeV cm$^{-2}$ s$^{-1}$. The GeV emission obtained is mainly due to the  EC of the seed photons from the disk or the corona (depending on the conditions in the jet). The predicted luminosity should be considered as an upper limit for comparison with the unidentified sources and in that sense EC emission in low-mass microquasars largely fails to explain the variable EGRET sources at large scale heights.\\

\begin{figure}[!t]
\centering
\resizebox{13cm}{!}{\includegraphics{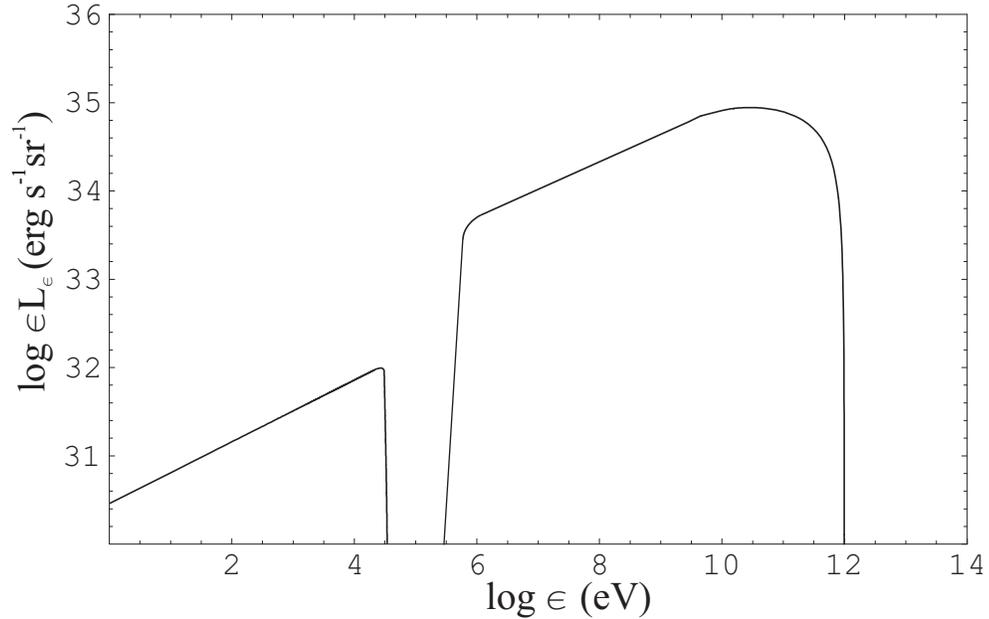}}
\caption{Spectral energy distribution of the synchrotron and SSC emission from the jet of an extreme microblazar, with a 10 G magnetic field, seen at $1^{\circ}$ from its axis, for $\Gamma=10$, $q_{\rm jet}=10^{-2}$, and a jet electron index p=2.3 with  $\gamma_{e}^{\rm min}=10$ and $\gamma_{e}^{\rm max}=10^5$.} \label{fig5-Lowmass}
\end{figure}

In contrast with high-mass systems where the external radiation energy density largely surpasses the magnetic one, SSC emission in low-mass systems is likely to dominate even for a modest field strength of $\sim$ 10 G in the jet. For this value the magnetic energy density compares with that of a K-M star. Applying the model developed by Bosch-Ramon et al. (2005) to a cylindrical jet, we can get luminosities similar to those expected from the EGRET halo sources (see
Fig.~\ref{fig5-Lowmass}). The luminosity per unit of frequency for the synchrotron radiation in the observer frame is:
\vspace{0.3cm}
\be
 L_{\nu}^{\rm syn}=D^{2+p}\int^{z_{\rm max}}_{z_{\rm min}}2\pi R s_{\nu} \tau_{\nu}dz, 
\ee

\vspace{0.3cm}

\noindent where $R$ is the radius of the jet, $s_{\nu}$ is the source function of the synchrotron emission from an isotropic particle  distribution and $\tau_{\nu}$ is the synchrotron optical depth of the jet\footnote{The expressions for $s_{\nu}$ and $\tau_{\nu}$ can be found in Pacholczyk (1970).}.\\

The resulting SSC emission has a luminosity in the jet frame given by:

\begin{eqnarray}
\epsilon L_{\epsilon}&=&\epsilon
\int^{z_{\rm max}}_{z_{\rm min}}
\int^{\epsilon_{\rm 0 max}}_{\epsilon_{\rm 0 min}}
\int^{\gamma_{\rm emax}}_{\gamma_{\rm emin}}
\Sigma U_{\rm syn}(\epsilon_0) n(\gamma_{\rm e}) \frac{dN_{p}}{dtd{\epsilon}}
\frac{\epsilon}{\epsilon_0}d\gamma 
d\epsilon_0 
dz,
\label{eq:Lic}
\end{eqnarray}

\vspace{0.4cm} 

\noindent where $\Sigma$ is the surface of a perpendicular jet slice, $\displaystyle \frac{dN_{p}}{dtd{\epsilon}}$ is the lab frame rate of IC interactions per final photon energy (see~\ref{rate}) and $U_{\rm syn}$, the synchrotron radiation density is given by:
\begin{eqnarray}
U_{\rm syn,\, \nu}&\approx&\frac{s_{\nu} \tau_{\nu}}{c}.
\end{eqnarray}

\vspace{0.4cm}

Fig.~\ref{fig5-Lowmass} shows the luminosity of the SSC emission as calculated from the observer frame. The integration of (\ref{eq:Lic}) is first performed in the co-moving frame and then transformed to the lab:
\begin{eqnarray}
\epsilon' L'_{\epsilon'}&=& D^{2+p} \epsilon' L_{\epsilon'}.
\label{eq:LicSRobs}
\end{eqnarray}
 
\vspace{0.4cm}

\noindent The energy of the scattered photon in the jet's reference frame,
$\epsilon$, is boosted to:

\be
 \epsilon'=D~\epsilon.
\ee

\vspace{0.4cm}

The synchrotron part of the spectrum is $\sim$ 100 -- 1000 times fainter than the near-IR to UV synchrotron emission recorded from XTE J1118+480 during the outburst\footnote{See Markoff et al. (2001) for a synchrotron model of this specific object.}, the adopted electron index in Fig.~\ref{fig5-Lowmass} being softer than that of XTE J1118+480 to avoid a too hard spectrum at high energies. The latter softens beyond 10 GeV because of the Klein-Nishina effect. One should keep in mind, however, that $\gamma\gamma$
absorption against the disk photons is not included in the calculation although it could efficiently limit the emerging gamma-ray flux (Bosch-Ramon et al. in preparation). Adiabatic losses in an expanding jet would also modify the result although not severely.\\ 

SSC emission in a low-mass microquasar appears then an interesting possibility to be investigated in greater detail, bearing in mind that modeling the high-energy radiation from these fascinating objects is severely limited by the highly uncertain choice of the jet bulk motion and magnetic field. These first results of SSC emission in LMMQs, make of these objects a promising parent population for the halo group of unidentified EGRET sources.

\chapter{\label{Conclusions}Conclusions}
\thispagestyle{empty}

\newpage
\thispagestyle{empty}
\phantom{.}

\newpage 

\vspace*{5.9cm}

\noindent The fact of being a scale-down version of quasars makes of microquasars a very interesting type of object, rich in an important variety of physical phenomena that were always difficult to study in detail in quasars considering their time scales and distances. For example, MQs have already provided an important insight into the connection between accretion disk instabilities and the formation of jets, since the characteristic times in the flow of matter onto a black hole are proportional to its mass, and therefore the accretion-ejection phenomena in quasars should last $10^5-10^7$ times longer than the analogous phenomena in MQs. In this way, variations on scales of tens of minutes in MQs have revealed events that had been difficult to observe in quasars.\\

Before the launch of the CGRO, only one AGN had been detected at high-energy gamma-rays. We are referring to the quasar 3C 273 that was observed in this wavelength by COS B (Swanenburg et al. 1978). After phases I (91/05 -- 92/11) and II (92/11 -- 93/09) of the CGRO, EGRET detected with a high degree of confidence, 33 AGNs in high-energy gamma rays ($E > 100$ MeV) (von Montigny et al. 1995, and references therein). The strong connection between this high-energy emission and the presence of relativistic jets has been widely discussed, for example in the mentioned paper from von Montigny at al. (1995) and many other similar references. Following the line of the quasar -- microquasar analogy, it looks reasonable at least to suspect the possibility that high gamma-ray emission could arise from MQs' jets as well.\\

In fact, the association proposed by Paredes et al. (2000) of the microquasar LS 5039 with the gamma-ray unidentified source 3EG J1824-1514, was a first observational step towards the identification of the gamma-ray emitting galactic jet sources.\\

This was the starting point for the research described in this thesis, also motivated by the significant number of unidentified gamma ray sources contained in the $3^{\rm rd}$ EGRET catalog. Many of these sources, as we saw in this text , are suspected to be of galactic origin because of their  spatial distribution. In order to establish microquasars as a parent population of some subset of unidentified gamma-ray sources we had first to look for a basic theoretical picture that could justify the gamma-ray production in different scenarios involving diverse conditions given in microquasars.\\

We started with a very simple leptonic toy model, in the context of high-mass microquasars, where the emission is due to inverse Compton process between relativistic leptons of the jet and the photon field of the high-mass stellar companion. We showed then that some variable unidentified EGRET sources in the galactic plane could be produced (with luminosities in the range $\sim 10^{35-37}$ erg s{$^{-1}$}) by microquasars with precessing jets. When the jet points towards the observer, gamma-ray emission could be detectable yielding a variable source with weak or undetectable counterpart at longer wavelengths. The ``microblazar'' case was studied in some detail, reaching the conclusion that some weak or distant sources could periodically appear in the detectors like EGRET due to variable Doppler boosting magnification.\\

The next step was to add some complexity to the leptonic model, by studying also the inverse Compton interactions with other external photon fields due to the accretion disk and the corona. We concluded that for HMMQs the predominant contribution comes from the scattering of the stellar photon field. This is explained by the higher luminosities  obtained in this way for MeV gamma-ray energies. For higher energies, where the luminosities due to the upscattered disk and coronal photons might be important, the absorption by pair creation makes it difficult for these gamma-rays to escape in many cases. At this point, the model was applied to the recurrent flaring events that have been recently reported for the galactic black hole candidate Cygnus X-1 at X-ray and soft gamma-ray energies, obtaining the observed luminosities and explaining the variability of these events through the precession of the jet.\\

We then turned towards hadronic models for the gamma-ray emission in HMMQs, where the jets are assumed to have a relativistic proton content. The jets of microquasars with massive, early-type stellar companions are exposed to the dense matter field of the stellar wind. We presented estimates of the gamma-ray emission expected from the jet-wind interaction (typical luminosities of $\sim 10^{34-36}$ erg s$^{-1}$, spectra harder than in leptonic cases). The proposed mechanism could explain some of the unidentified gamma-ray sources detected by EGRET on the galactic plane as well as predict possible neutrino detections since those particles are produced in the successive decays that follow the $p-p$ interaction.\\

In the case of the subset of unidentified EGRET sources whose spatial distribution forms a halo around the galactic center, we considered low-mass microquasars as their possible counterparts since these are old sources that, as we saw, can present large proper motions. We performed detailed calculations of the jet inverse Compton emission in the seed photon fields from the star, the accretion disk, and the hot coronal region, in different configurations of parameters such as jet Lorentz factors, powers, and angles with the line of sight. The conclusion was that unlike the HMMQs case, the external Compton emission largely fails to produce the required luminosities. Synchrotron-self-Compton emission appears as a promising alternative.\\ 

In order to test these models in a statistically significant number of sources, new generation of gamma-ray detectors is needed. In fact these satellites are already planned for the quite near future, which is the case of the AGILE and GLAST missions. Their sensitivity is expected to be about 10 to 100 times better than the EGRET one.\\ 

The variety of models proposed along the thesis was not only applied to the EGRET range of energies, but also to lower ones, like in the case of the soft gamma-ray emission from Cyg X-1, as well as to predict TeV gamma-ray production in the hadronic or the leptonic cases. The jets of microquasars could not only be the site of emission that would cover the electromagnetic spectra from radio to MeV, GeV or even TeV energy ranges, but as we proposed here they may also be a source of neutrinos. Furthermore, it has even been suggested that they may be important sites of particle acceleration in the interstellar medium\footnote{Heinz \& Sunyaev (2002) have discussed the possible contribution of X-ray binary jets to the production of galactic cosmic rays. They conclude that, whereas in terms of overall energetics such jets are still likely to inject less power into the ISM than supernovae, they may contribute a specific and detectable component to the cosmic ray spectrum. In particu
 lar, the shocks in the ISM associated with jets from X-ray binaries will be considerably more relativistic than those associated with the supernovae, and thus may be considerably more efficient for particle acceleration.}.\\

The existence of TeV emission in MQs might be soon confirmed by new and powerful instruments like MAGIC, HESS and VERITAS. Neutrino observations with IceCube and ANTARES will be crucial with respect to the hadronic model predictions.  

\vspace{0.9cm}
\begin{center}
$\clubsuit$
\end{center}
\vspace{0.9cm}

In such a rich context of physical processes to be explained, observations to be interpreted, diverse theories to be tested, we certainly foresee new projects to straightforwardly continue and immediately complement the work developed during this thesis period. To begin with, a deep and detailed study of the SSC emission in low-mass microquasars systems is desirable, after realizing the promising prospect this process could have to explain the observed luminosities in some cases. Different combinations of parameters as well as scenarios with and without adiabatic losses will be studied.\\

The generation of jets, either steady or transient, is also a pending issue that will be explored in the next stage of our research. This is a more difficult ground that involves the complex physics relating the accretion disk to energetic outflows of matter. The tight correlations observed in MQs between radio emission and changes in the X-ray state hopefully will provide some new hints to make progress in this direction. 

\begin{figure}[!b]
\begin{center}
\resizebox{13cm}{!}{\includegraphics{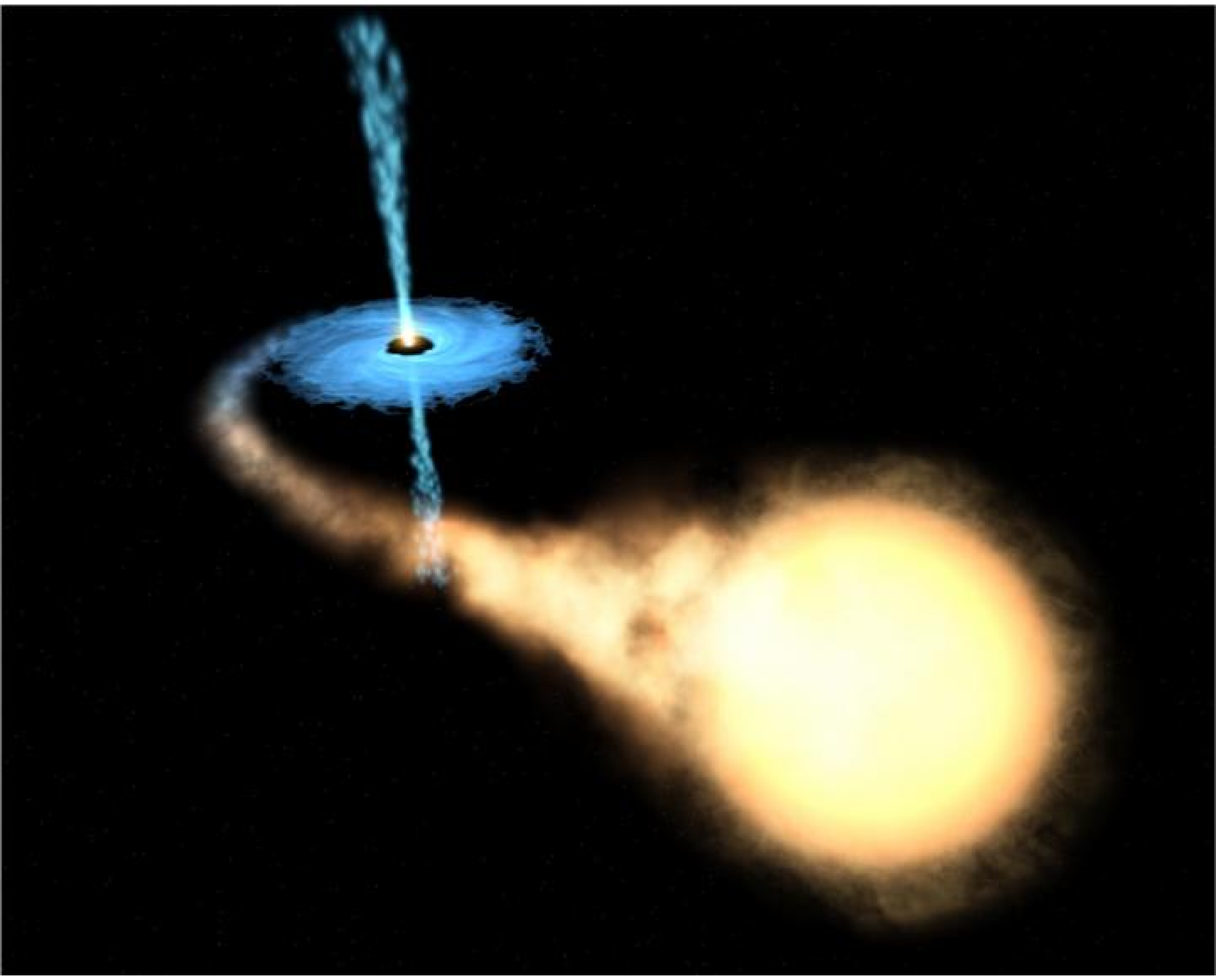}}
\end{center}
\textit{The central aim of this thesis was to propose diverse models that could support the idea that microquasars are possible gamma-ray emitters. In the different models developed for this work, gamma-rays are produced by leptonic or hadronic processes always related with the interaction of the jets with diverse photon and matter fields. The next stage of our research will be connected with the generation of these jets as well as with the different lines of research opened along the thesis period. The discovery of this local scale-down version of quasars is allowing a better understanding of the physics related with the accretion-ejection relativistic phenomena in the Universe.}
\end{figure}

\appendix

\chapter{\label{NS}Gamma-Rays from Magnetized Accreting Pulsars}
\thispagestyle{empty}

\newpage
\thispagestyle{empty}
\phantom{.}

\newpage
\vspace*{6cm}

In this appendix we do not deal with the gamma-ray emission from MQs but we rather concentrate on a specific unidentified EGRET source, 3EG J0542+2610, that seems originated in a completely different kind of scenario where, nonetheless, similar physics to that described in Chapter~\ref{Hadron} take place. After a multi-wavelength study of the surroundings of this source we show that the only known object within the
95\% confidence location contour of the source capable of
generating the observed gamma-ray emission is the Be/X-ray
transient A0535+26. We shall argue that the gamma-rays are
produced during the accretion disk formation and loss phases in
each orbit, through hadronic interactions between relativistic
protons accelerated in an electrostatic gap in the pulsar
magnetosphere and the matter in the disk.\\ 

\section{Gamma-Ray Source 3EG J0542+2610}

\begin{figure}[!t]
\begin{center}
\resizebox{12cm}{!}{\includegraphics{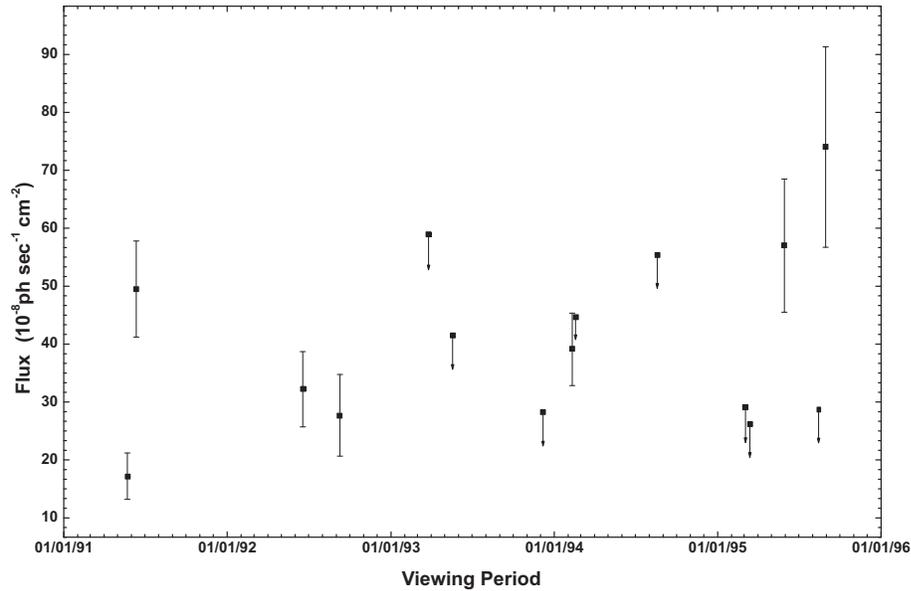}} 
\caption{\rm Flux
evolution of 3EG J0542+2610 through single viewing periods.}
\label{0535fig1}
\end{center}
\end{figure}

The best estimated position of the gamma-ray source 3EG J0542+2610
is at ($l$, $b$) $\approx$ (182.02,$-1.99$). Its 95\% confidence
location contour overlaps with the shell-type supernova remnant
(SNR) G180.0-1.7 (Romero et al. 1999). The
gamma-ray flux for the combined EGRET viewing periods is
$(14.7\pm3.2)\;10^{-8}$ ph cm$^{-2}$ s$^{-1}$. Since the flux is
highly variable on timescales of months, it should originate
in a compact object and not in the extended SNR. In Fig.~\ref{0535fig1} we
show the EGRET light curve. The source switches between periods of
clear detections and periods when only upper bounds to the flux
can be determined. The variability analysis of Torres et al.
(2001b) assigns to 3EG J0542+2610 a variability index $I=3.16$,
which means that the source variability level is $4.32\sigma$
above the average (spurious) variability of all known gamma-ray
pulsars. Tompkins' (1999) variability index for this source
($\tau=0.7$) also indicates that the source is variable.\\

If a pulsar origin is discarded due to the high variability and
the steep spectral index ($\Gamma=-2.67\pm0.22$), we are left with
two main possibilities: 1) the source is a background, unnoticed
gamma-ray blazar seen through the galactic plane, or 2) it is a
galactic compact object with an energy budget high enough as to
generate significant gamma-ray emission and, at the same time, does not have
the stable properties usually associated with isolated
pulsars.\\

\begin{figure}
\begin{center}
\resizebox{6cm}{!}{\includegraphics{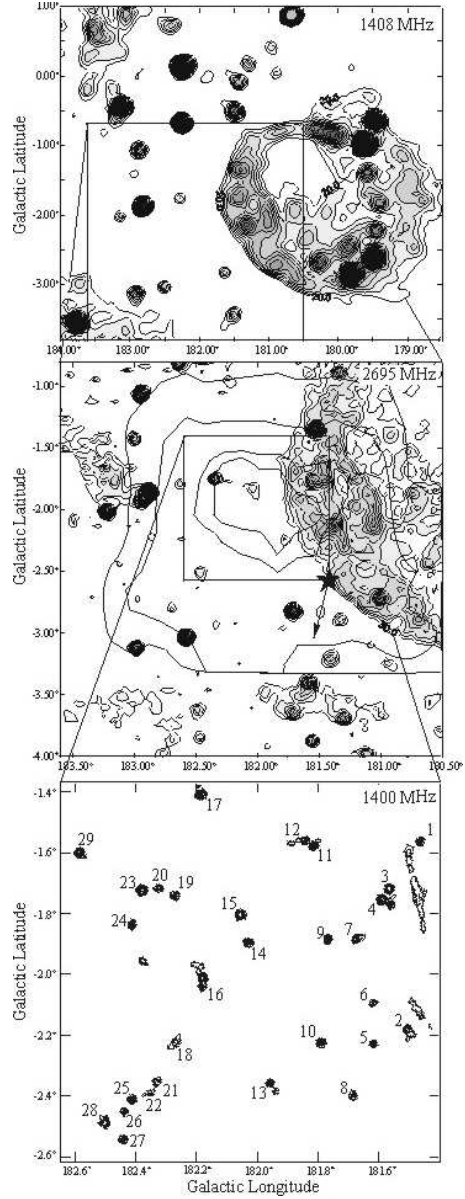}}
\caption{Radio maps at different resolutions of the field
containing the gamma-ray source 3EG J0542+2610. {\bf Upper panel}:
1408-MHz image obtained with Effelsberg 100-m single dish
telescope. The shell-type SNR G180.0-1.7 is clearly visible in the
map. Contours are shown in steps of 25 mJy beam$^{-1}$, starting
from 20 mJy beam$^{-1}$. {\bf Middle panel}: 2695-MHz map obtained
by the same telescope. Contours in steps of 30 mJy beam$^{-1}$,
starting from 30 mJy beam$^{-1}$. The position of A0535+26 is
marked by a star symbol. The arrow indicates the direction of the
proper motion. EGRET confidence location contours are superposed
to the radio image. {\bf Lower panel}: VLA image of the inner
region at 1.4 GHz. Contours in steps of 1 mJy beam$^{-1}$,
starting from 1 mJy beam$^{-1}$. The diffuse background emission
has been removed from the first two maps.} \label{0535fig2}
\end{center}
\end{figure}

In Fig.~\ref{0535fig2}, lower panel, we show a 1.4-GHz VLA map made with data
from the NVSS Sky Survey (Condon et al. 1998), where all
point-like radio sources within the 68\% confidence contour of 3EG
J0542+2610 can be seen. The measured characteristics of these 29
sources are listed in Table~\ref{radio}. Most of them have no entry in any
existing point source catalog. In those cases where we were able
to find positional counterparts at other frequencies we have
estimated the spectral indices, which are also shown in the table.
Most of these sources are very weak, at the level of a few mJy. No
strong (at Jy level), flat or nonthermal source is within the
location error box of the gamma-ray source. The strongest radio
source (No. 23 in our table) has a rather steep spectrum and is a
factor $\sim 10$ below the minimum flux density of firm gamma-ray
blazar identifications given by Mattox et al. (1997). The nature
of this source is not clear at present; it could be a background
weak radio quasar. The fact that it is not seen at X-rays seems to
argue against a galactic microquasar or any other kind of
accreting source.\\

\begin{table}[!t]\label{radio}
\begin{center}
\caption[]{Point radio sources within the inner location
probability contours of the gamma-ray source 3EG J0542+2610}
\vspace{0.8cm}
\begin{tabular}{ccccc}
\hline
\hline
Number & $l$ & $b$ & $S_{1.42\;{\rm GHz}}$ & Spectral\\ (on the
map) & (deg) & (deg) & (mJy) & index\\
\hline
\hline
\\
1&181.46&-1.56&18.00&\\
2&181.50&-2.18&13.15&$\alpha_{1.42}^{0.408}=-2.9$\\
3&181.56&-1.71&62.90&$\alpha_{1.42}^{0.408}=-1.3$\\
4&181.59&-1.75&78.67&$\alpha_{1.42}^{0.365}=-1.6$\\
&&&&$\alpha_{4.85}^{1.42}=-0.9$\\ 5&181.61&-2.22&7.65&\\
6&181.62&-2.09&7.39&\\ 7&181.67&-1.88&22.47&\\
8&181.68&-2.39&14.61&\\ 9&181.76&-1.88&25.00&\\
10&181.78&-2.22&35.00&$\alpha_{4.85}^{1.42}=0.2$\\
11&181.81&-1.58&27.00&\\ 12&181.84&-1.56&18.22&\\
13&181.95&-2.35&11.80&\\ 14&182.02&-1.89&20.32&\\
15&182.05&-1.80&53.40&\\ 16&182.18&-2.01&37.30&\\
17&182.18&-1.40&33.00&\\ 18&182.27&-2.22&8.70&\\
19&182.27&-1.73&18.88&\\
20&182.32&-1.71&9.91&$\alpha_{1.42}^{0.151}=-2.27$\\
21&182.33&-2.35&10.30&\\ 22&182.35&-2.38&6.00&\\
23&182.38&-1.72&225.40&$\alpha_{0.151}^{0.365}=-0.86$\\
&&&&$\alpha_{1.42}^{0.408}=-0.75$\\ 24&182.40&-1.83&18.72&\\
25&182.41&-2.41&20.67&\\ 26&182.43&-2.44&11.90&\\
27&182.44&-2.53&22.33&\\
28&182.50&-2.48&57.80&$\alpha_{1.42}^{0.408}=-1.33$\\
29&182.58&-1.59&41.86&\\
\\
\hline
\hline
\end{tabular}
\end{center}
\end{table}

At X-ray energies the most significant source within the EGRET
error box is the X-ray transient A0535+26, which is discussed in
the next section. This source does not present significant radio
emission and consequently it cannot be seen in our maps. We have
indicated its position with a star symbol in Fig.~\ref{0535fig2}, middle panel.
We also show in this figure the direction of the proper motion of
the system, as determined by Lee Clark \& Dolan (1999). The upper
panel shows the entire radio field as determined from radio
observations with Effelsberg 100-m single dish telescope at 1.408
GHz (data from Reich et al. 1997). The middle panel present an
enhanced image obtained at 2.695 GHz with the same telescope
(F\"urst et al. 1990), where the gamma-ray location probability
contours have been superposed (Hartman et al. 1999). We have
processed these large-scale images using the background filtering
techniques described by Combi et al. (1998).\\

\section{X-Ray Transient A0535+26}

A0536+26 is a Be/X-ray transient where the compact object is a
104s pulsar in an eccentric orbit around the B0III star HDE 245770
(Giovanelli \& Sabau Graziati 1992). Be stars are rapidly rotating
objects which eject mass irregularly forming gaseous disks on
their equatorial planes. If there is a compact companion in a
close orbit, accretion from the star can result in strong X-ray
emission. In the case of A0535+26, strong and recurrent X-ray
outbursts are observed with a period of 111 days, which has been
identified with the orbital period (Giovanelli \& Sabau Graziati
1992). It is generally agreed that these outbursts occur when the
accretion onto the neutron star increases at the periastron
passage. The average ratio of the X-ray luminosity at the
periastron to that of the apoastron is $\sim100$ (Janot-Pacheco et
al. 1987). In Table~\ref{0535} we list the main characteristics of the
A0535+26 system.\\

During a major outburst of A0535+26 in 1994, the BATSE instrument
of the Compton Gamma Ray Observatory detected a broad
quasi-periodic oscillation (QPO) in the power spectra of the X-ray
flux (Finger et al. 1996). The QPO component was detected during
33 days, with a central frequency that was well correlated with
both the hard X-ray flux and neutron star spin-up rate inferred
from pulse timing. Finger et al.'s (1996) observations are the
first clear evidence that an accretion disk is formed during giant
outbursts.\\

Using the simultaneous variations of the spin and the QPO
frequencies in the context of the beat frequency model (Alpar \&
Shaham 1985), Li (1997) has determined the evolution of the ratio
$\xi$ of the inner accretion disk radius to the Alfv\'en radius.
He found that during the initial rise of the outbursts $\xi$
quickly increased from $\sim 0.6$ to $\sim 1$, indicating a
transition of the accretion process from spherical accretion
before the outbursts to disk accretion during the high X-ray
luminosity phase.
Prior to the direct evidence for a transient accretion disk in
A0535+26, Motch et al. (1991) had already suggested, on the basis
of an analysis of the long-term X-ray, UV, and optical history of
the system, that two different types of interactions exist between
the Be star and the pulsar, accordingly with the dynamical state
of the highly variable circumstellar envelope. ``Normal" outbursts
would occur for high equatorial wind velocities ($\sim 200$ km
s$^{-1}$), whereas ``giant" outburst would result from lower
equatorial wind velocities ($\sim 20-80$ km s$^{-1}$) which allow
the formation of a transient accretion disk.\\

It is interesting to notice that, during the initial stage of the
1994 outburst, when QPOs were found in A0535+26, the gamma-ray
source 3EG J0542+2610 was detected by EGRET with a flux of
$(39.1\pm12.5)\;10^{-8}$ ph cm$^{-2}$ s$^{-1}$ (viewing period
321.1: February 8 - 15, 1994). But when A0535+26 was at the peak
of its X-ray luminosity, on February 18, the gamma-ray source was
not detected (viewing period 321.5). The gamma-ray emission seems
to have been quenched precisely when the accretion disk was
well-formed and maximally rotating. In the next section we present
a model that can account for the gamma-ray production in A0535+26
necessary to explain the EGRET source 3EG J0542+2610 and that is
in agreement with our present knowledge of the Be/X-ray transient
source.\\

\begin{table}\label{0535}
\begin{center}
\caption[]{Physical parameters for A0535+26 (from Janot-Pacheco et
al. 1987 and Giovanelli \& Sabau Graziati 1992)}
\vspace{0.8cm}
\begin{tabular}{l l}
\hline 
\hline
\\
Primary spectral type & B0III \cr Primary mass & 9 - 17
$M_{\odot}$ \cr Secondary mass & $<2.7\;M_{\odot}$\cr Primary mass
loss rate & $7.7\;10^{-7}$ $M_{\odot}$ yr$^{-1}$ \cr  Distance &
$2.6\pm0.4$ kpc \cr $L_X$ peak (average) & $7.5\pm2.4\;10^{36}$
erg s$^{-1}$ \cr X-ray pulse period & 104 s \cr Orbital period &
$111\pm0.5$ d \cr Orbital eccentricity & 0.3 - 0.8 \cr $L_X^{\rm
max}/L_X^{\rm min}$ & $\sim 100$ \cr 
\\
\hline
\hline
\end{tabular}
\end{center}
\end{table}

\section{The Model}

Our purpose in this section is to show that there exist a
plausible mechanism that could explain the gamma-ray emission of
the source 3EG J0542+2610 as originated in the X-ray transient
A0535+26. This mechanism should be capable of predicting the
observed gamma-ray flux, the variability in the lightcurve, the
fact that no gamma-ray emission was observed on February 18, 1994, when
A0535+26 was at the peak of the X-ray outburst, but also that it
was positively detected in the previous days when the X-ray flux
was rising, and finally the fact that A0535+26 is not a
non-thermal radio source. This latter restriction seems to suggest
a hadronic origin for the gamma rays. Otherwise, relativistic
electrons should also produce synchrotron radio emission.\\

\begin{figure}[!t]
\begin{center}
\resizebox{12cm}{!}{\includegraphics{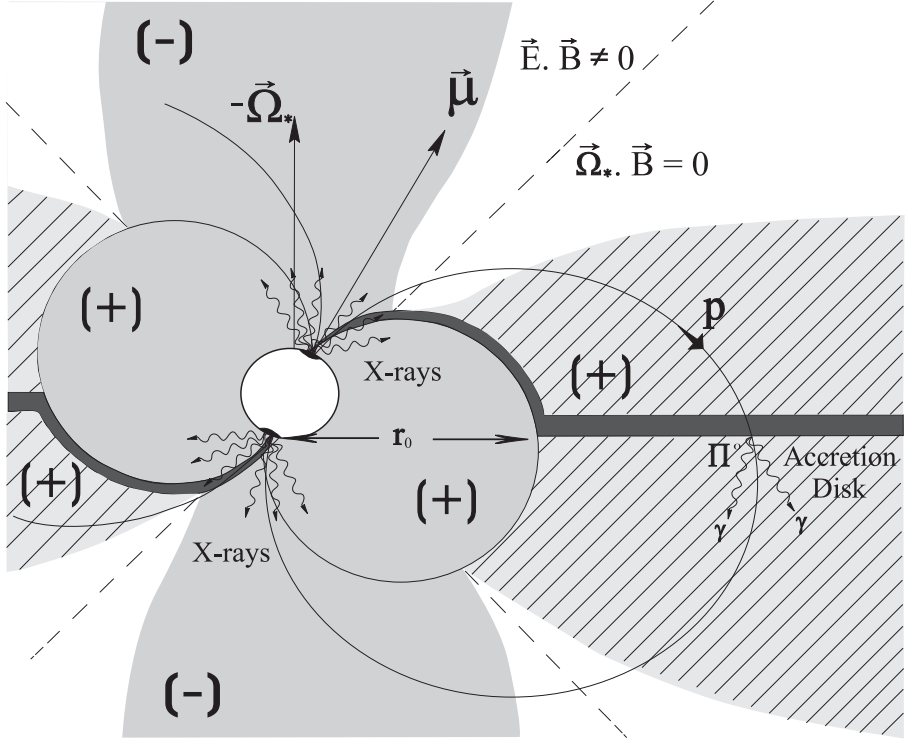}} 
\caption{Sketch of
the magnetosphere model for A0535+26 when $\Omega_{\rm
d}>\Omega_{*}$. The gray region co-rotates with the star, whereas
the hatched region co-rotates with the accretion disk. An empty
plasma gap where ${\bf E\cdot B}\neq 0$ exists around ${\bf
\Omega_{*}\cdot B}=0$. Protons from the stellar surface are
accelerated in this gap and collide with the disk, where they
produce $\pi^0$-decay gamma-rays. Adapted from Cheng \& Ruderman
(1991).} \label{0535fig3}
\end{center}
\end{figure}

Cheng \& Ruderman (1989, 1991) have studied the disturbances
produced in the magnetosphere of an accreting pulsar when the
Keplerian disk rotates more rapidly than the star. As in the case
of equal angular velocities, when $\Omega_{*}<\Omega_{\rm d }$,
inertial effects of electrons ($-$) and ions ($+$) lead to a
complete charge separation around the ``null surface" ${\bf
\Omega_{*}\cdot B}=0$. The equatorial plasma
between the inner accretion disk radius $r_0$ and the Alfv\'en
radius $r_{\rm A}$ co-rotates with the disk, whereas the rest of
plasma co-rotates with the star. An electrostatic gap with no
charge at all is then created around the ``null surface" (see Fig.~\ref{0535fig3}). In this gap ${\bf E\cdot B}\neq 0$ and a strong potential drop
is established (see Cheng \& Ruderman 1991 for details).\\

Cheng \& Ruderman (1989) have shown that the potential drop along
the magnetic field lines through the gap is:

\begin{eqnarray}
\Delta V_{\rm max}&\sim&\frac{B_{\rm s}R^3\Omega_{\rm d}(r_0)}{r_0
c }\nonumber\\ &\sim&4\;10^{14}
\beta^{-5/2}\left(\frac{M}{M_{\odot}}\right)^{1/7}\!\!
R_6^{-4/7}L_{37}^{5/7}B_{12}^{-3/7}\;\;{\rm V},
\end{eqnarray}

\vspace{0.4cm}

\noindent where $B_{\rm s}$ is the neutron star's surface dipole magnetic
field, $R$ is its radius, $r_0$ is the inner accretion disk
radius, $M$ is the compact star mass, and $L_{37}$ is the X-ray
luminosity in units of $10^{37}$ erg s$^{-1}$. The radius and
magnetic field in the second expression are in units of $10^6$ cm
and $10^{12}$ Gauss, respectively. The parameter
$\beta\equiv2r_0/r_{\rm A}$ is twice the ratio of the inner
accretion disk radius to the Alfv\'en radius. For the particular
case of A0535+26 we can adopt $\beta\sim 1$, according to the
estimates by Li (1997).\\

Protons entering into the gap are accelerated up to energies above
$E_p\sim e\Delta V_{\rm max}$ whereas inverse Compton and
curvature losses would limit the energy gain of electrons and
positrons to lower values. The maximum current that can flow
through the gap can be determined from the requirement that the
azimuthal magnetic field induced by the current cannot exceed that
of the initial field ${\bf B}$ (Cheng \& Ruderman 1989):

\begin{eqnarray}
J_{\rm max}&\sim& cB_{\rm s}R^3 r_0^{-2}\nonumber\\
&\sim&1.5\;10^{24}
\beta^{-2}\left(\frac{M}{M_{\odot}}\right)^{-2/7}\!\!\!\!
R_6^{1/7}L_{37}^{4/7}B_{12}^{-1/7}\;{\rm esu\;s^{-1}}.
\end{eqnarray}

\vspace{0.6cm}

\noindent As it can be seen from Fig.~\ref{0535fig3}, the proton current flux is directed
from the polar cap of the star, where the accreting material
impacts producing strong X-ray emission and abundant ions to be
captured by the gap, towards the accretion disk. Electrons move in
the opposite sense in order to cancel any net charge current flow,
keeping in this way ${\bf E \cdot B}\neq0$ in the gap.\\

The collision of the relativistic proton beam into the disk will
produce hadronic interactions with copious $\pi^0$ production
(Cheng et al. 1990). These $\pi^0$ will quickly decay into
gamma-rays that could escape only if the disk density is
sufficiently low. Otherwise, they will be absorbed by the matter
and then re-emitted as X-rays. The interaction of relativistic
protons with a thin hydrogen layer has been studied by Cheng et
al. (1990) in the context of their model for the Crab pulsar. The
$\pi^0$-decay gamma-rays will only escape insofar as the column
density of the disk would be $\Sigma\leq100$ g cm$^{-2}$. From
Novikov \& Thorne (1973), we have:

\vspace{0.3cm}

\begin{eqnarray}
\Sigma&\approx&7\;10^{2}
\beta^{-3/5}\left(\frac{M}{M_{\odot}}\right)^{-3/35}\!\!\!\!R_6^{-30/35}
L_{37}^{27/35}B_{12}^{-12/35}{\rm g\;cm^{-2}}.\label{mu}
\end{eqnarray}

\vspace{0.6cm}

\noindent In a system like A0535+26, the column density of the disk will
evolve with time following the variations in the X-ray luminosity.
All other physical parameters in Eq.~(\ref{mu}) remain constant
along the orbital period. If at some point near the periastron,
when the X-ray luminosity is close to its maximum, the column
density exceeds the value $\sim 100$ g cm$^{-2}$, the medium will
be no longer transparent to gamma-rays and the gamma-emission will
be quenched (we are assuming that the disk is seen from the
opposite side to that where the current impacts). This could
explain the fact that A0535+26 has not been detected by EGRET at
the peak of its X-ray outbursts on February 18, 1994. The
detection, instead, was clear a few days before, during the
previous viewing period.\\

The expected hadronic gamma-ray flux from A0535+26 on Earth
will be:

\vspace{0.3cm}

\begin{equation}
F(E_{\gamma})=\frac{1}{\Delta \Omega
D^2}\int^{\infty}_{E_{\gamma}}
\frac{d^{2}N_p(E_{\pi^0})}{dt\;dE_{\pi^0} }\;dE_p,\label{flux}
\end{equation}

\vspace{0.6cm}

where $D$ is the distance to the source, $\Delta \Omega$ is the
beaming solid angle and:

\vspace{0.3cm}

\be
\frac{d^{2}N_p(E_{\pi^0})}{dt\;dE_{\pi^0} }=\frac{2\,J_{\rm
max}}{e}\frac{\Sigma
<m_{\pi^0}>}{m_p}\frac{d\sigma(E_p,\,E_{\pi^0})}{dE_{\pi^0}} \ee

\vspace{0.4cm}

\noindent is the differential production rate of neutral pions from $p-p$
interactions. In this latter expression,
$d\sigma(E_p,\,E_{\pi^0})/dE_{\pi^0}$ is the differential
cross-section for the production of $\pi^0$-mesons of energy
$E_{\pi^0}$ by a proton of energy $E_p$ in a $p-p$ collision, and
$<m_{\pi^0}>$ is the mean multiplicity of $\pi^0$.\\

For calculation purposes we shall adopt the cross section given by
Dermer (1986a, b) and a beaming factor  $\Delta
\Omega/4\pi\approx0.3$ as used by Cheng et al. (1991). We shall
evaluate the total gamma-ray luminosity between 100 MeV and 20 GeV
of A0535+2610 at an early epoch $t$ of the disk formation, when
the X-ray luminosity ratio is $L_X^{\rm max}/L_X^{t}\sim 10$. If
the absorption feature observed near 110 keV is a cyclotron line,
the polar magnetic field of the neutron star results $9.5\;
10^{12}$ G (Finger et al. 1996). In our calculations we adopt this
value along with $\beta\sim 1$, $R_6=1$ and $M=1.4\;M_{\odot}$
(Janot-Pacheco et al. 1987). We then obtain that the column
density of the disk is $\sim42.5$ g cm$^{-2}$ at this stage and,
consequently, gamma-rays are not absorbed in the disk material.
The potential drop in the electrostatic gap results $\Delta V_{\rm
max}\sim2.5\;10^{13}$V whereas the proton current deposited into
the disk is $N_p=J_{\rm max}/e\sim4.8\;10^{32}$ s$^{-1}$. Using
Eq.~(\ref{flux}) for photons in EGRET's energy range we obtain:

\begin{equation}
F(100\;{\rm MeV}<E_{\gamma}<20\;{\rm
GeV})\approx5.1\;10^{-8}\;\;{\rm ph\;cm^{-2}\;s^{-1}}.
\end{equation}

\vspace{0.4cm}

This flux is consistent with the lower EGRET detections. As the
accretion onto the neutron star increases the gamma-ray flux also
increases, reaching a maximum when $\Sigma\approx100$ g cm$^{-2}$.
At this point the flux has grown by a factor $\sim 4$ (notice the
dependence on $J_{\rm max}$ in addition to $\Sigma$) and photon
absorption into the disk becomes important quenching the
radiation. The gamma-ray source, then, is not detected when the
peak of the X-ray luminosity occurs. The additional gamma-ray flux
expected from secondary electrons and positrons radiating in the
the strong magnetic fields anchored in the disk can explain even
higher EGRET fluxes (Cheng et al. 1991). A more detailed analysis
in this sense will be presented elsewhere, including spectral
considerations. The observed spectrum depends on the proton
injection spectrum at the gap. Strong shocks near the polar gap
can accelerate the protons up to a power law which will be
preserved through the propagation across the gap and imprinted
into the $\pi^0$ gamma-ray spectrum. Additionally, $e^+e^-$
cascades induced in the disk material can result in a relativistic
Bremsstrahlung contribution. Electromagnetic shower calculations
are in progress in order to determine this contribution (Romero et
al., in preparation).

\section{Discussion}

The model here outlined does not imply strict periodicity because
of the highly chaotic nature of the Be stellar winds, which can
significantly vary on short time-scales producing strong changes in
the accretion rate. Notwithstanding, the general prediction that
the peak of the gamma-ray emission should not be coincident with
the maximum X-ray luminosity during a given outburst can be used
to test the general scenario proposed here. At present, the poor
time resolution of the gamma-ray lightcurve does not allow
correlation studies. EGRET data for the best sampled X-ray
outburst (in February 1994) consist only of two viewing periods,
as it was mentioned, and in one of them the source was not
detected. In the future, however, new instruments like GLAST could
provide the tools for these kind of investigations.\\

TeV emission should be produced in the accretion disk according to
Eq.~(\ref{flux}), although degradation effects during the
propagation in the strong magnetic and photon fields around the
accretion disk could suppress much of it. The magnetic field at
the inner accretion disk radius is (e.g. Cheng et al. 1991):

\vspace{0.3cm}

\begin{equation}
B(r_0)=3\;10^5 \beta^{-3} B_{12}^{-5/7} R_6^{-9/7} L_{37}^{6/7}
(M/M_{\odot})^{-3/7}\;\;\;\;{\rm G} \label{B}
\end{equation}

\vspace{0.4cm}

\noindent which for the typical values of A0535+26 yields $B\sim10^4$ G. TeV
gamma-rays can be absorbed in the field through one-photon pair
production above the threshold given by (e.g. Bednarek 1993):

\vspace{0.2cm}

\begin{equation}
\chi\equiv \frac{E_{\gamma}B\sin{\theta}}{2m_e c^2 B_{\rm
cr}}=\frac{1}{15},
\end{equation}

\vspace{0.4cm}

\noindent where $\theta$ is the angle between the photon momentum and the
magnetic field and $ B_{\rm cr}$ is the critical magnetic field
given by  $B_{\rm cr}=m^2 c^3/ e\hbar\approx 4.4\;10^{13}$ G. This
process, then, suppress gamma-ray photons with energies higher
than $\sim300$ TeV in A0535+26. Since this value is well above the
energy of the protons that impact on the disk, we find that
absorption in the magnetic field should not occur.\\

However, TeV photons with lower energies should be absorbed by
two-photon pair production in the accretion disk X-ray
photosphere. Calculations by Bednarek (1993) show that the optical
depth quickly goes to values above 1 for disks with luminosities
$L_{37}\sim 1$. The opacity effects can reach even GeV energies
leading to a steepening in the spectrum respect to what is
expected from a pure pion-decay mechanism. The fact that the
observed spectrum in 3EG J0542+4610 has an index $\Gamma\sim -2.7$
seems to support the idea that important absorption is occurring
in the X-ray photosphere of this source. Additional spectral
modifications should be produced by the emission of secondary
pairs in the magnetic field close to the disk (Cheng et al. 1991,
Romero et al., in preparation).\\

In their original model, Cheng \& Ruderman (1989) suggested that
if the disk is sufficiently dense then the gamma-rays could be
produced only in a moving low-density ``window". This window would
collimate a pencil beam of gamma-rays aligned with the magnetic
axis. Consequently, a pulsed emission could exist with the same
period of the X-ray source (104 sec in the case of A0535+26).
Small changes in the gamma-ray period might be produced by the
radial motion of the ``window". It would be interesting to test
whether this pulses are present in A0535+26. However, since the
disk is a transient structure and only would excess the critical
density during a few days per orbit, the number of photon counts
in EGRET data are too low to allow a periodicity analysis as in
the case of isolated and stable gamma-ray pulsars. Instruments
with higher sensitivity like GLAST could sum up over several
viewing periods in order to look for these features.\\

\section{Conclusions}

We have shown that the Be/X-ray transient system A0535+26 can be
also a transient gamma-ray source under very reasonable
assumptions. The existence of a QPO phenomenon detected by BATSE
during the 1994 X-ray outburst provided direct evidence of the
formation of a transient accretion disk near the periastron
passage. In the beat frequency model for QPO the Keplerian orbital
frequency of the material at the inner edge of the accretion disk
should exceed the spin frequency of the neutron star. Cheng \&
Ruderman (1989) have shown that in such a circumstance an
electrostatic gap is open in the magnetosphere around the null
surface determined by ${\bf \Omega_*\cdot B=0}$. This gap can
accelerate protons up to energies of tens of TeV, producing a
hadronic current that impacts into the accretion disk generating
$\pi^0$ gamma-rays. The transient character of the disk makes the
high-energy gamma radiation highly variable, as was found in the
observed gamma-ray flux evolution of 3EG J0542+2610.\\

A specific prediction of the
model is the suppression of gamma-ray emission when the column
density of the disk exceeds a critical value, near the peak of the
X-ray luminosity. Future GeV and TeV observations of this source
with instruments of high temporal resolution, like GLAST or 5@5
(see Aharonian et al. 2001), could be used to test the proposed
model and, if it is basically correct, to probe the evolution of
the matter content on the accretion disk in this extraordinary
X-ray binary.

\newpage
\thispagestyle{empty}
\phantom{.}

\chapter{\label{radiative processes}Relevant Radiative Processes}
\thispagestyle{empty}

\newpage
\thispagestyle{empty}
\phantom{.}

\newpage
\vspace*{6cm}

The aim of this appendix is to provide some insight into the conceptual tools used in gamma-ray astrophysics and in particular in this thesis. We will
summarize the mechanisms that produce gamma rays in the astronomical events that were discussed along this thesis, and how these gamma rays can be affected once they are created. We will closely follow the presentation made by K.S. Cheng \& G.E. Romero in their recent book \textit{Cosmic Gamma-Ray Sources} (Cheng \& Romero 2004).\\

\section{Basic Concepts}

It is convenient to start by introducing as a basic concept
the number of particles incident per unit of surface
area per unit of solid angle per unit of time arriving
at a given, unspecified detector. We will call this
basic quantity the {\sl intensity} of particles, and
we will denote it by $I$. In general, we will use a
subscript to indicate the type of particles, e.g.
$I_{\gamma}$ and $I_{p}$ denote intensities of
gamma rays and protons, respectively.\\

Once the intensity is introduced, we can define the
particle flux as

\begin{equation}
F=\int_{\Omega} I \cos\theta d\Omega,
\end{equation}

\vspace{0.4cm}

\noindent where the angle $\theta$ is determined by the
direction of motion of the particles with respect to
the normal to the area, and the integration is
performed over the solid angle. For isotropic
radiation the flux is $F=\pi I$, and the number of
particles per unit of volume is
\begin{equation}
N=\frac{4\pi}{v}I.
\end{equation}

\vspace{0.4cm}

\noindent In most cases $v=c$ is a good approximation because we
deal with relativistic particles.\\

Normally, we will have particles with different
energies, so it is useful to introduce a particle
energy distribution $N(E)$ such that

\begin{equation}
N=\int^{\infty}_{0} N(E) dE.
\end{equation}

\vspace{0.4cm}

\noindent The number density of particles with energies greater
than $E$ is obtained just by integrating from $E$. In
a similar way, the integrated flux density is

\begin{equation}
F(>E)=\int^{\infty}_{E} F(E) dE.
\end{equation}

\vspace{0.4cm}

\noindent The luminosity of a source located at a distance $d$
that radiates isotropically is given by
\begin{equation}
L(>E)=4\pi d^2 \int^{\infty}_{E} F(E) dE,
\end{equation}

\vspace{0.4cm}

\noindent where $d$ is the distance to the source.\\

The energy density of the particles is

\begin{equation}
w=\int^{E_{\rm max}}_{E_{\rm min}} E N(E) dE.
\end{equation}

\vspace{0.4cm}

\noindent The energy flux is obtained from this expression just
by multiplying by $c/4\pi$, if we deal with
relativistic particles or photons.\\

Let us consider now that a flux of particles of type
$a$, with velocity $v_a$, interacts with some target
formed by particles of type $b$ within a volume $dV$.
The number of particle interactions of a given type,
$dN_i$, occurring in a time $dt$ in the volume $dV$
will be proportional to the number of particles $b$ in
the volume $dV$ and to the number of incident
particles that traverse
the cross section $dA$ of that volume in the time
$dt$:

\begin{equation}
     dN_i=d\sigma_i(n^0_b dV)(n_a v_a dt).
\end{equation}

\vspace{0.4cm}

\noindent In this expression $n^0_b$ and $n_a$ are the densities
of target and incident particles in a coordinate
system with the target at rest. The differential cross
section $d\sigma_i$ characterizes the number of
reactions of type $i$ occurring per unit of time in
unit volume for a unit flux density of incident
particles and unit density of the target. It is
measured in units of area, the standard unit being the
mb (i.e. $10^{-3}$ barn, 1 barn$=10^{-24}$ cm$^{2}$).\\

The total cross section for a given interaction
$\sigma_i$ is the sum over all possible momenta of the
resulting particles after the interaction. Both
$\sigma_i$ and $d\sigma_i$ are relativistic
invariants. The total cross section $\sigma_{\rm tot}$
is obtained by summing the cross sections of all
possible processes that occur upon the interaction of
particles $a$ and $b$, i.e.

\begin{equation}
\sigma_{\rm tot}=\sum_{i}\sigma_i.
\end{equation}

\vspace{0.4cm}

\noindent The relative probability of a given reaction channel
is given simply by $p=\sigma_i/\sigma_{\rm tot}$.\\

In the case of gamma-ray emission, if the generation
of the gamma rays is due to the interaction of
particles of type $i$ with a given intensity
$I_i(E_i,\vec{r})$ with a target of density
$n(\vec{r})$, we can write the intensity of the
radiation from the resulting gamma-ray source as

\begin{equation}
I_{\gamma}(E_{\gamma})=\int_{\vec{l}}
\int_{E_{\gamma}}^{\infty} n(\vec{r}) \sigma(E_i,
E_{\gamma}) I_i(E_i,\vec{r}) dE_i dr, \label{I_g}
\end{equation}

\vspace{0.4cm}

\noindent where $\vec{l}$ defines the direction along the line
of sight (i.e. $\vec{l}=\vec{r}/r$). The emissivity of
the gamma-ray source is defined as

\vspace{0.2cm}

\begin{equation}
q_{\gamma}(E_{\gamma}, \vec{r})\int_{E_{\gamma}}^{\infty} n(\vec{r})
\sigma(E_i, E_{\gamma}) I_i(E_i,\vec{r}) dE_i,
\end{equation}

\vspace{0.4cm}

\noindent in such a way that

\begin{equation}
I_{\gamma}(E_{\gamma})=\int_{\vec{l}}
q_{\gamma}(E_{\gamma},\vec{r}) dr.
\end{equation}

\vspace{0.6cm}

\section{Gamma--Ray Production}

\subsection{Thermal Mechanism}

Any body with a temperature different from zero emits
thermal radiation. For a perfect absorber in
thermodynamical equilibrium at temperature $T$ (i.e.
for a black body) the spectrum will be given by the
Planck formula:

\vspace{0.2cm}

\begin{equation}
I_{\rm BB}(E_{\rm ph})=\frac{2 E_{\rm
ph}^3}{(hc)^2}\left[ \frac{1}{\exp(E_{\rm ph}/kT)-
1}\right],
\end{equation}

\vspace{0.6cm}

\noindent where $h$ and $k$ are the Planck and Boltzmann's
constants, respectively. The corresponding number
density of photons per unit energy is:

\vspace{0.2cm}

\begin{equation}
n_{\rm BB}(E_{\rm ph})=\frac{(E_{\rm ph}/m_{\rm
e}c^2)^2}{m_{\rm e}c^2 (\pi^2 \lambda_{\rm c}^3)}\left[ \frac{1}{\exp(E_{\rm
ph}/kT)-1}\right].
\end{equation}

\vspace{0.6cm}

\noindent In this last expression, $\lambda_{\rm c}=(2\pi
m_{\rm e} c)^{-1} h$ is the Compton wavelength of the electron. The
maximum of the intensity occurs at

\begin{equation}
E_{\rm ph, \; max}\approx 4.7\times 10^{-10}
\left(\frac{T}{K} \right)\;\;{\rm MeV},
\end{equation}

\vspace{0.4cm}

\noindent and the average energy of the photons is

\begin{equation}
\left\langle E_{\rm ph}\right\rangle=2.7 kT\approx
2.3\times 10^{-10} \left(\frac{T}{K} \right)\;\;{\rm
MeV}.
\end{equation}

\vspace{0.4cm}

In order to have photons with average energies of 1
GeV, temperatures of $\sim10^{13}$ K are necessary.
These temperatures cannot be found in steady
astrophysical objects, but only in explosive events
and in the Big Bang. In addition, the photon density
in a source with $T\sim10^{13}$ K would be $\sim 3
\times 10^{34}$ cm$^{-3}$. Since the mean free path of
a photon in this radiation field is
$\lambda_{\gamma}\sim (n \sigma_{\gamma\gamma})^{-
1}<<$ 1 cm, the source would be self-absorbed by photon-photon pair creation.\\

Typical astrophysical gamma-ray sources in the
continuum are non-thermal sources where the gamma rays
are produced by the interaction of relativistic
particles with radiation or matter fields. We describe next the interactions that are present in our models.\\

\subsection{Non-Thermal Mechanism I: Particle--Field Interactions}

\vspace{0.6cm}

\subsubsection{\small{\textsl{SYNCHROTRON RADIATION}}}

A relativistic particle moving in a magnetic field
will emit photons within an angle $\theta\sim mc^2/E$
of its direction of motion. In a magnetic field $B$ an
electron moves along a helical path with an angular
frequency $\omega_B$ given by

\begin{equation}
\omega_B=\frac{eB}{m_{ e} c}\frac{m_{ e}c^2}{E}.
\end{equation}

\vspace{0.4cm}

\noindent The radiation spectrum of the electron is given by
(e.g. Ginzburg \& Syrovatskii 1964):

\begin{equation}
P(E)=\frac{\sqrt{3} e^3}{m_{\rm e} c^2} B_{\bot}
\frac{E}{E_{\rm c}} \int^{\infty}_{E/E_{\rm c}}
K_{5/3}(\eta) d\eta, 
\label{P}
\end{equation}

\vspace{0.4cm}

\noindent where $E=h\nu$ is the energy of the radiation, $B_{\bot}=B \sin \theta$,
$\theta$ is the pitch angle, and $K_{5/3}$ is a modified Bessel
function of the second kind. The characteristic energy
of the photons is given by

\begin{equation}
E_{\rm c}=\frac{3h}{4\pi}\frac{e B_{\bot}}{m_{\rm
e}c}\left(\frac{E}{m_{\rm e}c^2}\right)^2.
\end{equation}

\vspace{0.4cm}

\noindent The maximum of $P(E)$ occurs at $E_{\rm max}=1.9\times
10^{-11} B_{\bot} (E/{\rm GeV})^2$ GeV. We see then
that only for extremely energetic particles and strong
magnetic fields we can get gamma-ray photons from
synchrotron radiation.\\

The total energy rate loss by synchrotron radiation of
an electron moving in a field $B$ can be obtained by
integrating equation (\ref{P}). The result is

\begin{equation}
-\left(\frac{dE_e}{dt} \right)_{\rm
syn}=\frac{2}{3}c\left(\frac{e^2}{m_e c^2} \right)^2
B^2_{\bot} \gamma^2,
\end{equation}

\vspace{0.4cm}

\noindent where $\gamma=E_e/m_{\rm e} c^2$ is the Lorentz factor of
the particle.\\

Introducing the Thomson cross
section $\sigma_{\rm T}= 8\pi e^4/3m_e^2 c^4\approx
0.665\times 10^{-24}$ cm$^{-2}$ and averaging over an
isotropic pitch angle distribution, the expression for
the energy losses can be set in the following
convenient form:

\vspace{0.2cm}

\begin{equation}
-\left(\frac{dE_e}{dt} \right)_{\rm
syn}=\frac{4}{3}\sigma_{T} c w_{\rm mag}
\gamma^2=0.66\times 10^{3} B^2 \gamma^2\;\;\; {\rm
eV/s},\label{t_syn}
\end{equation}

\vspace{0.4cm}

\noindent where $w_{\rm mag}=B^2/8\pi$ is the magnetic energy
density and $B$ is measured in Gauss.\\

If we have a homogeneous and isotropic power-law
electron distribution given by
\begin{equation}
     N_e(E_e)dE_e=K_e E_e^{-p} dE_e
\end{equation}

\vspace{0.4cm}

\noindent in a random magnetic field, the resulting spectrum is
(Ginzburg \& Syrovatskii 1964):

\begin{equation}
     I(E_{\gamma})=a(p) \frac{e^3}{m_{\rm e} c^2}
\left(\frac{3e}{4 \pi m_{\rm e}^3 c^5} \right)^{(p-1)/2}
B^{(p+1)/2} K_e L E_{\gamma}^{-(p-1)/2}.
\end{equation}

\vspace{0.4cm}

\noindent In this expression $L$ is the characteristic size of
the emitting region and $a(p)$ is given by

\begin{equation}
     a(p)=\frac{2^{(p-1)/2} \sqrt{3}
\Gamma\left(\frac{3p1}{12}\right)
\Gamma\left(\frac{3p+19}{12}\right)
\Gamma\left(\frac{p+5}{4}\right) }{8\sqrt{\pi} (p+1)
\Gamma\left(\frac{p+7}{4}\right)}.
\end{equation}

\vspace{0.4cm}

\noindent $a(p)$ is 0.147, 0.103, 0.0852, and 0.0742 for $p=$
1.5, 2, 2.5, and 3, respectively. The emission is a
power law with index $\alpha=(p-1)/2$.\\

In the case of a homogeneous magnetic field, the
degree of polarization is

\begin{equation}
     \Pi_0(p)=\frac{p+1}{p+7/3},
\end{equation}

\vspace{0.4cm}

\noindent which typically yields values in the range 69-75 \%
for $p$ between 2 and 3. If the magnetic field has
random component $B_{\rm r}$, then the degree of
polarization will be

\begin{equation}
     \Pi(p)=\Pi_0 \frac{B_0^2}{B_0^2 + B_{\rm r}^2},
\end{equation}

\vspace{0.4cm}

\noindent where $B_{\rm r}$ is the random component. Additional
details can be found in Ginzburg \& Syrovatskii (1964,
1965), in Blumenthal \& Gould (1970) and in Longair (1992, 1997).

\vspace{0.6cm}

\subsubsection{\small{\textsl{INVERSE COMPTON (IC) INTERACTIONS}}}

The scattering of relativistic electrons on soft
photons can produce gamma rays. According to equation
(\ref{I_g}) the intensity of the radiation from this
process when the soft photon field has a density
$n_{\rm ph}(E_{\rm ph}, \vec{r})$ is

\begin{equation}
I^{\rm IC}_{\gamma}(E_{\gamma})=\int_{\vec{l}}
\int_{E_{\gamma}}^{\infty}\int^{\infty}_{0}
I_e(E_e,\vec{r}) \sigma(E_e, E_{\gamma}, E_{\rm ph})
n_{\rm ph}(E_{\rm ph},\,\vec{r})  dE_e dE_{\rm ph}
d\vec{r}.   \label{I_ic}
\end{equation}

\vspace{0.4cm}

We can introduce a parameter $\xi=E_e E_{\rm ph}/(m_e
c^2)^2$, such that for $\xi<<1$ the scattering is
classical. In such a limit the cross section can be
approximated by the Thomson cross section $\sigma_{\rm
T}$ and the average energy of the emerging photons will be

\begin{equation}
\left\langle E_{\gamma}
\right\rangle=\frac{4}{3}\left\langle E_{\rm
ph}\right\rangle \gamma^2,
\end{equation}

\vspace{0.4cm}

\noindent where $\left\langle E_{\rm ph}\right\rangle$ is the
average energy of the target photons. The energy
losses for an electron in a photon field of energy
density $w_{\rm ph}$ when $\xi<<1$ can be approximated
as (e.g. Ginzburg \& Syrovatskii 1964):

\vspace{0.2cm}

\begin{equation}
-\left(\frac{dE_e}{dt}\right)_{IC}= c\sigma_{\rm T}
w_{\rm ph} \gamma^2\approx 2\times10^{-14} w_{\rm ph}
\gamma^2\;\;\; {\rm eV/s}.
\end{equation}

\vspace{0.6cm}

\noindent Comparing with equation (\ref{t_syn}), we see that at
these energies the ratio of synchrotron to IC cooling
times is simply $t_{\rm IC}/t_{\rm syn}\approx w_{\rm
mag}/w_{\rm ph}$.\\

If the incident electron spectrum is a power law and the photon
field can be approximated by a monoenergetic distribution, then
we get from equation (\ref{I_ic}):

\begin{equation}
I^{\rm IC}_{\gamma}(E_{\gamma})=\frac{1}{2} n_{\rm ph} L
\sigma_{\rm T} (m_e c^2)^{1-p} \left(\frac{4}{3}\left\langle
E_{\rm ph}\right\rangle\right)^{(p-1)/2}K_e E_{\gamma}^{
-(1+p)/2}.
\end{equation}

\vspace{0.4cm}

\noindent Here, $L$ is the typical source dimension, and $\left\langle
E_{\rm ph}\right\rangle$ and $n_{\rm ph}$ are average values for
the photon energy and the photon density in the source (Ginzburg
\& Syrovatskii 1964). If the photon field is thermal
radiation, then we get from equation (\ref{I_ic}):

\begin{equation}
I^{\rm IC}_{\gamma}(E_{\gamma})= \left(r_0^2/4\pi^2 \hbar^3c^2\right)
L K_e(kT)^{(p+5)/2}F(p) E_{\gamma}^{-(p+1)/2},
\end{equation}

\vspace{0.4cm}

\noindent where $T$ is the temperature and

\vspace{0.2cm}

\begin{equation}
F(p)=\frac{2^{p+3}(p^2+4p+11)\Gamma\left[\frac{1}{2}(p+5)
\right]\zeta\left[\frac{1}{2}(p+5)\right]}
{(p+3)^2(p+1)(p+5)},
\end{equation}

\vspace{0.6cm}

\noindent where $\zeta$ is the Riemann function. For $p=1.5$, $2.0$ and $2.5$, $F(p)=3.91$, $5.25$, and $7.57$ respectively (Blumenthal \& Gould 1970).\\

If $\xi>>1$, then the electron gives most of its
energy to the photon: $E_{\gamma}\sim E_e$. The cross
section at these energies decreases drastically and
can be represented by the well-known Klein-Nishina
formula (e.g. Heitler 1954):

\vspace{0.2cm}

\begin{equation}
\sigma_{\rm KN}(E_e, E_{\rm ph})=\frac{3}{8}
\sigma_{\rm T} \frac{m_e c^2}{\left\langle E_{\rm
ph}\right\rangle} \ln\left[\left(\frac{2\gamma
E_{\gamma}}{m_e c^2}\right)+\frac{1}{2}\right].
\end{equation}

\vspace{0.4cm}

Here, $\gamma=E_e/m_e c^2$ is the electron Lorentz
factor, as usual. The electron energy losses are now
given by 

\begin{eqnarray}
     -\left(\frac{dE_e}{dt}\right)^{\rm KN}_{IC}&=&
\frac{3}{8}c\sigma_{\rm T} w_{\rm ph} \left(\frac{m_e
c^2}{\left\langle E_{\rm ph}\right\rangle}\right)^2
\ln\left(\frac{2E_e\left\langle E_{\rm
ph}\right\rangle}{m^2 c^4}+\frac{1}{2}\right)\nonumber
\\
     &\approx &10^{-14}w_{\rm ph}
\left(\frac{m_ec^2}{\left\langle E_{\rm
ph}\right\rangle}\right)^2
\ln\left(\frac{2E_e\left\langle E_{\rm
ph}\right\rangle}{m^2 c^4}\right)\;\;\;{\rm eV/s}.
\end{eqnarray}

\vspace{0.4cm}

A useful general expression for IC scattering by an
electron moving in a monoenergetic, isotropic photon
field has been obtained by Jones (1968), applying the so-
called ``head-on" approximation: the seed photons are
treated as coming from the direction opposite to the
electron velocity. The corresponding cross section can
be written as (e.g. Blumenthal \& Gould 1970):

\begin{equation}
     \sigma_{\rm IC}(x, \epsilon_{\rm ph},
\gamma)=\frac{3\sigma_{\rm T}}{4 \epsilon_{\rm ph}
\gamma^2} f(x), \end{equation}

\vspace{0.4cm}

\noindent where

\begin{equation}
     f(x)=\left[2x\ln x +x+1-2x^2 +\frac{(4
\epsilon_{\rm ph} \gamma x)^2(1-x)}{2(1+4 \epsilon_{\rm ph}
\gamma x)} \right] P(1/4\gamma^2, 1, x), \label{fx}
\end{equation}

\vspace{0.4cm}

\noindent being $\epsilon_{\rm ph}=E_{\rm ph}/m_e c^2$ the
target photon energy and $x$ a function of the energy
$\epsilon_{\gamma}=E_{\gamma}/m_e c^2$ of the
scattered photons given by

\vspace{0.2cm}

\begin{equation}
      x=\frac{\epsilon_{\gamma}}{4 \epsilon_{\rm
                    ph}\gamma^2(1-
\epsilon_{\gamma}/\gamma)}.
\end{equation}

\vspace{0.6cm}

The function $P$ in equation (\ref{fx}) is 1 for
$1/4\gamma^2\leq x\leq 1$ and 0 otherwise. It
constrains the cross section to the physical case,
where the energy of the scattered photons cannot be
lower than  that of the seed photons or higher than the energy of electron. The maximum
energy of the scattered photons is $\sim (4
\epsilon_{\rm ph} \gamma_{\rm max}^2)/ (1+ 4
\epsilon_{\rm ph} \gamma_{\rm max})$, with
$\gamma_{\rm max}$ the maximum Lorentz factor for the
electrons.\\

\subsection{Non-Thermal Mechanism II: Particle--Matter Interactions}

\vspace{0.6cm}

\subsubsection{\small{\textsl{HADRONIC GAMMA--RAY EMISSION}}}

\subsubsection{$\pi^0$-decays from proton-proton
interactions}

The dominant $\pi$-producing channels in hadronic
interactions are ($E_{\rm thr}\sim m_{\pi} c^2$):

\begin{eqnarray}
p+p &\rightarrow& p+p+ a \pi^0 + b (\pi^+ + \pi^-)
\label{pi0}\\
p+p &\rightarrow& p+ n+\pi^+ + a \pi^0 + b (\pi^+ + \pi^-)\\
p+p &\rightarrow& n+n + 2\pi^+ + a \pi^0 + b (\pi^+ + \pi^-) \label{pi+},
\end{eqnarray}

\vspace{0.4cm}

\noindent where $a$ and $b$ are positive integers. Neutral pions decay into gamma rays with a proper lifetime of only
$9\times10^{-17}$ s. The gamma-ray emissivity generated through
such decays at a source with a proton spectrum
$I_p(E_p)=(c/4\pi)N_p(E_p) $ given by a power law:

\begin{equation}\label{powerlawp}
   I_p(E_p)=K E_p^{-\Gamma}
\end{equation}

\vspace{0.4cm}

\noindent is

\begin{equation}\label{Fgam}
   q_{\gamma}(E_{\gamma})=2\int^{\infty}_{E_{\pi}^
   {\rm min}}
   \frac{q_{\pi}(E_{\pi})}{\sqrt{E_{\pi}^{2}-
   m_{\pi}^{2} c^4}}
   \;dE_{\pi},
\end{equation}

\vspace{0.4cm}

\noindent where

\begin{equation}\label{Epi}
E_{\pi}^{\rm
min}(E_{\gamma})=E_{\gamma}+\frac{m_{\pi}^{2}
c^4}{4E_{\gamma}}, \end{equation}

\vspace{0.4cm}

\noindent and

\begin{equation}\label{Fpi}
q_{\pi}(E_{\pi})=4\pi \int^{E^{\rm max}_{p}}_{E^{\rm
min}_{p}} I_p(E_p)
\frac{d\sigma_{\pi}(E_{\pi},\;E_{p})}{dE_{\pi}}
\;dE_{p}.
\end{equation}

\vspace{0.4cm}

\noindent Here, $d\sigma_{\pi}(E_{\pi},\;E_{p})/dE_{\pi}$ is the
differential cross section for the production of
$\pi^0$-mesons of energy $E_{\pi}$ by a proton of
energy $E_{p}$ in a $p-p$ collision.\\ 

The gamma-ray emissivity can be approximated at high
energies by 

\begin{equation}
      q_{\gamma}(E_{\gamma}, \vec{r})\approx 4\pi
                      \sigma_{pp}
n(\vec{r})\frac{2 Z^{\Gamma}_{p\rightarrow\pi^0}}{\Gamma}
I_p(E_{\gamma}) \eta,
\end{equation}

\vspace{0.4cm}

\noindent where $Z^{\Gamma}_{p\rightarrow\pi^0}$ are the so-called spectrum
weighted moments (Gaisser 1990, Drury et al. 1994) and is $\sim$
0.275, 0.26, 0.245 for $\Gamma \sim$ 1.6, 1.8, 2.0 respectively,
$\eta\sim 1.5$ is a parameter that takes into account the
contribution of nuclei other than protons into the gamma-ray
production, and the cross section for inelastic $p-p$ interactions
can be approximated at $E_p\approx10 E_{\gamma}$, with
$E_{\gamma}>1$ GeV, by

\vspace{0.2cm}

\begin{equation} 
\sigma_{pp}(E_p)\approx 30
\times \left[0.95 + 0.06 \log \left(\frac{E_p}{\rm
GeV}\right)\right]\label{sigma_pp}\;\;\;{\rm mb}.
\end{equation}

\vspace{0.4cm}

The spectral gamma-ray intensity (photons per unit of
time per unit of energy-band) is

\begin{equation}
I_{\gamma}(E_{\gamma})=\int_V  q_{\gamma}(E_{\gamma},
\vec{r}) d^3\vec{r}, \label{I2}
\end{equation}

\vspace{0.4cm}

\noindent where $V$ is the interaction volume. The photon flux
observed at the Earth
from a source at a distance $d$ then results

\vspace{0.2cm}

\begin{equation}
F_{\gamma}(E_{\gamma})=\frac{I_{\gamma}(E_{\gamma})}{4 \pi d^2}.
\end{equation} 

\vspace{0.4cm}

\noindent Since the injection proton spectrum was a power
law, we can also expect a power law spectrum for the observed
gamma rays: 

\begin{equation}\label{powerlawg}
   F_{\gamma}(E_{\gamma})\propto E_{\gamma}^{-\Gamma}.
\end{equation}

\vspace{0.4cm}

For energies lower than $E_p\sim10$ GeV (i.e. which
corresponds to $E_{\gamma}\sim 1$ GeV) this simple
approach is not possible and we must use more complex parameterizations of the cross
section (see Blattnig et al. 2000).\\ 

As indicated by the reaction chains (\ref{pi0}) -
(\ref{pi+}), electron-positron pairs from the decay of the charged pions are also created
in $p-p$ interactions. These pairs can in turn
generate gamma rays through relativistic
Bremsstrahlung, IC, and/or synchrotron processes
depending on the environment. The total number of
pions generated in a $p-p$ interaction depends on
the energy: $N_{\pi}(E_p)=N_0 E_p^{\delta}$, with one
third each of $\pi^0$, $\pi^+$, and $\pi^-$. The total
energy that goes to pions is $E_{\pi}^{\rm
tot}=N_{\pi} \left\langle E_{\pi}\right\rangle=k E_p$,
with $k=k_0 E_p^{\alpha}$. Typical values for the
parameters are (e.g. Ginzburg \& Syrovatskii 1964):
$N_0=3.3$, $\delta=1/4$, $k_0=1/3$, and $\alpha=0$.
Hence, $\left\langle E_{\pi} \right\rangle=k
E/N_{\pi}(E_p)\approx0.1 E_p^{3/4}$. If we have an
incident proton spectrum with a power law intensity
$I_p(E_p)=K_p E_p^{-\Gamma}$ interacting with a medium
of particle density $n$, the pion emissivity can be
approximated by

\begin{equation}
     q_{\pi}(E_{\pi})dE_{\pi}=\sigma_{pp} n N_0
E_p^{\delta} I(E_p) dE_p=\sigma_{pp} n K_{\pi}^{-
\Gamma_{\pi}} dE_{\pi},
\end{equation}

\vspace{0.4cm}

\noindent where

\begin{equation} \Gamma_{\pi}=\frac{\Gamma+\alpha-
     2\delta}{1+\alpha-\delta}
\end{equation}

\vspace{0.2cm}

\noindent and

\begin{equation}
K_{\pi}=K_p\frac{N_0}{1+\alpha
\delta}\left(\frac{k_0}{N_0}\right)^{\Gamma_{\pi}-1}.
\end{equation}

\vspace{0.6cm}

The decays $\pi^{\pm}\rightarrow\mu^{\pm}+\nu$ and
$\mu^{\pm}\rightarrow e^{\pm}+\nu+\bar{\nu}$ will lead
to the injection of leptons and neutrinos in the
source. The electron emissivity can be approximated by
(Ginzburg \& Syrovatskii 1964, see also Dermer 1986c
for a more general expression):

\vspace{0.2cm}

\begin{equation}
     q_e(E_e)dE_e=\sigma_{pp} n K_e(\Gamma) K_{\pi}
E_e^{-\Gamma} dE_e,
\end{equation}

\noindent where

\begin{equation}
     K_e(\Gamma)=\left(\frac{m_{\mu}}{m_{\pi}}\right)^
     {\Gamma-
1}\frac{2(\Gamma+5)}{\Gamma(\Gamma+2)(\Gamma+3)}.
\end{equation}

\vspace{0.4cm}

\noindent Here, $ m_{\pi}c^2\approx140 $ MeV and $
m_{\mu}c^2\approx106 $ MeV.
The mean energies of the produced secondaries in the
lab frame are $\left\langle
E_{e^{\pm}}\right\rangle=E_{\pi^{\pm}}/4$,
$\left\langle E_{\nu,
\bar{\nu}}\right\rangle=E_{\pi^{\pm}}/4$, and
$\left\langle E_{\gamma}\right\rangle=E_{\pi^0}/2$.
Since we have equal luminosities for the different
pions, then $L_{\gamma}\sim L_{\nu, \bar{\nu}}\sim
L_{\pi}/2$.

\vspace{0.6cm}

\subsubsection{\small{\textsl{ELECTRON--POSITRON ANNIHILATION}}}

Electron-positron annihilation can be an important
source of gamma rays through the reaction
$e+e^+\rightarrow\gamma+\gamma$. When
the two particles are at rest the energy of the
resulting photons is simply $E_{\gamma}=m_e c^2=0.511$
MeV. Line radiation at this energy is usually referred
as annihilation radiation. If one of the leptons moves
at high velocity when it collides with the other at
rest, then one of the photons will have a high energy
whereas the other will have an energy $\sim 0.511$
MeV.  The cross section for pair annihilation of an
electron with energy $\gamma m_e c^2$ with a  positron
at rest is

\begin{eqnarray}
     \sigma_{e^{\pm}}&=&\frac{\pi
r_0^2}{\gamma+1}\left[\frac{\gamma^2+4\gamma+1}{\gamma
^2-1} \ln(\gamma+\sqrt{\gamma^2-1})-
     \frac{\gamma+3}{\sqrt{\gamma^2-1}} \right]\\
     &\approx&\frac{\pi r_0^2}{\gamma}[\ln(2\gamma)-
     1]\;\;\;{\rm
for}\; \gamma>>1\\
     &\approx&\frac{\pi r_0^2}{\beta}\;\;\;{\rm for}\;
\beta<<1,\;\;\; \beta=\frac{v}{c}.
\end{eqnarray}

\vspace{0.4cm}

\noindent Expressed in terms of the center-of-mass frame (cm)
this becomes

\begin{eqnarray}
\sigma(e^{\pm}_{\rm cm})=\frac{\pi r_0^2}{4\beta_{\rm
cm}\gamma_{\rm
cm}^2}\times\;\;\;\;\;\;\;\;\;\;\;\;\;\;\;\;\;\;
\;\;\;\;\;\;\;\;\;\;\;\;\;\;\;\;\;\;
\;\;\;\;\;\;\;\;\;\;\;\;\;\;\;\;\;\;\hspace{3cm}
&&\\ \nonumber
\times\left[\frac{1}{\beta_{\rm cm}}
\left(2+\frac{2}{\gamma_{\rm cm}^2}-
\frac{1}{\gamma_{\rm cm}^4}\right)
\ln\left(\frac{1+\beta_{\rm
cm}}{1-\beta_{\rm cm}}\right) -2-\frac{2}{\gamma_{\rm
cm}^2}\right],&&
\end{eqnarray}

\vspace{0.6cm}

\noindent where all quantities
are referred to the center-of-mass system of the
colliding particles.\\

Interpolating between the non-relativistic and ultra-relativistic limits
of the corresponding pair annihilation rates, a simple
expression with an
accuracy within 14 per cent can be found (Coppi \&
Blandford 1990):

\begin{equation}
R_{e^{\pm}}\approx\frac{3}{8}\frac{\sigma_{\rm T}
c}{x} [\ln x + x^{-1/2}],
\end{equation}

\vspace{0.4cm}

\noindent where $x=\gamma_{e^{+}}\gamma_{e^{-}}$.
Then, the annihilation luminosity for a plasma with
particle density given by
$N_{e^{\pm}}(E_{e^{\pm}})dE_{e^{\pm}}$ is:

\vspace{0.2cm}

\begin{equation}
L_{\gamma}^{({\rm ann})}=\int (E_{e^{+}}+E_{e^{-}})\:
R_{\pm}\: N_{e^{+}}(E_{e^{+}})N_{e^{-}}(E_{e^{-
}})\;dE_{e^{+}}dE_{e^{-}}dV. 
\end{equation}

\vspace{0.4cm}

Electron-positron annihilation can also occur with the
emission of a single photon, but in this case the
electron must be bound to an atom. The one-photon
annihilation cross section when the electron is bound
to an atom with charge $eZ$ is

\begin{eqnarray}
     \sigma^{1\;\rm ph}_{e^{\pm}}&=&\frac{4 \pi Z^5
\alpha^4
r_0^2}{\beta\gamma(\gamma+1)^2}\left[\gamma^2
+\frac{2\gamma}{3}+\frac{4}{3}
-\frac{\gamma+2}{\beta\gamma}
\ln[(1+\beta)\gamma] \right]\\
           &\approx&\frac{4 \pi Z^5 \alpha^4
               r_0^2}{\gamma}\;\;\;{\rm
for}\; \gamma>>1\\
           &\approx&\frac{4\pi Z^5 \alpha^4
               r_0^2\beta}{3}\;\;\;{\rm
for}\; \beta<<1,
\end{eqnarray}

\vspace{0.4cm}

\noindent where as before $\alpha$ is the fine structure
constant and the energy of the positron is $\gamma m_e
c^2$.\\

Three or more photons can be produced in the
annihilation of free electron-positron pairs, but the
cross section is down by a factor $\sim\alpha^{i-
2}\sim (1/137)^{i-2}$ from the free two photon
annihilation case, where $i$ is the number of
resulting photons.

\section{Gamma-ray absorption processes}

Gamma rays, once created at the source can be absorbed
by photon and matter fields either in the same source
or in the medium between the source and the detector.
We worked along the thesis with one of the main mechanisms that results in the absorption of gamma rays: pair creation in photon-photon
interaction.\\ 

If gamma rays with an initial intensity
$I^0_{\gamma}(E_{\gamma})$ are injected into a medium
of particle density $n$, the intensity after
traversing a distance $x$ will be

\vspace{0.2cm}

\begin{equation}
I_{\gamma}(E_{\gamma})=  I^0_{\gamma}(E_{\gamma}) e^{-
\tau}, 
\end{equation}

\noindent where
\begin{equation}
\tau=\sigma n x
\end{equation}

\vspace{0.4cm}

is the optical depth and $\sigma$ is the cross
section of the relevant interaction for the photons.
The quantity $\lambda_{\gamma}= (\sigma n)^{-1}$ is
the mean free path of the photon in the medium. The
probability for a photon to interact after traveling
a distance $L$ is $1-e^{-L/\lambda}$.\\

\subsection{Photon-Photon pair creation}

A gamma-ray photon of energy $E_{\gamma, 1}$ can
produce an electron-positron pair in a collision with a
photon of energy $E_{\gamma, 2}$ if $E_{\gamma,
1}E_{\gamma, 2}>(m_e c^2)^2$. The pair creation cross
section is (Dirac 1930):

\begin{equation}
     \sigma_{\gamma\gamma}(E_{\gamma, 1}, \;E_{\gamma,
2})=\frac{\pi r_0^2}{2}(1-\beta^2)\left[2\beta(\beta^2-
2)+(3\beta^4)\ln\left(\frac{1+\beta}{1-\beta}\right)
\right], 
\end{equation}

\vspace{0.4cm}

\noindent where $r_0$ is as before the classical radius of the
electron and

\begin{equation}
\beta=\left[1-\frac{(m_e c^2)^2}{E_{\gamma, 1} E_{\gamma,2}}\right]^{1/2}.
\end{equation}

\vspace{0.4cm}

\noindent The outgoing electron (positron) has an energy $m_e
c^2/\sqrt{1\beta^2}$ in the center-of-mass system. It is also useful to express the cross
section in terms of the total energy square in the center of
mass frame ($s=\left( p_{\gamma, 1}+p_{\gamma, 2}\right)^{2}$):

\begin{equation}
\sigma \left( s\right) =\frac{\pi r_{0}^{2}}{2}\left(
1-v^{2}_s\right) \left[
2v_s\left( v^{2}_s-2\right) +\left( 3-v^{4}_s\right) \ln \left( \frac{1+v_s}{1-v_s}
\right) \right],
\end{equation}

\vspace{0.4cm}

\noindent where $p_{\gamma}$ is the four-momentum of photon and

\vspace{0.2cm}

\begin{equation}
v_s=\left[1-\frac{4\left( m_{e}c^{2}\right) ^{2}}{s}\right] ^{1/2}. 
\end{equation}

\vspace{0.4cm}

\noindent The threshold of pair creation is $s$=$4m_{\rm e}^2 c^4$.\\

If a gamma ray of energy $E_{\gamma}$ should traverse a
region of size $R$ with a photon field of number
density $N(E_{\rm ph}, r)dE_{\rm ph}$, the optical
depth is given by:

\vspace{0.2cm}

\begin{equation}
     \tau_{\gamma\rm
ph}(E_{\gamma})=\int^{\infty}_{0}\int^{R}_{0} N(E_{\rm
ph}, r)\sigma_{\gamma\gamma}(E_{\gamma}, E_{\rm ph}) dr
\;dE_{\rm ph}.
\end{equation}

\vspace{0.4cm}

\noindent Gould \& Schr\'eder (1967) present estimates of the
absorption by a blackbody photon gas and a power-law
photon spectrum.\\

In the case of a gamma-ray source with an intrinsic
luminosity $L_{\gamma}$, the intrinsic $\gamma\gamma$-attenuation will be determined by the compactness parameter $l$, which is defined as the ratio of the
intrinsic luminosity to the source radius $R$, and the
mean photon energy $\left\langle
E_{\gamma}\right\rangle$:

\begin{equation}
     \tau_{\gamma\gamma}\approx \sigma_{\gamma\gamma}
n_{\gamma} R=\frac{\sigma_{\gamma\gamma}}{4\pi c
\left\langle E_{\gamma}\right\rangle} l.
\end{equation}

\vspace{0.4cm}

\noindent For $\left\langle E_{\gamma}\right\rangle=1$ MeV, we
have $\tau\sim 1.7 \times 10^{-31} l$, where $l$ is
measured in erg s$^{-1}$ cm$^{-1}$. We see, then, that
compact and luminous gamma ray sources can be self-absorbed through pair production. If the radiation is
beamed, the inferred (e.g. through variability
observations) luminosity might be significantly larger
than the intrinsic value ($L_{\rm app}=D^n L_{\rm
int}$, with $3\leq n\leq4$, where $D$ is the
Doppler factor\footnote{The Doppler factor is defined
as $D=[\Gamma(1-\beta\cos\phi)]^{-
1})$, where $\Gamma$ is the bulk Lorentz
factor of the radiating plasma, $\beta$ is its
velocity in units of $c$ and $\phi$ is the angle between the beamed radiation and the line of sight.}). In such a case the opacity
constraints are not so strong.

\newpage
\thispagestyle{empty}
\phantom{.}

\chapter{\label{superluminal}Superluminal Motion}
\thispagestyle{empty}

\newpage
\thispagestyle{empty}
\phantom{.}

\newpage
\vspace*{6cm}

\begin{figure}[!h] 
\centering 
\resizebox{10cm}{!}{\includegraphics{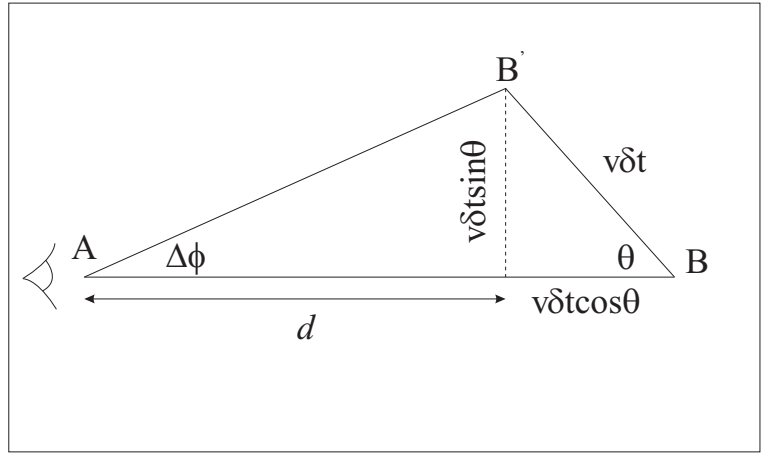}}
\caption{ {The geometry assumed in the model. The observer at A sees a radio source moving from B to B' at a speed apparently exceeding the speed of light, $c$. This effect can happen if $v \rightarrow c$ and the angle to the line of sight, $\theta$, is small, but non-zero.}}\label{superluminal-fig}
\end{figure}

\noindent Many explanations have been proposed to understand the existence of these apparently \textit{superluminal}, $v>c$, motions in radio sources. The most reliable is that apparent superluminal velocities are attributable to bulk relativistic motion along the line of sight to the continuum source. We will expand this idea following the model  presented in Blandford et al. (1977). We shall use Fig. \ref{superluminal-fig} to a better understanding of the situation.

\vspace{0.2cm}

Let us consider that the source is at point B at time $t_1$. Its emission is detected by an observer at point A at time $t'_1$ after the light has crossed the distance AB. Later, at $t_2=t_1+\delta t$, the source has moved a distance $v\delta t$. Observations are made again, recorded at $t'_2$, taking into account the light-travel time between the observer and the new position of the source. The distance between A and B is $d+v \: \delta t\: \cos\theta$ while the distance between A and B' can be approximated by $d$ if the angle to the line of sight, $\theta$, is small.\\

The angular separation between B and B' is given by,

\begin{equation}
 \Delta \phi = \frac{v\: \delta t \: \sin\theta}{d}
\end{equation} 

\vspace{0.4cm}

\noindent and the times of the observations recorded by the observer are,

\begin{eqnarray}
 \lefteqn{t'_1 = t_1+\frac{d+v \: \delta t \: \cos \theta}{c}}\nonumber\\
 & &t'_2 = t_2+\frac{d}{c}
\end{eqnarray} 

\vspace{0.4cm}

\noindent Then, the measured interval between the observations is:

\begin{eqnarray}
  \lefteqn{\Delta t \: = t'_2 - t'_1 = t_2 - t_1 - \frac{v \: \delta t \: \cos \theta}{c}} \nonumber\\
  & & = \delta t \:(1 - \beta \: \cos\theta)
\end{eqnarray} 

\vspace{0.4cm}

\begin{figure}[!t] 
\centering 
\resizebox{11cm}{!}{\includegraphics{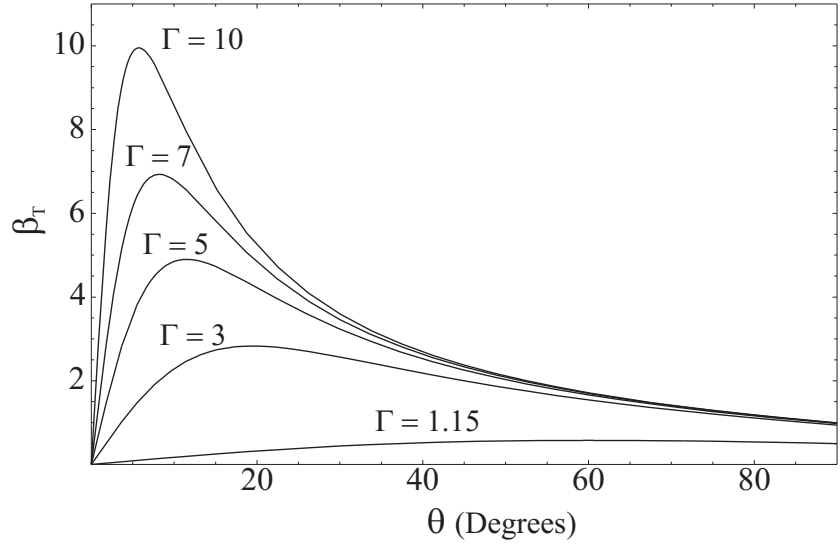}}
\caption{ {The apparent transverse velocity $\beta_{\rm T} = v_{\rm T} / c$ of a source moving at an angle $\theta$ to the observer's line of sight as a function of the Lorentz factor $\Gamma$. It can be seen how, for $\Gamma \gg 1$, the apparent transverse velocity can exceed the speed of light.}}\label{superlumcasos}
\end{figure}

\noindent where $\beta = v/c$. Therefore, the transverse velocity inferred by the observer is

\vspace{0.3cm}

\begin{equation}
 \beta_{\rm T} = \frac{v_{\rm T}}{c} = \frac{d}{c}\frac{\Delta\phi}{\Delta t} = \frac{v \sin\theta}{c(1-\beta \cos\theta)} = \frac{\beta \sin\theta}{1-\beta \cos\theta} 
 \label{final}
\end{equation} 

\vspace{0.5cm}

\noindent We can notice that $\beta_{\rm T} \rightarrow \infty$ if $\theta$ is small and $v \rightarrow c$.\\

Making equal to zero the differentiation of eq~\ref{final} with respect to $\theta$, we obtain at what value of $\theta$, $\beta_{\rm T}$ is maximized: $\theta_{\rm max} = \cos^{-1}\beta$. Inserting this result into eq~\ref{final} we get that $\beta_{\rm T}^{\rm max} = \beta \Gamma$, where $\Gamma$ is the Lorentz factor, $\Gamma = (1-\beta^2)^{-1/2}$. This is another way to realize that $\beta_{\rm T}$ can be arbitrarily high, since if $\beta \rightarrow 1$, $\beta \approx \Gamma$. \\ 

We show in Fig.~\ref{superlumcasos}, the measured transverse velocity $\beta_{\rm T}$ as a function of the angle to the line of sight, $\theta$, for different values of the Lorentz factor, $\Gamma$.\\

The situation of an approaching source have been analyzed up to here. The case of a receding source yields the following result:

\vspace{0.5cm}

\begin{equation}
 \beta_{\rm T_{receding}} = \frac{v_{\rm T_{receding}}}{c} = \frac{d}{c}\frac{\Delta\phi}{\Delta t} = \frac{v \sin\theta}{c(1+\beta \cos\theta)} = \frac{\beta \sin\theta}{1+\beta \cos\theta} 
\end{equation} 

\newpage
\thispagestyle{empty}
\phantom{.}

\chapter{List of Publications}
\thispagestyle{empty}

\newpage
\thispagestyle{empty}
\phantom{.}

\newpage
\vspace*{6cm}

{\Large \scshape Publications}
\vspace{0.4cm}

\begin{itemize}

\item ``Variable gamma-ray emission from the Be/X-ray transient A0535+26?''\\
Gustavo E. Romero, M.M. Kaufman Bernad\'o, Jorge A. Combi, Diego F. Torres, A\&A, \textbf{376}, 599, 2001.\\
\textbf{astro-ph/0107411}     

\item ``Precessing microblazars and unidentified gamma-ray sources''\\
M.M. Kaufman Bernad\'o, G.E. Romero, I.F. Mirabel, A\&A Letters, 385, L10, 2002.\\
\textbf{astro-ph/0202316}

\item ``Recurrent microblazar activity in Cygnus X-1?''\\
G.E. Romero, M.M. Kaufman Bernad\'o, I.F. Mirabel, A\&A Letters, \textbf{393}, L61, 2002.\\
\textbf{astro-ph/0208495}

\item ``CHANDRA / VLA Follow-up of TeV J2032+4131, the Only Unidentified TeV Gamma-ray Source''\\
Yousaf Butt, Paula Benaglia, Jorge Combi, Michael Corcoran,
Thomas Dame, Jeremy Drake, Marina Kaufman Bernad\'o, Peter Milne, Francesco
Miniati, Martin Pohl, Olaf Reimer, Gustavo Romero, Michael Rupen, ApJ, \textbf{597}, 494, 2003.\\
\textbf{astro-ph/0302342}

\item ``Hadronic gamma-ray emission from windy microquasar''\\
G.E. Romero, D.F. Torres, M.M. Kaufman Bernad\'o, I.F. Mirabel, A\&A Letters, \textbf{410}, L1, 2003.\\
\textbf{astro-ph/0309123}

\item ``Unidentified $\gamma$ - ray sources off the Galactic plane as low-mass microquasars?''\\
Grenier, I.A., Kaufman Bernad\'o, M.M., Romero, G.E., Ap\&SS, 2004, in press.\\
\textbf{astro-ph/0408215}

\end{itemize}

{\Large \scshape Proceedings}
\vspace{0.4cm}

\begin{itemize}

\item ``Variable gamma-ray emission from microblazars''\\
M.M. Kaufman Bernad\'o, G.E. Romero, I.F. Mirabel, Proceedings of the 4th Microquasar Workshop, eds. Durouchoux, Fuchs \& Rodriguez, published by the Center for Space Physics: Kolkata, p. 148, 2002.\\
\textbf{astro-ph/0207578}

\item ``The Association of High-Mass Microblazars with Variable Gamma-Ray Sources''\\
Marina Kaufman Bernad\'o, Gustavo E. Romero, I. Felix Mirabel,  Proceedings of the International Symposium on Astrophysics Research and on the Dialogue between Science and Religion, Vatican Observatory, eds. C. D. Impey \& C. E. Petry, published by University of Notre Dame press, p. 145, 2002.\\
\textbf{astro-ph/0302039}

\item ``CHANDRA / VLA follow-up of TeV J2032+4131''\\
Butt, Y., Benaglia P., Corcoran, M., Dame, T., Drake, J., Kaufman Bernadó, M., Milne, P., Miniati, F., Pohl, M., Reimer, O., Romero, G., Rupen, M., Proceedings of the American Astronomical Society 201st Meeting, Bulletin of the AAS, \textbf{34}, 613, 2003.

\item ``CHANDRA / VLA  follow-up of  TeV J2032+4131, the Only Unidentified TeV Gamma-ray Source''\\
Butt, Y., Benaglia P., Corcoran, M., Dame, T., Drake, J., Kaufman Bernad\'o, M., Milne, P., Miniati, F., Pohl, M., Reimer, O., Romero, G., Rupen, M., Proceedings of the American Astronomical Society Meeting 201,  HEAD \#7, Session 10.22, Bulletin of the AAS, \textbf{35}, 2003.

\item ``Cosmic ray acceleration by stellar associations? The case of Cygnus OB2.''\\
Butt, Y., Benaglia P., Corcoran, M., Dame, T., Drake, J., Kaufman Bernad\'o, M., Milne, P., Miniati, F., Pohl, M., Reimer, O., Romero, G., Rupen, M., Proceeding of the 2nd VERITAS Symposium on the TeV Astrophysics of Extragalactic Sources, New Astronomy Reviews, 2003.\\
\textbf{astro-ph/0306243} 

\item ``Unidentified Gamma-Ray Sources and Microquasars''\\
G.E. Romero, I.A Grenier, M.M. Kaufman Bernad\'o, I.F. Mirabel, D.F. Torres, ESA-SP (Proceeding of the fifth INTEGRAL workshop), \textbf{552}, 703, 2004.\\
\textbf{astro-ph/0402285}

\end{itemize}

\end{document}